\pdfoutput=1

\documentclass[
PhD, % Change to PhD to change text at bottom of title page
DIC=off,
12pt,
bibliography=totoc, % Include bibliography in contents
listof = totoc, % Include lists of figures and tables in contents
index= totoc, % Include index in contents
onehalfspacing, % onehalfspacing or doublespacing
a4paper %letterpaper
]
{icldt}

\usepackage[square, numbers, sort&compress]{natbib}
\usepackage[%CVSon, % CVS version in footers
	        todos, % show todos (requires todonotes package)
            hyperref % Use hyperref for links
            ]{ihformat}
            
\makeatletter
\newcommand\appendix@numberline[1]{}%\autodot\ } %#1} %\appendixname\  }
\g@addto@macro\appendix{%
  \addtocontents{toc}{
    \let\protect\numberline\protect\appendix@numberline}%
}
\makeatother
\begin{document}

% Title page is input not included to remove extra page.
% % % % % % % % % % % % % % % % % % % % % % % 
% title.tex - Ian Huston
% $Id: title.tex,v 1.1 2009/12/17 17:33:39 ith Exp $
% % % % % % % % % % % % % % % % % % % % % % % 
% Redefine CVSRevision for this section
\renewcommand{\CVSrevision}{\version$Id: title.tex,v 1.1 2009/12/17 17:33:39 ith Exp $}

\setlength{\tabcolsep}{0in}
\newcommand{\isep}{-2 pt}
\newcommand{\lsep}{-0.5cm}
\newcommand{\psep}{-0.6cm}
\renewcommand{\labelitemii}{$\circ$}
\college{Queen Mary, }
\department{Physics and Astronomy}
\supervisor{Karim Malik and David Mulryne}
\title{Induced gravitational waves -- beyond linear cosmological perturbation theory}
\author{Rapha{\" e}l Picard}
\date{September 2025}% cth added
%Note Actual title format is done in icldt.cls
% 
%
\hypersetup{pdftitle={Stick it},pdfauthor={Raphael Picard}}
% 

% Change as appropriate
\declaration{%
I, Rapha{\" e}l Picard, confirm that the research included within this thesis is my own work or that where it has been carried out in collaboration with, or supported by others, that this is duly acknowledged below and my contribution indicated. Previously published material is also acknowledged below. 
\newline
\newline
I attest that I have exercised reasonable care to ensure that the work is original, and does not to the best of my knowledge break any UK law, infringe any third party's copyright or other Intellectual Property Right, or contain any confidential material. 
\newline
\newline
I accept that the College has the right to use plagiarism detection software to check the electronic version of the thesis.
\newline
\newline
 I confirm that this thesis has not been previously submitted for the award of a degree by this or any other university. The copyright of this thesis rests with the author and no quotation from it or information derived from it may be published without the prior written consent of the author.
\newline
\newline
Details of collaboration and publications: Part of this work is done in collaboration with Karim Malik, Matthew Davies, Luis Padilla and David Mulryne. I have made a major contribution to the original research presented in this thesis. It is based on the following papers, all of which have been published: \newline
\begin{itemize} 
	\item \underline{Induced gravitational waves: the effect of first order tensor perturbations} \\
	          R. Picard and K. A. Malik, \textit{JCAP10(2024)010}, \\
	          Arxiv:2410.17819 [astro-ph.CO]
	\item \underline{Effects of scalar non-Gaussianity on induced scalar-tensor 
    gravitational waves} \\
	          R. Picard and M. W. Davies, \textit{JCAP02(2025)037}, \\
	          ArXiv:2311.14513  [astro-ph.CO]
        \item \underline{Suppression of the induced gravitational wave background due to third-order} \\
        \underline{perturbations} \\
	          R. Picard, L. E. Padilla, K. A. Malik and D. J. Mulryne, \textit{Under review by JCAP}, \\
	          Arxiv:2509.07811 [astro-ph.CO]
\end{itemize}
\vfill
Signature: Rapha{\" e}l Picard
\newline
Date: 19/09/2025}

\maketitle

% Dedication page
\cleardoublepage          % Start on a new right-hand page (book style). Use \clearpage if you don’t care.
\thispagestyle{empty}     % Removes page number
\vspace*{\fill}           % Push text to vertical center
\begin{center}
    À ma mère, Virginie Picard, trop tôt partie.
\end{center}
\vspace*{\fill}
% % % % % % % % % % % % % % % % % % % % % % % % % 
% Abstract
% % % % % % % % % % % % % % % % % % % % % % % % % 
% abstract.tex - Ian Huston
% $Id: abstract.tex,v 1.2 2009/12/17 17:41:41 ith Exp $
% % % % % % % % % % % % % % % % % % % % % % % % % 
% Redefine CVSRevision for this section
\renewcommand{\CVSrevision}{\version$Id: abstract.tex,v 1.2 2009/12/17 17:41:41 ith Exp $}
\chapter*{Abstract}
\label{ch:abstract}
\addcontentsline{toc}{chapter}{Abstract}
\section*{}
\singlespacing

This thesis focuses on gravitational waves (GWs) that arise beyond linear order in cosmological perturbation theory. In recent years, scalar-induced GWs have attracted significant attention because they may serve as the observational signature of primordial black holes (PBHs) formed in the early universe. The formation of PBHs requires large density perturbations, which can naturally emerge in some models of inflation. When these large density fluctuations couple, they act as a source for scalar-induced GWs at second order. In this work, we extend the existing formalism by including linear tensor fluctuations as an additional source term. This gives rise to two new classes of second-order GWs: those sourced by scalar–tensor couplings (scalar–tensor induced GWs) and those quadratic in tensor modes (tensor–tensor induced GWs). We find that the scalar–tensor contribution becomes significant if first-order tensor modes are enhanced, whilst the tensor–tensor contribution remains subdominant. Moreover, we demonstrate that the spectrum of scalar–tensor induced GWs exhibits an unphysical enhancement in the UV limit when the primordial scalar power spectrum is insufficiently peaked. To investigate whether this can be resolved, we study third-order induced GWs and their correlation with primordial GWs. We find that this new contribution suppresses the overall signal but does not cancel the unphysical enhancement. Possible explanations for this behaviour are discussed and left for future work. Finally, we explore the effect of primordial scalar non-Gaussianity on the spectrum of scalar–tensor induced GWs, building on previous results showing its impact on scalar-induced GWs.     

% % % % % % % % % % % % % % % % % % % % % % % % %

% % % % % % % % % % % % % % % % % % % % % % % % % 
% Acknowledgements
% % % % % % % % % % % % % % % % % % % % % % 
% acknowledgements.tex - Ian Huston
% $Id: acknowledgements.tex,v 1.2 2009/12/17 17:41:41 ith Exp $
% % % % % % % % % % % % % % % % % % % % % % 
% Redefine CVSRevision for this section
\renewcommand{\CVSrevision}{\version$Id: acknowledgements.tex,v 1.2 2009/12/17 17:41:41 ith Exp $}
\chapter*{Acknowledgements}
\label{ch:acknowledgements}
\addcontentsline{toc}{chapter}{Acknowledgements}

I would like to thank the Cosmology and Relativity group at Queen Mary for welcoming me into their research community: it is an experience I will never forget.

First and foremost, I owe immense gratitude to my two supervisors, Karim Malik and David Mulryne, from whom I have learned a great deal. Karim never refused a meeting and, perhaps at times too generously, always placed complete confidence in me. I even forgive him for making me go to third order in perturbation theory; his unwavering support is something I never took for granted. David `always busy' Mulryne adopted me in my final year of PhD and has also been a constant source of encouragement and help. 

Within the department, I would like to thank Chris Clarkson for his guidance and for giving me the opportunity to travel to Cape Town to meet his collaborators, in particular Roy Maartens. Writing a paper with Matthew, entirely independently, was both a fun and memorable experience. I am especially grateful to Charles and Laura, who supported me enormously throughout my PhD and encouraged me to think more deeply and approach my research with rigour. I also greatly enjoyed collaborating with Luis, whose remarkable programming skills made it possible to complete my final project. 

I began my PhD alongside Stefano, Pritha, and Chris, and it is safe to say the experience would not have been the same without them. I am truly grateful to have made such wonderful friends and created memories for a lifetime during this journey. 
%I would also like to give special thanks to Lucy.  

I would like to thank my friends who, despite still having no real idea what I actually do (“what's up with the planets bro?”), have made this whole experience infinitely better -- especially on weekends. Merci à Nico, Emma, Nour, Anais, EJ, Ellie, Saub, Beber, Ruben, Tara, Elise, Laure(s), Philou, Neidex, Lara, Lotus, Josh, Jake, Jason, Max(s) and Lucy.  

Finally, I would like to thank my family for their constant support, particularly my dad and my brother, for always making home truly feel like home. I am immensely grateful for the people who were caring for my mother in her final moments:  Magda, Ubah, Amal and Joahir. À mes deux grand-mères que j’aime fort, je vous remercie pour vos appels et votre soutien constant, mais surtout pour les bons plats qui m’ont donné de l’énergie jusqu’à la fin. Thank you to Lou for being such a wonderful addition to the household for a year, and to my entire Picard and Le Rouzic families for their love and support.

% % % % % % % % % % % % % % % % % % % % % % % % % 

% % % % % % % % % % % % % % % % % % % % % % % % % 
% Table of contents and figures
% % % % % % % % % % % % % % % % % % % % % % % % % 
%\setcounter{tocdepth}{0}
\tableofcontents
%\listoffigures %cth commented
%\listoftables %cth commented
% % % % % % % % % % % % % % % % % % % % % % % % % 

% Set one half spacing now that thesis proper is starting.
\onehalfspacing
% % % % % % % % % % % % % % % % % % % % % % % % % 

% Include files for each chapter here using 
% \include{relative-path-to-file}
% 
% Example
% % % % % % % % % % % % % % % % % % % % % % % % % % % % 
% chapter.tex - Ian Huston
% Sample chapter layout
% % % % % % % % % % % % % % % % % % % % % % % % % % % % 
% Redefine CVSRevision for this section. 
% If you don't want to use CVS tags comment out this line
\renewcommand{\CVSrevision}{\version$Id: chapter.tex,v 1.3 2009/12/17 18:16:48 ith Exp $}

% % % % % % % % % % % % % % % % % % % % % % % % % % % % % % % % 
% =========================================================== %
% % % % % % % % % % % % % % % % % % % % % % % % % % % % % % % % 
\chapter{Introduction}
\label{Ch_intro}
% % % % % % % % % % % % % % % % % % % % % % % % % % % % % % % % 
% =========================================================== %
% % % % % % % % % % % % % % % % % % % % % % % % % % % % % % % % 
Cosmology, the study of the Universe, is a science that has evolved over thousands of years. From the geocentric models of ancient Greece to the heliocentric model proposed by Copernicus in the 16th century, new ideas, theories, and observations have steadily reshaped our understanding of the cosmos. Since Isaac Newton’s seminal work on gravity \cite{Newton:1687eqk}, it has been recognised as a fundamental force essential to any accurate description of the Universe, and as such, cosmology and gravitational theory are deeply interconnected. Hence, in 1915, when Albert Einstein published his theory of gravity, known as the theory of General Relativity (GR) \cite{Einstein:1915ca, Einstein:1916vd}, it brought a profound transformation to the science of cosmology \cite{Friedman:1922kd, lemaitre}. GR redefines gravity not as a fundamental force, but as the curvature of a dynamic fabric called spacetime, which is distorted by the presence of mass or energy. Two years after introducing his theory of gravity, Einstein applied it to the cosmos \cite{Einstein:1917ce}, marking the beginning of a new era in cosmology: relativistic cosmology. Meanwhile, GR gained widespread acceptance as a successful theory of gravity, as it consistently passed experimental and observational tests \cite{Einstein:1915bz, eddington, Shapiro, Will:2018bme}.

Einstein initially envisioned a static Universe, sustained by a mysterious repulsive force he introduced into his equations, known as the cosmological constant, to counterbalance the gravitational attraction of matter. However, in 1929, Edwin Hubble made a groundbreaking discovery: by measuring the velocities of nearby and distant galaxies, he showed that the Universe is expanding \cite{Hubble:1929ig}. This became known as Hubble’s Law and was revolutionary as it marked a turning point in the history of cosmology, establishing it as an empirical science, one in which the Universe itself became the laboratory. In the final decade of the 20th century, it was found that this expansion is in fact accelerating \cite{SupernovaSearchTeam:1998fmf, SupernovaCosmologyProject:1998vns}. Interestingly, the accelerated expansion of the Universe can be modelled theoretically by using the cosmological constant that Einstein had originally introduced.

During the 20th century, two other seminal discoveries fundamentally shaped modern cosmology: the inference of dark matter and the detection of the cosmic microwave background (CMB). While the nature of dark matter remains unknown, the formation of the CMB is well understood, though the details of its ultimate origin are not. 

Dark matter is estimated to constitute approximately 27\% of the Universe’s total energy density, yet no widely accepted theory has identified its fundamental nature. What is firmly established is that it does not interact through the forces described by the Standard Model of particle physics or only very weakly; its existence is inferred solely from its gravitational effects. The first indications of dark matter date back to the 1930s \cite{Zwicky:1933gu, Zwicky:1937zza, Smith:1936mlg}, when observations of the Coma Cluster revealed that the luminous matter was insufficient to account for the observed galaxy velocity dispersions. In the latter half of the twentieth century, additional and more compelling evidence emerged on galactic scales \cite{Einasto:1974dra, 1973A&A....26..483R, Rubin:1980zd}. Rotation curves of stars demonstrated that orbital velocities at large radii remain unexpectedly high, a phenomenon that can be explained by invoking non-luminous matter. Further support has come from measurements of gravitational lensing, such as those conducted in the Sloan Digital Sky Survey (SDSS) \cite{SDSS:2005sxd}, and from the analysis of anisotropies in the CMB \cite{Planck:2015fie, Planck:2018vyg, WMAP:2003cmr, WMAP:2008lyn}.

The CMB was first predicted as the relic radiation of a hot and dense early Universe by extrapolating the dynamical equations of cosmology backwards in time \cite{PhysRev.73.803}. Its accidental discovery in 1965 \cite{1965ApJ...142..419P}, followed by its interpretation as the thermal afterglow of the Big Bang \cite{Dicke:1965zz}, marked a major milestone in modern cosmology. This background radiation is often referred to as the first light of the Universe. Observations of the CMB reveal that its spectrum is an almost perfect blackbody \cite{firas}, providing strong evidence that the Universe was in near-perfect thermal equilibrium at early times. Furthermore, the CMB exhibits an extremely uniform temperature. This striking isotropy naturally raises the question: why is the Universe so remarkably homogeneous and isotropic on the largest scales? Widely separated regions of the CMB should never have been in causal contact in the hot big bang models of the time, so why did they look so alike? This ultimately gave rise to the theory of cosmic inflation \cite{Starobinsky:1979ty,Guth:1980zm,Linde:1981mu}. Inflation, is postulated as a brief epoch of exponential expansion in the very early Universe, and it offers an elegant resolution to several fundamental problems in cosmology, including the isotropy/homogeneity problem. In its simplest form, inflation is driven by a scalar field, the inflaton, slowly rolling down a potential. Even more strikingly, inflation provides a mechanism that links quantum fluctuations of this scalar field in the early Universe to the temperature variations observed in the CMB. These fluctuations, first detected by the \textsc{cobe} satellite in 1992~\cite{cobe}, revealed a rich structure that continues to shape modern cosmology \cite{ACT:2025tim}. The analysis of these primordial perturbations is carried out within the framework of cosmological perturbation theory, which allows tiny deviations from homogeneity to be systematically studied and provides the foundation for understanding the formation of cosmic structure.

In the 21st century, a new groundbreaking mechanism for detecting gravitational waves (GWs) emerged. Einstein's theory of GR predicts the existence of GWs: perturbations of spacetime which travel at the speed of light. GWs arise as wave-like solutions to the perturbed Einstein field equations and are generated by accelerating masses. They are characterised by two independent polarisations, which distort space in transverse directions as they pass. The first strong indirect evidence for such waves came from observations of a binary pulsar discovered in 1974 \cite{Hulse:1974eb}. GR predicts that as the two neutron stars orbit each other, they should emit GWs, carrying away orbital energy and causing the orbital period to gradually shrink. Hulse and Taylor measured the arrival times of radio pulses from the pulsar with extreme precision and found that the orbital period was decreasing exactly as predicted by GR if GWs were being emitted continuously. This was another successful milestone in the confirmation of the theory of GR.  

\begin{figure}
    \centering
    \includegraphics[width=0.99\linewidth]{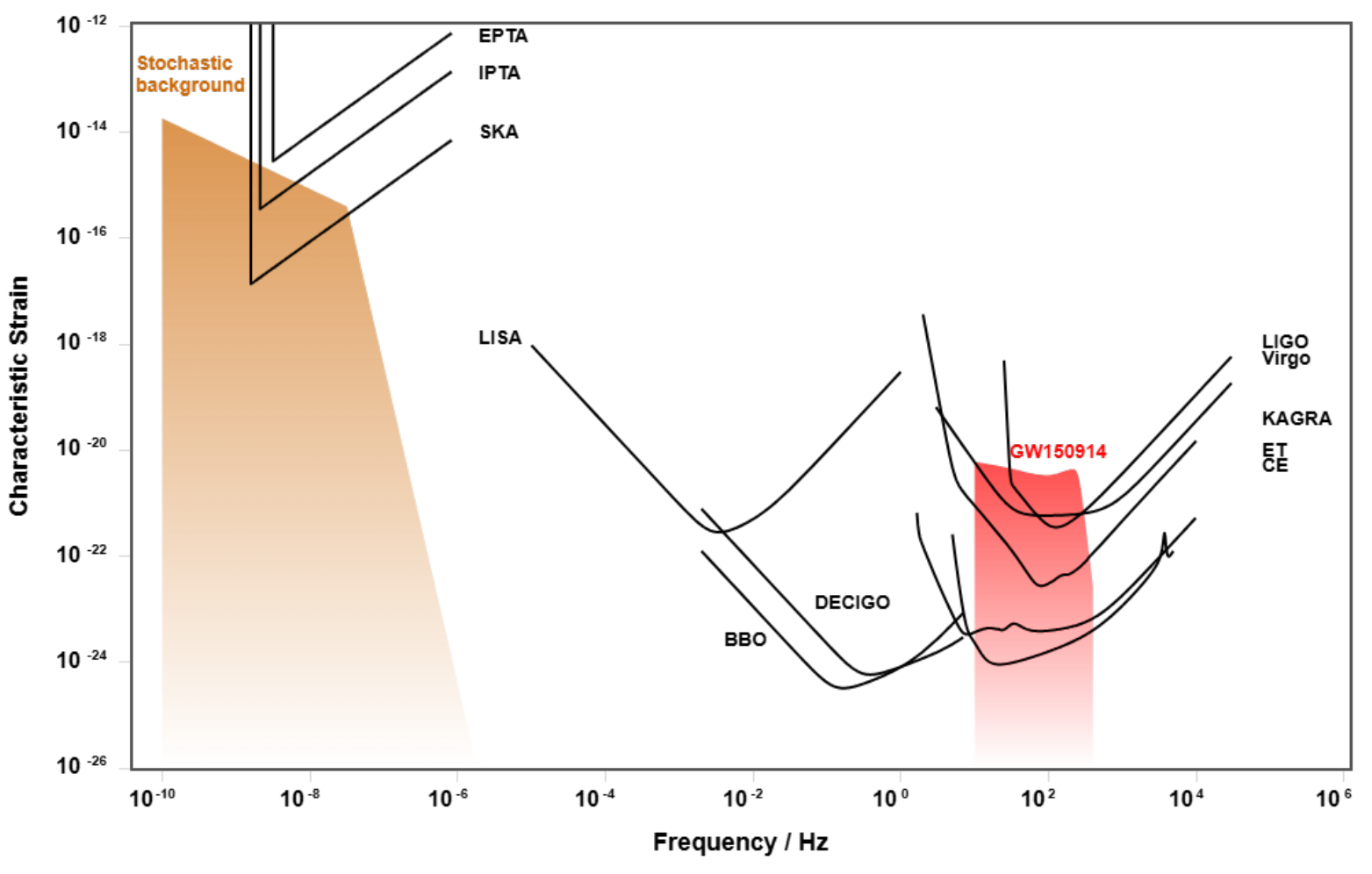}
    \caption{\footnotesize{Plot of the characteristic strain against frequency. We have plotted the sensitivity curves of current ground-based detectors (sensitive to higher frequencies), i.e. the LVK collaboration. Future ground-based detectors include the Cosmic Explorer \cite{cosmicexplorer} and the Einstein Telescope \cite{einsteintelecope}. At intermediate frequencies, future space-based detectors include the Laser Interferometer Space Antenna (LISA) \cite{LISA:2017pwj}, the DECi-hertz Interferometer Gravitational wave Observatory (DECIGO) \cite{decigo} and the Big Bang Observer (BBO) \cite{bbo, Crowder:2005nr}. At lower frequencies, PTA experiments detect GWs by monitoring the arrival times of radio pulses from many millisecond pulsars and looking for tiny, correlated deviations across the sky caused by passing GWs between Earth and the pulsars. Shaded areas correspond to the first GW detection from two black hole mergers (red) and the stochastic background which would come from unresolved sources (orange). \textit{Credit}: This plot was generated via the \href{http://gwplotter.com/}{online GW tool plotter} (the website was last accessed on the 09/09/2025), see Ref.~\cite{Moore:2014lga} for details. }}
    \label{fig:gwfrequencies}
\end{figure}

It was not until 2015 that the first direct detection of GWs occurred \cite{LIGOScientific:2016aoc}. The Laser Interferometer Gravitational-Wave Observatory (LIGO) and Virgo collaboration announced the observation of a signal from the merger of two black holes. This was followed by numerous further detections of binary black hole mergers. In 2017, the Virgo interferometer joined the network, and soon after, a joint detection of a binary neutron star merger was announced \cite{LIGOScientific:2017ync}. This event was a watershed moment, since it was the first time that GWs were observed together with an electromagnetic counterpart, including a short gamma-ray burst. This marked the beginning of multimessenger astrophysics. The Kamioka Gravitational-Wave Detector (KAGRA) has since joined the network, and together, the three collaborate as the LVK collaboration \cite{KAGRA:2013rdx}. The future of GW astrophysics is promising, with a new generation of detectors expected to come online over the next two decades. These instruments will not only achieve greater sensitivity but will also extend coverage to different frequency ranges, opening a different window on the universe. For a comprehensive overview of future prospects in GW physics, see Ref.~\cite{Bailes:2021tot}. Fig.~\ref{fig:gwfrequencies} is a plot of the characteristic strain against frequency and it provides a summary of current and planned detectors, together with the frequency ranges to which they are (or will be) sensitive. The characteristic strain quantifies the strength of a GW signal at each frequency, representing the cumulative stretching and squeezing of spacetime over the duration of the signal. For a detector, a lower sensitivity curve indicates better sensitivity, while for a signal, a higher characteristic strain corresponds to a stronger, louder signal. Consequently, signals that rise well above the detector’s sensitivity curve are easier to detect. Loosely speaking, the area between the signal curve and the detector sensitivity curve provides an intuitive measure of the signal-to-noise ratio.

\begin{figure}
    \centering
    \includegraphics[width=0.99\linewidth]{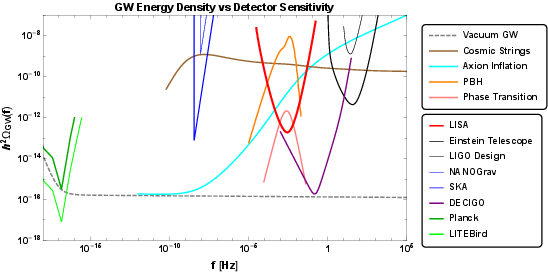}
    \caption{\footnotesize{Spectral density of GWs against frequency. The spectral density is a measure of how much energy GWs carry. Some typical cosmological sources of GWs are plotted with different detectors also represented. We note that the grey dashed line corresponds to GWs generated during inflation, and it assumes we can extrapolate the Planck data \cite{Planck:2018jri} to all scales, whilst the cyan line represents GWs produced in a different model of inflation (axion inflation) \cite{Garcia-Bellido:2016dkw}. We further note that the authors have included at very low frequencies the sensitivity of the Planck \cite{Planck:overview} and the future LiteBIRD telescope \cite{LiteBIRD:2022cnt}: GWs interact with the CMB by imprinting a distinct curl-like polarisation pattern, known as B-modes \cite{Kamionkowski:2015yta}. \textit{Credit}: This plot was taken from Ref.~\cite{LISACosmologyWorkingGroup:2022jok}.}}
    \label{fig:lisaplotGW}
\end{figure}

So far, we have solely discussed GW astrophysics. However, the first direct detection of GWs reinvigorated the field and opened the exciting possibility of pursuing GW cosmology. Recently, pulsar timing array (PTA) experiments have reported data that tantalizingly suggests the presence of a stochastic gravitational-wave background (SGWB) \cite{NANOGrav:2023gor,EPTA:2023sfo,EPTA:2023fyk,EPTA:2023xxk,Xu:2023wog,Reardon:2023gzh,Zic:2023gta}. Although the origin of this SGWB remains uncertain, these results mark a milestone for researchers who have long anticipated a background of cosmological origin. The observed signal is consistent with an unresolved population of supermassive black hole binaries, as described by the Hellings–Downs correlation curve \cite{hellings}. At the same time, the NANOGrav collaboration has explored cosmological interpretations of their data \cite{NANOGrav:2023hvm}, fitting models involving different cosmological sources which include: vacuum fluctuations \cite{Guzzetti:2016mkm}, topological defects \cite{Jenkins:2020ctp, Deng:2016vzb}, phase transitions \cite{Jedamzik:1999am, Rubin:2000dq, Davoudiasl:2019ugw} and scalar-induced GWs, which we will discuss further throughout this thesis. They emphasise that current evidence should not yet be taken as confirmation of new physics. Fig.~\ref{fig:lisaplotGW} shows different kinds of cosmological sources and detectors which are sensitive to these sources. 

Gravitational waves will then offer a powerful and transformative new way to probe the Universe, opening an exciting era in observational cosmology with promising prospects for future discoveries. Because gravity is so weakly interacting, GWs generated in the very early Universe, such as those potentially produced during inflation, can propagate essentially unimpeded, offering a direct window into the first moments of cosmic history, long before the release of the CMB at around 380,000 years after the Big Bang. In this thesis, we will focus on a particular class of GWs: so-called induced GWs. In the framework of cosmological perturbation theory, these GWs appear naturally after perturbing the Einstein equations. 

To do this consistently, it is essential to first study the non-perturbed Einstein equations, which we refer to as the background equations. Chapter~\ref{Chap:background} is therefore dedicated to extracting and solving these background equations. We begin by introducing the key equations and mathematical tools from GR that are needed to describe the Universe on scales where the Universe is homogeneous and isotropic. We then solve the Einstein equations in a cosmological context, highlighting fundamental relations and their solutions. Finally, we introduce several important parameters and epochs of cosmic evolution, such as inflation, that will play a central role in later discussions.

In Chapter~\ref{Chap:CPT} we introduce the perturbative framework that will be used throughout the remainder of this thesis and present the tools needed to discuss first, second, and third-order perturbations. As a starting point, it is instructive to study first-order perturbations in Minkowski spacetime, since this provides a natural way to introduce GWs. We then extend the discussion to a cosmological setting by solving the Einstein equations at linear order in perturbation theory. In doing so, we obtain solutions for the three different classes of perturbations: scalar, vector, and tensor modes. These solutions describe how perturbations propagate in the Universe, given suitable initial conditions.  Furthermore, we introduce the concept of the power spectrum, which provides a statistical description of cosmological perturbations via their correlations. We conclude the chapter by examining how first-order perturbations can \emph{induce} higher-order perturbations, in particular how they act as a source for GWs at second-order in cosmological perturbation theory.

Chapter~\ref{Chap:SOGWs} is dedicated to solving the second-order tensor equation, i.e.\ the GW equation of motion. Under specific initial conditions, namely Gaussian and adiabatic initial conditions, we investigate how first-order perturbations induce GWs at second order in a radiation-dominated Universe. GWs induced by scalar perturbations have been extensively discussed in the literature; hence, our focus is on how the inclusion of first-order tensor perturbations modifies the picture. In particular, we develop the framework to include first-order tensor modes as a source term and discuss implications of these contributions for future GW surveys. We also highlight the behaviour of the source term that couples first-order scalars and tensors in the UV sector, showing that it leads to an unphysical enhancement of the GW signal.

Chapter~\ref{Chap:NG} examines how changing the initial conditions of scalar perturbations affects the GW observable (the spectral density). In particular, we show that when the scalar perturbations are non-Gaussian, a new source term arises, and we analyse its distinctive features.

Finally, in Chapter~\ref{Chap:Third} we study GWs induced at third order in cosmological perturbation theory. In particular, we investigate how the correlation of third-order tensor modes with primordial tensors leads to a non-trivial contribution to the SGWB. These correlations appear at the same order as the correlation of second-order induced GWs. The motivation behind this chapter is to explore whether these additional contributions can remedy the shortcoming of the scalar–tensor coupling at second order, namely its behaviour in the UV limit. To this end, we provide a detailed analysis of the new source terms in the UV sector and conclude that no such cancellation occurs. 

We close the thesis by outlining potential directions for future work.

\vfill \noindent
\textit{Conventions:} We adopt the Einstein summation convention and use the `mostly plus' metric \cite{Misner:1973prb}. Greek characters denote spacetime indices ($0,1,2,3$) whilst Latin characters denote spatial ones ($1,2,3$). Unless explicitly stated otherwise, a dot over a function refers to a derivative with respect to cosmic time $t$, whilst a dash refers to a derivative with respect to conformal time $\eta$. We work in natural units, i.e.~$c=1$.
% % % % % % % % % % % % % % % % % % % % % % % % % % % % 
% chapter.tex - Ian Huston
% Sample chapter layout
% % % % % % % % % % % % % % % % % % % % % % % % % % % % 
% Redefine CVSRevision for this section. 
% If you don't want to use CVS tags comment out this line
\renewcommand{\CVSrevision}{\version$Id: chapter.tex,v 1.3 2009/12/17 18:16:48 ith Exp $}

% % % % % % % % % % % % % % % % % % % % % % % % % % % % % % % % 
% =========================================================== %
% % % % % % % % % % % % % % % % % % % % % % % % % % % % % % % % 
\chapter{Cosmology at the background level}
\label{Chap:background}
% % % % % % % % % % % % % % % % % % % % % % % % % % % % % % % % 
% =========================================================== %
% % % % % % % % % % % % % % % % % % % % % % % % % % % % % % % %
As a starting point, we review the formalism of cosmology at the background level. Given the breadth of the field, our discussion is restricted to concepts and results most relevant to this work. We begin with the framework provided by General Relativity (GR), which supplies the mathematical tools necessary for a precise description of our universe. From the Einstein equations and the conservation of the energy-momentum tensor, we can extract three key equations that govern the evolution of the universe and its matter content: the Friedmann, acceleration, and energy-conservation equations. We then solve these equations in different cosmological settings. Next, we provide an overview of the current standard model of cosmology, the $\Lambda$CDM model, and finally introduce the epoch of inflation, a hypothesised period of accelerated expansion in the very early universe, along with some of the primary motivations for postulating such an epoch.

% % % % % % % % % % % % % % % % % % % % % % % % % % % % % % % % 
% % % % % % % % % % % % % % % % % % % % % % % % % % % % % % % % 
\section{General relativity: a (very) brief recap}
In this section, we briefly recall some key objects and tensors which form the basis of GR and which will be used continuously throughout this work. This section follows Ref.~\cite{Carroll:2004st}, to which we refer the reader for further details. In GR, gravity appears as the curvature of spacetime, which necessitates mathematical tools to describe that geometric structure. 

Spacetime is modelled as a 4D manifold with a pseudo-Riemannian signature. Our starting point is the central object of GR, the metric tensor $g_{\mu \nu}$, a symmetric 4D dynamical tensor which we use to construct the line-element between two infinitesimal events in spacetime
\begin{equation}
    ds^2=g_{\mu \nu}dx^{\mu}dx^{\nu} \, .
\end{equation}
The metric tensor is the GR counterpart of the Newtonian gravitational field, and it contains all the geometrical information about the spacetime of interest. The laws of physics usually take the form of differential equations, and hence we must be able to take derivatives appropriately on the manifold. To do this, we introduce the Christoffel symbols, which allow us to connect the tangent spaces of tensors at all points on the manifold. They are defined as
\begin{equation}\label{defChristofell}
    \Gamma^\rho_{\mu \nu} = \frac{1}{2} g^{\rho \lambda} \left( \partial_\mu g_{\nu \lambda} + \partial_\nu g_{\mu \lambda} - \partial_\lambda g_{\mu \nu} \right) \, .
\end{equation}
These objects allow us to construct derivatives in a covariant way, so-called covariant derivatives $\nabla _\mu$, which act on a general tensor $T^{\mu_1\mu _2...\mu _k}_{\nu_1\nu _2...\nu _l}$
\begin{equation}
    \begin{split}
        \nabla _\sigma T^{\mu_1\mu _2...\mu _k}_{\nu_1\nu _2...\nu _l} = &\, \,\partial _\sigma T^{\mu_1\mu _2...\mu _k}_{\nu_1\nu _2...\nu _l} \\
        &+ \Gamma ^{\mu _1}_{\sigma \lambda}  T^{\lambda\mu _2...\mu _k}_{\nu_1\nu _2...\nu _l}+ \Gamma ^{\mu _2}_{\sigma \lambda}  T^{\mu _1\lambda...\mu _k}_{\nu_1\nu _2...\nu _l}+...+ \Gamma ^{\mu _k}_{\sigma \lambda}T^{\mu_1\mu _2...\lambda}_{\nu_1\nu _2...\nu _l}\\
        &- \Gamma ^{\lambda}_{\sigma \nu _1}  T^{\mu _1\mu _2...\mu _k}_{\lambda\nu _2...\nu _l}- \Gamma ^{\lambda}_{\sigma \nu _2}  T^{\mu _1\mu _2...\mu _k}_{\nu _1\lambda...\nu _l}-...-\Gamma ^{\lambda}_{\sigma \nu _l}T^{\mu_1\mu _2...\mu _k}_{\nu_1\nu _2...\lambda} \, .
    \end{split}
\end{equation}
The Christoffel symbols allow us to construct the Riemann (intrinsic) curvature tensor
\begin{equation} \label{defRiemann}
        R^\rho_{ \sigma\mu\nu} = \partial_\mu \Gamma^\rho_{\nu\sigma} - \partial_\nu \Gamma^\rho_{\mu\sigma} + \Gamma^\rho_{\mu\lambda} \Gamma^\lambda_{\nu\sigma} - \Gamma^\rho_{\nu\lambda} \Gamma^\lambda_{\mu\sigma}\, ,
\end{equation}
whereas the name indicates, this tensor contains all the information about the curvature of the manifold/spacetime. It vanishes when the spacetime is flat. Finally, the Einstein equations relate components of this tensor to the matter content of the spacetime
\begin{equation} \label{Einsteineq}
    G_{\mu \nu} = \frac{8\pi G}{c^4} \, T_{\mu \nu} \, ,
\end{equation}
where $G_{\mu \nu}$ is the Einstein tensor, $T_{\mu \nu}$ tells us what matter content fills up this spacetime and $ \frac{8\pi G}{c^4}$ is a constant of proportionality\footnote{Just for the master equation, we have restored the units.}. The Einstein tensor is defined as
\begin{equation}
    G_{\mu \nu} = R_{\mu \nu} - \frac{1}{2}Rg_{\mu \nu} \, , 
\end{equation}
where  $R_{\mu \nu} = R^\rho_{\mu\rho\nu}$ is the Ricci tensor and $R = g^{\mu\nu}R_{\mu\nu}$ the Ricci scalar. The Bianchi identities tell us that the covariant derivative of the Einstein tensor vanishes ($\nabla _{\mu}G^{\mu \nu}=0$), which directly implies
\begin{equation}\label{energymomentumconsevration}
    \nabla _{\mu}T^{\mu \nu} = 0 \, ,
\end{equation}
known as the energy-momentum conservation equation.

Loosely speaking, the Einstein equations in Eq.~\eqref{Einsteineq} can be read as \textit{curvature} $\approx$ \textit{matter}. Given an energy-momentum tensor, we can solve the Einstein field equations to determine the $10$ components of the metric tensor. As previously noted, the metric tensor encapsulates the geometric structure of spacetime. Solving for the metric allows us to compute various physical and geometric quantities, such as spacetime curvature, tidal effects, and particle trajectories via the geodesic equation, amongst others. In the following section, we will do exactly this in the context of cosmology, i.e. we need to find a metric to describe the geometry of the universe given a suitable energy-momentum tensor which describes the matter content of the universe. 

% % % % % % % % % % % % % % % % % % % % % % % % % % % % % % % % 
% % % % % % % % % % % % % % % % % % % % % % % % % % % % % % % % 
\section{The Friedmann-Lema\^itre-Robertson-Walker metric and the perfect fluid}

Observations from large-scale surveys indicate that the matter in our universe is, to a good approximation, both distributed homogeneously and isotropically on large scales. These observations underpin the \textit{Cosmological Principle}, which is fundamental to building any cosmological model. They also enable us to make well-justified assumptions about the form of the metric tensor that describes the geometry of the universe.

The Cosmological Principle is based on two fundamental properties:
\begin{enumerate}
\item \textit{Homogeneity}: The universe has no preferred location in spacetime. Because matter is distributed uniformly on large scales, our position is not special, and thus the metric must be independent of spatial position. Therefore, the metric will only depend on time. 
\item \textit{Isotropy}: The universe has no preferred direction in spacetime. Observations should appear the same regardless of the direction we look, so the metric must be rotationally symmetric around any point.
\end{enumerate}
A metric which fulfils these requirements is the Friedmann-Lema\^itre-Robertson-Walker (FLRW) metric defined as:
\begin{align}\label{FLRWflat}
    \begin{split}
       ds^2 &= -dt^2 + a^2(t)g_{ij}dx^idx^j \, ,\\
       ds^2 &= -dt^2 + a^2(t) \left (\frac{dr^2}{1-Kr^2} + r^2 d\theta^2 + r^2 \sin^2 \theta d\phi^2  \right ) \, .
    \end{split}
\end{align}
We have chosen to work in spherical coordinates where $r\in [0;\infty )$ is the radial coordinate, $\theta \in [0;\pi ]$ is the polar coordinate and $\phi \in [0;2\pi]$ is the azimuthal coordinate. $K=0,\pm 1$ is the curvature of the universe and $a(t)$ is the scale factor. GR, when constrained by homogeneity and isotropy, allows spacetime to have one of three possible spatial curvatures: positive (spherical geometry), negative (hyperbolic geometry), or zero (flat or Euclidean geometry). The non-vanishing components of the metric $g_{\mu \nu}$ can be expressed as 
\begin{equation}\label{FLRWflatcomponent}
    g_{00}=-1 \quad \text{and} \quad g_{ij}=a^2(t)\begin{bmatrix}
\frac{1}{1-Kr^2} & 0 & 0 \\
0  & r^2 & 0 \\
0 & 0 & r^2 \sin ^2\theta
\end{bmatrix} \, .
\end{equation}
The scale factor is the only remaining degree of freedom we need to solve for via the Einstein equations, but first, we must specify the matter content of the universe. 

On very large scales, the matter content of our universe can be modelled as a fluid, and the most general energy-momentum tensor we can write down is \cite{Peter:2013avv}
\begin{equation}\label{generalfluid}
    T_{\mu \nu} = (\rho +P) u_{\mu}u_{\nu} + Pg_{\mu \nu} +2 q_{(\mu}u_{\nu )} + \pi _{\mu \nu}\, ,
\end{equation}
where $\rho$ is the density, $u_{\mu}$ the 4-velocity , $P$ the pressure, $q_{\mu}$ the the energy flux relative to $u^u$ and $\pi _{\mu \nu}$ is the anisotropic stress of the fluid. The Cosmological Principle also implicitly selects a preferred frame of reference: the comoving frame. In this frame, observers move along with the cosmic fluid, meaning that matter is at rest relative to the coordinates. If matter were to have motion relative to the observer, it would introduce a preferred direction, thereby violating isotropy. These conditions imply $q_u=0$ and $\pi \munu=0$. Eq.~\eqref{generalfluid} therefore becomes
\begin{equation}\label{perfectfluid}
    T_{\mu \nu} = (\rho +P) u_{\mu}u_{\nu} + Pg_{\mu \nu}  ,
\end{equation}
which is the energy-momentum tensor of a perfect fluid. It follows that in this frame $u^i=\mathbf{0}$ and hence the velocity vector $u^\mu=(1,0,0,0)$, such that $u_\mu u^\mu=-1$.

We now have an expression for the metric tensor and for the energy-momentum tensor: we can now proceed to solving the Einstein equations.

% % % % % % % % % % % % % % % % % % % % % % % % % % % % % % % % 
% % % % % % % % % % % % % % % % % % % % % % % % % % % % % % % % 
\section{The cosmological equations in different settings}\label{sec:scalefactor}

In this section, we derive the fundamental equations governing cosmological dynamics. These include two equations from distinct components of the Einstein field equations, commonly known as the Friedmann equations, and a third equation arising from the conservation of energy-momentum, referred to as the conservation equation. To close the system, we also specify an equation of state appropriate for a perfect fluid. We will solve these equations during a radiation domination (RD), matter domination (MD) and dark energy domination epoch. 

The LHS of the Einstein equations in Eq.~\eqref{Einsteineq} depends exclusively on the metric tensor, which we argued takes the form of Eq.~\eqref{FLRWflat}, and the RHS depends on the metric tensor and the energy-momentum tensor for a perfect fluid defined in Eq.~\eqref{perfectfluid}. We are left with computing the Einstein tensor. The non-vanishing connection coefficients are
\begin{equation}
    \Gamma ^0_{ij}=a\dot{a}g _{ij}\, , \quad \Gamma ^i_{0j}=\frac{\dot{a}}{a}g ^i_{j}\, , \quad \Gamma ^i_{jk}= \frac{1}{2}g ^{il} \left (\partial _j g_{lk} + \partial _k g_{lj} - \partial _l g_{jk} \right ) \, .
\end{equation}
From these, we can get components of the Ricci tensor
\begin{equation}
    R_{00}= -3\frac{\ddot{a}}{a} \, , \quad \text{and} \quad R_{ij} = (a\ddot{a} +2\dot{a}^2 + 2K) g_{ij} \, ,
\end{equation}
and hence the Ricci scalar
\begin{equation}
    R = \frac{6}{a^2} (a\ddot{a} + \dot{a}^2 +K) \, .
\end{equation}
The non-vanishing Einstein tensor components then read
\begin{equation}
    G_{00}= \frac{3}{a^2}(\dot{a}^2+K) \, , \quad \text{and} \quad G_{ij} = -g_{ij} (2a\ddot{a}+\dot{a}^2+K) \, .
\end{equation}
The Einstein equations, Eq.~\eqref{Einsteineq}, relate components of the Einstein-tensor $G_{00}$ and $G_{ij}$ to components of the energy-momentum tensor $T_{00}$ and $T_{ij}$ respectively. Again, we highlight that as a consequence of the Cosmological principle, the $G_{0i}$ and $T_{0i}$ components vanish. The two non-trivial equations we get are 
\begin{equation} \label{FriedmannEqs}
    \left  ( \frac{\dot{a}}{a}\right )^2 =\frac{\kappa}{3}\rho - \frac{K}{a^2}  \quad \text{and} \quad  \frac{\ddot{a}}{a}  = -\frac{\kappa}{6}(\rho+3P) \,  ,
\end{equation}
\begin{comment}
    or using conformal time\footnote{In the following sections we will work exclusively with conformal time $\eta$ rather than cosmic time $t$. The relationship between the time derivative of a quantity $X$ with respect to cosmic time and conformal time is $X\conf=a\dot{X}$, whilst $X ^{\prime \prime} -\hubble X \conf = a^2 \ddot{X}$. Here, for clarity, we work with both cosmic and conformal time, but if one is not specified it should be clear how to switch between the two.}
\begin{equation} \label{FriedmannEqsconf}
    \left  ( \frac{a \conf}{a}\right )^2 =\frac{a^2}{3M_{Pl}^2} \rho \quad \text{and} \quad \left ( \frac{a \conf}{a}\right ) \conf = -\frac{a^2}{6M_{Pl}^2}(\rho+3P) \,  ,
\end{equation}
\end{comment}
which are known as the Friedmann equations, and we defined $\kappa=8\pi G$. The first Friedmann equation comes from the time-time component of the Einstein equations ($G_{00}=\kappa T_{00}$) and the second from taking the trace of the spatial part of the Einstein equations ($G^i_{\, \,i}=\kappa T^i_{\, \,i}$). The quantity 
\begin{comment}
    \begin{equation}
    H= \frac{\dot{a}}{a}\, , \quad \text{or} \quad \hubble = \frac{a \conf}{a} =aH \, ,
\end{equation}
\end{comment}
\begin{equation}
    H \equiv \frac{\dot{a}}{a}\, , 
\end{equation}
is known as the Hubble parameter. It plays a central role in cosmology as it quantifies the rate of change of the scale factor. The first Friedmann equation in Eq.~\eqref{FriedmannEqs} relates this expansion rate to the energy density of the universe’s matter content. In other words, knowing the matter density allows us to determine the expansion rate of the universe, and vice versa. The second Friedmann equation, often called the acceleration equation, describes the conditions under which the universe undergoes an accelerated or decelerated expansion. To reduce the number of unknowns, we make a standard assumption in cosmology: the fluid is barotropic, meaning its pressure depends only on its energy density. We can then relate the pressure to the density via an equation of state
\begin{equation}\label{barotropicfluid}
    P=w\rho \, ,
\end{equation}
where $w$ is known as the equation of state parameter. It can depend on time in the following way
\begin{equation}
    \dot{w} = -3H (1+w) (\cs  -w)\, ,
\end{equation}
where we have further defined the speed of sound, $\cs$,
\begin{equation}\label{speedofsound}
    \dot{P} = \cs \dot{\rho} \, .
\end{equation}
We see that when $w=\cs$ then $w$ is constant\footnote{The actual definition of the speed of sound is $\cs \equiv \partial P/\partial\rho$ at a constant entropy, $S$. However, in FLRW spacetimes, $S=0$ at the background level. Furthermore, in FLRW spacetimes, we have established that the background quantities must exclusively depend on time, hence the less formal definition we have given. We will come back to this in the context of cosmological perturbation theory.}. Since the universe has undergone extended periods of time dominated by radiation ($w=1/3$), matter ($w=0$), and dark energy ($w=-1$), each well-described by a constant equation of state parameter, it is a good approximation to assume that this parameter remains constant within each era. Henceforth, $w$ is independent of time. 

The Friedmann equations in Eq.~\eqref{FriedmannEqs} form a closed system that can be solved once the equation of state parameter is specified. However, it is often more convenient to introduce the energy density conservation equation, which follows from the time component of the covariant conservation law in Eq.~\eqref{energymomentumconsevration}. This takes the form  
\begin{equation}\label{backgroundconservation}
    \dot{\rho} + 3H \left( \rho + P \right) = 0 \, .
\end{equation}  
Together, we now have three equations governing the background dynamics and matter content of the universe, all of which depend only on cosmic time. As stated, the equations are not fully independent: Eq.~\eqref{backgroundconservation} can actually be derived from the Friedmann equations. In fact, it is redundant as a consequence of the Bianchi identities. Nevertheless, the conservation equation is a useful first-order ODE that provides a direct relation between the energy density and the scale factor. Once this relation is obtained, the first Friedmann equation can be integrated to yield the time evolution of the scale factor $a(t)$.

We can now proceed to solve the Friedmann equations. By substituting Eq.~\eqref{barotropicfluid} into the conservation equation \eqref{backgroundconservation}, we arrive at
\begin{equation}\label{densityscaling}
    \rho \propto a^{-3(1+w)} \, .
\end{equation}
The results for different epochs of the universe are summarised in Table.~\ref{tablecosmoparameters}. For RD, MD and $\Lambda$ domination, we have taken $K=0$. Interestingly, when $w=-1/3$, we find from Eq.~\eqref{backgroundconservation} that $\rho \propto a^{-2}$, suggesting that the curvature term can be interpreted as a fluid with a constant equation of state. This value also marks the critical threshold in the acceleration equation (the second Friedmann equation): for $w<-1/3$, the universe undergoes accelerated expansion.

\begin{comment}
    \begin{table}[h!]
\centering
\begin{tabular}{|>{\centering\arraybackslash}p{2cm}|
                >{\centering\arraybackslash}p{2cm}|
                >{\centering\arraybackslash}p{2cm}|
                >{\centering\arraybackslash}p{2cm}|
                >{\centering\arraybackslash}p{2cm}|
                >{\centering\arraybackslash}p{2cm}|}
\hline
\textbf{Period} & \textbf{w} & \textbf{b} & \boldmath$\rho$ & \boldmath$ a(\eta)$ & \boldmath{$a(t)$} \\
\hline
RD & $1/3$ & $0$ & $\propto a^{-3}$ & $\propto \eta $ & $\propto t ^{1/2}$ \\
\hline
MD & $0$ & $1$ & $\propto a^{-4}$ & $\propto \eta ^2$ & $\propto t^{2/3}$ \\
\hline
DE & $-1$ & $-2$ & $\propto \text{const}$ &  $\times$ &$\propto e^{Ht}$   \\
\hline
\end{tabular}
\caption{Different values of cosmological parameters and relationships for different epochs. The value $w=-1$ is a special case which will be explore further in the next chapter.}
\end{table}

\end{comment}
\begin{table}[h!]
\centering
\begin{tabular}{|>{\centering\arraybackslash}p{2cm}|
                >{\centering\arraybackslash}p{2cm}|
                >{\centering\arraybackslash}p{2cm}|
                >{\centering\arraybackslash}p{2cm}|
                >{\centering\arraybackslash}p{2cm}|}
\hline
\textbf{Species} & \textbf{w} & \textbf{b} & \boldmath$\rho (a)$ & \boldmath{$a(t)$} \\
\hline
Radiation & $1/3$ & $0$ & $\propto a^{-3}$  & $\propto t ^{1/2}$ \\
\hline
Matter & $0$ & $1$ & $\propto a^{-4}$  & $\propto t^{2/3}$ \\
\hline
$\Lambda$ & $-1$ & $-2$ & $\propto \text{const}$ &  $\propto e^{Ht}$   \\
\hline
Curvature & $-1/3$ & $\times$ & $\propto a^{-2}$ & $\propto t$ \\
\hline
\end{tabular}
\caption{\footnotesize{Different values of cosmological parameters and relationships between key variables for different epochs.}}
\label{tablecosmoparameters}
\end{table}

An equivalent way to parametrise the equation of state is via the parameter $b$, defined as a function of $w$, which is ($w\neq -\frac{1}{3}$)
\begin{equation}\label{defb}
    b=\frac{1-3w}{1+3w} \Leftrightarrow w=\frac{1-b}{3(1+b)} \, .
\end{equation}
This parameter will prove useful when solving linear equations of motion in perturbation theory.

\begin{figure}
    \centering
    \includegraphics[width=\linewidth]{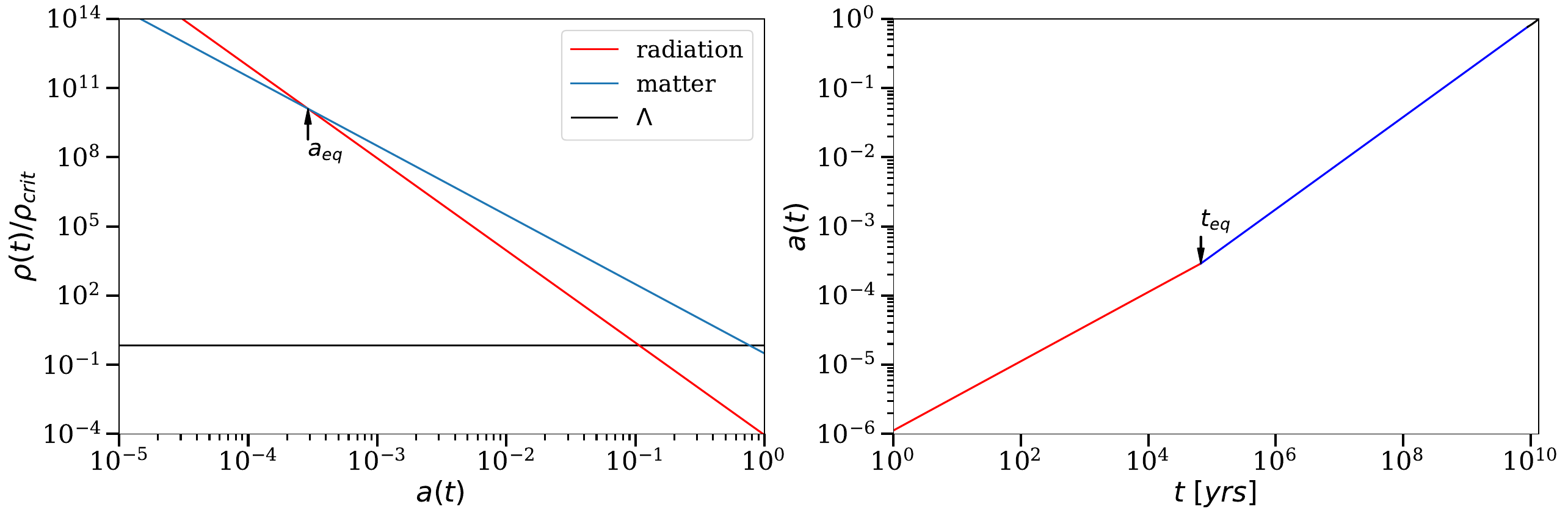}
    \caption{\footnotesize{(\textit{Left}) Plot of density (normalised to the critical density, introduced in the next sub-section) against the scale factor for different values of $w$.  (\textit{Right}) How the scale factor evolves throughout the history of our universe, normalised such that $a(t_0)=1$, where $t_0$ is the age of our universe. We have labelled on the two plots $a_{eq}$ and $t_{eq}$, which respectively correspond to the value of the scale factor and time at the time of radiation-matter equality. For both plots, we use the values in Eq.~\eqref{omegaLCDM}.}}
    \label{fig:scalefactor}
\end{figure}

The relationship between the density and the scale factor having been established, we can now solve for the scale factor. Substituting the scaling in Eq.~\eqref{densityscaling} into the first Friedmann equation, we arrive at
\begin{comment}
    \begin{equation}
    a(t)\propto t^{\frac{2}{3(1+w)}} \, , \quad \text{or} \quad  a(\eta) \propto \eta ^ {\frac{2}{1+3w}} \propto \eta ^{1+b}
\end{equation}
\end{comment}
\begin{equation}
    a(t)\propto t^{\frac{2}{3(1+w)}} \, , 
\end{equation}
for $w \neq -1 $. When $w=-1$ we see that the ratio $\dot{a}/a$ becomes constant and hence we get an exponential expansion
\begin{comment}
    \begin{equation}
    a(t)\propto e^{Ht} \, , \quad \text{or} \quad  a(\eta) \propto -\frac{1}{\eta}
\end{equation}
\end{comment}
\begin{equation}
    a(t)\propto e^{Ht} \, ,
\end{equation}
where $H$ is constant. As expected, $\ddot{a}>0$, corresponding to an accelerating expansion. In Fig.~\ref{fig:scalefactor}, we have plotted the `history of the scale factor' throughout different periods of the universe. 

\begin{comment}
    Finally, we need to consider what happens around the so-called `equality scale'. This is the time, $t_{eq}$, where matter and radiation are equal: $\rho_r (a_{eq}) = \rho _m (a_{eq}) = \frac{1}{2}\rho_{eq}(a_{eq})$. To find the value of $a_{eq}$ we then need to solve our background equations of motion for a universe filled of both radiation and matter.

    For a universe dominated by both radiation and matter we can show that the acceleration equation takes the form
\begin{equation}
    a ^{\prime\prime} = \frac{2\pi G}{3} \rho_{eq} a_{eq}^3 \, ,
\end{equation}
which has for solution
\begin{equation}
    a(\eta) = a_{eq} \left [ \left ( \frac{\eta}{\eta_*} \right ) ^2 + 2\left ( \frac{\eta}{\eta_*} \right ) \right ] \, ,
\end{equation}
with
\begin{equation}
    \eta_* = \left ( \frac{\pi G}{3}\rho _{eq} a^2_{eq} \right )^{-\frac{1}{2}} = \frac{\eta _{eq}}{\sqrt{2}-1} \, .
\end{equation}
\end{comment}
Often in cosmology, it is convenient to re-express the FLRW metric in Eq.~\eqref{FLRWflat} in the following way
\begin{equation}\label{FLRWconformal}
        ds^2 = a^2(\eta) \left ( -d\eta ^2 +g_{ij}dx^idx^j  \right ) \, ,
\end{equation}
which corresponds to a rescaling of the time variable in the FLRW metric. In this new coordinate system, the time coordinate, $\eta$, is called conformal time. It is defined via
\begin{equation}
    d\eta = \frac{dt}{a(t)} \, , \quad \eta (t) = \int _0^{t} \frac{1}{a(t^{\prime})} {\rm }t^{\prime} \, .
\end{equation}
This transformation is particularly useful as it makes the metric conformal to Minkowski spacetime, simplifies equations and is used extensively during inflation. In this new coordinate system, the components of the metric tensor read
\begin{align}
    \begin{split}
        g_{00} &= -a^2(\eta) \, , \quad g^{00} = -a^{-2}(\eta)\, , \\
        g_{ij} &= a^2(\eta)\delta _{ij} \, , \quad g^{ij} = a^{-2}(\eta)\delta ^{ij} \, ,
    \end{split}
\end{align}
and now the velocity 4-vector
\begin{equation}
    u^\mu = \left (\frac{1}{a(\eta)},0,0,0 \right ) \, , \quad u_\mu = \left (-a(\eta),0,0,0 \right ) \, ,
\end{equation}
hence the components of the perfect fluid from Eq.~\eqref{perfectfluid} take the form
\begin{align}\label{perfectlfluidconformal}
    \begin{split}
        T_{00} &= a^2(\eta)\rho \, , \quad T^{00} = a^{-2}(\eta)\rho\, , \\
        T_{ij} &= a^2(\eta)P\delta _{ij} \, , \quad T^{ij} = a^{-2}(\eta)P\delta ^{ij} \, .
    \end{split}
\end{align}
The Friedmann equations (with $K=0$) become
\begin{equation} \label{FriedmannEqsconf}
    \hubble ^2 =\left  ( \frac{a \conf}{a}\right )^2 =\frac{a^2\kappa}{3} \rho \quad \text{and} \quad \hubble \conf = -\frac{a^2\kappa}{6}(\rho+3P) \, ,
\end{equation}
whilst the energy conservation equation becomes
\begin{equation}
    \rho \conf + 3\hubble (\rho+P) = 0.
\end{equation}
As a result, for $w\neq -\frac{1}{3}$, we have the useful following relationships
\begin{equation}\label{afuncb}
    a(\eta) \propto \eta ^{\frac{2}{1+3w}} = \eta ^{1+b} \quad \text{and} \quad \hubble = \frac{2}{\eta(1+3w)}=\frac{1+b}{\eta} \, .
\end{equation}
They will be used extensively throughout this work. 

% % % % % % % % % % % % % % % % % % % % % % % % % % % % % % % % 
% % % % % % % % % % % % % % % % % % % % % % % % % % % % % % % % 
\section{Critical density and current parameters in a $\Lambda$CDM universe}
An alternative way of expressing the first Friedmann equation is to divide it by a factor of $H^2$, to arrive at the dimensionless equation
\begin{equation}
    1 = \frac{8\pi G }{3H^2} \rho - \frac{K}{H^2a^2}\ .
\end{equation}
By defining $\rho _{crit} \equiv  \left (3H^2/8\pi G \right ) $, the above equation becomes
\begin{equation}\label{defcriticaldensity}
     \Omega -1 \equiv  \frac{\rho}{\rho _{crit}}-1= \frac{K}{a^2H^2}  \, .
\end{equation}
We see that if $\Omega =1$ then $K=0$ and $\rho = \rho_{crit}$. The critical density, $\rho _{crit}$, is then the density the universe would need to have to be perfectly flat. If we further split $\rho$ into its constituents ($\rho_r$ density of radiation, $\rho_m$ density of matter, $\rho_{\Lambda}$ density of dark energy and $\rho_{K}$ the curvature density) we arrive to 
\begin{equation}
    H^2 =   \frac{8\pi G}{3} \left ( \rho_r + \rho_m  + \rho_{\Lambda} + \rho_{K} \right ) \, ,
\end{equation}
from which follows
\begin{equation}\label{densityscaling}
    \frac{H^2}{H_0^2} = \Omega _{r,0}a^{-4} + \Omega _{m,0}a^{-3} + \Omega _{\Lambda,0}+ \Omega _{K,0}a^{-2} \, ,
\end{equation}
where the subscript ``$0$'' means present time. We have defined
\begin{equation}
     \Omega _i \equiv  \frac{\rho _i}{\rho _{crit}}
\end{equation}
where $i$ refers to a constituent of the universe, and we have used the relationships between density and the scale factor in Table.~\ref{tablecosmoparameters}. The precise value of the Hubble constant, $H_0$, remains a matter of significant debate, a discrepancy widely known as the Hubble tension. This tension stems from the persistent discrepancy between the value inferred from observations of the CMB by the Planck satellite \cite{Planck:2018vyg} and those obtained from present-day measurements of Type~Ia supernovae \cite{Pan-STARRS1:2017jku}. For a recent compilation of $H_0$ determinations across different experiments, the reader is referred to Figure~9 of Ref.~\cite{DESI:2024mwx} and references therein. It is therefore convenient to introduce a dimensionless number $h$ and express Hubble's constant as
\begin{equation}
    H_0= h \times 100 \, \text{km } \text{s}^{-1} \text{ Mpc}^{-1} \, , 
\end{equation}
and substitute this into Eq.~\eqref{densityscaling} to factor the uncertainty in the value of $H_0$. Therefore, values of the cosmological parameters such as $ \Omega _{m,0}$ are often quoted in surveys as $ \Omega _{m,0}h^2$. From the Planck satellite \cite{Planck:2018vyg} and surveys of galaxies such as the Sloan Digital Sky Survey \cite{SDSS:2006lmn}, the density parameters today are determined to be 
\begin{equation}\label{omegaLCDM}
    \Omega _{K,0} < |0.01| \, , \text{ } \Omega _{r,0} = 9.4\times10^{-5}  \, ,\text{ }  \Omega _{m,0} = 0.31  \,  \text{ and } \Omega _{\Lambda,0} = 0.69 \, .
\end{equation}
The Universe is then to a good approximation spatially flat ($K \approx 0$) and its energy budget is dominated by dark energy and matter. The present-day matter density parameter is decomposed as
\begin{equation}
    \Omega_{m,0} = \Omega_{\mathrm{CDM},0} + \Omega_{b,0},
\end{equation}
with $\Omega_{\mathrm{CDM},0} \approx 0.27$ for cold dark matter (CDM) and $\Omega_{b,0} \approx 0.04$ for baryonic matter. Remarkably, about $95\%$ of the Universe’s energy content remains of unknown physical nature; there is no widely accepted theory for the origin of $\Lambda$ or what matter CDM is made of, despite strong observational evidence for both \cite{SupernovaSearchTeam:1998fmf,SupernovaCosmologyProject:1998vns,Rubin:1980zd,Einasto:1974dra,1973A&A....26..483R,SDSS:2005sxd}. Of interest in this work, primordial black holes (PBHs) have recently gained traction as a viable CDM candidate as they could represent a non-negligible fraction of CDM. These objects result from the collapse of large density perturbations in the very early universe \cite{Zeldovich:1967lct,Hawking:1971ei,Carr:1974nx}, prior to nucleosynthesis, making them viable candidates for CDM \cite{Chapline:1975ojl,Meszaros:1975ef}. Their observational signatures and constraints on their abundance will be discussed further in Chapter \ref{Chap:SOGWs}. 
 
The values of cosmological parameters listed above define the so-called $\Lambda$CDM model: a ``\textit{Euclidean universe dominated today by non-baryonic cold dark matter and a cosmological constant, with initial perturbations generated by inflation in the very early universe}" \cite{Dodelson:2003ft}. Before turning to a detailed discussion of these perturbations in Chap.~\ref{Chap:CPT}, we first provide a brief overview of inflation.   
% % % % % % % % % % % % % % % % % % % % % % % % % % % % % % % % 
% % % % % % % % % % % % % % % % % % % % % % % % % % % % % % % % 
\section{Inflation}\label{inflation}
Although our work does not focus directly on inflation, it is arguably an important (postulated) epoch in the history of the universe. As discussed in the introduction, inflation was originally proposed to address several shortcomings of pre–$\Lambda$CDM cosmological models, and was formulated as a theory involving a brief period of exponential expansion \cite{Starobinsky:1979ty,Guth:1980zm,Linde:1981mu}. One of the key motivations for inflation arises from the observed near-homogeneity and isotropy of the CMB, with temperature fluctuations on the order of $\Delta T / T \approx 10^{-5}$. Such uniformity suggests that different regions of the CMB were once in causal contact. However, in the absence of inflation, regions separated by more than roughly $1.5^\circ$ on the sky could not have been causally connected at the time of formation of the CMB. A way to view this is to introduce $\chi _p$, the comoving particle horizon, defined as
\begin{equation}\label{particlehorizon}
    \chi _p = \int _{t_i=0}^{t} \frac{\rm t \conf}{a(t\conf)} \, ,
\end{equation}
where $t_i=0$ is the time of the initial singularity (big-bang). If two points are separated by a distance greater than $\chi_p$, they have never been in causal contact. This is precisely the case for the CMB, where points separated by more than $1.5^\circ$ have a comoving distance much larger than the comoving particle horizon \cite{Baumann:2009ds}. Inflation elegantly resolves this so-called horizon problem by stretching a small, causally connected region to encompass the entire observable universe. As highlighted earlier, the universe appears to be nearly spatially flat, which is a feature that inflation naturally explains by dynamically driving the universe towards flatness. In what follows, we will briefly describe the dynamics of inflation.

There are various models of inflation; arguably the simplest consists of a scalar field called the inflaton, $\varphi$, which rolls down a potential $V(\varphi)$. The corresponding energy-momentum tensor is
\begin{equation}
    T \munu = \partial _\mu \varphi \partial _\nu \varphi - \left ( \frac{1}{2} \partial _ \alpha \varphi \partial ^ \alpha \varphi + V (\varphi) \right ) g \munu \, .
\end{equation}
By comparing the above energy-momentum tensor with that of the perfect fluid in Eq.~\eqref{perfectlfluidconformal}, we get the following correspondence\footnote{Usually, it is natural to work in conformal time coordinates during inflation; however, for our purposes, it is not necessary.}:
\begin{equation}
    \rho _{\varphi} = \frac{\dot{\varphi} ^{2}}{2} + V(\varphi) \, , \quad P _{\varphi} = \frac{\dot{\varphi} ^{2}}{2} - V(\varphi) \, .
\end{equation}
Furthermore, we recall that when the equation of state parameter $w=-1$, we get an exponential expansion. From the above equation, we see that this is the case if $\tfrac{\dot{\varphi} ^{2}}{2} < V(\varphi)$. Hence, inflation is a period of accelerated expansion characterised by a slowly varying scalar field which evolves in its potential. The corresponding Friedmann equations are\footnote{It is convenient during inflation to use the Planck mass $M_{Pl}$ rather than $\kappa$, since these processes happen at high energy scales. $M_{Pl}^{-2}=8\pi G$.}
\begin{equation}
    H^2 = \frac{1}{3M_{Pl}^2} \left (  \frac{\dot{\varphi} ^{2}}{2}  +  V(\varphi) \right ) \, , \quad \frac{\ddot{a}}{a} = \frac{1}{3M_{Pl}^2} \left (  V(\varphi) - \dot{\varphi} ^{2} \right ) \, ,
\end{equation}
and the energy conservation equation reads
\begin{equation}
    \ddot{\varphi} + 3H \dot{\varphi}  +  \frac{dV(\varphi)}{d\varphi} =0 \, .
\end{equation}
These equations can be solved analytically or numerically, once we specify a potential. The fact that the potential energy of the scalar field is larger than the kinetic energy of the scalar field can be encoded in the so-called slow-roll parameter  $\epsilon$, defined as
\begin{equation}
    \epsilon = - \frac{\dot{H}}{H^2} = \frac{3 \times \tfrac{\dot{\varphi}^2}{2}}{\tfrac{\dot{\varphi}^2}{2}+V(\varphi)} \ll 1 \quad \text{if} \quad \frac{\dot{\varphi} ^{2}}{2} < V(\varphi) \, ,
\end{equation}
and hence we see that the condition for accelerated expansion becomes $\epsilon \ll 1$. 

It is useful to re-express the comoving particle horizon in Eq.~\eqref{particlehorizon} as
\begin{equation}
    \chi _p = \int _{\ln a_i=0}^{\ln a} \frac{1}{a\conf H(a\conf)} \rm \ln a \conf \, ,
\end{equation}
where $(aH)^{-1}$ is the comoving Hubble radius and determines whether two points in spacetime are able to communicate at some time $t$. As such, the particle horizon can be seen as the integral over past comoving Hubble scales, at a given time $t$. If the comoving Hubble radius increases, it can bring previously causally disconnected points into causal contact over time. However, the observed uniformity of the CMB implies that widely separated regions must have already been in causal contact. Inflation then resolves this by introducing an earlier epoch during which the comoving Hubble radius decreases, allowing regions that were once in causal contact to be stretched far apart, beyond the comoving Hubble radius. To see this, we consider
\begin{equation}
    \frac{\rm}{\rm t}(aH)^{-1} = -\frac{a(H^2 + \dot{H})}{aH^2} \, ,
\end{equation}
and notice that $\epsilon <1$, the condition for an inflating universe, implies a decreasing comoving Hubble radius. We also remark that when we defined the critical density through equation Eq.~\eqref{defcriticaldensity}, the quantity $\Omega -1$ is inversely proportional to the square of the comoving Hubble radius
\begin{equation}
    |\Omega - 1| \propto (aH)^{-2} \, .
\end{equation}
It turns out that having a period where the comoving Hubble decreases also solves the flatness problem, by forcing the quantity $|\Omega - 1|$ to be extremely close to zero by the end of inflation.

Finally, we would like to emphasise that, in addition to addressing key cosmological problems, inflation naturally connects the quantum fluctuations of the inflaton field, $\delta \varphi$, to the large-scale structure observed in the universe today. In other words, these fluctuations serve as the primordial seeds from which all cosmic structure arises. Although we will not delve into this topic, the connection becomes evident through cosmological perturbation theory, which we will introduce in the next chapter.

% % % % % % % % % % % % % % % % % % % % % % % % % % % % 
% chapter.tex - Ian Huston
% Sample chapter layout
% % % % % % % % % % % % % % % % % % % % % % % % % % % % 
% Redefine CVSRevision for this section. 
% If you don't want to use CVS tags comment out this line
\renewcommand{\CVSrevision}{\version$Id: chapter.tex,v 1.3 2009/12/17 18:16:48 ith Exp $}

% % % % % % % % % % % % % % % % % % % % % % % % % % % % % % % % 
% =========================================================== %
% % % % % % % % % % % % % % % % % % % % % % % % % % % % % % % % 
\chapter{Cosmological perturbation theory}
\label{Chap:CPT}
% % % % % % % % % % % % % % % % % % % % % % % % % % % % % % % % 
% =========================================================== %
% % % % % % % % % % % % % % % % % % % % % % % % % % % % % % % % 
In this section, we introduce the formalism necessary to study GWs in a cosmological context. As a starting point, it is instructive to examine how GWs propagate in Minkowski spacetime. This serves two purposes: it provides a simple introduction to perturbation theory, and it illustrates the role of coordinate freedom, highlighting how to choose a convenient gauge when analysing metric perturbations. We then introduce the cosmological perturbation theory framework, order by order, and outline the equations required for the rest of this work\footnote{The Einstein and conservation equations become increasingly lengthy at higher orders, and listing them all can be counterproductive. Therefore, we will focus only on the equations relevant for the upcoming chapters.}. We will also introduce the power spectrum, which characterises the distribution and correlations of perturbations in the universe.

% % % % % % % % % % % % % % % % % % % % % % % % % % % % % % % % 
% % % % % % % % % % % % % % % % % % % % % % % % % % % % % % % % 
\section{Gravitational waves in Minkowski spacetime}\label{sectionminkwoski}

General relativity predicts wave-like solutions. To understand this, it's helpful to first examine GWs in a flat, non-expanding spacetime. We will then show in one of the following sections that, in a cosmological setting, tensor modes behave like GWs in Minkowski spacetime on small scales. In this section, we follow closely Ref.~\cite{Caprini:2018mtu}. We start by perturbing the Minkowski metric as
\begin{equation} \label{linearminkmetric}
    g_{\mu \nu} = \eta _{\mu \nu} +h_{\mu \nu} \, .
\end{equation}
Our objective is to find an equation of motion for $h_{\mu \nu}$, the metric perturbation. To this end, we will show that $h \munu$ obeys the wave equation $\Box h \munu =0$, where $\Box = \eta ^{\mu \nu} \partial _\mu \partial _\nu$ ($\Box$ is referred to as the d'Alembertian operator). Our only requirement is that $|h\munu |\ll 1$, i.e. the components of the tensor $h\munu$ are small, which allows us to work with the linearised Einstein equations and not consider terms $\mathcal{O}(h^2)$. To illustrate this, we consider taking the inverse of the metric perturbation $h^{\mu \nu}$:
\begin{equation}
    h^{\mu \nu}=g^{\mu \alpha} g^{\nu \beta} h_{\alpha \beta} = \eta^{\mu \alpha} \eta^{\nu \beta} h_{\alpha \beta} + \mathcal{O}(h^2) \, . 
\end{equation}
In order to satisfy $g_{\mu \alpha}g^{\nu \alpha}=\delta _{\mu}^{\nu}$ we need
\begin{equation}
    g^{\mu \nu} = \eta ^{\mu \nu} - h^{\mu \nu} \, ,
\end{equation}
to keep everything linearised. Expanding the metric tensor as we have done in Eq.~\eqref{linearminkmetric}, allows us to separate the Einstein equations in the following way
\begin{equation}
    G \munu ^{(0)} = \kappa T \munu ^{(0)} \, , \quad \text{and} \quad G \munu ^{(1)} = \kappa T \munu ^{(1)} \, .
\end{equation}
The Einstein equations with the superscript ``$0$'' are called the background equations, and the equations with the superscript ``$1$'' are referred to as the linearised Einstein equations; they contain only terms linear in $h\munu$. In Minkowski space $T\munu ^{(0)}=0$ and we assume nothing is sourcing the GWs, i.e. $T\munu ^{(1)}=0$. Therefore, to extract an equation of motion for $h\munu$, we need to consider
\begin{equation}
    G \munu ^{(1)} = 0.
\end{equation}
Accordingly, we compute the Christoffel symbols as defined in Eq.~\eqref{defChristofell}, which at linear order take the simple form
\begin{equation}
    \Gamma ^{\rho \, \, (1)}_{\mu \nu} = \frac{1}{2} \eta ^{\rho \lambda} \left ( h_{\nu \lambda,\mu} + h_{\mu \lambda,\nu} - h_{\mu \nu ,\lambda} \right ) \, ,
\end{equation}
where we have introduced the notation $h_{\nu \lambda,\mu} = \partial _{\mu} h_{\nu \lambda}$. We can then compute the Ricci tensor from contracting Eq.~\eqref{defRiemann},
\begin{align}
    \begin{split}
        R^{(1)} \munu&= \partial_\rho \Gamma^\rho \munu - \partial_\nu \Gamma^\rho_{\rho \mu} + \mathcal{O}(h^2) \\
        &= \frac{1}{2} \left ( \partial ^\lambda \partial _\mu h_{\nu \lambda} + \partial ^\lambda \partial _\nu h_{\mu \lambda} - \Box h \munu - \partial _\mu \partial _\nu h \right ) \, , 
    \end{split}
\end{align}
where $h=h^\mu _{ \, \, \mu}$, is the trace of the metric perturbation. From the equation above follows the Ricci scalar
\begin{equation}
    R^{(1)} = \partial ^\mu \partial ^\nu h \munu - \Box h \, .
\end{equation}
Finally, the linearised Einstein equations in vacuum read
\begin{equation}\label{minkGpre}
    G\munu ^{(1)} =  \partial ^\lambda \partial _\mu h_{\nu \lambda} + \partial ^\lambda \partial _\nu h_{\mu \lambda} - \Box h \munu - \partial _\mu \partial _\nu h - \eta \munu \partial ^\alpha \partial ^\beta h_{\alpha \beta} + \eta \munu \Box h =0 \, .
\end{equation}
The form of the above equation is not very instructive, so we introduce the trace-reversed perturbation, $\overline{h}\munu$, defined as
\begin{equation}\label{deftracereversed}
    \overline{h} \munu = h \munu -\frac{1}{2}\eta \munu h \, ,
\end{equation}
such that $\overline{h}=-h$. Substituting the above relationship into Eq.~\eqref{minkGpre}, the linearised Einstein equation becomes 
\begin{equation}\label{linearisedeinsteinhbar}
    \partial ^\lambda \partial _\mu \overline{h}_{\nu \lambda} + \partial ^\lambda \partial _\nu \overline{h}_{\mu \lambda} - \Box \overline{h} \munu - \eta \munu \partial ^ \alpha \partial ^\beta \overline{h}_{\alpha \beta} =0\, .
\end{equation}
Although more compact, it is not yet clear that the perturbation $\overline{h} \munu$ obeys the wave equation. In fact, we see from the above equation that if $\partial ^\mu \overline{h}_{\mu \nu}=0$ then we are left with $\Box \overline{h} \munu =0$, which is the wave equation describing the propagation of the massless perturbed variable $\overline{h} \munu$. There is one nuance in GR that we have not yet discussed: the freedom to choose a coordinate system.

If we consider an infinitesimal change of coordinates $x^{\mu} \rightarrow x^{\prime \mu }= x^{\mu} + \xi ^{\mu}$ with the condition $|\xi ^{\mu}|\ll1$, it then induces a change in the perturbed variables such that\footnote{In this section the $\prime$ signifies a different coordinate system and not a conformal time derivative.} 
\begin{subequations}
    \begin{align}
       &h \munu (x)\rightarrow h^{\prime}\munu (x\conf)= h \munu (x) - \pounds_{\xi}\eta \munu = h \munu (x) - \partial _\mu \xi _\nu - \partial _\nu \xi _\mu \, , \\
       &h \rightarrow h\conf = h-2\partial_{\alpha}\xi ^\alpha \, , 
    \end{align}
\end{subequations}
and\footnote{We have used the Lie derivative defined by its action on a general tensor as
\begin{equation} \label{liederivative}
    \pounds_{\xi} T^{\mu_1...\mu _p}_{\nu_1...\nu _q} = \xi ^\alpha \partial _\alpha  T^{\mu_1...\mu _p}_{\nu_1...\nu _q} -  \sum _{i=1}^p T^{\mu_1...\alpha ...\mu _p}_{\nu_1...\nu _q} \partial _\alpha \xi ^{\mu _i} + \sum _{j=1}^q T^{\mu_1...\mu _p}_{\nu_1...\beta ...\nu _q} \partial _{\nu _j} \xi ^{\beta} \, ,
\end{equation}}
\begin{equation}\label{hbargauge}
    \overline{h} \munu (x)\rightarrow \overline{h}\conf \munu (x \conf) = \overline{h} \munu (x) - \pounds_{\xi}\eta \munu = h \munu (x) - \partial _\mu \xi _\nu - \partial _\nu \xi _\mu + \eta \munu \partial_{\alpha}\xi ^\alpha  \, .
\end{equation}
where we have truncated everything to linear order. The coordinate transformations leave the Einstein equations invariant, and therefore, the equations of motion. Consequently, there is an infinite number of ways to pick $h\munu$ (or $\overline{h}\munu$), since the linearised equations do not fix the perturbed variables uniquely. We then have the freedom to pick a set of coordinates to our advantage: this freedom is referred to as a choice of gauge. We already hinted that $\partial ^\mu \overline{h}_{\mu \nu}=0$ seems like an intuitive choice, since it reduces Eq.~\eqref{linearisedeinsteinhbar} to the wave equation for the trace-reversed metric perturbation. In fact, this gauge choice is called the Lorentz Gauge, and in this gauge (or in these coordinates) Eq.~\eqref{linearisedeinsteinhbar} becomes 
\begin{equation}\label{hbarwaveequation}
    \Box \overline{h} \munu =(-\partial_t^2+\nabla ^2)\overline{h}\munu=0 \, .
\end{equation}
We note that we can always pick the Lorentz gauge since, according to Eq.~\eqref{hbargauge}, 
\begin{equation}
    \partial _\mu \overline{h} ^{\mu \nu} (x)\rightarrow \partial _\mu \conf \overline{h} ^{\prime \mu \nu} (x \conf)= \partial _\mu \overline{h} ^{\mu \nu} (x)- \Box \xi ^\nu \, ,
\end{equation}
and we see that if we want $\partial _\mu \conf \overline{h} ^{\prime \mu \nu} (x \conf)=0$ (i.e. for the Lorentz gauge to hold in our new coordinate system) then we require $\partial _\mu \overline{h}^{\mu \nu}=\Box \xi ^{\nu}$, which is always possible since the d'Alembertian is invertible. The Lorentz gauge allows us to reduce the ten degrees of freedom in $\overline{h}\munu$ to six.

Is there still freedom to choose coordinates? Suppose we know the trace-reversed metric perturbation is in the Lorentz gauge (which we showed was always possible), then according to the transformation of $\overline{h}^{\mu \nu}(x)$ under infinitesimal coordinate transformation in Eq.~\eqref{hbargauge}, 
\begin{equation}
    \partial _\mu \overline{h} ^{\mu \nu} (x)\rightarrow \partial _\mu \conf \overline{h} ^{\prime \mu \nu} (x \conf)= - \Box \xi ^\nu \, .
\end{equation}
For $\partial _\mu \conf \overline{h} ^{\prime \mu \nu} (x \conf)=0$ then we need  $\Box \xi ^\nu=0$. Therefore, it is possible to change coordinates and arrive in a new coordinate system where the Lorentz gauge is still satisfied if $\Box \xi ^\nu=0$. Similarly to when we picked the Lorentz gauge because we had the freedom to do so, we can further pick a coordinate system of our choosing which still respects the Lorentz gauge.

The wave equation for the trace-reverse perturbation in Eq.~\eqref{hbarwaveequation} can be solved with the ansatz
\begin{equation}\label{gwminkansatz}
    \overline{h} \munu (\mathbf{x},t)= \Re \left ( \int \frac{\rm  \mathbf{k}}{(2\pi)^{\frac{3}{2}}} A \munu e^{i(\mathbf{k}\cdot \mathbf{x}-\omega t)} \right ) \, ,
\end{equation}
where $\mathbf{k}$ is the wave vector, $\omega$ the angular frequency and $A \munu$ is a constant tensor we are yet to determine. The wave equation implies that the propagating degrees of freedom travel at the speed of light ($k=|\omega|$), whilst the Lorentz gauge implies $k_\mu A^{\mu \nu}$, i.e. it puts restrictions on the form of $A^{\mu \nu}$ (on top of the fact that is symmetric by virtue of the metric being symmetric). To completely determine $A^{\mu \nu}$, we use our residual gauge freedom, which was that the infinitesimal transformation generator $\xi ^\mu$ obeys the wave equation. We can pick the ansatz
\begin{equation}
    \xi _\mu(\mathbf{x},t)= \int \frac{\rm  \mathbf{k}}{(2\pi)^{\frac{3}{2}}} A _\mu e^{i(\mathbf{k}\cdot \mathbf{x}-\omega t)} \, ,
\end{equation}
and are free to choose $A _\mu$ to our liking. In order to connect with what follows, we choose an $A _\mu$ which satisfies
\begin{equation}
    A^{\mu i}=A^{i\mu } =0\, , \quad \text{and} \quad A^{\mu}_{\,\, \mu} =0 \, .
\end{equation}
This gauge choice, combined with the Lorentz gauge, is called the transverse-traceless gauge. The four degrees of freedom we have just eliminated reduce the total degrees of freedom to two. Furthermore, looking back at the relationship between our original metric perturbation $h\munu$ and the trace-reversed perturbation $\overline{h}\munu$ in Eq.~\eqref{deftracereversed}, we see that the two perturbation variables are equal in the transverse-traceless gauge.   

It is instructive to consider a wave travelling in the $z$-direction\footnote{This will be a viable choice when we discuss cosmological gravitational waves.} to fully determine the structure of $A\munu$. To satisfy the wave-equation the wave-vector $k_{\mu}$ becomes
\begin{equation}
    k_{\mu} = [-\omega ,0,0,\omega] \, ,
\end{equation}
and the transverse-traceless condition implies 
\begin{equation}
    A\munu = 
A \begin{bmatrix}
0 & 0 & 0 & 0 \\
0 & 1 & 0 & 0 \\
0 & 0 & -1 & 0 \\
0 & 0 & 0 & 0
\end{bmatrix}
+B \begin{bmatrix}
0 & 0 & 0 & 0 \\
0 & 0 & 1 & 0 \\
0 & 1 & 0 & 0 \\
0 & 0 & 0 & 0
\end{bmatrix} \, .
\end{equation}
We can therefore express the GW solution in Eq.~\eqref{gwminkansatz} for one plane wave travelling in the $z$-direction as
\begin{equation}
    \overline{h} \munu (\mathbf{x},t)=h_{ij} (z,t)= \sum_{\lambda=+,\times} A_{\lambda} \cos [\omega t - kz] \eps _{ij}^{\lambda} \, ,
\end{equation}
where we have defined the plus ($+$) and cross ($\times$) polarisations 
\begin{equation} 
\eps _{ij}^{+} =\begin{bmatrix}
1 & 0 & 0  \\
0 & -1 & 0  \\
0 & 0 & 0 
\end{bmatrix}
\, , \quad \text{and} \quad  \eps _{ij}^{\times} =\begin{bmatrix}
0 & 1 & 0  \\
1 & 0 & 0  \\
0 & 0 & 0 
\end{bmatrix}\, .
\end{equation}

Before moving on to perturbations in an FLRW background, we conclude this section by noting that, in Minkowski spacetime, a component of the Riemann tensor can be directly related to the transverse-traceless perturbation via \cite{Caprini:2018mtu}
\begin{equation}\label{riemannandh}
    R_{i0j0}^{(1)}=-\frac{1}{2} \ddot{h}_{ij} \, .
\end{equation}
This expression illustrates how gravitational waves induce curvature and reveals how they locally perturb the spacetime through which they propagate.

% % % % % % % % % % % % % % % % % % % % % % % % % % % % % % % % 
% % % % % % % % % % % % % % % % % % % % % % % % % % % % % % % % 
\section{Perturbations in an isotropic and homogeneous universe}

In the previous chapter, we assumed that the universe is homogeneous and isotropic on large scales. However, it clearly possesses structures: on smaller scales, we observe stars and planets, then galaxies and galaxy clusters. At first glance, this seems to contradict our most fundamental assumption. The FLRW metric in Eq.~\eqref{FLRWflat} depends only on time through the scale factor and not on spatial coordinates. How can we reconcile this with the observed structure? We address this using cosmological perturbation theory, treating our universe as close to FLRW, a homogeneous and isotropic background with small perturbations that evolve on this background depending on both time and space. In what follows, we draw closely from the setup established in Ref.~\cite{Malik:2008im} (with different notations) and use the \href{https://www.xact.es/xPand/}{\textit{xPand}}~\cite{Pitrou:2013hga} package to extract the different cosmological equations we require.

Tensorial quantities can be split up as \cite{Malik:2008im}
\begin{equation}\label{tensorsplit}
    \mathbf{T}(x^\mu) = \mathbf{T}_0(\eta) + \delta \mathbf{T}(x^\mu) \, ,
\end{equation}
where $\mathbf{T}_0(\eta)$ is the `background tensor', depending only on time, and $\delta \mathbf{T}(x^\mu)$ are perturbations which depend on space and time. Perturbations can then be further expanded as a power series
\begin{equation}
    \delta \mathbf{T}(x^\mu) = \sum _0^{\infty} \frac{\eps ^{(n)}}{n!} \delta \mathbf{T}^{(n)}(x^\mu) \, ,
\end{equation}
$(n)$ being the order of the perturbation and $\eps$ is assumed to be small. When dealing with linear perturbation theory, we only keep terms linear in $\eps$. When going to second order in perturbation theory, we deal only with terms involving $\eps ^2$, and so on. We will not keep track of this parameter; however, one can imagine it comes attached to every perturbative quantity we use.
% % % % % % % % % % % % % % % % % % % % % % % % % % % % % % % %
\subsection{Perturbing the Einstein equations to linear order}

Applying the formalism of Eq.~\eqref{tensorsplit} to the metric tensor, we have 
\begin{equation}\label{metricexpansion}
    g\munu = g^{(0)} \munu + g^{(1)} \munu + \frac{1}{2}g^{(2)} \munu + \frac{1}{6}g^{(3)} \munu + \, ... \, ,
\end{equation}
where $g^{(0)} \munu$ is given by Eq.~\eqref{FLRWconformal} with $K=0$ and we are now working in conformal time. Since we have not included $\eps$ in the expansion, we assume $g^{(1)} \munu$ to be small, $g^{(2)} \munu$ to be smaller and so on. Physically, the above expression is an attempt to represent the `real metric' of our universe by an expansion \cite{Malik:2008yp}. In this section, we will focus exclusively on the term $ g^{(1)} \munu$ and then, when needed, we will just `add' higher order perturbations as the formalism we develop at linear order will be similar at higher orders. The line element up to linear order is
\begin{equation}\label{metriclinearflrw}
    ds^2 = a^2(\eta) \left [ - (1+2A^{(1)}) d\eta ^2 + 2B_i^{(1)}dx^id\eta + (\delta _{ij} + H_{ij}^{(1)}) dx^i dx^j  \right ] \, .
\end{equation}
$\delta _{ij}$ is the Kronecker delta, $A^{(1)}(\eta,x^i)$, $B_i^{(1)}(\eta,x^i)$ and $H_{ij}^{(1)}(\eta,x^i)$ represent the scalar, vector, and tensor perturbations of the temporal, cross, and spatial parts of the metric, respectively. All are functions of both time and space, and as for the scale factor, their evolution can be found by solving the Einstein equations (see Sec.~\ref{sec:scalefactor}). We also note that by turning these perturbations off, we return to the conformal FLRW metric in Eq.~\eqref{FLRWconformal}. From now on, we will drop the superscript which indicates a perturbation is first-order to reduce clutter.

We can further decompose the vector and tensor perturbations using the Helmholtz decomposition theorem. As we will see, the advantage of doing such a decomposition at first order in perturbation theory lies in the fact that scalar, vector and tensor perturbations decouple, i.e. the respective equations of motion for the different types of perturbations are independent. This is referred to as the scalar-vector-tensor (SVT) decomposition. The vector perturbation can be split up into the sum of the divergence of a scalar (curl-free part) and a divergenceless vector (solenoidal part):
\begin{equation}
    B^i = \partial ^i B + \overline{B^i} \, , \quad \text{with} \quad \partial_i\overline{B^i} =0 \, .
\end{equation}
The overline on a vector indicates it is divergenceless. The tensor perturbation can analogously be split into
\begin{equation}
    H_{ij}(\eta,x^i) = -2C\delta_{ij} + 2\partial _i \partial _j E + \partial _i \overline{E}_j + \partial _j\overline{E}_i + 2\overline{h}_{ij} \, ,
\end{equation}
where here the overline on the tensor indicates it is transverse and traceless (TT). All the quantities we have decomposed are three-dimensional, because we have sliced the FLRW spacetime into hypersurfaces of constant time, and then we study how these spatial slices evolve. This convenient approach is known as the (3+1) decomposition  \cite{Arnowitt:1962hi}. We can then track how the perturbations evolve in time on these spatial hypersurfaces. To summarise, our metric up to linear order takes the form
\begin{align}
    \begin{split}
        g_{00} &= -a^2(\eta) \big ( 1+\overbrace{2A}^{\delta g_{00}}\big ) \, , \\
        g_{0i} &=  a^2(\eta)\big  (  \partial _i B + \overline{B}_i \big ) =\delta g_{0i} \, ,\\
        g_{ij} &= a^2(\eta) \big ( \delta _{ij} -  \underbrace{2C\delta_{ij} + 2\partial _i \partial _j E + \partial _i \overline{E}_j + \partial _j\overline{E}_i + 2\overline{h}_{ij} }_{\delta g_{ij}}\big ) \, .
    \end{split}
\end{align}
There are ten degrees of freedom: four coming from the scalars ($A,\,B,\, C$ and $D$), four from the vectors ($\overline{B}_i$ and $\overline{E}_i$) and two from the tensor perturbation ($\overline{h}_{ij}$). We are yet to identify which are physical degrees of freedom. For completeness, we give the expressions for the inverse components of the metric tensor (such that $g_{\mu \alpha} g^{\nu \alpha}=\delta _\mu ^\nu$)
\begin{align}
    \begin{split}
        g^{00} &= -a^{-2}(\eta) \left ( 1-2A\right ) \, , \\
        g^{0i} &= g_{i0} = a^{-2}(\eta) \left (  \partial ^i B + \overline{B^i} \right ) \, ,\\
        g^{ij} &= a^{-2}(\eta) \left ( \delta ^{ij} +  2C\delta^{ij} - 2\partial ^i \partial ^j E - \partial ^i \overline{E}^j - \partial ^j\overline{E}^i - 2\overline{h}^{ij} \right ) \, .
    \end{split}
\end{align}

The energy-momentum tensor of a perfect fluid in Eq.~\eqref{perfectfluid} reads
\begin{equation}
    \delta T \munu = (\delta \rho + \delta P) u \back _\mu u \back _\nu +  \delta P g \back \munu  + (\rho + P) (u\back _\mu \delta u _\nu+ u\back _\nu \delta u _\mu) + P\delta g\munu + a^2\pi \munu \, ,
\end{equation}
at linear level\footnote{Again, every tensorial quantity has been expanded following Eq.~\eqref{tensorsplit}. We have
\begin{align}
    \begin{split}
        &\rho = \rho \back  + \delta \rho + \frac{1}{2}\rho \second + \frac{1}{6} \rho \third +...\, , \\
        &P = P\back  + \delta P + \frac{1}{2}P \second + \frac{1}{6} P \third +... \, , \\
        &u^\mu = u^{\mu \, (0)}  + \delta  u^\mu + \frac{1}{2} u^{\mu \, (2)} + \frac{1}{6} u^{\mu \, (3)} +... \, , \\
        &\text{etc...}
    \end{split}
\end{align}
We will drop the superscript for the background pressure and density as it is obvious when we are referring to them or perturbed quantities.}. In the above, we have introduced $\delta \rho$, the first-order density perturbation, $\delta P$, the first-order pressure perturbation, $\delta u _{\mu}$, the first-order velocity perturbation, $\pi \munu$, the anisotropic stress tensor and here $\delta g\munu$ is the linear perturbed metric. Using the timelike normalisation of the velocity, we have
\begin{align}
    \begin{split}
        \delta u^\mu &= \frac{1}{a(\eta)}(-A,v^i) \, , \\
        \delta u_\mu &= a(\eta)(-A,v_i + \partial_i B +\overline{B}_i) \, ,
    \end{split}
\end{align}
where $v^i$ is the linear velocity perturbation, and, like the vectorial metric perturbations, we decompose it as
\begin{equation}
    v_i=\partial _i v + \overline{v}_i \, .
\end{equation}
Finally, the anisotropic stress can also be decomposed into scalar, vector and tensor degrees of freedom. Its non-vanishing components read
\begin{equation}
    \pi _{ij} = \left ( \partial _i \partial _j-\frac{1}{3}\delta _{ij} \nabla ^2 \right ) \pi + \frac{1}{2} \left (\partial _i \overline{\pi}_j + \partial _j \overline{\pi}_i \right ) + \overline{\pi}_{ij} \, ,
\end{equation}
since we have the freedom to set $\pi_{00}=\pi_{0i}=0$ and to make it orthogonal to the velocity vector $\pi \munu u^\mu =0$ \cite{Peter:2013avv}. In what follows, we will turn off the anisotropic stress at all orders, only restoring it briefly to clarify certain points. Hence, up to leading order, the energy-momentum tensor reads
\begin{align}
    \begin{split}
        T_{00}&= a^2(\eta) \big (\rho + \overbrace{\delta \rho + 2\rho A }^{\delta T_{00}} \big ) \, , \\
        T_{0i}&=-a^2(\eta) \big (\rho (\partial_iv+\overline{v}_i+\partial_iB+\overline{B}_i) +P (\partial_iv+\overline{v}_i)\big )=\delta T_{0i} \, , \\
        T_{ij}&= a^2(\eta) \big ( P\delta _{ij} + \underbrace{\delta P\delta _{ij} + 2P (-2C\delta_{ij} + 2\partial _i \partial _j E + \partial _i \overline{E}_j + \partial _j\overline{E}_i + 2\overline{h}_{ij})+\pi _{ij}}_{\delta T_{ij}} \big ) \, , 
    \end{split}
\end{align}
and $\delta T^{\mu \nu}$ at linear order can be computed via $\delta T^{\mu \nu}=g^{\mu \alpha}g^{\nu \beta}T_{\alpha \beta}=g_{(0)}^{\mu \alpha}g_{(0)}^{\nu \beta}\delta T_{\alpha \beta}+\delta g^{\mu \alpha}g_{(0)}^{\nu \beta}T^{(0)}_{\alpha \beta}+ g_{(0)}^{\mu \alpha}\delta g^{\nu \beta}T^{(0)}_{\alpha \beta}$. There are again ten degrees of freedom: Four scalar ($\delta \rho , \,\delta P , \, v$ and $\pi$), four vectorial ($\overline{v}_i$ and $\overline{\pi}_i$) and two tensorial ($\overline{\pi}_{ij}$).

At the background level, we established a relationship between the pressure and density via the equation of state parameter $w$ in Eq.~\eqref{barotropicfluid}. Pressure in general is a function of both density and entropy; $P=P(\rho,S)$. A small perturbation of the pressure leads to
\begin{equation}
    \delta P = \left.\frac{\partial P}{\partial \rho} \right |_S \delta \rho + \left.\frac{\partial P}{\partial S} \right |_{\rho} \delta S \, ,
\end{equation}
where $\delta S$ is an entropy perturbation. If the latter vanishes, we arrive at
\begin{equation}
    \delta P = \left.\frac{\partial P}{\partial \rho} \right |_S \delta \rho = \cs \delta \rho \, ,
\end{equation}
at leading order. Pressure perturbations with vanishing entropy perturbations are called adiabatic perturbations, and these will be used throughout the remainder of this work.

We therefore have a total of ten degrees of freedom we need to solve for with the Einstein equations. As in the case of perturbing Minkowski spacetime, however, we need to identify which degrees of freedom are physical, i.e. gauge invariant. We will do exactly this in the next section by considering small coordinate transformations of the FLRW background and choosing a suitable gauge to carry out the rest of our work.

% % % % % % % % % % % % % % % % % % % % % % % % % % % % % % % %
\subsection{Coordinate dependence and the Newtonian Gauge}\label{sectiongauge}
The foundations of gauge-invariant cosmological perturbation theory can be found in Ref.~\cite{Bardeen:1980kt}
and for a comprehensive discussion on gauge invariance in cosmology, we refer the reader to Ref.~\cite{Malik:2012dr}. 

As in Sec.~\ref{sectionminkwoski}, where we perturbed around a Minkowski background, here we have perturbed the metric around an FLRW background. In fact, when we wrote down the perturbed metric in Eq.~\eqref{metriclinearflrw}, we already made a choice about the coordinates we are using, and we have assumed that this spacetime (which we refer to as the physical spacetime) is `close' to a FLRW spacetime (which here is the background spacetime) by making the perturbations small. To compare the two spacetimes, we need to introduce a map, generated by a vector field $\xi ^\mu$, between the `background manifold' and the `physical manifold'. $\xi ^\mu$ is called the gauge generator and defines a map between the background manifold and the perturbed manifold; this map is called a gauge. A tensor $\mathbf{T}$, once the gauge generator has been specified, transforms as \cite{Malik:2008yp,Christopherson:2009fp}
\begin{equation}
    \mathbf{T}  \rightarrow e^{\pounds _{\xi}} \mathbf{T} \, ,
\end{equation}
with 
\begin{equation}
    \xi ^\mu = \xi ^\mu _{(1)} + \frac{1}{2}\xi ^\mu _{(2)} + \frac{1}{6}\xi ^\mu _{(3)} + \mathcal{O}(\eps ^{(4)}) \, ,
\end{equation}
where at each perturbative order $(n)$, we can further SVT decompose $\xi ^\mu _{(n)}$ as $\xi ^\mu _{(n)} = \left  (T_{(n)},\partial ^i L _{(n)} + \overline{L}^i _{(n)}\right )$. Order by order, tensorial quantities transform as 
\begin{subequations}
    \begin{align}
        \mathbf{T}_0 & \rightarrow \mathbf{T}_0 \,,\\
        \mathbf{T}_1 & \rightarrow\mathbf{T}_1 + \pounds _{\xi _1} \mathbf{T}_0  \,, \label{tensorgauge1}\\
        \mathbf{T}_2 & \rightarrow\mathbf{T}_2 + \pounds _{\xi _2} \mathbf{T}_0 + \pounds _{\xi _1}^2 \mathbf{T}_0 + 2\pounds _{\xi _1} \mathbf{T}_1   \,, \label{tensorgauge2}\\
        \mathbf{T}_3 & \rightarrow \mathbf{T}_3 + \left ( \pounds _{\xi _3} + \pounds _{\xi _1}^3 + \frac{3}{2} \pounds _{\xi _1} \pounds _{\xi _2} + \frac{3}{2}\pounds _{\xi _2} \pounds _{\xi _1}  \right ) \mathbf{T}_0  + 3 \left ( \pounds _{\xi _1} ^2 + \pounds _{\xi _2} \right ) \mathbf{T}_1  + 3  \pounds _{\xi _1} \mathbf{T}_2 \label{tensorgauge3}  \,.
    \end{align}
\end{subequations}

At linear order, using Eq.~\eqref{tensorgauge1} and the definition of the Lie derivative in Eq.~\eqref{liederivative}, we summarise below how metric SVT decomposed linear-order perturbations transform
\begin{align*}
&A \rightarrow A + \hubble T + T\conf \, , \quad & B &\rightarrow B + L \conf - T \, , \quad & C &\rightarrow C - \hubble T \, , \\
&E \rightarrow E + L \, , \quad & \overline{B}^i &\rightarrow \overline{B}^i + \overline{L}^{i\prime} \, , \quad & \overline{E}^{i} &\rightarrow \overline{E}^{i} + \overline{L}^{i} \, , \\
&\overline{h}_{ij} \rightarrow \overline{h}_{ij} \, ,
\end{align*}
and the functions from the energy-momentum tensor
\begin{align*}
&\delta \rho \rightarrow \delta \rho + \rho \conf T \, , \quad & \delta P &\rightarrow \delta P+P\conf T \, , \quad & v &\rightarrow v - L \conf \, , \\
&\overline{v}_i \rightarrow \overline{v}_i - \overline{L}_i \conf\, , \quad & \pi _{ij} &\rightarrow \pi _{ij} \, .
\end{align*}
$\overline{h}_{ij}$ and $\pi _{ij}$ are automatically gauge invariant, i.e. they pick up no gauge artefacts ($T$ and $L^i$) when submitted to coordinate change\footnote{It is important to note, as we will show later, $\overline{h}_{ij}$ is gauge invariant at first-order, but this is not the case for $\overline{h}_{ij}^{(n)}$, when $n>1$.}. This result may be regarded as a corollary of the Stewart–Walker lemma \cite{Stewart:1974uz}. Specifically, from Eq.~\eqref{tensorgauge1} we see that if $\mathbf{T}_0$ vanishes, is constant, or consists of Kronecker deltas, then the corresponding linear perturbation is gauge invariant.

Knowing how the perturbations change under a shift in coordinates, we can equivalently: pick a suitable gauge from the onset so that the governing equations do not depend on gauge artefacts or work in an arbitrary gauge and cancel out the gauge artefacts by construction. For example, we can construct variables which do not pick up gauge artefacts when they transform, such as
\begin{subequations}
    \begin{align}
            &\Psi =  C - \hubble (B-E \conf) \, , \label{psibardeen} \\
            &\Phi = A + \hubble ( B-E\conf ) + (B-E\conf )\conf \, ,  \label{phibardeen}\\
            &\overline{\Phi} \, ^i =  \overline{E} \, ^{i\prime}- \overline{B} \, ^i \, .
    \end{align}
\end{subequations}
$\Psi$ and $\Phi$ are know as the Bardeen potentials \cite{Bardeen:1980kt}. Now the ten degrees of freedom we originally had have been reduced to six, removing all gauge artefacts; their physical interpretation is yet to be discussed. We can construct similar quantities for the matter perturbations
\begin{subequations}
    \begin{align}
        & \delta \rho ^N = \delta \rho + \rho \conf (B-E \conf) \, , \\
        & \delta P ^N = \delta P+ P \conf (B-E \conf) \, , \\
        & V = v+ E \conf \, , \\
        & \overline{V}_i = \overline{v}_i + \overline{B}_i\, , 
    \end{align}
\end{subequations}
where the superscript ``$N$'' stands for Newtonian, as we shall next see. 

Throughout this work, it will be convenient to work in the conformal Newtonian gauge, also referred to as the conformal longitudinal gauge. In this gauge, we have
\begin{equation}
    B=0 \, , \quad E=0 \, , \quad \text{and} \quad \overline{E}\, ^i=(0,0,0) \, ,
\end{equation}
so that the other metric perturbations reduce to the Bardeen potentials
\begin{equation}
    A=\Phi \quad \text{and} \quad C=\Psi \, ,
\end{equation}
as defined in Eqs.~\eqref{psibardeen} and \eqref{phibardeen}. The matter variables also become independent of the gauge artefacts. In this gauge, the linear equations of motion resemble the Newtonian equation of motion \cite{Bardeen:1980kt,Mukhanov:1990me}.

Having picked a gauge, we now move on to extracting and solving different components of the Einstein equations and the energy-momentum tensor. 
% % % % % % % % % % % % % % % % % % % % % % % % % % % % % % % % 
% % % % % % % % % % % % % % % % % % % % % % % % % % % % % % % % 
\section{First-order equations}\label{firstorderNEWTON}
In the conformal Newtonian gauge, the metric defined in Eq.~\eqref{metriclinearflrw} reduces to
\begin{equation}\label{metriclinearnewtonian}
    ds^2 = a^2(\eta) \left [ - (1+2\Phi) d\eta ^2 - 2\Phi _id\eta dx^i + (\delta _{ij} - 2\Psi \delta _{ij} + 2h_{ij}) dx^i dx^j  \right ] \, .
\end{equation}
We have dropped the overline indicating a vector is divergence-free and a tensor is TT to reduce clutter, and we will keep on doing so throughout the rest of this work. Below we list different components of the Einstein tensor, energy-momentum tensor and governing equations\footnote{All the linear Einstein and conservation equations in a general gauge can be found in Ref.~\cite{Peter:2013avv} and \cite{Malik:2008im}. However, both authors have slight different conventions and notation.}. We have
\begin{subequations}
    \begin{align}
        &G^{(1)}_{00}= - 6\hubble \Psi \conf + 2\nabla ^2\Psi \, , \label{G00newton1}\\
        &G^{(1)}_{0i}= 2\hubble \partial _i \Phi + 2 \partial _i \Psi \conf + \frac{1}{2} \nabla ^2 \Phi _i +\hubble ^2 \Phi _i +2\hubble \conf \Phi _i \, , \label{G0inewton1} \\
        \begin{split}
            &G^{(1)}_{ij}= \partial _i \partial _j \Psi - \partial _i \partial _j \Phi + \delta _{ij} \big [2\Psi ^{\prime \prime} + 4\hubble \Psi \conf -\nabla ^2 \Psi + \nabla ^2 \Phi + 2 (2\hubble \conf + \hubble ^2)(\Phi + \Psi)\\
            &+ 2 \hubble \Phi \conf \big ] + \frac{1}{2} (\partial _i \Psi ^{\prime } _j+ \partial _j \Psi _i ^{\prime }) + \hubble (\partial _i \Psi  _j+ \partial _j \Psi _i) +h ^{\prime \prime} _{ij} + 2\hubble h \conf _{ij} - \nabla ^2 h_{ij} \\
            & -2(2\hubble \conf + \hubble ^2) h_{ij} \label{Gijnewton1} \, , 
        \end{split} 
    \end{align}
\end{subequations}
and 
\begin{subequations}
    \begin{align}
        &T^{(1)}_{00}=  a^2 \left ( \delta \rho + 2\Phi \right ) \, , \\
        &T^{(1)}_{0i}= -a^2 (\rho + P) \left (\partial _i V + V_i \right ) +a^2\rho \Phi _i \, , \\
        &T^{(1)}_{ij}= a^2\left ( 2Ph_{ij} + \delta P \delta _{ij} -2P\delta_{ij}\Psi +  \left [ \left ( \partial _i \partial _j-\frac{1}{3}\delta _{ij} \nabla ^2 \right ) \pi + \frac{1}{2} \left (\partial _i \overline{\pi}_j + \partial _j \overline{\pi}_i \right ) + \overline{\pi}_{ij} \right ]  \right ) \, .
    \end{align}
\end{subequations}
We previously mentioned that first-order scalar, vector and tensor perturbations evolve independently, and we will show this now by extracting different parts of the Einstein equations.

First, we will extract the vectorial degrees of freedom, for which we find two equations coming from the Einstein equations and one from the energy-momentum conservation equation. It is useful to consider $G^{0 \, (1)}_{\, \, i}=\kappa T^{0 \, (1)} _{\, \, i}$ to arrive at an equation which relates the velocity perturbation to the metric perturbation. By subtracting the divergence-free part of the equation, we arrive at 
\begin{equation}
    \nabla ^2 \Phi _i = -2\kappa \rho a^2 (1+w)V_i \, .
\end{equation}
We can then extract a vector evolution equation from $G^{(1)}_{ij}=\kappa T^{(1)}_{ij}$ for the perturbation $\Phi _i$ by projecting out the vectorial degrees of freedom via an operator defined as
\begin{equation}\label{vectorextractorrealspace}
    V^{ij}_a  = \frac{1}{\nabla ^2} \left ( \delta ^i_a - \frac{\partial ^i \partial_a}{\nabla ^2} \right ) \partial ^j \, ,
\end{equation}
where the action of $\nabla ^{-2}$ on a general tensor $\mathbf{T}$ is defined such that $\nabla^{-2}(\nabla ^2\mathbf{T})=\mathbf{T}$. We can check that the operator works by considering its action on terms in Eq.~\eqref{Gijnewton1}, for example
\begin{equation}
    V^{ij}_a \left ( \partial _i \partial _j \Psi + \partial _i \Psi  _j+ \partial _j \Psi _i \right ) = \Psi _a \, .
\end{equation}
We got rid of the scalar degrees of freedom and kept only the vectorial ones. Applying the operator to the spatial part of the Einstein equations, i.e. $ V^{ij}_a G_{ij}^{(1)}=\kappa V^{ij}_a T_{ij}^{(1)}$, we get
\begin{equation}
    \Phi _i \conf + 2\hubble \Phi _i = \kappa a^2 \pi _i \, .
\end{equation}
Finally, the last equation comes from the space component of the energy-momentum conservation equation, $\nabla _\mu T^{\mu i} _{(1)}$, which leads to
\begin{equation}
    V_i \conf + (1-3\cs ) \hubble V_i = -\frac{1}{2}\frac{1}{\rho (1+w)} \nabla ^2 \pi _i \, .
\end{equation}
The vector part of the anisotropic stress $\pi _i$ thus sources $\Phi _i$ and the divergence sources velocity perturbations. In general, we will turn off the anisotropic stress, and we see that if $\pi _i = \mathbf{0}$, then there are no propagating vector modes and 
\begin{equation}
    \Phi _ i \propto a^{-2} \, , \quad V _ i \propto a^{-(1-3 \cs)} = a^{-(1-3 w)} \, ,
\end{equation}
hence, the vectors are diluted in an expanding universe. For this reason, we will not consider linear vector perturbations in the rest of our work.

Secondly, we can extract a tensor evolution equation by extracting via a TT operator the tensor degrees of freedom from the spatial part of the Einstein equations. From $ \Lambda ^{ij}_{ab} G_{ij}^{(1)}=\kappa \Lambda ^{ij}_{ab} T_{ij}^{(1)}$, where $\Lambda ^{ij}_{ab}$ is defined as \cite{Domenech:2019quo}
\begin{equation}\label{tensorextractorrealspace}
    \Lambda ^{ij}_{ab}  =  \left (\delta ^i_a - \frac{\partial ^i \partial _a}{\nabla ^2} \right )\left (\delta ^j_b - \frac{\partial ^j \partial _b}{\nabla ^2} \right ) -\frac{1}{2}\left (\delta _{ab} - \frac{\partial _a \partial _b}{\nabla ^2}\right )\left (\delta ^{ij} - \frac{\partial ^i \partial ^j}{\nabla ^2} \right )\,,
\end{equation}
we arrive at
\begin{equation} \label{hevolution}
    h_{ij} ^{\prime \prime} + 2 \hubble h_{ij} \conf - \nabla ^2 h_{ij} = \kappa a^2 \pi _{ij} \, .
\end{equation}
We will proceed to solve this equation in the next section.

Finally, from the various components of the Einstein equations, several combinations can be projected out, yielding scalar equations. We are interested in particular equalities that we will use in the following chapters. We can first extract the symmetric trace-free part of the spatial linear Einstein equations via the operator
\begin{equation}\label{scalarextractorrealspace}
    D^{ij} = \partial ^i \partial ^j -\frac{1}{3}\delta ^{ij}\nabla ^2 \, ,
\end{equation}
which gives rise to
\begin{equation}
    \Psi - \Phi = \pi \, .
\end{equation}
We note that in the case of vanishing anisotropic stress, we have
\begin{equation}\label{noanis}
    \Psi = \Phi \, ,
\end{equation}
which will greatly simplify equations when it is the case. The time-time component of the Einstein equations contains only scalar degrees of freedom, and it relates the density perturbation to metric fluctuations
\begin{equation} \label{density1}
    a^2\kappa \delta \rho = 2\nabla ^2 \Psi - 6\hubble ^2 \Phi - 6\hubble \Psi \conf \, .
\end{equation}
We can also take the divergence of the $0-i$ component of the Einstein equations, which leads to a relationship between the velocity perturbation and the metric fluctuations
\begin{equation}\label{velocity1}
   V =- \frac{2}{3(1+w)\hubble } \left ( \Phi + \frac{\Psi \conf}{\hubble}  \right ) \, .
\end{equation}
Finally, the trace of the spatial part of the Einstein equations leads to
\begin{equation}
    6\Psi ^{\prime \prime} + 6\hubble ( 2\Psi \conf + \Phi \conf ) + 2 (\nabla ^2 \Phi - \nabla ^2 \Psi ) -18w\hubble ^2 \Phi -3\kappa wa^2 \delta \rho =0 \, .
\end{equation}
In the absence of anisotropic stress we can use Eq.~\eqref{noanis} and we can substitute Eq.~\eqref{density1} to get a closed form equation for $\Psi$ (or $\Phi$)
\begin{equation} \label{psievolution}
    \Psi ^{\prime \prime} + 3(1+w)\hubble \Psi \conf -\cs \nabla ^2 \Psi =0 \, .
\end{equation}
We notice that we have written all the scalar perturbation equations as functions $\Psi$ and/or $\Phi$; these relations will be useful in the following chapters. Furthermore, once the solution of the above equation is known, we can then get evolution equations for $V$ and $\rho$. In the next section, we proceed to solve this equation and the tensor equation in Fourier space.

% % % % % % % % % % % % % % % % % % % % % % % % % % % % % % % % 
% % % % % % % % % % % % % % % % % % % % % % % % % % % % % % % % 
\section{Solutions to the linear scalar and tensor equations of motion}

Eqs.~\eqref{hevolution} and \eqref{psievolution} are both second-order, linear, partial differential equations, and we can find solutions by switching to Fourier space. Our conventions are set out in App.~\ref{appFourierspace}. The tensor equation becomes
\begin{equation}\label{hevolutionfourier}
     h_{\lambda}(\eta ,\mathbf{k}) ^{\prime \prime} + 2 \hubble h_{\lambda}\conf (\eta ,\mathbf{k})  +k ^2 h_{\lambda}(\eta ,\mathbf{k}) = 0 \, ,
\end{equation}
when we switch off the anisotropic stress. $\lambda$ is the polarisation of the GW wave. We note that when in Fourier space, the polarisations of the tensor modes decouple. The scalar equation reduces to
\begin{equation}\label{psievolutionfourier}
    \Psi ^{\prime \prime} (\eta ,\mathbf{k}) + 3(1+w)\hubble \Psi  \conf (\eta ,\mathbf{k})+\cs k ^2 \Psi (\eta ,\mathbf{k}) =0 \, .
\end{equation}
The advantage of going to Fourier space lies in the fact that now both equations have become second-order ordinary differential equations. We shall proceed to solve them for a range of $k$ values, which evolve independently. The magnitude of the comoving wavevector $k$ has units of inverse length and is related to the comoving wavelength of a perturbation as
\begin{equation}
    k=\frac{2\pi}{\lambda} \, .
\end{equation}
Perturbations with long wavelengths correspond to the $k\rightarrow 0$ limit (the so-called IR limit), and perturbations with short wavelengths correspond to the $k\rightarrow \infty$ limit (the so-called UV limit). In Sec.~\ref{inflation} we introduced the comoving Hubble radius, $(aH)^{-1}$, which in conformal coordinates is $\mathcal{H}^{-1}$. There are two different regimes with which we are concerned when analysing solutions to equations of motion in cosmology:
\begin{itemize}
    \item $\lambda > \hubble ^{-1}$ or $k<\hubble$, the wavelength is larger than the comoving Hubble radius and 
    \item $\lambda < \hubble ^{-1}$ or $k>\hubble$ the wavelength is smaller than the comoving Hubble radius.
\end{itemize}
As we shall see, on super-Hubble scales (scales larger than the Hubble radius), the perturbations remain effectively constant. We say they “freeze” because points separated by more than the comoving Hubble radius cannot communicate causally via gravity. This happens because the gradient terms in the evolution equations Eq.~\eqref{hevolutionfourier} and Eq.~\eqref{psievolution} become negligible compared to the Hubble friction (or damping) term in the perturbation equations. Conversely, on sub-Hubble scales (smaller than the Hubble radius), perturbations can oscillate and be damped due to gravitational effects and the pressure of the cosmic fluid. With this in mind, we can now proceed to solving the equations presented above.

% % % % % % % % % % % % % % % % % % % % % % % % % % % % % % % % 
\subsection{Tensor solution}

\begin{figure}
    \centering
    \includegraphics[width=0.6\linewidth]{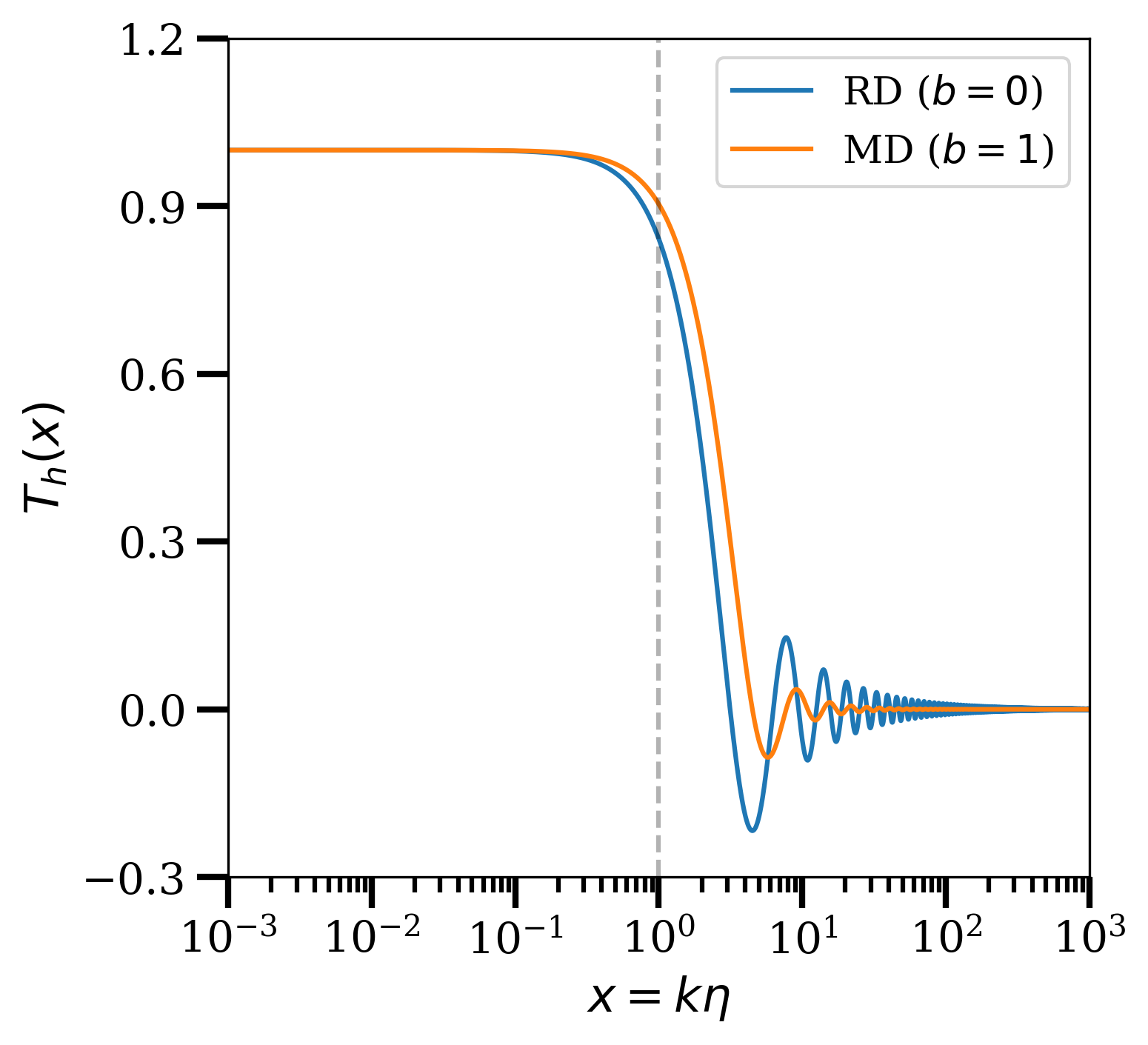}
    \caption{\footnotesize{Plot of the transfer function for tensor modes in Eq.~\eqref{tensortransfer} against $x=k\eta$, normalised by their initial value. We have plotted the behaviour of tensor modes in an RD universe (blue line), MD universe (orange line) and the $x=1$ line (dashed, faded black). Modes are constant and around $x\approx 1$ start to decay and oscillate.}}
    \label{fig:tensor_evolution}
\end{figure}

In this section, we proceed with solving the tensor equation Eq.~\eqref{hevolutionfourier}. By introducing the variable 
\begin{equation}
    x=k\eta \,,
\end{equation}
and the function $y(x)$ defined via $h(x)=x^{-\frac{1}{2}-b}y(x)$ (we drop the momentum and polarisation dependence for now to reduce clutter), the equation becomes
\begin{equation}
    \frac{d^2y}{dx^2} + \frac{1}{x}\frac{dy}{dx} + \left (1-\frac{\left (\frac{1}{2}+b \right )^2}{x^2} \right ) y(x) =0 \, ,
\end{equation}
where we recall $b$ was previously defined in Eq.~\eqref{defb}. We recognise the Bessel equation (see App.~\ref{appBessel}), and it has the solution 
\begin{equation}
    h_\lambda (\mathbf{k},\eta) = x^{-\frac{1}{2}-b} \left (A_{\mathbf{k}}^{\lambda } J_{\frac{1}{2}+b}(x) + B_{\mathbf{k}}^{\lambda } Y_{\frac{1}{2}+b}(x) \right ) \, ,
\end{equation}
where $J_n(z)$ and $Y_n(z)$ are Bessel functions of the first and second-kind respectively. As $x \to 0$, the Bessel function of the second kind exhibits a divergent behaviour, which allows us to set $B_{\mathbf{k}}^{\lambda} = 0$. Consequently, the general solution reduces to a constant. The final solution is then 
\begin{equation}
    h_\lambda (\mathbf{k},\eta) = h_{\mathbf{k}}^{\lambda } \times x^{-\frac{1}{2}-b}  J_{\frac{1}{2}+b}(x) \, ,
\end{equation}
where $h_{\mathbf{k}}^{\lambda }$ is the initial value of the perturbation, which we will come to shortly. The perturbation $ h_\lambda (\mathbf{k},\eta)$ has been split into two parts: an initial value $h_{\mathbf{k}}^{\lambda }$ which depends on the wave-vector $\mathbf{k}$ and the polarisation $\lambda$, and a part which describes how the mode decays as it re-enters the horizon, called the transfer function, which depends only the scalar momenta and conformal time. We can re-express this as
\begin{equation}\label{tensor1sol}
     h_\lambda (\mathbf{k},\eta) = h_{\mathbf{k}}^{\lambda } \times T_h(x) \, ,
\end{equation}
where 
\begin{equation}\label{tensortransfer}
    T_h(x)\equiv \frac{j_{b}(x)}{x^b}\, .
\end{equation}
When going from the Bessel function to the spherical Bessel function we have discarded an overall factor of $\sqrt{2/\pi}$ as this does not affect the solution to Eq.~\eqref{hevolutionfourier}. We have plotted the transfer functions in Fig.~\ref{fig:tensor_evolution}.
The tensor modes are constant on super-Hubble scales and, as they do not oscillate, we cannot yet interpret them as GWs. However, as they enter the horizon, they start to oscillate. This should be expected since if we look at the metric in Eq.~\eqref{metriclinearnewtonian} on small scales (i.e. sub-Hubble), we can ignore the scale factor, scalar and vector fluctuations \cite{Domenech:2023fuz}, and arrive at 
\begin{equation}\label{metriclinearnewtonian}
    ds^2 =  - d\eta ^2  + (\delta _{ij} + 2h_{ij}) dx^i dx^j   \, .
\end{equation}
The above metric is identical to the Minkowski metric in Eq.~\eqref{linearminkmetric}, where we showed that the spatial part of the metric satisfies the wave equation, i.e. tensor modes at linear order on sub-Hubble scales behave as freely propagating GWs. 
% % % % % % % % % % % % % % % % % % % % % % % % % % % % % % % % 
\subsection{Scalar solution}\label{sec:scalarsol}
We now proceed to solving the equation of motion for $\Psi (\eta,\mathbf{k})$ in Eq.~\eqref{psievolutionfourier}, and we recall $\cs =w$ when the equation of state is constant. Here we distinguish between a universe in MD and a general equation of state $w>0$. 

\begin{figure}
    \centering
    \includegraphics[width=0.6\linewidth]{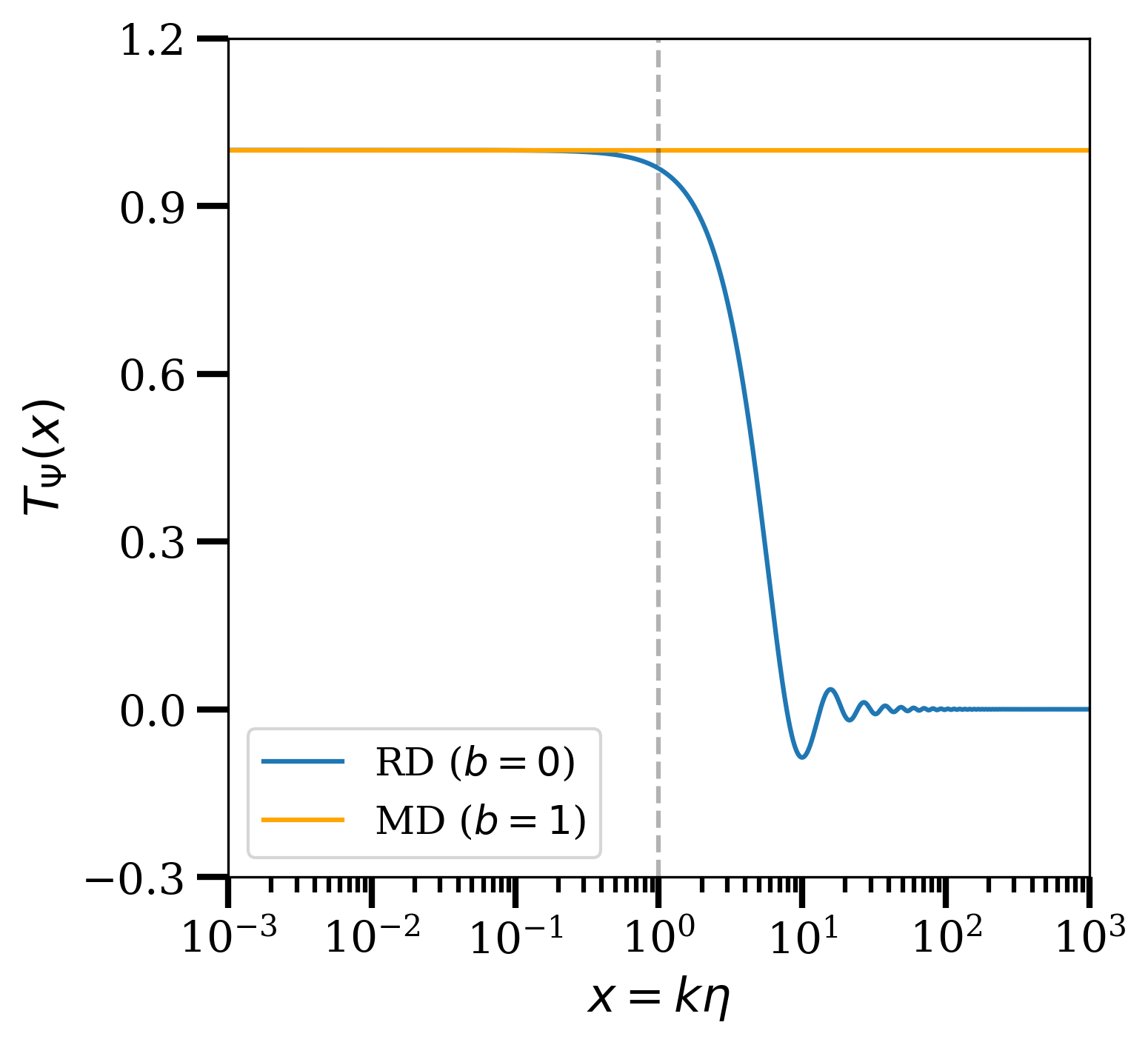}
    \caption{\footnotesize{Plot of the transfer functions for scalar modes against $x=k\eta$, normalised by their initial value. We have plotted the behaviour of scalar modes in an RD universe (blue line), MD universe (orange line) and the $x=1$ line (dashed, faded black). Modes are constant and around $x\approx 1$ start to decay and oscillate only in an RD universe.}}
    \label{fig:scalar_evolution}
\end{figure}

When in an MD universe, the equation reduces to
\begin{equation}
    \Psi ^{\prime \prime} (\eta , \mathbf{k}) + \frac{6}{\eta} \Psi ^{\prime }(\eta ,  \mathbf{k}) = 0 \, ,
\end{equation}
which has the solution 
\begin{equation}
     \Psi  (\eta ,  \mathbf{k}) = A_{\mathbf{k}} x^{-5} + B_{\mathbf{k}} \, .
\end{equation}
Dropping the decaying term (which blows up as $k \rightarrow0$) we conclude that the perturbation is constant during MD and fix it to its initial value
\begin{equation}
     \Psi  (\eta ,  \mathbf{k}) = \Psi_{\mathbf{k}} \, .
\end{equation}
When not in MD, we can define the variable $\tilde{x}=x\sqrt{w}$ and define $y(\tilde{x})$ via $\Psi(\tilde{x})=\tilde{x}^{-\frac{3}{2}-b}y(\tilde{x})$, then Eq.~\eqref{psievolutionfourier} reduces to 
\begin{equation}
    y^{\prime \prime} (\tilde{x}) + \frac{1}{\tilde{x}}y^{\prime } (\tilde{x}) + \left (1- \left (\frac{3+2b}{2} \right ) ^2 \frac{1}{\tilde{x}^2}\right ) y (\tilde{x}) =0 \, .
\end{equation}
Its solution is made up of Bessel functions,
\begin{equation}
    \Psi (\eta, \mathbf{k}) = A_{\mathbf{k}} \tilde{x}^{-\frac{3}{2}-b}J_{b+\frac{3}{2}}(\tilde{x}) + B_{\mathbf{k}} \tilde{x}^{-\frac{3}{2}-b}Y_{b+\frac{3}{2}}(\tilde{x}) \, ,
\end{equation}
and after setting the initial conditions by taking the limit $x \rightarrow 0$, 
\begin{equation}\label{psitransfer}
    \Psi (\eta, \mathbf{k})=\Psi_{\mathbf{k}}\frac{2^{2+b}\Gamma \left (\frac{5}{2} +b\right )}{w^{\frac{1+b}{2}}x^{1+b}\sqrt{\pi}}j_{b+1}(x\sqrt{w}) \equiv \Psi_{\mathbf{k}} T_{\Psi}(x)\, .
\end{equation}
In Fig.~\ref{fig:scalar_evolution} we have plotted the scalar transfer functions during MD and RD.

\section{Initial conditions}
In our discussion on inflation (Sec.~\ref{inflation}), we mentioned that scalar fluctuations generated during inflation serve as the initial conditions for all the scalar/density perturbations observed in our universe today. In this section, we will provide a bridge between $\Psi$ and scalar fluctuations produced during inflation.

We begin by defining a quantity, $\zeta$, the curvature perturbation in the uniform density gauge\footnote{The uniform density gauge is defined by setting density perturbations to vanish, i.e. $\delta \rho =0$.}\cite{Deruelle:1995kd,Martin:1997zd}
\begin{equation}
    \zeta = - C - \hubble  \frac{\delta \rho}{\rho \conf} \, , 
\end{equation}
and note it is gauge invariant. We can express it in terms of the Bardeen potential $\Psi$, using Eq.~\eqref{density1} and arrive at
\begin{equation}\label{zetadef}
    \zeta = - \Psi - \frac{2}{3(1+w)} \left ( \Psi + \frac{\Psi \conf}{\hubble} + \frac{1}{3} \left ( \frac{k}{\hubble} \right ) ^2 \Psi \right ) \, ,
\end{equation}
where we have switched to Fourier space. Taking a conformal time derivative 
\begin{equation}
    \zeta \conf = \frac{2}{3(1+w)}  \frac{1}{3} \left ( \frac{k}{\hubble} \right ) ^2 \left ( \Psi \hubble + \Psi \conf \right ) \approx 0 \, ,
\end{equation}
if and only if $k/\hubble ^2 \ll 1$ and $\left ( \Psi \hubble + \Psi \conf \right )$ does not diverge (we showed this was the case since we neglected the decaying mode in the transfer function Eq.~\eqref{psitransfer}). The quantity $\zeta$ we have introduced is then constant on super-Hubble scales for adiabatic initial conditions (vanishing entropy perturbations). We also notice that from the definition in Eq.~\eqref{zetadef}, and knowing $\Psi \conf \approx 0$ from Eq.~\eqref{psitransfer}; on super-Hubble scales we can establish a relationship between $\zeta$ and $\Psi$ 
\begin{equation}
    \zeta = - \left ( \frac{3+3w}{5+3w} \right ) \Psi \, .
\end{equation}
When solving the equation of motion for $\Psi$ in the previous section, we could have used the above relation when fixing the initial condition $\Psi_{\mathbf{k}}$. 

Similarly, we could have also introduced the gauge-invariant quantity, $\mathcal{R}$, which is the comoving curvature perturbation\footnote{In the comoving (orthogonal) gauge the velocity of the matter fluid is zero.} defined via \cite{Mukhanov:1990me}
\begin{equation}
    \mathcal{R}= C - \frac{\hubble \left (\hubble A + C \conf \right )}{\hubble \conf - \hubble ^2}  = \Psi + \frac{2}{3(1+w)} \left ( \Psi + \frac{\Psi \conf}{\hubble} \right )\, ,
\end{equation}
and notice on super-Hubble scales we have $\mathcal{R}=-\zeta$, which means that we can use both these variables interchangeably\footnote{Throughout the literature some researchers use $\mathcal{R}$ while they mean $\zeta$ and vice versa. In the rest of this work, we will use the definitions we have presented above, but the reader should be aware that this can change from reference to reference, including the authors! This is why we gave the definitions of both $\zeta$ and $\mathcal{R}$.}. To summarise, the initial conditions of the scalar perturbations are encoded in gauge invariant variables $\zeta _{\mathbf{k}}$ and $\mathcal{R} _{\mathbf{k}}$, which are constant on super-Hubble scales and therefore allow us to relate them to the variable $\Psi$ at later times. 

There are two main advantages of using the variables introduced and not $\Psi$. Indeed, on super-Hubble scales, they are always constant even when the equation of state parameter is not, which is not the case for $\Psi$. On super-Hubble scales, we would have to track how $\Psi$ changes as the universe goes through successive periods of different dominating species. From now on we will use $\mathcal{R}$, and therefore we re-write Eq.~\eqref{psitransfer} as
\begin{equation}\label{curlyR}
    \Psi (\eta , \mathbf{k}) = \left ( \frac{3+3w}{5+3w} \right ) T_{\Psi}(x) \mathcal{R} _{\mathbf{k}} \, .
\end{equation}
Furthermore, $\mathcal{R}$ has the advantage of being directly proportional to the perturbed inflaton when in the flat-slicing gauge\footnote{The flat-slicing gauge is defined by setting the scalar perturbations $C=B=0$.}. This is a convenient gauge to use during inflation. If $\delta \varphi$ is the scalar perturbation of the inflaton, then we have \cite{Peter:2013avv} 
\begin{equation}
    \mathcal{R} = C-\hubble \frac{\delta \varphi}{\varphi \prime} = -\hubble \frac{\delta \varphi}{\varphi \prime} \, ,
\end{equation}
where the second equality holds when in the flat-slicing gauge. Therefore, the variable $\mathcal{R}$ gives us a direct link between scalar perturbations produced during inflation and the scalar perturbations which evolve later in the universe.

Inflation predicts different values of $\mathcal{R}$, which will depend on the theory of inflation one uses. As we shall see in the next chapter, certain theories predict large values of the comoving curvature perturbation, which will be of interest to us. The same is true for $h_{\mathbf{k}}^{\lambda}$, where we did not have to introduce an auxiliary field since the tensor mode is constant on super-horizon scales regardless of the dominating background species.

% % % % % % % % % % % % % % % % % % % % % % % % % % % % % % % % 
% % % % % % % % % % % % % % % % % % % % % % % % % % % % % % % % 
\section{The power spectrum and the spectral density}

In the previous section, we mentioned that inflation predicts values of $\mathcal{R}_{\mathbf{k}}$ and $h_{\mathbf{k}}^{\lambda}$. Rather than track individual perturbations, in cosmology, we track the statistics of these perturbations. In fact, since these perturbations are originally produced by quantum fluctuations, when they become super-Hubble, they can be replaced by stochastic fields with Gaussian statistics \cite{Polarski:1995jg}. In Fourier space, two-point correlations of these fields are called the power spectrum, an object we will define shortly. If perturbations obey a Gaussian distribution, then the statistics of these perturbations are completely defined by the power spectrum, and we do not require higher-order statistics. The timeline of an individual perturbation goes as follows: quantum fluctuations are produced during inflation (either matter or metric perturbations) and are stretched beyond the cosmological horizon by virtue of the comoving Hubble radius shrinking; they become frozen, and then when the comoving Hubble radius starts to grow again during RD, it encompasses perturbations that start to evolve under the effects of gravity. We track how these perturbations evolve via the solutions to their equations of motion and make predictions on their statistics at later times. 

%%%%%%%%%%%%%%%%%%%%%%%%%%%%%%%%%%%%%%%%%%%%%%%%%%%%%%%%%%%%
\subsection{Power spectrum}
We begin by defining two-point correlation function, $\xi (\mathbf{x},\mathbf{r})$, as 
\begin{equation}
    \xi (\mathbf{x},\mathbf{r}) = \langle f(\mathbf{x})f(\mathbf{x}+\mathbf{r}) \rangle \, , 
\end{equation}
where $f(\mathbf{x})$ is a classical random field at location $\mathbf{x}$ and $f(\mathbf{x+r})$ is the same field evaluated at $\mathbf{x+r}$. The two-point correlation function is then the expectation value of the product of two fields. This can be extended to higher-order statistics; the N-point correlation function then represents the expectation value of the product of N fields. Homogeneity and isotropy, or rather, statistical homogeneity and isotropy, implies that the translated and rotated function $\zeta(\mathbf{x},\mathbf{r})$ remains the same, and hence the two-point function should only depend on the distance $r=|\mathbf{r}|$. This implies
\begin{equation}
    \xi (\mathbf{x},\mathbf{r}) = \xi (r) \, .
\end{equation}
We can express the field as a Fourier integral
\begin{equation}
    f(\mathbf{x}) = \int \frac{d^3\mathbf{k}}{(2\pi)^{\frac{3}{2}}} f_{\mathbf{k}} e^{i\mathbf{k}\cdot \mathbf{x}} \, .
\end{equation}
By considering the two-point function in Fourier space
\begin{equation}
    \langle f_{\mathbf{k}}f_{\mathbf{k}\conf} \rangle = \int \frac{d^3\mathbf{x}}{(2\pi)^{\frac{3}{2}}} \int \frac{d^3\mathbf{x}^{\prime}}{(2\pi)^{\frac{3}{2}}} \langle f(\mathbf{x}) f(\mathbf{x}^{\prime})\rangle e^{-i\mathbf{k}\cdot \mathbf{x}} e^{-i\mathbf{k}\cdot \mathbf{x}\conf} \, ,
\end{equation}
and using the fact that the integral is invariant under the shift $\mathbf{x}\conf= \mathbf{x} + \mathbf{r}$, we arrive at 
\begin{equation}
    \langle f_{\mathbf{k}}f_{\mathbf{k}\conf} \rangle = \delta (\mathbf{k}+\mathbf{k}\conf) P_{f}(k) \, ,
\end{equation}
where we have defined the power spectrum as
\begin{equation}
    P_{f}(k) = \int d^3 \mathbf{r} \, \xi (r) e^{-i\mathbf{k}\cdot \mathbf{r}}\, .
\end{equation}
Or, taking the Fourier inverse, we have
\begin{align}
    \begin{split}
        \xi (r) &= \int \frac{\rm ^3 \mathbf{k}}{(2\pi)^3} \, P_{f}(k) e^{-i\mathbf{k}\cdot \mathbf{r}} = \int _0 ^\infty \rm   k \,  \frac{k^2}{2\pi ^2}  P_{f}(k) \frac{\sin (kr)}{kr}= \int \rm \log  k \,  \frac{k^3}{2\pi ^2} P_{f}(k) \frac{\sin (kr)}{kr} \\
        &= \int \rm \log  k \, \mathcal{P}_{f}(k) \frac{\sin (kr)}{kr}\, ,
    \end{split}
\end{align}
where we have defined the dimensionless power spectrum in Fourier space   
\begin{equation}
    \langle f_{\mathbf{k}}f_{\mathbf{k}\conf} \rangle = \delta (\mathbf{k}+\mathbf{k}\conf) \frac{2\pi ^2}{k^3} \mathcal{P}_{f}(k) \, .
\end{equation}
If $f(x)=\mathcal{R}(x)$ then we have 
\begin{equation}\label{defpsR}
    \langle \mathcal{R}_{\mathbf{k}}\mathcal{R}_{\mathbf{k}\conf} \rangle = \delta (\mathbf{k}+\mathbf{k}\conf) \frac{k^3}{2\pi ^2} \mathcal{P}_{\mathcal{R}}(\eta,k) \, ,
\end{equation}
and if $f(x)=h_{ij}(x)$ then we have
\begin{equation}\label{defpsh}
    \langle h^{\lambda }_{\mathbf{k}}h^{\lambda \conf }_{\mathbf{k}\conf} \rangle = \delta (\mathbf{k}+\mathbf{k}\conf) \delta ^{\lambda \lambda \conf} \frac{k^3}{2\pi ^2} \mathcal{P}_{h,\lambda}(\eta,k) \, ,
\end{equation}
where $\delta ^{\lambda \lambda \conf}$ indicates different polarisations do not correlate.

Finally, if $\mathcal{R}_{\mathbf{k}}$ and $h^{\lambda }_{\mathbf{k}}$ are drawn from a Gaussian distribution, then we can break down any N-point function into products of two-point functions using Wick's theorem\footnote{We refer the reader to \href{https://nms.kcl.ac.uk/eugene.lim/AdvCos/lecture2.pdf}{lecture notes} by Eugene A. Lim for a detailed explanation. Last accessed by the author on 15/02/2023.}\cite{Durrer:2008eom}. Since the N-point odd correlator of Gaussian perturbations vanishes, if we consider, for example, $N=4$, we have 
\begin{equation}\label{4pointdefscalar}
    \langle \mathcal{R}_{\mathbf{k}_1} \mathcal{R}_{\mathbf{k}_2} \mathcal{R}_{\mathbf{k}_3}\mathcal{R}_{\mathbf{k}_4}\rangle = \langle \mathcal{R}_{\mathbf{k}_1} \mathcal{R}_{\mathbf{k}_2} \rangle \langle \mathcal{R}_{\mathbf{k}_3}\mathcal{R}_{\mathbf{k}_4}\rangle + \langle \mathcal{R}_{\mathbf{k}_1} \mathcal{R}_{\mathbf{k}_3} \rangle \langle \mathcal{R}_{\mathbf{k}_2}\mathcal{R}_{\mathbf{k}_4}\rangle + \langle \mathcal{R}_{\mathbf{k}_1} \mathcal{R}_{\mathbf{k}_4} \rangle \langle \mathcal{R}_{\mathbf{k}_2}\mathcal{R}_{\mathbf{k}_3}\rangle \, .
\end{equation}

%%%%%%%%%%%%%%%%%%%%%%%%%%%%%%%%%%%%%%%%%%%%%%%%%%%%%%%%%%%%
\subsection{The spectral density}\label{sec:spectraldensity}

In our discussion on GWs propagating in Minkowski spacetime, we concluded with Eq.~\eqref{riemannandh}, which shows that GWs can curve spacetime. An interpretation of this is that GWs can be treated analogously to energy, in the sense that they curve spacetime, and so in theory we should be able to construct an energy-momentum tensor which represents how GWs curve spacetime. The difficulty of this in GR is that there is no notion of localised energy, and consequently of energy in the form of gravitational radiation. Indeed, Einstein's equivalence principle states that we can always go to a local inertial frame where gravitational radiation fields vanish (see Ref.~\cite{Misner:1973prb} Sec. 20.4 for a discussion on this). As such, we cannot localise the energy carried out by a GW and so instead we average over some macroscopic region containing multiple wavelengths. In this section, we will first look at the definition of the energy-momentum tensor for GWs in Minkowski spacetime, and then we will argue that this also holds in an FLRW background on short scales. Finally, we will introduce the spectral density $\Omega _{GW}(\eta,k)$, which characterises the SGWB in cosmology. 

Because GWs are oscillatory, their linear contribution to the Einstein tensor averages to zero over several wavelengths; only quadratic terms, involving products of the perturbation and its derivatives, will survive the averaging and define a non-vanishing effective energy–momentum tensor. We therefore need to go to second order in perturbation theory over our Minkowski background. The energy-momentum tensor for GWs in Minkowski spacetime reads \cite{Misner:1973prb}
\begin{equation}\label{gwtensorminkowski}
    t\munu = \frac{1}{4\kappa} \left \langle \partial _\mu h_{ij } \partial _\nu h^{ij }  \right \rangle \, ,
\end{equation}
where we recall that the tensor perturbations are TT and the brackets represent an average over several wavelengths. The energy density of GWs is then 
\begin{equation} \label{rhomink}
    \rho_{GW}=t_{00}=\frac{1}{4\kappa} \left \langle  \dot{h}_{ij }  \dot{h}^{ij}  \right \rangle \, .
\end{equation}

We previously argued that tensor modes in cosmology can only be interpreted as gravitational waves on small (sub-Hubble) scales, since on large (super-Hubble) scales they remain approximately constant. Moreover, as shown in Eq.~\eqref{metriclinearnewtonian}, on small scales the perturbed FLRW metric in Newtonian gauge closely resembles Minkowski spacetime, which is a reasonable approximation at detector scales. When deriving the energy-momentum tensor for GWs on a curved background like FLRW, one employs a procedure to separate long-wavelength (absorbed into the background) and short-wavelength modes, known as the Isaacson prescription \cite{Isaacson:1968hbi,Isaacson:1968zza}. This approach requires the curvature scale to be much larger than the wavelength of the gravitational waves, motivating the focus on short-wavelength perturbations. For these reasons, we can write the energy density in a FLRW spacetime analogously as
\begin{equation}\label{gwtensorminkowski}
    \rho _{GW} = \frac{M_{Pl}^2}{a^{2}(\eta)} \left \langle  h_{ij} \conf  h^{\prime ij}   \right \rangle  \approx \frac{M_{Pl}^2}{a^{2}(\eta)} \left \langle  \partial _k h_{ij} \partial^k  h^{ij}   \right \rangle\, ,
\end{equation}
where in the second equality we used the approximation for a freely propagating wave\footnote{The factor of 4 in Eq.~\eqref{rhomink} disappeared to match with our conventions, see the metric in Eq.~\eqref{metriclinearnewtonian}.}. In Fourier space, the GW energy density can be related to the time-averaged power spectrum such that
\begin{equation}\label{densityGWs}
    \rho _{GW}(\eta , k) = \frac{M_{Pl}^2}{4} \frac{k^2}{a^2(\eta)} \sum _{\lambda} \overline{\mathcal{P}^{\lambda}_{h}(\eta , k)} \, .
\end{equation}
where we recall $\lambda$ is the polarisation of the GW, the overline denotes a time average, and the superscript $(11)$ indicates that the power spectrum of two linear-order tensor modes.

Finally, although the power spectrum is well defined for tensor modes, the central object to characterise the SGWB is the spectral density defined as \cite{Moore:2014lga}
\begin{equation}\label{omegalinear}
    \Omega _{GW}(\eta,k)=\frac{\rho _{GW}}{\rho _{crit}} = \frac{1}{3}\frac{k^2}{\mathcal{H}^2 (\eta)} \sum _{\lambda}\overline{\mathcal{P}^{\lambda}_{h}(\eta , k)}= \frac{1}{3}\frac{k^2}{\mathcal{H}^2 (\eta)}\overline{\mathcal{P}_{h}(\eta , k)}\, .
\end{equation}
In later sections, when we go beyond the linear approximation for gravitational waves, the spectral density will be the primary observable of interest. We close this section by noting that the derivation presented here applies specifically to first-order tensor modes.

% % % % % % % % % % % % % % % % % % % % % % % % % % % % % % % % 
% % % % % % % % % % % % % % % % % % % % % % % % % % % % % % % % 
\section{Beyond linear order: second order equations of motion}

In the following chapters, we will examine the evolution of second-order tensor perturbations. Hence, in this section, we will set up the framework to do such calculations. One of the main differences with linear theory is that at higher orders in perturbation theory, scalar, vector and tensor degrees of freedom no longer evolve independently. We will show this by extracting different parts of the second-order spatial Einstein equations. The origins of second-order perturbation theory can be found in Ref.~\cite{Tomita:1967wkp} and was later rediscovered in Refs.~\cite{Matarrese:1992rp,Matarrese:1993zf,Matarrese:1997ay}. 

Following the same expansion of the metric we had at first-order in Eq.~\eqref{metriclinearflrw}, we can expand the metric to second-order as \cite{Malik:2008im} 
\begin{align}
    \begin{split}
        ds^2 &= a^2(\eta) \big [ - (1+2A^{(1)} + A^{(2)}) d\eta ^2 + 2 \left ( B_i^{(1)}+\frac{1}{2}B_i^{(2)}\right )dx^id\eta + (\delta _{ij}  \\
        &+ H_{ij}^{(1)}+ \frac{1}{2}H_{ij}^{(2)}) dx^i dx^j  \big ] \, ,
    \end{split}
\end{align}
and then further split the perturbations into scalar, vector and tensor degrees of freedom. In the conformal Newtonian-Gauge
\begin{subequations}\label{metricdefsecondorder}
    \begin{align}
        g_{00} &= -a^2(\eta)(1+2\Phi ^{(1)} + \Phi ^{(2)}) \, , \\
        g_{0i} &= a^2(\eta)\frac{1}{2}B_i^{(2)}=g_{i0}\, , \\
        g_{ij} &= a^2(\eta) \left ( (1-2\Psi ^{(1)} - \Psi ^{(2)})\delta _{ij}  +2h^{(1)}_{ij}+h^{(2)}_{ij} \right )\, ,
    \end{align}
\end{subequations}
where we have disregarded first-order vector perturbations since we showed in Sec.~\ref{firstorderNEWTON} that they are diluted away by the expansion of the universe. Like at background and linear order, we have to specify a relationship between pressure and density. In the Newtonian gauge, it can be related to familiar quantities \cite{Carrilho:2015cma}
\begin{equation}
    \delta P^{(2)} = c_s^2 \delta \rho ^{(2)} + \frac{2}{\rho ^{\prime}} \delta \rho \delta P^{\prime} - \left (\frac{P^{\prime \prime}}{\rho ^{\prime ^2}} - \frac{P^{\prime}\rho ^{\prime \prime}}{\rho ^{\prime 3}} \right )\delta \rho ^2 - \frac{2c_s^2}{\rho ^{\prime}} \delta \rho \delta \rho ^{\prime} \, .
\end{equation}
In the case of a constant equation of state then
\begin{equation}
    \delta P^{(2)} = c_s^2 \delta \rho ^{(2)} = w  \delta \rho ^{(2)} \, .
\end{equation}

We also saw at linear order, that we can extract evolution equations for $\Psi(\eta,k)$, $\Phi(\eta,k)$, $B_i(\eta,k)=\Psi _i(\eta ,k)$ and $h_{ij}(\eta , k)$ by projecting out different parts of the spatial Einstein equations via the scalar, vector and tensor operators defined in Eq.~\eqref{scalarextractorrealspace}, Eq~\eqref{vectorextractorrealspace} and Eq.~\eqref{tensorextractorrealspace} respectively. We can then relate them to other perturbations by using different components of the Einstein and conservation equations. 

Since we are ultimately interested in tensor perturbations, we provide the spatial part of the Einstein equations. The spatial part of the Einstein tensor is
\begin{equation}\label{gij2}
    \begin{split}
        G_{ij}^{(2)} &= h_{ij}^{\prime \prime (2)} + 2 \mathcal{H}h_{ij}^{\prime (2)} - h_{ij}^{(2)} (2\mathcal{H}^2 + 4\mathcal{H}^{\prime}) -\nabla ^2 h_{ij}^{(2)} + \delta _{ij} (\Phi ^{(2)} + \Psi ^{(2)})(2\mathcal{H}^2 + 4\mathcal{H}^{\prime}) \\
        &+2\delta _{ij} \Phi ^{\prime (2)}\mathcal{H} +4 \delta _{ij} \Psi ^{\prime (2)}\mathcal{H} + 2 \delta _{ij} \Psi ^{\prime \prime (2)} + \delta _{ij} \nabla ^2 \Phi ^{(2)} - \delta _{ij} \nabla ^2 \Psi ^{(2)}  \\
        &- \partial _i \partial _j \Phi ^{(2)} + \partial _i \partial _j \Psi ^{(2)} - 4 h_i^{m\prime}h_{jm}^{\prime} +  \delta _{ij} \left ( 3h_{mk}^{\prime} h^{mk \prime} + 4 h^{mk}h_{mk}^{\prime \prime} +8 \mathcal{H} h^{mk}h_{mk}^{\prime}  \right ) \\
        &- \Phi \left (4 h_{ij}^{\prime \prime} + 8 \mathcal{H}h_{ij}^{\prime} - 8\mathcal{H}^2 h_{ij} - 16\mathcal{H}^{\prime} h_{ij} \right ) - \delta _{ij} \left (8\mathcal{H}^2 + 16 \mathcal{H}^{\prime} \right ) \Phi ^{(2)} - 16 \delta _{ij} \mathcal{H} \Phi \Phi ^{\prime} \\
        &+\Phi ^{\prime} \left (8 \mathcal{H}h_{ij} - 2h_{ij}^{\prime} \right ) + \delta _{ij} \left ( 16 \mathcal{H}^{\prime} - 8\mathcal{H}^2 \right )\Phi \Psi - 8 \delta _{ij} \mathcal{H} \Psi \Phi ^{\prime} + 2 h_{ij}^{\prime} \Psi ^{\prime} + 24 \mathcal{H} h_{ij} \Psi ^{\prime} \\
        &-16 \delta _{ij} \mathcal{H} \Phi \Psi ^{\prime} - 4 \delta _{ij} \Phi ^{\prime} \Psi ^{\prime} + 2 \delta _{ij} \Psi ^{\prime 2} + 12h_{ij} \Psi ^{\prime \prime} - 8 \delta _{ij} \Phi \Psi ^{\prime \prime} - 4 \Psi \nabla ^2 h_{ij} + 4 h_i^m \partial _m \partial _j \Psi \\
        &+ \left ( 4 h_{ij} - 4\delta _{ij} \Phi \right )\nabla ^2 \Phi - \left ( 8 h_{ij} + 4\delta _{ij} \Psi \right ) \nabla ^2 \Psi + 4 h_j^m \partial _m \partial _i \Psi - 2 \partial _k h_{ij} \partial ^k \Phi \\
        & - 2 \delta _{ij} \partial _k \Phi \partial ^k \Phi - 6 \partial _k h_{ij} \partial ^k \Psi - 4 \delta _{ij} \partial _k \Psi \partial ^k \Psi + 4 h^{mk} \partial _m \partial _k h_{ij} - 4 h^{mk} \delta _{ij} \partial _m \partial _k \Phi \\
        & - 4 \partial _m h_{jk} \partial ^k h_i^m + 4 \partial _k h_{jm} \partial ^ k h_i^m - 4h^{mk}\delta _{ij} \nabla ^2 h_{mk} + 2 \delta _{ij} \partial _k h_{md} \partial ^d h^{mk} \\ 
        & - 3\delta _{ij} \partial _d h_{mk} \partial ^d h^{mk} + 2 \partial ^m \Phi \partial _i h_{jm} + 2 \partial ^m \Psi \partial _i h_{jm} - 4 h^{mk} \partial _i \partial _k h_{jm} + 2 \partial _i h^{mk} \partial _j h_{mk} \\
        &+2 \partial ^m \Psi \partial _j h_{im} +2 \partial ^m \Phi \partial _j h_{im} + 2\partial _i \Phi \partial _j \Phi - 2\partial _i \Psi \partial _j \Phi - 2\partial _i \Phi \partial _j \Psi +6 \partial _i \Psi \partial _j \Psi \\ 
        &-4h^{mk} \partial _j \partial _k h_{im} + 4 h^{mk} \partial _j \partial _i h_{mk} + 4\Phi \partial _j \partial _i \Phi + 4 \Psi \partial _j \partial _i \Psi \, ,
    \end{split}
\end{equation}
whilst the energy-momentum tensor is given by
\begin{eqnarray}\label{tij2}
    T_{ij}^{(2)} &=& a^2 \bigg ( \delta _{ij} \left [ \frac{1}{2}P^{(2)} -2c_s^2\delta \rho \Psi -w\rho \Psi ^{(2)} \right]  + 2c_s^2\delta \rho  h_{ij} +(1+w)\rho \partial _i V \partial _j V \nonumber \\
    &+&  w\rho h_{ij}^{(2)}\bigg )\, .
\end{eqnarray}
We note here that we have not simplified the terms using background and linear-order equations of motion; these are the `raw' Einstein equations in the Newtonian gauge and after SVT decomposition. We proceed to project out different parts of interest. 

% % % % % % % % % % % % % % % % % % % % % % % % % % % % % % % % 
\subsection{Second order induced scalars} \label{sec:inducedscalars}
To find the scalar equation of motion for both $\Psi ^{(2)} (\eta,k)$ and $\Phi ^{(2)} (\eta,k)$ we can take the trace of $G_{ij}^{(2)}=\kappa T_{ij}^{(2)}$ and extract the symmetric trace-free part via the operator defined in Eq.~\eqref{scalarextractorrealspace}. Whilst this operator was straightforward to use at linear order, we find that at second-order in perturbation theory it is useful to go directly to Fourier space to circumvent the chain rule. Indeed, acting with the real space operator in Eq.~\eqref{scalarextractorrealspace} on a term like $\delta _{ij} \nabla^2 \Psi ^{(2)} \subset G_{ij}^{(2)}$ in Eq.~\eqref{gij2} is trivial, however, a term like $-16\delta _{ij}\mathcal{H}\Phi \Psi \conf$ would give a multitude of terms due to the chain rule. We find it is easier to define the projector, $D^{ij}$, defined via 
\begin{equation}\label{scalarextractor}
    D^{ij}S_{ij}(\mathbf{\eta ,x^{\prime}}) = \int \frac{d ^3 \mathbf{k^{\prime}}}{(2\pi)^{\frac{3}{2}}} e^{i\mathbf{k^{\prime}}\cdot \mathbf{x}} \int \frac{d ^3 \mathbf{x^{\prime}}}{(2\pi)^{\frac{3}{2}}} \left (-k^{\prime i}k^{\prime j}+\frac{1}{3}k^{\prime 2}\right ) e^{-i\mathbf{k^{\prime}}\cdot \mathbf{x^{\prime}}} S_{ij}(\mathbf{\eta ,x^{\prime}}) \, ,
\end{equation}
where $S_{ij}(\mathbf{\eta ,x^{\prime}})$ is a term in the spatial part of the Einstein equations expressed as a Fourier integral. Applying the above operator to the spatial part of the Einstein equations, the surviving terms are 
\begin{equation}\label{eqanisrealspace}
    \nabla ^2 \nabla ^2 \Phi ^{(2)} - \nabla ^2 \nabla ^2 \Psi ^{(2)} = -3 D^{ij} \left ( \Pi _{ij}^{(2),ss}  +  \Pi _{ij}^{(2),st} +  \Pi _{ij}^{(2),tt} \right ) \, ,
\end{equation}
where we have defined
\begin{subequations}
    \begin{align}
        \begin{split}
            \Pi^{(2),ss}_{ij}  &=  \kappa a^2 \rho (1+w) \partial _i V \partial_j V + \partial _i \Psi \partial _j \Phi -3 \partial _i \Psi \partial _j \Psi - \partial _i \Phi \partial _j \Phi +\partial _i \Phi\partial _j \Psi \\
            &-2\Phi\partial _i  \partial _j \Phi -2\Psi\partial _i  \partial _j \Psi \, , 
        \end{split}
        \\ 
        \begin{split}
            \Pi^{(2),st}_{ij}  &= 2\cs a^2 \kappa h_{ij} \delta \rho + h_{ij}^{\prime\prime} \Phi + 4 h_{ij}^\prime \mathcal{H} \Phi - 4 h_{ij} \mathcal{H}^2 \Phi - 8 h_{ij} \mathcal{H}^\prime \Phi + h_{ij}^\prime \Phi^\prime - 4 h_{ij} \mathcal{H} \Phi^\prime\\
            &- h_{ij}^\prime \Psi^\prime - 12 h_{ij} \mathcal{H} \Psi^\prime - 6 h_{ij} \Psi^{\prime\prime} + 2 \Psi \partial_m \partial^m h_{ij} - 2 h_{ij} \partial_m \partial^m \Phi + 4 h_{ij} \partial_m \partial^m \Psi \\
            &- 2 h_j^m \partial_m \partial_i \Psi - 2 h_i^m \partial_m \partial_j \Psi + \partial_m h_{ij} \partial^m \Phi + 3 \partial_m h_{ij} \partial^m \Psi - \partial^m \Phi \partial_i h_{jm}\\
            &- \partial^m \Psi \partial_i h_{jm} - \partial^m \Phi \partial_j h_{im} - \partial^m \Psi \partial_j h_{im} \, , 
        \end{split}
        \\
        \begin{split}
            \Pi^{(2),tt}_{ij}  &= 2 h_i^{m\prime} h_{j m}^{\prime} - 2 h^{m n} \partial_n \partial_m h_{ij} + 2 \partial_m h_j^n \partial_n h_i^m - 2 \partial_n h_j^m \partial_n h_i^m + 2 h^{m n} \partial_i \partial_n h_{j m}\\
            &- \partial_i h^{m n} \partial_j h_{m n} + 2 h^{m n} \partial_j \partial_n h_{i m} - 2 h^{m n} \partial_j \partial_i h_{m n} \, . \label{rsanistt}
        \end{split}
    \end{align}
\end{subequations}
We now have terms quadratic in linear perturbations sourcing the difference between $\Phi ^{(2)}$ and $\Psi ^{(2)}$, even in the absence of second-order anisotropic stress. We have separated the contributions into scalar-scalar (superscript ``ss''), scalar-tensor (superscript ``st'') and tensor-tensor (superscript ``tt'') source terms, as this will be useful in the following chapters. We recall that we have disregarded first-order vector perturbations. We also note that `pure' second-order vector and tensor perturbations do not source second-order scalar perturbations, in the same way that at linear order, scalar, vector and tensor perturbations evolve independently. 

We can also take the trace of the spatial part of the Einstein equations, which leads to
\begin{equation}\label{eqtracescalarsrealspace}
    \begin{split}
        &3\Psi ^{(2) \prime \prime} + 6 \hubble \Psi ^{(2) \prime} + 6 \hubble \conf \Psi ^{(2)} + 3 \hubble ^2 \Psi ^{(2)} + 3 \hubble \Phi ^{(2)\prime}+ 6 \hubble \conf \Phi ^{(2)}  + 3\hubble ^2 \Phi ^{(2)} + \nabla ^2 \Phi ^{(2)}   \\
        &- \nabla ^2 \Psi ^{(2)} + 3\kappa w a^2 \rho \Psi ^{(2)}- \frac{3}{2} \cs \kappa a^2 \rho ^{(2)} = S^{\Psi^{(2)}_{ss}} + S^{\Psi^{(2)}_{st}} + S^{\Psi^{(2)}_{tt}} \, , 
    \end{split}
\end{equation}
where we have further defined
\begin{subequations}
    \begin{align}
        \begin{split}
            S^{\Psi^{(2)}_{ss}}  &= 12 \mathcal{H}^2 \Phi^2 + 24 \mathcal{H}^\prime \Phi^2 + 24 \mathcal{H} \Phi \Phi^\prime - 6 c_s^2 \kappa  a^2 \delta \rho \Psi + 12 \mathcal{H}^2 \Phi \Psi + 24 \mathcal{H}^\prime \Phi \Psi\\
            &+ 12 \mathcal{H} \Phi^\prime \Psi + 24 \mathcal{H} \Phi \Psi^\prime + 6 \Phi^\prime \Psi^\prime - 3 (\Psi^\prime)^2 + 12 \Phi \Psi^{\prime\prime} + 4 \Phi \nabla ^2 \Phi \\
            &+ 4 \Psi \nabla ^2 \Psi + \kappa a^2  \rho (1+w) \partial_m V \partial^m V + 2 \partial_m \Phi \partial^m \Phi + 2 \partial_m \Psi \partial^m \Phi \\
            &+ 3 \partial_m \Psi \partial^m \Psi \, , 
        \end{split}
        \\ 
        \begin{split}
            S^{\Psi^{(2)}_{st}}  &= 6 h^{mn}\partial _m \partial _n \Phi - 4  h^{mn}\partial _m \partial _n \Psi \, , 
        \end{split}
        \\
        \begin{split}
            S^{\Psi^{(2)}_{tt}}  &= -\frac{5}{2} h_{mn}^\prime h^{mn\prime} - 6 h_{mn} h^{\prime\prime mn} - 12 \mathcal{H} h_{mn} h^{mn\prime} + 4 h_{mn} \partial_k \partial_n h_m^{k} + 4 h_{mn} \nabla ^2 h^{mn} \\
            &- \partial_n h_{mk} \partial^{k} h^{mn} + \frac{3}{2} \partial_k h_{mn} \partial^{k} h^{mn}\, . \label{rstracestt}
        \end{split}
    \end{align}
\end{subequations}
We see that to proceed and solve the above equation for $\Psi ^{(2)}$, we need to substitute in the relation between $\rho ^{(2)}$ and other perturbation variables, which is given by the time-time component of the second-order Einstein equations
\begin{equation}
    \begin{split}
        &\frac{1}{2} \kappa a^2 \delta\rho^{(2)}=-\frac{1}{2} h_{mn}' h^{mn\prime} - 4 h_{mn} h^{mn\prime} \mathcal{H}  - 2 \kappa a^2 \delta\rho \Phi - \kappa a^2 \rho  \Phi ^{(2)} - 12 \mathcal{H} \Psi \Psi' + 3 (\Psi')^2 \\
        &- 3 \mathcal{H} \Psi ^{(2)\prime} + 4 \Phi  \partial^m \partial_m \Psi + 8 \Psi  \partial^m \partial_m \Psi + \nabla ^2 \Psi ^{(2)} - \kappa a^2 (1+w) \delta\rho  \partial_m V^m \partial^m V^m \\
        &+ 3 \partial_m \Psi \partial^m \Psi - 2 h_{mn} \partial^n \partial^m \Psi + 2 h_{mn} \nabla ^2 h^{mn} - \partial_n h_{mk} \partial^k h^{mn} + \frac{3}{2} \partial_k h_{mn} \partial^k h^{mn} \, .
    \end{split}
\end{equation}
Finally if we substitute the relationship between $\Phi ^{(2)}$ and $\Psi ^{(2)}$ from Eq.~\eqref{eqanisrealspace}, we arrive at a closed form equation for $\Psi ^{(2)}$.

The equations presented above are all the scalar equations we will need at second order in perturbation theory, and we now proceed to see how second-order vector perturbations are sourced by first-order fluctuations.

% % % % % % % % % % % % % % % % % % % % % % % % % % % % % % % % 
\subsection{Second order induced vectors}

In Eq.~\eqref{vectorextractorrealspace} we defined a vector extractor in real space, which was useful at linear order; however, once again, at higher orders in perturbation theory it is useful to work in Fourier space. We can build an extractor (see Ref.~\cite{Lu:2007cj} for example) $V^{ij}_a(\mathbf{x})$, defined by its action on a source term $S_{ij}(\mathbf{\eta ,x^{\prime}})$ as
\begin{equation}\label{vectorextractor}
    V^{ij}_a(\mathbf{x})S_{ij}(\mathbf{\eta ,x^{\prime}}) =-i \sum _{s=+,-} \int \frac{ \rm ^3 \mathbf{k^{\prime}}}{(2\pi)^{\frac{3}{2}}} e^s_a(\mathbf{k^{\prime}}) e^{i\mathbf{k^{\prime}}\cdot \mathbf{x}} \frac{1}{k^{\prime 2}} \int \frac{\rm  ^3 \mathbf{x^{\prime}}}{(2\pi)^{\frac{3}{2}}} k^{\prime i}e_s^j(\mathbf{k^{\prime}})^* e^{-i\mathbf{k^{\prime}}\cdot \mathbf{x^{\prime}}} S_{ij}(\mathbf{\eta ,x^{\prime}}) \, ,
\end{equation}
where $ e^s_a(\mathbf{k^{\prime}})$ is the parity vector, with parity index $a$ and $S_{ij}(\mathbf{\eta ,x^{\prime}})$ is expressed as a Fourier integral.

Applying the vector extractor to the spatial part of the Einstein equations 
\begin{equation}\label{eqinducedvectorsrealspace}
    B^{(2)\prime}_a + 2\mathcal{H}B^{(2)}_a= -4 V ^{ij}_{a} \left (S_{ij}^{B^{(2)}_{ss}} +  S_{ij}^{B^{(2)}_{st}} + S_{ij}^{B^{(2)}_{tt}} \right ) \, ,
\end{equation}
where again we have split the source terms as 
\begin{subequations}
    \begin{align}
        \begin{split}
            S_{ij}^{B^{(2)}_{ss}}  &= \kappa (1 + w) a^2 \rho \, \partial_i V \partial_j V + \partial_i \Psi \partial_j \Phi - 3 \partial_i \Psi \partial_j \Psi + \partial_i \Phi (-\partial_j \Phi + \partial_j \Psi) \\
            &- 2 \Phi \partial_j \partial_i \Phi - 2 \Psi \partial_j \partial_i \Psi \, , 
        \end{split}
        \\ 
        \begin{split}
            S_{ij}^{B^{(2)}_{st}}  &= 2 \kappa a^2 h_{ij} P + 2 h''_{ij} \Phi + 4 h'_{ij} \mathcal{H} \Phi - 4 h_{ij} \mathcal{H}^2 \Phi - 8 h_{ij} \mathcal{H}' \Phi + h'_{ij} \Phi' - 4 h_{ij} \mathcal{H} \Phi' \\
            &- h'_{ij} \Psi' - 12 h_{ij} \mathcal{H} \Psi' - 6 h_{ij} \Psi'' + 2 \Psi \partial^m \partial_m h_{ij} - 2 h_{ij} \partial^m \partial_m \Phi + 4 h_{ij} \partial^m \partial_m \Psi \\
            &- 2 h_{j}^{m} \partial_m \partial_i \Psi - 2 h_{i}^{m} \partial_m \partial_j \Psi + \partial^m h_{ij} \partial_m \Phi + 3 \partial^m h_{ij} \partial_m \Psi - \partial^m \Phi \partial_i h_{jm} \\
            &- \partial^m \Psi \partial_i h_{jm} - \partial^m \Phi \partial_j h_{im} - \partial^m \Psi \partial_j h_{im}\, , 
        \end{split}
        \\
        \begin{split}
            S_{ij}^{B^{(2)}_{tt}}  &= 2 h_{i}^{m \prime} h_{j m}^{\prime} - 2 h^{m n} \partial_n \partial_m h_{ij} + 2 \partial^m h_{j}^{n} \partial_n h_{i m} - 2 \partial^n h_{j}^{m} \partial_n h_{i m} + 2 h^{m n} \partial_i \partial_n h_{j m} \\
            &- \partial_i h^{m n} \partial_j h_{m n} + 2 h^{m n} \partial_j \partial_n h_{i m} - 2 h^{m n} \partial_j \partial_i h_{m n}\, . \label{rsvectortt}
        \end{split}
    \end{align}
\end{subequations}
Although at linear order we discarded vector perturbations because they are diluted away by the expansion of the universe, above we show that second-order vector perturbations are sourced by combinations of first-order scalars and tensors which are non-vanishing at first-order, and so we might have to take second-order vector perturbations into account. 

We now move on to induced gravitational waves at second order.

% % % % % % % % % % % % % % % % % % % % % % % % % % % % % % % % 
\subsection{Second order induced tensors}
In this subsection, we will build a transverse-traceless operator and look at how second-order tensor modes can be sourced by first-order fluctuations. This will be the basis of our work in the next Chapters. 

Applying the transverse-traceless operator in real space, as defined in Eq.~\eqref{tensorextractorrealspace}, to quadratic combinations of linear terms would generate numerous terms due to the chain rule. To avoid this complexity, we again define the operator $\Lambda ^{ij}_{ab}$ in Fourier space \cite{Ananda:2006af,Baumann:2007zm}
\begin{equation}\label{tensorextractor}
    \Lambda ^{ij}_{ab}(\mathbf{x})S_{ij}(\mathbf{\eta ,x^{\prime}}) = \sum _{\lambda=R,L} \int \frac{\rm  ^3 \mathbf{k^{\prime}}}{(2\pi)^{\frac{3}{2}}} \eps ^{\lambda ^{\prime}}_{ab}(\mathbf{k^{\prime}}) e^{i\mathbf{k^{\prime}}\cdot \mathbf{x}}  \int \frac{\rm  ^3 \mathbf{x^{\prime}}}{(2\pi)^{\frac{3}{2}}}\eps _{\lambda ^{\prime}}^{ij}(\mathbf{k^{\prime}})^* e^{-i\mathbf{k^{\prime}}\cdot \mathbf{x^{\prime}}} S_{ij}(\mathbf{\eta ,x^{\prime}}) \, ,
\end{equation}
where $ \eps ^{\lambda ^{\prime}}_{ab}(\mathbf{k^{\prime}})$ is the polarisation of the GW.

Applying Eq.~\eqref{tensorextractor} to the spatial part of the second-order spatial Einstein equations, we arrive at 
\begin{equation}\label{GW2eqraw}
    h_{ab}^{\prime \prime (2)} + 2\mathcal{H}h_{ab}^{\prime (2)} - \nabla ^2h_{ab}^{(2)}= \Lambda ^{ij}_{ab} \left (S_{ij}^{h^{(2)}_{ss}}+ S_{ij}^{h^{(2)}_{st}}+S_{ij}^{h^{(2)}_{tt}}\right )\, ,
\end{equation}
where $S_{ij}^{h^{(2)}_{ss}}$ corresponds to a source term made of quadratic linear scalar fluctuations, $S_{ij}^{h^{(2)}_{st}}$ is a source term which couples first-order scalar and tensor fluctuations and $S_{ij}^{h^{(2)}_{tt}}$ a term quadratic in first-order tensor fluctuations. These source terms read
\begin{subequations}
    \begin{align}
        \begin{split}
            S_{ij}^{h^{(2)}_{ss}} &= 2(1+w) \kappa a^2  \rho \, \partial_i V \, \partial_j V  - 2 \partial_i \Phi \, \partial_j \Phi + 2 \partial_i \Psi \, \partial_j \Phi + 2 \partial_i \Phi \, \partial_j \Psi - 6 \partial_i \Psi \, \partial_j \Psi \\
            &- 4 \Phi \, \partial_j \partial_i \Phi - 4 \Psi \, \partial_j \partial_i \Psi \, ,  \label{sourceh2SS}
        \end{split}
        \\
        \begin{split}
            S_{ij}^{h^{(2)}_{st}} &=4 c_s \kappa \mathcal{H}^2 h_{i j} \delta \rho + 4 h_{i j}^{\prime\prime} \Phi + 8 h_{i j}^{\prime} \mathcal{H} \Phi - 8 h_{i j} \mathcal{H}^2 \Phi - 16 h_{i j} \mathcal{H} \mathcal{H}' \Phi + 2 h_{i j}^{\prime} \Phi' \\
            &- 8 h_{i j} \mathcal{H} \Phi' - 2 h_{i j}^{\prime} \Psi' - 24 h_{i j} \mathcal{H} \Psi' - 12 h_{i j} \Psi^{\prime\prime} + 4 \Psi \nabla ^2 h_{i j}  - 4 h_{i j} \nabla ^2 \Phi \\
            &+ 8 h_{i j} \nabla ^2 \Psi  - 4 h_{j}^{m}  \partial_{m} \partial_{i} \Psi  - 4 h_{i}^{m}  \partial_{m} \partial_{j} \Psi  + 2 \partial_{m} h_{i j}   \partial^{m} \Phi  \\
            &+ 6 \partial_{m} h_{i j}  \partial^{m} \Psi - 2 \partial^{m} \Phi   \partial_{i} h_{j m} - 2  \partial^{m} \Psi  \partial_{i} h_{j m}  - 2  \partial^{m} \Phi  \partial_{j} h_{i m}  \\
            &- 2  \partial^{m} \Psi  \partial_{j} h_{i m}  \, , \label{sourceh2ST}
        \end{split}
        \\
        \begin{split}
            S_{ij}^{h^{(2)}_{tt}} &= - 4h^{mn} \partial _m \partial _n h_{ij} + 4\partial _n h_{jm} \partial ^m h_i^n - 4\partial _n h_{jm} \partial ^n h^m_i + 8 h^{nm} \partial _i \partial _ m h_{jn} \\& 
            +4h^{m\prime}_i h_{jm}^{\prime}+ 2 \partial _i h^{mn} \partial _j h_{mn}\, . \label{sourceh2TT}
        \end{split}
    \end{align}
\end{subequations}
In the following chapter, we will further simplify these source terms by employing the linear equations of motion. At this stage, our goal has been to demonstrate how GW are induced at second order.

Finally, we specify that at second order it is possible to build gauge invariant quantities, like we did at linear order. For example, in Ref.~\cite{Malik:2003mv}, the authors build a gauge-invariant definition of the second-order curvature perturbation on uniform density hypersurfaces. However, our work in the next chapters is concerned with tensor perturbations at second and third order, and so it is outside the scope of this work to use such gauge-invariant quantities for scalar perturbations. Nevertheless, since we are interested in tensors, it is necessary to point out that tensor modes beyond linear theory are no longer gauge invariant. As we just showed, first-order scalar, vector and tensor modes mix at second-order. Hence, under infinitesimal coordinate transformation at second-order using Eq.~\eqref{tensorgauge2}, it will also be the case. The second-order tensor transforms as 
\begin{equation}\label{h2gaugetrans}
    \begin{split}
        &h^{(2)}_{ab} \rightarrow h^{(2)}_{ab} + \Lambda _{ab}^{ij}  \big ( 2 h_{ij}^{\prime} T + 4 h_{ij} \mathcal{H} T + 2 L^{m} \partial_{m} h_{ij} + 2 h_{j}^{\ m} \partial_{m} \partial_{i} L + 2 h_{i}^{\ m} \partial_{m} \partial_{j} L 
        \\
        &+ \partial_{m} \partial_{j} L \partial^{m} E_{i} + \partial_{m} \partial_{i} L \partial^{m} E_{j} + 2 \partial_{m} h_{ij} \partial^{m} L + \tfrac{1}{2} \partial_{m} \partial_{j} L \partial^{m} L_{i} + \tfrac{1}{2} \partial_{m} \partial_{i} L \partial^{m} L_{j} \\
        &+ 2 \partial_{m} \partial_{j} L \partial^{m} \partial_{i} E + 2 \partial_{m} \partial_{j} L \partial^{m} \partial_{i} L + 2 \partial_{m} \partial_{i} L \partial^{m} \partial_{j} E + \partial_{m} \partial_{j} L \partial_{i} E^{m}+ 2 \mathcal{H} T \partial_{i} E_{j} \\
        &+ T \partial_{i} E^{\prime}_{j} + \partial^{m} E_{j} \partial_{i} L_{m} + \tfrac{1}{2} \partial^{m} L_{j} \partial_{i} L_{m} + 2 h_{j m} \partial_{i} L^{m} + 2 \partial_{m} \partial_{j} E \partial_{i} L^{m} + \tfrac{3}{2} \partial_{m} \partial_{j} L \partial_{i} L^{m}\\
        &+ 2 \mathcal{H} T \partial_{i} L_{j} - 2 \Psi \partial_{i} L_{j} + \tfrac{1}{2} T \partial_{i} L^{\prime}_{j} B_{j} \partial_{i} T + \tfrac{1}{2} L_{j} \partial_{i} T + L^{m} \partial_{i} \partial_{m} E_{j} + L^{m} \partial_{i} \partial_{m} E_{j}  \\
        &+ \tfrac{1}{2} L^{m} \partial_{i} \partial_{m} L_{j}+ \tfrac{1}{2} L^{m} \partial_{i} \partial_{m} L_{j} + \partial_{i} T \partial_{j} B + L^{m} \partial_{j} \partial_{m} E_{i} + L_{i}^{\ m} \partial_{j} \partial_{m} E^{m}+ 2 \mathcal{H} T \partial_{j} E_{i} \\
        &+ T \partial_{j} E^{\prime}_{i} + T \partial_{j} E^{\prime}_{i} + \tfrac{1}{2} \partial_{i} T \partial_{j} L + E^{m} \partial_{j} L_{i m} + \tfrac{1}{2} L^{m} \partial_{j} L_{i m} + L^{m} \partial_{j} L_{i m} + \tfrac{3}{2} L^{m} \partial_{j} L_{i m} \\
        &+ 2 \mathcal{H} T \partial_{j} L_{i} - 2 \Psi \partial_{j} L_{i} + \tfrac{1}{2} T \partial_{j} L^{\prime}_{i} + B_{i} \partial_{j} T + \tfrac{1}{2} L_{i} \partial_{j} T + B_{i} \partial_{j} T + \tfrac{1}{2} L_{i} \partial_{j} T + B_{i} \partial_{j} T \\
        &+ \tfrac{1}{2} L_{j} \partial_{i} T - T \partial_{j} T + L^{m} \partial_{j} \partial_{m} E_{i}+ \partial_{m} L \partial_{j} \partial_{m} E_{i} + \tfrac{1}{2} L^{m} \partial_{j} \partial_{m} L_{i} + \tfrac{1}{2} \partial_{m} L \partial_{j} \partial_{m} L_{i} \\
        &+ 2 L^{m} \partial_{j} \partial_{m} \partial_{i} E + 2 \partial_{m} L \partial_{j} \partial_{m} \partial_{i} E + L^{m} \partial_{j} \partial_{m} \partial_{i} L + \partial_{m} L \partial_{j} \partial_{m} \partial_{i} L + 4 \mathcal{H} T \partial_{j} \partial_{i} E \\
        &+ 2 T \partial_{j} \partial_{i} E + 4 \mathcal{H} T \partial_{j} \partial_{i} L - 4 \Psi \partial_{j} \partial_{i} L + T \partial_{j} \partial_{i} L + \partial_{j} \partial_{i} L^{(2)} \big ) \, .
    \end{split}
\end{equation}
We note that, as is the case at first order, we can pick a gauge and then the quantity we work with is fine. However, the gauge dependence of second-order tensors exposes a problem in the definition of the spectral density, since the observable should, in principle, be independent of the chosen coordinates. We will discuss this further in the following Chapter.

In this chapter, we developed the formalism for cosmological perturbation theory, both at linear and non-linear orders. At linear order, we examined in detail how to perturb the FLRW universe and extracted parts of the Einstein equations of interest in the conformal Newtonian gauge. We then solved the equations of motion for the curvature perturbation $\Psi(\eta, \mathbf{x})$ and the tensor perturbation $h_{ij}(\eta, \mathbf{x})$ in Fourier space. We also introduced the observational framework for analysing these perturbations through the power spectrum and spectral density. Finally, when going to second-order in perturbation theory, we illustrated how first-order scalar and tensor fluctuations can generate second-order scalar, vector and tensor perturbations. In the following chapter, we will look closely into the second-order induced gravitational waves by solving Eq.~\eqref{GW2eqraw} and then studying its observational imprints by computing the spectral density.
\usetikzlibrary{calc}
\usetikzlibrary{arrows.meta}

% % % % % % % % % % % % % % % % % % % % % % % % % % % % % % % % 
\chapter{Second order induced gravitational waves}
\label{Chap:SOGWs}

In the previous chapter, we revealed how first-order scalar and tensor perturbations can act as sources for second-order perturbations. Among these, we are particularly interested in induced gravitational waves (iGWs), since they are expected to contribute to the SGWB and may be detectable by upcoming GW observatories. Unlike their linear counterpart, second-order iGWs are sourced, even in the absence of anisotropic stress. In this chapter, our focus will then be on solving Eq.~\eqref{GW2eqraw}, which governs the dynamics of iGWs at second order in cosmological perturbation theory. After obtaining a solution to the equation of motion, we will compute the corresponding power spectrum for a general equation of state parameter $w$, adiabatic perturbation, in the absence of anisotropic stress and with Gaussian initial conditions. We will then specialise in the RD era to evaluate the associated spectral density. We begin with the case of scalar-induced gravitational waves (SIGWs), those sourced by quadratic combinations of first-order scalar fluctuations. This serves as a natural starting point; not only have SIGWs received sustained attention in the literature, but they also provide a clean framework for developing the formalism and methodology. Building on this foundation, we will later incorporate first-order tensor modes as an additional source.

% % % % % % % % % % % % % % % % % % % % % % % % % % % % % % % %
\section{Scalar-induced gravitational waves}
% % % % % % % % % % % % % % % % % % % % % % % % % % % % % % % %
The prospect of SIGWs was first studied in Refs.~\cite{Ananda:2006af,Baumann:2007zm} as a new way to probe the scalar power spectrum on scales not accessible to large-scale structure surveys or the CMB. For now, we turn off first-order tensor perturbations in Eq.~\eqref{GW2eqraw} and look at how to solve the second-order iGW equation with only the source term made of pure scalar fluctuations. We will then extract the power spectrum and compute the spectral density in an RD universe. 

Using first-order and background equations, we can further simplify the source term in Eq.~\eqref{sourceh2SS}, so that
\begin{equation}
     h_{ab}^{\prime \prime (2)} + 2\mathcal{H}h_{ab}^{\prime (2)} - \nabla ^2h_{ab}^{(2)}= \Lambda ^{ij}_{ab} S_{ij}^{h^{(2)}_{ss}} \, ,
\end{equation}
with
\begin{equation}
    S_{ij}^{h^{(2)}_{ss}} = \frac{8}{3(1+w)}\left [(\partial _i\Psi +\frac{\partial _i \Psi ^{\prime}}{\mathcal{H}})(\partial _j \Psi +\frac{\partial _j \Psi ^{\prime}}{\mathcal{H}})\right ] +4 \partial _i \Psi \partial _j \Psi \,,
\end{equation}
where we have used the equation relating velocity perturbations to the Bardeen potentials in Eq.~\eqref{velocity1} and used the fact that $\Psi = \Phi$. The prospects of having scalar anisotropic stress at linear order were studied in Ref.~\cite{Baumann:2007zm} and shown to be negligible. As is the case in linear order perturbation theory, it is convenient to solve this equation in Fourier space
\begin{equation}\label{sigwseqfour}
     h_{\mathbf{k},\lambda}^{(2)\prime \prime} + 2\mathcal{H}h_{\mathbf{k},\lambda}^{(2)\prime}  +k^2h_{\mathbf{k},\lambda}^{(2)} =  4\epsilon ^{ij}_{\lambda}(\mathbf{k})^{*}S_{ij}^{h^{(2)}_{ss}}(\eta , \mathbf{k}) \equiv 4 S_{\lambda}^{h^{(2)}_{ss}}(\eta , \mathbf{k}) \, ,
\end{equation}
where
\begin{equation}\label{sssourcefourier}
    \begin{split}
        &S_{\lambda}^{h^{(2)}_{ss}}(\eta , \mathbf{k}) = \avant ^2 \epsilon ^{ij}_{\lambda}(\mathbf{k})^{*} \int \frac{\rm ^3 \mathbf{p}}{\four}  \bigg (\frac{2}{3(1+w)} \bigg [ \left ( p_i T_{\Psi} (p\eta) + p_i \frac{ T_{\Psi}\conf (p\eta)}{\hubble} \right ) \\
        &\times \left ( p_j T_{\Psi} (|\mathbf{k}-\mathbf{p}|\eta) + p_j \frac{ T_{\Psi}\conf (|\mathbf{k}-\mathbf{p}|\eta)}{\hubble} \right )\bigg ]  + p_ip_j  T_{\Psi} (p\eta)  T_{\Psi} (|\mathbf{k}-\mathbf{p}|\eta)  \bigg ) \mathcal{R}_{\mathbf{p}}\mathcal{R}_{\mathbf{k}-\mathbf{p}} \, .
    \end{split}
\end{equation}
In the above we have used the fact that the polarisation tensors are transverse ($\epsilon ^{ij}_{\lambda}(\mathbf{k})^{*}k_i=\epsilon ^{ij}_{\lambda}(\mathbf{k})^{*}k_j=0$) and we decomposed the first-order scalar fluctuations $\Psi(\eta,k)$ according to Eq.~\eqref{curlyR}. We recall that the transfer functions depend only on the scalar momenta and on time, whereas the initial scalar perturbations, $\mathcal{R}_{\mathbf{k}}$, are functions of the wavevector. At this stage, we have not yet specified an equation of state, and therefore, the explicit form of the transfer functions remains undetermined.

We can proceed and solve Eq.~\eqref{sigwseqfour} using Green's method. We have already solved this equation at linear order when there was no source term. Here, it is more convenient to introduce the variable $z=a(\eta)h_{\mathbf{k},\lambda}$, so that the equation becomes
\begin{equation}
    z ^{\prime \prime} + \left ( k^2 - \frac{a^{\prime \prime}(\eta)}{a(\eta)} \right ) z = 4a(\eta)S_{\lambda}^{h^{(2)}_{ss}}(\eta , \mathbf{k}) \, ,
\end{equation}
and if we further parametrise $a(\eta)$ in terms of $b$ (see Eq.~\eqref{afuncb}), on the LHS we have
\begin{equation}
    z ^{\prime \prime} + \left ( k^2 - \frac{b(1+b)}{\eta ^2} \right ) z = 4a(\eta)S_{\lambda}^{h^{(2)}_{ss}}(\eta , \mathbf{k}) \, .
\end{equation}
The homogeneous equation of motion has two linearly independent solutions
\begin{equation}
    z^{\text{homo}} (\eta , k) = A \sqrt{\eta} \,  J_{\tfrac{1}{2}+b} (\eta  k) + B \sqrt{\eta} \, Y_{\tfrac{1}{2}+b} (\eta  k) \, .
\end{equation}
For initial conditions $h(\eta _i)=0$ and $h\conf(\eta _i)=0$, where $\eta _i$ is the initial time, we arrive at the Green's function 
\begin{equation}\label{greenstensors}
    G^h_k(\eta , \bareta) = \theta (\eta - \bareta) \frac{(k\eta)(k \bareta)}{k} \left (y_b(k\eta)j_b(k\bareta)- j_b(k\eta)y_b(k\bareta)   \right ) \, , 
\end{equation}
where we have reduced the Bessel functions to spherical Bessel functions. The solution then reads
\begin{equation}
    z = 4\int ^{\infty}_{\eta _i} \rm \bareta \, G^h_k(\eta , \bareta) a(\bareta) S_{\lambda}^{h^{(2)}_{ss}}(\bareta , \mathbf{k}) \, ,
\end{equation}
and consequently, the general solution to Eq.~\eqref{sigwseqfour} is
\begin{equation}\label{GW2soleq}
     h^{(2)ss}_{\lambda}(\eta , \mathbf{k}) =4\int ^{\infty}_{\eta _i} \rm \bareta \, G^h_k(\eta , \bareta) \frac{a(\bareta)}{a(\eta)} S_{\lambda}^{h^{(2)}_{ss}}(\bareta , \mathbf{k}) \, .
\end{equation}
The source term in Eq.~\eqref{sssourcefourier} can be rewritten in a more compact form as
\begin{equation}
    S_{\lambda}^{h^{(2)}_{ss}}(\eta , \mathbf{k}) = \avant ^2  \int \frac{\rm ^3 \mathbf{p}} {\four} \, Q_{\lambda}^{h^{(2)}_{ss}} (\mathbf{k},\mathbf{p}) f^{h^{(2)}}_{ss}(\bareta , p , |\mathbf{k}-\mathbf{p}|)\mathcal{R}_{\mathbf{p}}\mathcal{R}_{\mathbf{k}-\mathbf{p}} \, ,
\end{equation}
where\footnote{The factor of $\eps ^{ij}_{\lambda}(\mathbf{k}) ^*$ comes from the TT projection operator in Eq.~\eqref{tensorextractor}.}
\begin{equation}
    Q_{\lambda}^{h^{(2)}_{ss}}(\mathbf{k},\mathbf{p})= \eps ^{ij}_{\lambda}(\mathbf{k}) ^*  p_{i}p_{j} \, ,
\end{equation}
and
\begin{align}
    \begin{split}
    f^{h^{(2)}}_{ss}(\bareta , p , |\mathbf{k}-\mathbf{p}|)&=\frac{2}{3(1+w)} \bigg [ \bigg ((T_{\Psi}(\eta |\mathbf{k}-\mathbf{p}|) + \frac{T^{\prime}_{\Psi}(\eta |\mathbf{k}-\mathbf{p}|)}{\mathcal{H}}\bigg )  \bigg ( T_{\Psi}(\eta p) \\
    &+ \frac{T^{\prime}_{\Psi}(\eta p)}{\mathcal{H}} \bigg )  \bigg ] 
    +T_{\Psi}(\eta |\mathbf{k}-\mathbf{p}|)T_{\Psi}(\eta p)\, .
    \end{split}
\end{align}
Finally, putting all this together, we arrive at the solution of the SIGW equation
\begin{equation}\label{solsigws}
    h^{(2)ss}_{\lambda}(\eta , \mathbf{k}) = 4\int \frac{{\rm } ^3 \mathbf{p}}{(2\pi)^{\frac{3}{2}}} Q_{\lambda}^{h^{(2)}_{ss}}(\mathbf{k},\mathbf{p})\mathcal{I}^{h^{(2)}}_{ss}(\eta , p , |\mathbf{k}-\mathbf{p}|) \mathcal{R} _{\mathbf{p}}\mathcal{R} _{\mathbf{k}-\mathbf{p}} \,,
\end{equation}
where we have further defined
\begin{equation}
    \mathcal{I}^{h^{(2)}}_{ss}(\eta , p , |\mathbf{k}-\mathbf{p}|) = \left (\frac{3+3w}{5+3w} \right )^2 \int _{\eta _i}^{\infty} {\rm }  \bareta \, \frac{a(\bareta)}{a(\eta)} G^h_k(\eta , \bareta)f^{h^{(2)}}_{ss}(\bareta , p , |\mathbf{k}-\mathbf{p}|) \, .
\end{equation}
The solution in Eq.~\eqref{solsigws} contains three parts: The polarisation part encoded in the function $Q_{\lambda}^{h^{(2)}_{ss}}(\mathbf{k},\mathbf{p})$, a kernel $\mathcal{I}^{h^{(2)}}_{ss}(\eta , p , |\mathbf{k}-\mathbf{p}|)$ which contains all the time dependence of the SIGW solution and the initial values of the scalar perturbations, $\mathcal{R} _{\mathbf{p}}$ \& $\mathcal{R}_{\mathbf{k}-\mathbf{p}}$.

\begin{figure}[h!]
  \centering
  \begin{tikzpicture}

    %First DIAGRAM
    \coordinate (A) at (0,0);
    \coordinate (B) at (3,2.5);
    \coordinate (C) at (4,0);

    % Label the corners
    %\node[below left] at (A) {A};
    %\node[below right] at (B) {B};
    %\node[above] at (C) {C};

    \draw[thick,-{Stealth}, blue] (A) -- (C)
    node[midway, below, sloped] {$\mathbf{k}$};
    \draw[-{Stealth}, black] (A) -- (B)
    node[midway, above, sloped] {$\mathbf{p}$};
    \draw[-{Stealth}, black] (B) -- (C)
    node[midway, above, sloped] {$\mathbf{k}-\mathbf{p}$};

    %SECOND DIAGRAM
    \coordinate (D) at (6.5,0);
    \coordinate (E) at (9.5,2.5);
    \coordinate (F) at (10.5,0);

    % Label the corners
    %\node[below left] at (A) {A};
    %\node[below right] at (B) {B};
    %\node[above] at (C) {C};

    \draw[dashed,thick,-{Stealth}, blue] (D) -- (F)
    node[midway, below, sloped] {$\mathbf{k\conf}$};
    \draw[dashed,-{Stealth}, black] (D) -- (E)
    node[midway, above, sloped] {$\mathbf{p\conf}$};
    \draw[dashed,-{Stealth}, black] (E) -- (F)
    node[midway, above, sloped] {$\mathbf{k\conf}-\mathbf{p\conf}$};

    \begin{scope}[shift={(0,-4)}] % shift down by 4 units
        \coordinate (A) at (0,0);
        \coordinate (B) at (3,2.5);
        \coordinate (C) at (4,0);

        \draw[thick,-{Stealth}, blue] (A) -- (C)
        node[midway, above, sloped] {$\mathbf{k}$};
        \draw[-{Stealth}, black] (A) -- (B)
        node[midway, above, sloped] {$\mathbf{p}$};
        \draw[-{Stealth}, black] (B) -- (C)
        node[midway, above, sloped] {$\mathbf{k}-\mathbf{p}$};

        \coordinate (M) at (0,-0.1);
        \coordinate (N) at (4,-0.1);
        \coordinate (O) at (1,-2.5);

        % Label the corners
        %\node[below left] at (M) {M};
        %\node[below right] at (N) {N};
        %\node[above] at (O) {O};

        \draw[dashed,thick,-{Stealth}, blue] (N) -- (M)
        node[midway, below, sloped] {$\mathbf{k\conf}$};
        \draw[dashed,-{Stealth}, black] (N) -- (O)
        node[midway, below, sloped] {$\mathbf{p\conf}$};
        \draw[dashed,-{Stealth}, black] (O) -- (M)
        node[midway, below, sloped] {$\mathbf{k\conf}-\mathbf{p\conf}$};

    \end{scope}

    \begin{scope}[shift={(6.5,-4)}] % shift down by 4 units
        \coordinate (A) at (0,0);
        \coordinate (B) at (3,2.5);
        \coordinate (C) at (4,0);

        \draw[thick,-{Stealth}, blue] (A) -- (C)
        node[midway, below, sloped] {$\mathbf{k}$};
        \draw[-{Stealth}, black] (A) -- (B)
        node[midway, above, sloped] {$\mathbf{p}$};
        \draw[-{Stealth}, black] (B) -- (C)
        node[midway, above, sloped] {$\mathbf{k}-\mathbf{p}$};

        \coordinate (M) at (0.15,0.05);
        \coordinate (N) at (2.95,2.3);
        \coordinate (O) at (3.85,0.1);

        % Label the corners
        %\node[below left] at (M) {M};
        %\node[below right] at (N) {N};
        %\node[above] at (O) {O};

        \draw[dashed,thick,-{Stealth}, blue] (O) -- (M)
        node[midway, above, sloped] {$\mathbf{k\conf}$};
        \draw[dashed,-{Stealth}, black] (N) -- (M)
        node[midway, below, sloped] {$\mathbf{k\conf}-\mathbf{p\conf}$};
        \draw[dashed,-{Stealth}, black] (O) -- (N)
        node[midway, below, sloped] {$\mathbf{p\conf}$};

    \end{scope}

  \end{tikzpicture}
  
  \caption{\footnotesize{(\textit{Top Row}). Two gravitational waves of momenta $\mathbf{k}$ and $\mathbf{k\conf}$ (solid blue) are being induced by first-order scalar modes (solid black) of momenta $\mathbf{p}$ \& $\mathbf{k}-\mathbf{p}$ and $\mathbf{p\conf}$ \& $\mathbf{k\conf}-\mathbf{p\conf}$ respectively. (\textit{Bottom Row}). When we correlate the two scalar-induced GWs, there are two non-trivial setups. The first to the left corresponds to the second term in Eq.~\eqref{4pointSIGWs} and the second setup to the third term.}} 
  \label{fig:sigwscontraction}
\end{figure}
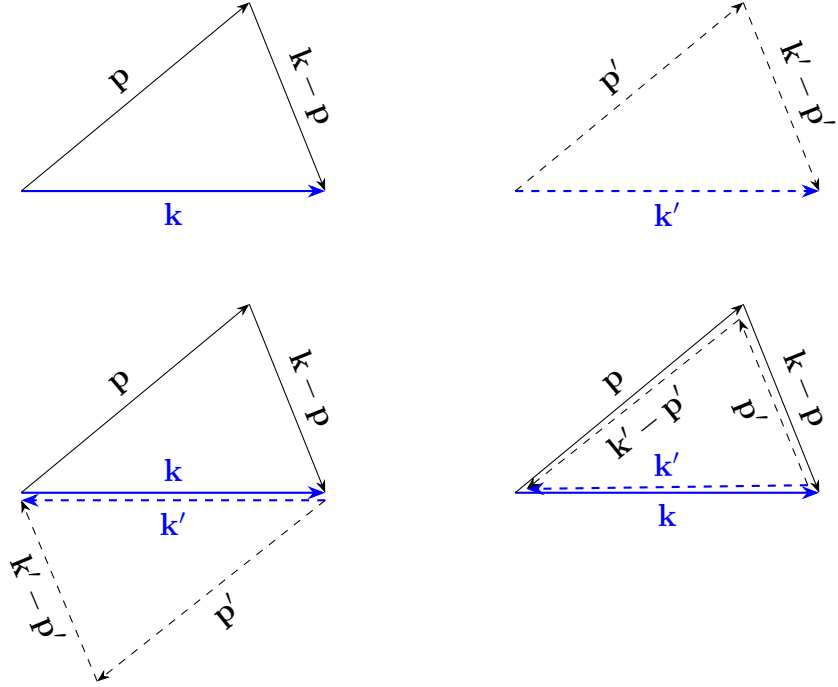

Having obtained the solution for SIGWs, we can now proceed to compute the power spectrum of these iGWs as defined in Eq.~\eqref{defpsh}, which, for clarity, reads
\begin{equation}\label{psigwsdef}
    \langle h^{(2)ss}_{\mathbf{k},\lambda}h^{(2)ss}_{\mathbf{k^{\prime}},\lambda \conf}\rangle = \delta ^{(3)} (\mathbf{k} + \mathbf{k^{\prime}}) \delta ^{\lambda \lambda ^{\prime}} P_{\lambda}^{h^{(2)}_{ss}}(\eta, k) \, .
\end{equation}
Substituting the solution from Eq.~\eqref{solsigws} into the above, we arrive at
\begin{equation}
    \begin{split}
        \langle h^{(2)ss}_{\mathbf{k},\lambda}h^{(2)ss}_{\mathbf{k^{\prime}},\lambda \conf}\rangle &= \frac{16}{(2\pi)^3}\int \rm  ^3 \mathbf{p} \int \rm  ^3 \mathbf{p \conf} \,  Q_{\lambda}^{h^{(2)}_{ss}}(\mathbf{k},\mathbf{p}) Q_{\lambda \conf}^{h^{(2)}_{ss}}(\mathbf{k\conf},\mathbf{p\conf})\mathcal{I}^{h^{(2)}}_{ss}(\eta , p , |\mathbf{k}-\mathbf{p}|)\\
        &\times \mathcal{I}^{h^{(2)}}_{ss}(\eta , p \conf , |\mathbf{k}\conf-\mathbf{p}\conf|)  \langle \mathcal{R} _{\mathbf{p}} \mathcal{R} _{\mathbf{k}-\mathbf{p}}\mathcal{R} _{\mathbf{p \conf}} \mathcal{R} _{\mathbf{k\conf}-\mathbf{p\conf}} \rangle \, .
    \end{split}
\end{equation}
For now, we are concerned with Gaussian initial conditions, which means that $\mathcal{R}_{\mathbf{k}}$ is drawn from some Gaussian distribution and so the four-point scalar function can be broken down according to Eq.~\eqref{4pointdefscalar}:
\begin{equation}\label{4pointSIGWs}
    \begin{split}
        \langle \mathcal{R} _{\mathbf{p}} \mathcal{R} _{\mathbf{k}-\mathbf{p}}\mathcal{R} _{\mathbf{p \conf}} \mathcal{R} _{\mathbf{k\conf}-\mathbf{p\conf}} \rangle &= \langle \mathcal{R} _{\mathbf{p}} \mathcal{R} _{\mathbf{k}-\mathbf{p}}\rangle \langle\mathcal{R} _{\mathbf{p \conf}} \mathcal{R} _{\mathbf{k\conf}-\mathbf{p\conf}} \rangle +\langle \mathcal{R} _{\mathbf{p}} \mathcal{R} _{\mathbf{p \conf}} \rangle \langle \mathcal{R} _{\mathbf{k}-\mathbf{p}} \mathcal{R} _{\mathbf{k\conf}-\mathbf{p\conf}} \rangle \\
        &+\langle \mathcal{R} _{\mathbf{p}}\mathcal{R} _{\mathbf{k\conf}-\mathbf{p\conf}}\rangle \langle  \mathcal{R} _{\mathbf{k}-\mathbf{p}} \mathcal{R} _{\mathbf{p \conf}}  \rangle \, .
    \end{split}
\end{equation}
The definition of the two-point function in Eq.~\eqref{defpsR} implies that the only non-trivial contributions are the last two terms. Of these last two terms, we can show using 
\begin{equation}
    Q_{\lambda}^{h^{(2)}_{ss}}(\mathbf{k},\mathbf{p}) = Q_{\lambda}^{h^{(2)}_{ss}}(\mathbf{k},\mathbf{k}-\mathbf{p}) \, ,
\end{equation}
and that the arguments of the transfer functions are symmetric in the shift $p \rightarrow |\mathbf{k}-\mathbf{p}|$ due to the convolution \cite{Espinosa:2018eve}, that
\begin{equation}
        \langle \mathcal{R} _{\mathbf{p}} \mathcal{R} _{\mathbf{k}-\mathbf{p}}\mathcal{R} _{\mathbf{p \conf}} \mathcal{R} _{\mathbf{k\conf}-\mathbf{p\conf}} \rangle = 2 \delta (\mathbf{p}+\mathbf{p\conf}) \delta (\mathbf{k}+\mathbf{k\conf}-\mathbf{p}-\mathbf{p\conf}) P_{\mathcal{R}}(p)P_{\mathcal{R}}(|\mathbf{k}-\mathbf{p}|) \, .
\end{equation}
See Fig.\ref{fig:sigwscontraction} for a geometrical interpretation. 
Substituting in the above relation and integrating over the internal momentum $\mathbf{p\conf}$, we arrive at
\begin{equation}
    \begin{split}
        \langle h^{(2)ss}_{\mathbf{k},\lambda}h^{(2)ss}_{\mathbf{k^{\prime}},\lambda \conf}\rangle &= \frac{32}{(2\pi)^3}\int \rm  ^3 \mathbf{p} \,  Q_{\lambda}^{h^{(2)}_{ss}}(\mathbf{k},\mathbf{p}) Q_{\lambda \conf}^{h^{(2)}_{ss}}(\mathbf{-k},\mathbf{-p})\left ( \mathcal{I}^{h^{(2)}}_{ss}(\eta , p , |\mathbf{k}-\mathbf{p}|) \right )^2\\
        &\times \delta (\mathbf{k}+\mathbf{k\conf}) P_{\mathcal{R}}(p)P_{\mathcal{R}}(|\mathbf{k}-\mathbf{p}|) \, ,
    \end{split}
\end{equation}
and hence from Eq.~\eqref{psigwsdef} we can read off the power spectrum of SIGWs to be
\begin{equation}
    P^{\lambda}_{h^{(2)}_{ss}}(\eta, k) = \frac{32}{(2\pi)^3}\int \rm  ^3 \mathbf{p} \,  Q_{\lambda}^{h^{(2)}_{ss}}(\mathbf{k},\mathbf{p}) Q_{\lambda }^{h^{(2)}_{ss}}(\mathbf{-k},\mathbf{-p}) \mathcal{I}^{h^{(2)}}_{ss}(\eta , p , |\mathbf{k}-\mathbf{p}|) ^2 P_{\mathcal{R}}(p)P_{\mathcal{R}}(|\mathbf{k}-\mathbf{p}|) \, .
\end{equation}
We are now left with the task of evaluating the product of the polarisation functions and the kernel. Since the kernel depends on the equation-of-state parameter, we begin by computing the polarisation product in full generality. For this purpose, it is convenient to switch to spherical coordinates, where the homogeneity and isotropy of the background can be exploited. Conventions and useful properties of the polarisation tensors are summarised in App.~\ref{appFourierspace}. We obtain
\begin{equation}
    Q_{\lambda}^{h^{(2)}_{ss}}(\mathbf{k},\mathbf{p}) Q_{\lambda }^{h^{(2)}_{ss}}(\mathbf{-k},\mathbf{-p}) = \eps ^{ij}_{\lambda}(\mathbf{k}) ^*  p_{i}p_{j} \eps ^{mn}_{\lambda}(\mathbf{-k}) ^*  p_{m}p_{n} =  \eps ^{ij}_{\lambda}(-\mathbf{k})  p_{i}p_{j} \eps ^{mn}_{\lambda}(\mathbf{k})   p_{m}p_{n} \, , 
\end{equation}
and for both polarisations ($\lambda=R,L$) this is evaluated to be
\begin{equation}
    Q_{\lambda}^{h^{(2)}_{ss}}(\mathbf{k},\mathbf{p}) Q_{\lambda }^{h^{(2)}_{ss}}(\mathbf{-k},\mathbf{-p}) = \frac{1}{4}p^4\sin ^4\theta_p \, .
\end{equation}
At this point, it is convenient to introduce the rescaled dimensionless scalar momenta $v$ and $u$ defined as
\begin{equation} \label{uvcoordinates}
    v=\frac{p}{k}, \quad u=\frac{|\mathbf{k}-\mathbf{p}|}{k} \,,
\end{equation}
where the following relationships are helpful and used extensively throughout the rest of our work:
\begin{equation}
    \sin^2{\theta _p} = 1-\frac{(1+v^2-u^2)^2}{4v^2} = 1- \cos^2{\theta _p}\, .
\end{equation}
The integrand measure changes as
\begin{equation}
    \int \rm ^3\mathbf{p}=k^3 \int_0^{\infty}\rm v \int_{|v-1|}^{1+v}\rm u \, uv  \int_0^{2\pi}\rm \phi _p \,.
\end{equation}
The power spectrum now reads 
\begin{equation}
    \begin{split}
        P_{h^{(2)}_{ss}}(\eta, k) &= k^3 \frac{32}{(2\pi)^2}\int ^{\infty}_0 \rm  v \int ^{v+1}_{|v-1|}\rm u \, \, (uv)  \frac{k^4}{4}v^4  \left ( 1-\frac{(1+v^2-u^2)^2}{4v^2} \right )^2 \\
        &\times  \mathcal{I}^{h^{(2)}}_{ss}(x,v,u) ^2 P_{\mathcal{R}}(kv)P_{\mathcal{R}}(ku) \, ,
    \end{split}
\end{equation}
and, noting that the above expression is the same for each polarisation, the dimensionless time-averaged power spectrum\footnote{We recall 
\begin{equation*}
       \mathcal{P}_{h^{(2)}} = \sum _{\lambda} \frac{k^3}{2\pi ^2}  P^{\lambda}_{h^{(2)}}\,.
\end{equation*}} is therefore 
\begin{equation}\label{psSIGWsfinal}
    \begin{split}
        \overline{\mathcal{P}_{h^{(2)}_{ss}}(\eta, k)} &= 8 \int ^{\infty}_0 \rm  v \int ^{v+1}_{|v-1|}\rm u \,  \,  \left ( \frac{4v^2-(1+v^2-u^2)^2}{4uv} \right )^2 k^4 \overline{\mathcal{I}^{h^{(2)}}_{ss}(x,v,u) ^2}
        \\
        &\times \mathcal{P}_{\mathcal{R}}(kv)\mathcal{P}_{\mathcal{R}}(ku) \, .
    \end{split}
\end{equation}
Before proceeding to compute the spectral density, we note that the above expression holds for all values of $b\neq -\tfrac{1}{3}$. To see this, we give the full expression of the kernel
\begin{equation}\label{kernelSIGWsfinal}
    k^2\mathcal{I}^{h^{(2)}}_{ss}(x,v,u) =  \left (\frac{3+3w}{5+3w} \right )^2 \frac{1}{x^b} \int _{x _i}^{x} {\rm }  \barx \, \, \barx ^{2+b} \left ( y_b(x)j_b(\barx)- j_b(x)y_b(\barx) \right )f^{h^{(2)}}_{ss}(\barx ,v , u) \, ,
\end{equation}
where we have introduce $\barx = k \bareta$, recall $x=k\eta$ and
\begin{align}\label{fSIGWseek}
    \begin{split}
    f^{h^{(2)}}_{ss}(\barx , v , u)&=\frac{1+b}{2+b} \bigg [ \bigg ((T_{\Psi}(\barx u) + \frac{\barx}{(1+b)}\frac{\partial T_{\Psi}(\barx u)}{\partial \barx}\bigg )  \bigg ( T_{\Psi}(\barx v) + \frac{\barx}{(1+b)}\frac{\partial T_{\Psi}(\barx v)}{\partial \barx} \bigg ) \bigg ]  \\
    &  +T_{\Psi}(\barx u)T_{\Psi}(\barx v)\, .
    \end{split}
\end{align}
The spectral density depends on the Hubble parameter and therefore requires specifying an equation of state parameter. Therefore, the time-averaged power spectrum is the most general expression we can provide that remains independent of the dominant species. Secondly, recall that in Sec.~\ref{sec:spectraldensity} the spectral density was defined for linear tensor perturbations. In the following section, we will examine the validity of this definition for the second-order tensor modes of interest, in connection with our earlier discussion on the gauge dependence of such perturbations. Finally, it should be understood from Eq.~\eqref{psSIGWsfinal} that SIGWs carry information on the primordial power spectrum $\mathcal{P}_{\mathcal{R}}(k)$, at a particular scale $k$. We have not yet specified the primordial scalar power spectrum, which depends on the particular inflationary model under consideration.

% % % % % % % % % % % % % % % % % % % % % % % % % % % % % % % %
\subsection{The gauge issue and the present spectral density}

Our goal is to compute the spectral density of SIGWs. Recall that the spectral density was introduced in Eq.~\eqref{omegalinear}. If one were to naively replace the linear tensor modes by their second-order counterparts, while accounting for the numerical factors arising from the metric expansion in the Newtonian gauge (Eq.~\eqref{metricdefsecondorder}), the corresponding observable for SIGWs can be defined as
\begin{equation}\label{spectraldensityss}
    \Omega _{GW} (\eta , k) = \frac{1}{12} \frac{k^2}{\hubble ^2 (\eta)}  \overline{\mathcal{P}_{h^{(2)}_{ss}}(\eta , k)} \, .
\end{equation}
with $\mathcal{P}_{h^{(2)}_{ss}}=\sum _{\lambda} \mathcal{P}_{h^{(2)}_{ss}}^{\lambda}(\eta , k)$. At linear order, when deriving the density of GWs, we argued that tensor modes correspond to GWs on small scales, since they oscillate when they are sub-Horizon and the background metric reduces to Minkowski. Moreover, because tensor modes are gauge invariant at first order, there is no ambiguity in identifying them as physical degrees of freedom. At second order, however, tensor modes are no longer gauge invariant, as shown in Eq.~\eqref{h2gaugetrans}. This presents a difficulty, as it implies that the observable defined above acquires a dependence on the choice of gauge, which is unphysical. This became known as the `gauge issue'.

The first computations of SIGWs in Ref.~\cite{Ananda:2006af} were in an RD universe, and in Ref.~\cite{Baumann:2007zm}, in an RD and MD universe. They were both carried out in the Newtonian gauge. It was not until Ref.~\cite{Hwang:2017oxa}, that the authors showed that in an MD universe, the spectral density changes with the choice of gauge. Subsequently, there have been a plethora of papers which have computed the spectral density of GWs in different gauges for an RD and MD universe \cite{Gong:2019mui, Tomikawa:2019tvi, Inomata:2019yww, Yuan:2019fwv, Chang:2020iji, Chang:2020tji, Ali:2020sfw, DeLuca:2019ufz, Domenech:2020xin} or see Ref.~\cite{Domenech:2021ztg} for a review. A full resolution of the gauge issue lies beyond the scope of this work. Nevertheless, we summarise several key points from the literature to justify the validity of using Eq.~\eqref{spectraldensityss} in the Newtonian gauge when the waves are induced in an RD universe.
 
When deriving the solution to the equation of motion for $\Psi(\eta,\mathbf{k})$ in Sec.~\ref{sec:scalarsol}, we distinguished between the RD and MD cases. In particular, during MD the solution remains constant, in contrast to RD where it decays. Consequently, if one were to compute the spectral density of SIGWs generated in an MD universe, the resulting expression would be unreliable since the source terms would not decay, and the approximation of freely propagating waves would no longer hold \cite{Sipp:2022kmb}. There are ways to address this issue. For example, if the MD era occurs prior to RD (as during reheating), the evolution of $\Psi(\eta,\mathbf{k})$ must be carefully tracked through the MD–RD transition \cite{Kohri:2018awv, Inomata:2019ivs, Inomata:2019zqy}. For simplicity, however, in this work we restrict to GWs sourced in an RD universe, where the decay of the source terms ensures that the generated waves behave as freely propagating modes. Additionally, Ref.~\cite{Domenech:2020xin} showed that the induced GW spectrum is invariant under a set of reasonable gauge transformations. What is meant by this is that once the source term becomes inactive, the gauge artefacts decay on small scales and the observable becomes independent of said artefacts. The authors argue that this is the case for the Newtonian gauge. For those reasons, we compute the spectral density of SIGWs (and in fact second-order tensors sourced by other perturbations) using Eq.~\eqref{spectraldensityss}. Using Eq.~\eqref{afuncb} we arrive at 
\begin{equation}\label{spectraldensityRD}
    \Omega _{GW} (\eta , k) = \frac{1}{12} x^2 \times   \overline{\mathcal{P}^{(22)}_{h^{(2)}_{ss}}(\eta , k)} \, .
\end{equation}

% % % % % % % % % % % % % % % % % % % % % % % % % % % % % % % %
\subsection{Primordial black holes and current constraints on the scalar power spectrum}\label{constraintsps}

The power spectrum of SIGWs depends directly on the primordial power spectrum of scalar perturbations, $\mathcal{P}_{\mathcal{R}}(k)$, and therefore encodes valuable information about it. Importantly, both the amplitude and the shape of $\mathcal{P}_{\mathcal{R}}(k)$ are subject to constraints arising from observational data as well as theoretical considerations. The recent surge of interest in SIGWs within cosmology is largely motivated by their potential connection to PBH formation in the early universe. As we highlighted earlier, PBHs could constitute a non-negligible fraction of CDM. This can be encoded in the parameter $f_{\text{PBH}}$, which is defined as
\begin{equation}
    f_{\text{PBH}} = \frac{\rho _{\text{PBH}}}{\rho _{\text{CDM}}} \, ,
\end{equation}
where $\rho _{\text{PBH}}$ is the density of PBHs in the universe and $\rho _{\text{CDM}}$ the density of CDM. There are different ways to produce PBHs in the early universe, see Ref.~\cite{Green:2014faa} for examples, but throughout this work we are concerned with large density perturbations collapsing under their own weight due to gravity during RD \cite{Carr:1974nx}. In what follows, we give a brief overview of the formation process, highlighting key properties and numerical values related to our work, following closely Ref.~\cite{Byrnes:2021jka}.

\begin{figure}
    \centering
    \includegraphics[width=0.8\linewidth]{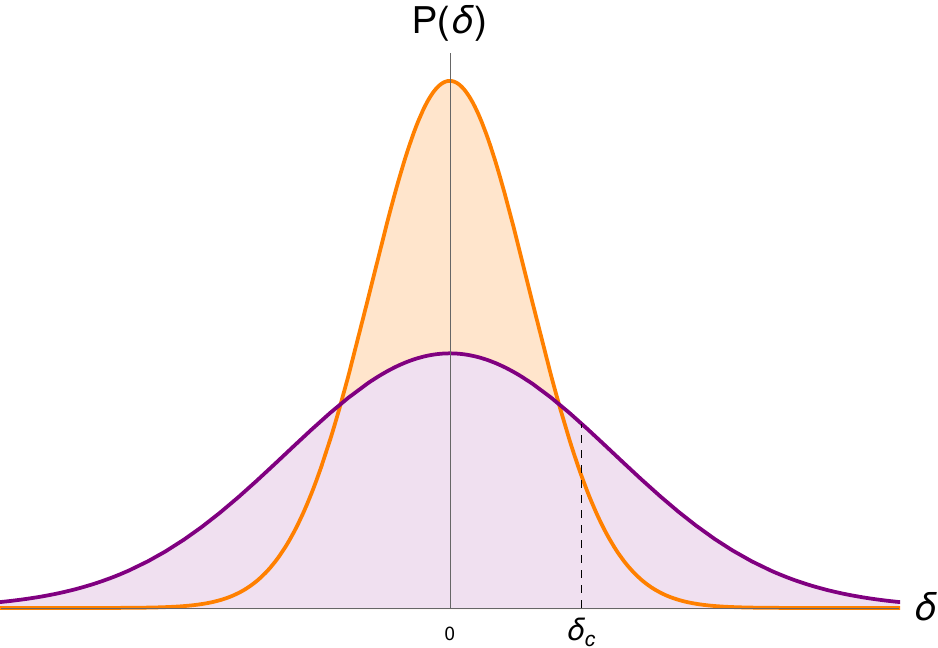}
    \caption{\footnotesize{The PDF of $\delta$. A larger variance, or equivalently, a higher amplitude of the power spectrum, increases the fraction of the PDF lying above the threshold $\delta_c$, thereby enhancing the probability of PBH formation.}}
    \label{fig:PDFPBH}
\end{figure}

The key characteristic of a PBH is its mass, which is directly related to the cosmic time ($t$) at which it formed and the scale ($k$) of the overdensity which collapses. In the early universe, if a region (or Hubble patch) contains a sufficiently large overdensity, gravity can overcome the radiation pressure and collapse. If the overdensity is big enough, it can form a PBH. The mass of the overdensity is related to how much mass there is inside the comoving horizon as the perturbation becomes sub-Hubble ($k\approx\hubble$), since above that scale causal contact is impossible and therefore gravity cannot act. The overdensity $\delta \rho$ has to be approximately $0.45$ times bigger than the background density $\rho$ \cite{Escriva:2021aeh}. Mathematically, we define the dimensionless critical density $\delta _{\text{c}}$, as the criterion for PBH formation during RD 
\begin{equation}
    \delta  = \frac{\delta \rho}{\rho} \bigg | _{k=\hubble} \geq \delta _{\text{c}} \approx 0.45 \, .
\end{equation}
What remains to be determined is how many fluctuations fulfil this criterion. A way to estimate this is via the Press-Schechter formalism \cite{press_schecter}, which estimates the fraction of the universe which has collapsed into PBHs at the time of creation (or at some particular scale since both are related), $\beta$, which is given by
\begin{equation}
    \beta = \int _{\delta _c}^{\infty} \rm \delta \,  \, P(\delta) \, ,
\end{equation}
where $P(\delta)$ is a probability distribution function (PDF). We further note that $f_{\text{PBH}} \approx \tfrac{a_{\text{eq}}}{a_{\text{form}}}\beta$, where $a_{\text{eq}}$ is the scale factor at time of radiation-matter equality and $a_{\text{form}}$ the scale factor at the time of formation of the PBH. Since the scalar perturbations are drawn from a Gaussian distribution, large variances ($\sigma ^2_{\delta}$) imply a bigger fraction $\beta$ or, equivalently, $f_{\text{PBH}}$. This is illustrated in Fig.~\ref{fig:PDFPBH}. If we assume $\mathcal{R}\sim \delta$ and the variance $\sigma ^2_{\delta}\sim \mathcal{P}_{\mathcal{R}}$, then in the limit where $\beta$ is small we approximately have
\begin{equation}
    \beta \approx \frac{\sigma _{\delta}}{\sqrt{2\pi}\delta _\text{c}} \exp{\left (-\frac{\delta _\text{c}^2}{2\sigma _{\delta}^2} \right )} \Leftrightarrow \mathcal{P}_{\mathcal{R}} \approx \frac{\delta _\text{c}^2}{\ln (1/\beta)} \, .
\end{equation}

\begin{figure}
    \centering
    \includegraphics[width=0.8\linewidth]{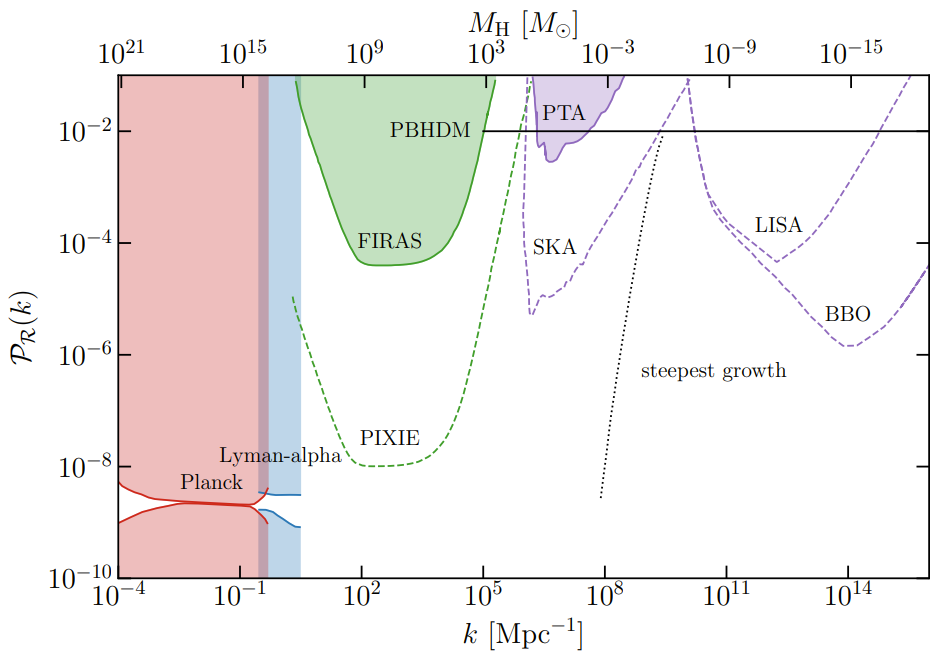}
    \caption{\footnotesize{Current constraints on $\mathcal{P}_{\mathcal{R}} (k)$ come from the CMB temperature angular power spectrum \cite{Planck:2018jri} (solid red), CMB spectral distortions \cite{Fixsen:1996nj} (solid green), the Lyman-alpha forest \cite{Bird:2010mp} (solid blue) and PTA data \cite{EPTA:2015qep, NANOGrav:2015aud, Shannon:2015ect} (solid magenta). The solid black line corresponds to  $\mathcal{P}_{\mathcal{R}} (k)=10^{-2}$. The dotted black line corresponds to the possible steepest growth of the power spectrum \cite{Byrnes:2018txb, Carrilho:2019oqg} in single-field inflation. Future constraints will come from PIXIE \cite{Kogut:2011xw} (dashed green) and proposed GW detectors (dashed magenta) such as SKA \cite{Weltman:2018zrl, Dror:2019twh}, LISA \cite{LISACosmologyWorkingGroup:2022jok}, and BBO \cite{Harry:2006fi}. This figure was reproduced with permission from Ref.~\cite{Green:2020jor}.}}
    \label{fig:psconstriants}
\end{figure}

On cosmological scales, the scalar power spectrum takes the form \cite{Planck:2018jri, Baumann:2009ds}  
\begin{equation}
    \mathcal{P}_{\mathcal{R}} (k) = \mathcal{A}_{\mathcal{R}} \left ( \frac{k}{k_*} \right ) ^{n_s-1} \, ,
\end{equation}
where $k_*$ is some pivot scale. Planck constrains on CMB scales ($k_*=0.05 \, \text{Mpc}^{-1}$) the amplitude $ \mathcal{A}_{\mathcal{R}}\approx 2\times 10^{-9}$ and $n_s\approx 0.96$ \cite{Planck:2018vyg}. A solar mass PBH, $M_{\text{PBH}}=M_{\odot}$, corresponds to a scale, time and scale factor of \cite{Byrnes:2021jka}
\begin{equation}
    k \sim 10^7 \, \text{Mpc}^{-1}\, , \, \, \, t\sim 10^{-6}\, \text{s} \, , \, \text{and}\, \,a_{\text{form}} \sim 10^{-8} \, a_{\text{eq}} \, ,
\end{equation}
which would result in a minuscule $\beta$ and $f_{\text{PBH}}$ if we extrapolated the power spectrum measured by the Planck satellite to these scales of interest. However, \textit{a priori}, there is no reason this should hold at smaller scales. In fact we see that if we want  $f_{\text{PBH}}=1$ then we need $\mathcal{P}_{\mathcal{R}} (k) \approx 10^{-2}$. This implies that if we want a significant fraction of the CDM to be in the form of PBHs, then we need a theory of inflation that enhances the curvature perturbation $\mathcal{R}$ on smaller scales. In Fig.~\ref{fig:psconstriants} we present current and future constraints on $\mathcal{P}_{\mathcal{R}} (k)$.     

\newpage

\subsection{SIGWs in a RD universe for a peaked input power spectrum}

We are now ready to compute the spectral density of SIGWs generated in an RD universe, i.e. with $w=\tfrac{1}{3}$ or $b=0$, provided we give an input primordial scalar spectrum. In an RD universe, the kernel in Eq.~\eqref{kernelSIGWsfinal} now reads
\begin{equation}
    k^2\mathcal{I}^{h^{(2)}}_{ss}(x,v,u) = \frac{4}{9} \times \frac{1}{x} \int _{0}^{x\rightarrow \infty} {\rm }  \barx \, \, \barx \sin (x-\barx)f^{h^{(2)}}_{ss}(\barx ,v , u) \, ,
\end{equation}
where we have chosen $\eta _i = 0$, i.e. we consider modes which start their evolution before horizon crossing. Furthermore, we are interested in SIGWs today or when they are deep inside the horizon, so we take the upper limit of the integral as $\eta \rightarrow \infty$, for fixed $\mathbf{k}$. The function $f^{h^{(2)}}_{ss}(\barx ,v , u)$, from Eq.~\eqref{fSIGWseek}, is explicitly given by 
\begin{equation}
    \begin{split}
    f^{h^{(2)}}_{ss}(\barx ,v , u) &= \frac{1}{2 u^3 v^3 \barx^6}
    \Bigg(
    54 v \cos\left(\tfrac{v \barx}{\sqrt{3}}\right) \barx
    \left( 9u \cos\left(\tfrac{u \barx}{\sqrt{3}}\right) \barx
    + \sqrt{3}(-9 + u^2 \barx^2)\sin\left(\tfrac{u \barx}{\sqrt{3}}\right)\right)\\
    &+ 27\Big(
    2\sqrt{3} u \cos\left(\tfrac{u \barx}{\sqrt{3}}\right) \barx (-9 + v^2 \barx^2)
    + \big(54 - 6(u^2+v^2)\barx^2 + u^2 v^2 \barx^4\big)\\
    &\times    \sin\left(\tfrac{u \barx}{\sqrt{3}}\right)
    \Big)\sin\left(\tfrac{v \barx}{\sqrt{3}}\right)
    \Bigg) \, .
    \end{split} 
\end{equation}
The integral is made out of products of oscillating and decaying functions and, therefore, naturally involves trigonometric integrals, defined as
\begin{equation}\label{SICIfunc}
    Si(z) = \int _0^z \frac{\sin t}{t} dt \, , \quad Ci(z) = -\int _z^{\infty} \frac{\cos t}{t} dt \, .
\end{equation}
Importantly, both these functions have finite limits at infinity:
\begin{equation}
    \lim_{z\to\infty} Si(z) = \pm\frac{\pi}{2} \, , \quad \lim_{z\to\infty} Ci(z) =0\, . 
\end{equation}
Putting all this together, we arrive at \cite{Kohri:2018awv, Espinosa:2018eve}
\begin{equation}\label{SSfullkernel}
    \begin{split}
        k^4\overline{\mathcal{I}_{ss}^{h^{(2)}}(x\rightarrow \infty,v,u)^2} &= \frac{1}{2} \left ( \frac{3(u^2+v^2-3)}{8xu^3v^3}\right )^2 \bigg [\left ( -4uv+(u^2+v^2-3) \log \bigg |\frac{3-(u+v)^2}{3-(u-v)^2} \bigg |\right )^2\\
    &+\pi ^2 (u^2+v^2-3)^2 \Theta (v+u-\sqrt{3})  \bigg]\, ,
    \end{split}
\end{equation}
where we have used the fact that 
\begin{equation}
    \overline{\sin x}=\overline{\cos x}=0, \quad \overline{\sin ^2 x}=\overline{\cos ^2 x}=\frac{1}{2}\,.
\end{equation}
Finally, the spectral density of SIGWs in an RD universe, according to Eq.~\eqref{spectraldensityRD}, is
\begin{equation}\label{omegaSIGWsfinal}
    \begin{split}
        \Omega _{ss}(\eta, k) &= \frac{4}{3} \int ^{\infty}_0 \rm  v \int ^{v+1}_{|v-1|}\rm u \,  \,  \left ( \frac{4v^2-(1+v^2-u^2)^2}{4uv} \right )^2 \overline{\mathcal{I}_{ss}^2(v,u)}
        \\
        &\times \mathcal{P}_{\mathcal{R}}(kv)\mathcal{P}_{\mathcal{R}}(ku) \, ,
    \end{split}
\end{equation}
where we have defined $\overline{\mathcal{I}_{ss}^{h^{(2)}}(v,u)^2}=x^2\overline{\mathcal{I}_{ss}^{h^{(2)}}(x\rightarrow \infty,v,u)^2}$. We are just left to specify the input power spectrum $\mathcal{P}_{\mathcal{R}}(k)$.

\begin{figure}
    \centering
    \includegraphics[width=0.8\linewidth]{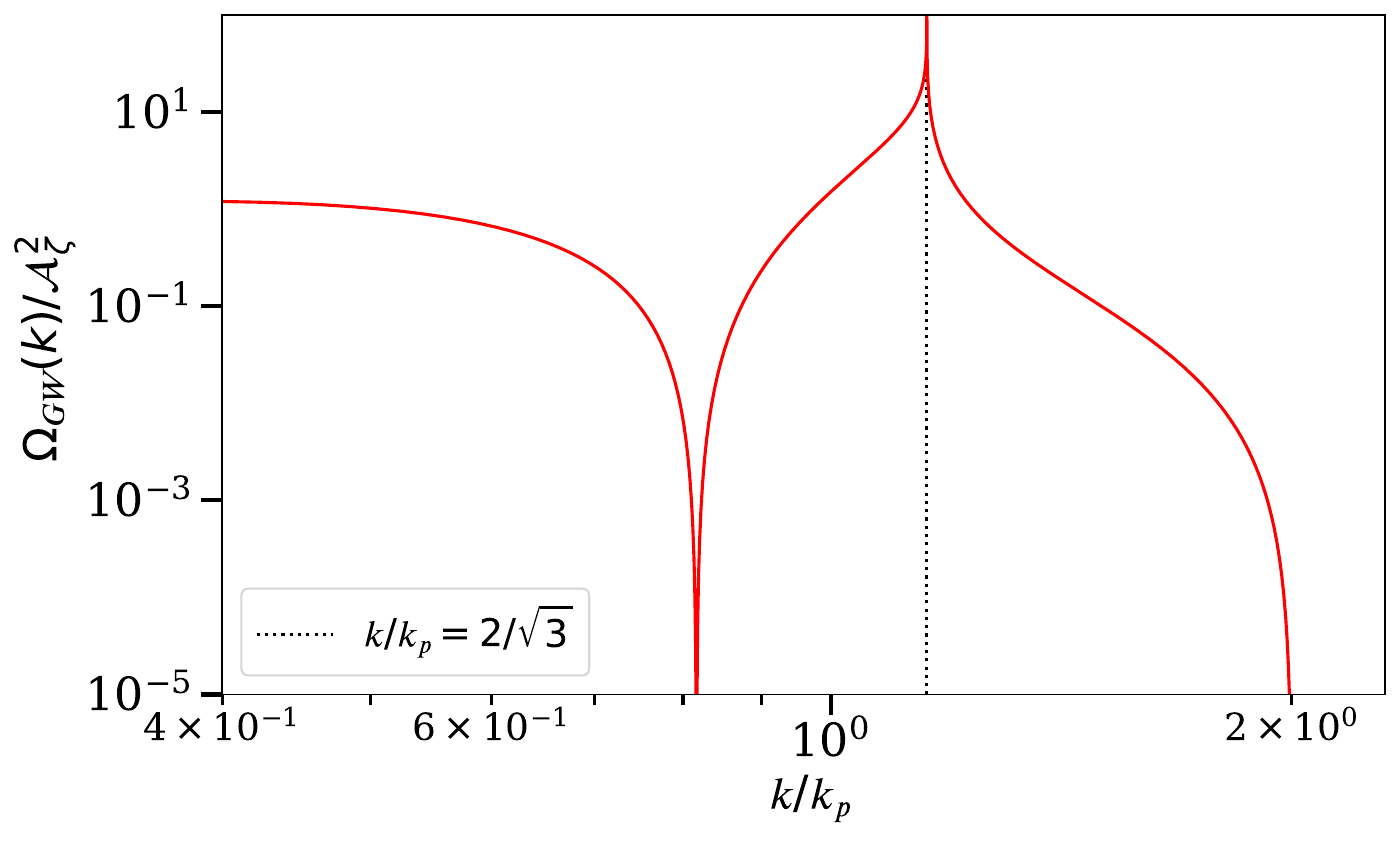}
    \caption{\footnotesize{The spectral density $ \Omega _{ss}(k)$ of SIGWs, normalised by $\mathcal{A}_{\mathcal{R}}$,  against a range of $k$ around some pivot scale $k_\mathcal{R}$. There are two notable features: the sharp cut-off at $k=2k_\mathcal{R}$ due to the presence of the Heaviside function in Eq.~\eqref{omegaSIGWsfinal} and the sharp peak at $\tilde{k}=\tfrac{2}{\sqrt{3}}$ which is due to the presence of the logarithmic term in the kernel Eq.~\eqref{SSfullkernel}.}}
    \label{fig:DiracDeltaSS}
\end{figure}

We will first consider the case of a Dirac delta peak for the input scalar power spectrum. Although `not physical' (i.e.~there is no such thing as an infinitely sharp peak in nature), the resulting spectral density tells us a lot about the properties and shape of resulting spectra in general. Furthermore, some multifield models of inflation predict ultra-sharp peaks at small scales \cite{Cai:2019jah, Kawasaki:1997ju, Frampton:2010sw, Kawasaki:2012wr, Inomata:2017okj, Pi:2017gih, Cai:2018tuh, Chen:2019zza, Chen:2020uhe}. The monochromatic power spectrum is given by
\begin{equation}\label{diracdeltascalars}
    \mathcal{P}_{\mathcal{R} }(k)= \mathcal{A}_{\mathcal{R}}\delta \left (\log  \frac{k}{k_{\mathcal{R}}} \right )\,,
\end{equation}
with $\mathcal{A}_{\mathcal{R}}$ the amplitude of the power spectrum and $k_{\mathcal{R}}$ the location of the scalar peak in $k$ space, respectively. In our case, the above can equally be expressed as
\begin{equation}
    \mathcal{P}_{\mathcal{R} }(kv)= \mathcal{A}_{\mathcal{R}} \frac{k_{\mathcal{R}}}{k} \delta (v-\frac{k_{\mathcal{R}}}{k})\, ,
\end{equation}
i.e. we are putting all the power at one scale. The same holds for $\mathcal{P}_{\mathcal{R} }(ku)$, if we swap $v\leftrightarrow u$. The way we have expressed the power spectrum makes it straightforward to use the sifting property of the Dirac delta function; we just have to take the limits of integration of Eq.~\eqref{psSIGWsfinal} into account:
\begin{equation}\label{krangedirac}
    \begin{cases}|1-v|<u<1+v\\ 0<v<\infty\end{cases}    =
    0<k<2k_{\mathcal{R}} \, ,
\end{equation}
which can be encoded in a Heaviside step function. By defining $\tilde{k}=k/k_p$, where here $k_p=k_\mathcal{R}$, Eq.~\eqref{omegaSIGWsfinal} becomes
\begin{equation}
    \Omega _{ss}(\eta, k) = \frac{3}{4} \mathcal{A}_{\mathcal{R}}^2\tilde{k}^{-2}(\frac{\tilde{k}^2-4}{4})^2 \overline{\mathcal{I}_{ss}^{h^{(2)}}(v=\tilde{k}^{-1},u=\tilde{k}^{-1})^2}\Theta(2k_p-k)
\end{equation}
for a Dirac delta input power spectra. In Fig.~\ref{fig:DiracDeltaSS} we have plotted the spectral density of SIGWs against some fiducial values of $k$ normalised by $k_p$. 
Through the kernel in Eq.~\eqref{SSfullkernel}, the spectral density is characterised by a logarithmic term. This term introduces a logarithmic divergence when the wave number, $k$, approaches a specific value, namely $k = \frac{2}{\sqrt{3}}k_p = 2c_s k_p$. This divergence is a distinctive feature of SIGWs produced from a sharply peaked input power spectrum in an RD universe. It results in the significant enhancement of the GW signal at this particular scale due to a resonance effect. There is also a sharp cut-off at $k=2k_p$ due to the presence of the Heaviside step function.

It is more realistic to consider an input peak with a finite width. To do this, we model the power spectrum as a Log-normal peak 
\begin{equation}
\label{lognorm_def}
    \mathcal{P}_{\mathcal{R} }= \frac{\mathcal{A}_{\mathcal{R} }}{\sqrt{2\pi}\sigma}\exp \left ( {-\frac{\log ^2(k/k_{\mathcal{R} })}{2\sigma ^2}}\right )\,,
\end{equation}
with $\sigma$ controlling the width of the peak and normalised such that $\int ^{\infty}_{- \infty}\rm \log k \mathcal{P}_{\mathcal{R} } = \mathcal{A}_{\mathcal{R} }$. We note that as $\sigma \rightarrow 0$ we recover the Dirac delta case. For a detailed discussion of SIGWs from a Log-normal peak, see Ref.~\cite{Pi:2020otn}. We will consider $\sigma =0.1 \, , 0.5\,,1 $ as these specific values will become useful when we add first-order tensors as a source term in the next section. For models of inflation corresponding to $\sigma \approx 1$ or larger, see Refs.~\cite{Espinosa:2018eve, Braglia:2020eai, Ando:2018nge, Ando:2017veq, Inomata:2017okj, Garcia-Bellido:1996mdl, Yokoyama:1998pt, Kohri:2012yw, Clesse:2015wea, Cheng:2016qzb, Kannike:2017bxn, Garcia-Bellido:2017mdw, Cheng:2018yyr, Inomata:2018cht, Palma:2020ejf, Fumagalli:2020adf}. The scale at which we centre the scalar power spectrum ($k_{\mathcal{R}}$) is chosen so that the physical frequency is centred in the LISA band: $k=f \times (2\pi a_0)$, with $f_{\mathcal{R}}=3.4$ mHz. We take $\mathcal{A}_{\mathcal{R}}\approx 2.1\times10^{-2}$, which corresponds to an enhancement of $\mathcal{O}(10^{7})$ of the Planck 2018 data \cite{Planck:2018vyg}, motivated by the enhancement needed to the power spectrum in order to produce primordial black holes \cite{Motohashi:2017kbs}, as we discussed in the previous section.

The integrands are now calculated numerically, using the numerical integration package \texttt{vegas+}~\cite{Lepage:2020tgj}. We transform to the coordinate system $(s,t)$ to perform these integrals\footnote{We do not write out the integral fully in terms of $s$ and $t$, but one can imagine every variable $v$ and $u$ now becomes a function $v(s,t)$ and $u(s,t)$.}, where
\begin{equation}\label{stcoordinates}
    s=u-v \, , \quad t = u+v-1 \, .
\end{equation}
This has the advantage of making the integration region rectangular. With these variables, the integral becomes
\begin{equation}
    \int _0^{\infty} {\rm } v  \int _{|1-v|}^{1+v} {\rm } u = \frac{1}{2}  \int _0^{\infty} {\rm } t  \int _{1}^{-1} {\rm } s \, .
\end{equation}
\begin{figure}
    \centering
    \includegraphics[width=0.8\linewidth]{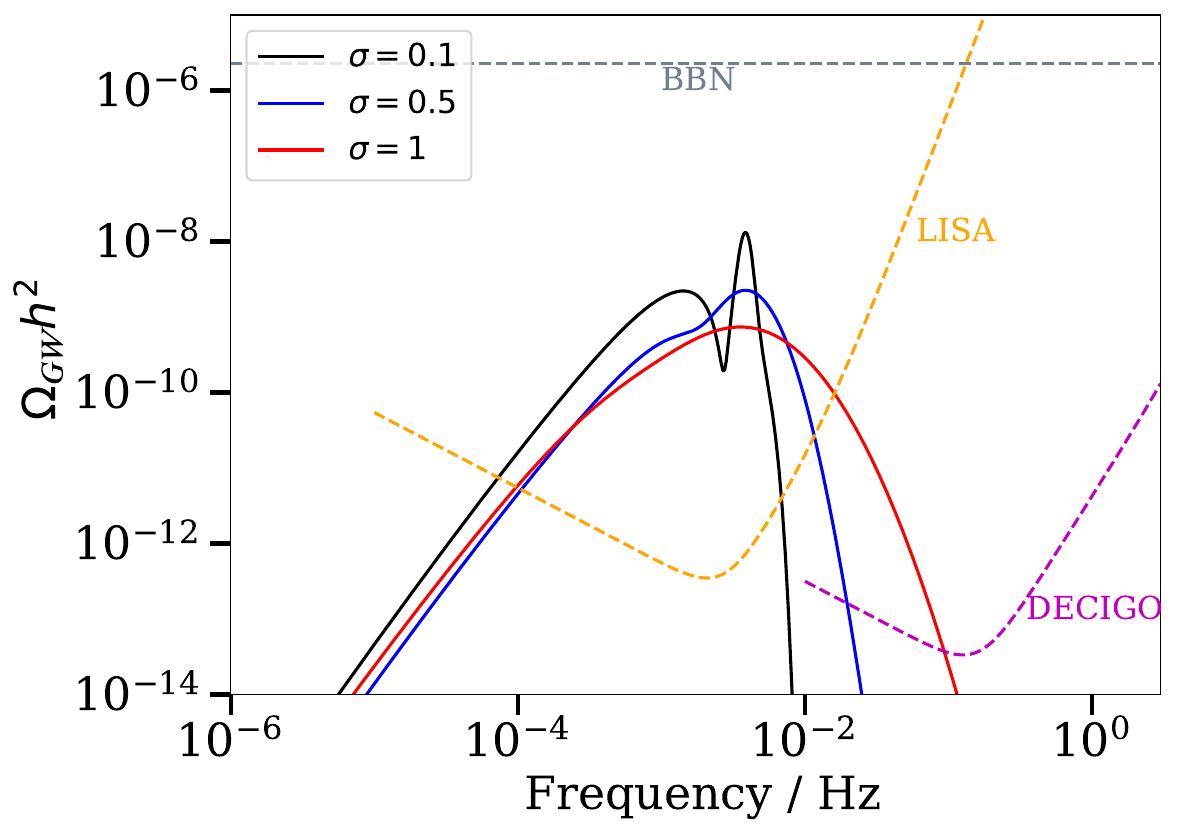}
    \caption{\footnotesize{Plots of the current spectral density of SIGWs against frequency. We have considered different (finite) widths of $\sigma$, with $\sigma =0.1$ in solid black, $\sigma =0.5$ in solid blue and $\sigma =1$ in solid red. We have also included the noise power spectral densities curves \cite{Moore:2014lga} for LISA \cite{Sathyaprakash:2009xs} and DECIGO \cite{Kawamura:2020pcg}, as well as bounds on primordial gravitational waves from BBN \cite{LISACosmologyWorkingGroup:2022jok, Caprini:2018mtu, Allen:1996vm, Cyburt:2015mya}.}}
    \label{fig:LogNormalSS}
\end{figure}
\newpage
In Fig.~\ref{fig:LogNormalSS}, we have plotted the current-day spectral density\footnote{The present-day spectral density, $\Omega (\eta _0,k)$, is related to the spectral density at time of creation,  $\Omega (\eta ,k)$, by a dilution factor \cite{Domenech:2021ztg, Domenech:2023fuz} $\mathcal{N}=1.62\times10^{-5}$ such that
\begin{equation*}
    \Omega (\eta _0,k)h^2 = \mathcal{N}\Omega (\eta ,k)\, ,
\end{equation*}To reduce clutter we continue using $\Omega (\eta ,k)$ instead of $\Omega (\eta _0 ,k)$.} of SIGWs for different values of the spectral width parameter $\sigma$. As expected, the closest scenario to a Dirac delta function (i.e., having $\sigma = 0.1$) gives a clear resonant peak with a sharp, high-frequency cut-off. As $\sigma$ increases, the defining features of the sharply peaked input spectrum are progressively washed out. The resonant peak diminishes and eventually disappears for $\sigma = 1$, and the frequency range of the spectrum extends beyond the original cut-off. Another distinctive feature of SIGWs generated in a RD universe, is the low frequency tail of the spectrum which decays as $(f/f_p)^3$ \cite{Cai:2019cdl, Yuan:2019wwo} where $f_p=f_{\mathcal{R}}$ (for the case of very peaked input power spectrum it decays as $(f/f_p)^3\log  ^2(f/f_p)$ \cite{Domenech:2023fuz}). This can be taken as a template for distinguishing SIGWs from other sources. For a general review on SIGWs, the interested reader can refer to Ref.~\cite{Domenech:2021ztg} and references therein. We conclude by specifying that the integrand in Eq.~\eqref{omegaSIGWsfinal} is well behaved in the IR ($k\rightarrow 0$ then $v\sim u \rightarrow \tfrac{1}{k}$) and UV ($k\rightarrow \infty$ then $v\rightarrow1$ and $u\rightarrow 0$ or vice-versa) limits, i.e. the integrand is finite in both those limits, regardless of the input primordial scalar power spectrum.

Having established the methodology for deriving the spectral density of SIGWs and examined the characteristic features of the spectrum arising from a peaked input power spectrum in an RD universe, we now proceed to incorporate linear tensor perturbations into our analysis. As we showed before, this adds two source terms: one made of terms quadratic in first-order tensor perturbations and one which couples first-order scalars and tensors. 

% % % % % % % % % % % % % % % % % % % % % % % % % % % % % % % % 
% % % % % % % % % % % % % % % % % % % % % % % % % % % % % % % % 
\section{Adding first-order tensors as a source of second-order gravitational waves}

In the previous section, we showed that an enhancement in the primordial scalar power spectrum, $\mathcal{P}_{\mathcal{R}}$, leads to a corresponding enhancement in the spectrum of SIGWs, $\mathcal{P}^{h^{(2)}_{ss}}(\eta, k)$, proportional to $\mathcal{P}_{\mathcal{R}}^2$. Once first-order tensor perturbations are included, however, additional contributions arise: one proportional to $\mathcal{P}_{h}^2$ and another to $\mathcal{P}_{h}\mathcal{P}_{\mathcal{R}}$. In this section, we derive the expressions for the spectral density of induced gravitational waves sourced by both first-order tensor and scalar modes, and then compare these results with the case of purely SIGWs.

The source terms made of first-order scalar-tensor and tensor-tensor perturbations shown in Eq.~\eqref{sourceh2ST} and Eq.~\eqref{sourceh2TT} can be further simplified using linear equations of motion from Sec.~\ref{firstorderNEWTON} 
\begin{subequations}
    \begin{align}
        \begin{split}
            S_{ij}^{h^{(2)}_{st}} &= 8\Psi\nabla^2h_{ij} + 8\partial _ch_{ij}\partial ^c \Psi + 4h_{ij} (   \mathcal{H}(1+3c_s^2) \Psi^{\prime} +(1-c_s^2)\nabla^2\Psi )\, ,
        \end{split}
        \\
        \begin{split}
            S_{ij}^{h^{(2)}_{tt}} &= - 4h^{cd} \partial _c \partial _d h_{ij} + 4\partial _d h_{jc} \partial ^c h_i^d
    - 4\partial _d h_{jc} \partial ^d h^c_i + 8 h^{dc} \partial _i \partial _ c h_{jd} \\& 
    +4h^{c\prime}_i h_{jc}^{\prime}+ 2 \partial _i h^{cd} \partial _j h_{cd}\, . 
        \end{split}
    \end{align}
\end{subequations}
We can follow the exact same procedure that we used for SIGWs to compute the spectral density. Namely, one of the advantages of Green's method is that the solution is independent of the form of the source term, and so we can just use the same solution from Eq.~\eqref{GW2soleq}, and just change the source term. We split the solution as
\begin{equation}\label{GW2split}
      h^{(2)}_{\lambda}(\eta , \mathbf{k})= h^{(2)ss}_{\lambda}(\eta , \mathbf{k})  + h^{(2)st}_{\lambda}(\eta , \mathbf{k})  + h^{(2)tt}_{\lambda}(\eta , \mathbf{k}) \, , 
\end{equation}
where $h^{(2)ss}_{\lambda}(\eta , \mathbf{k})$ corresponds to the SIGWs solution computed in the previous section, and $h^{(2)st}_{\lambda}(\eta , \mathbf{k})$ and $h^{(2)tt}_{\lambda}(\eta , \mathbf{k})$ correspond to the scalar-tensor solution and tensor-tensor solution. The new solutions are explicitly given by
\begin{subequations}
    \begin{align}
        h^{(2)st}_{\lambda}(\eta , \mathbf{k}) &= 4 \sum _{\lambda _1} \int \frac{{\rm } ^3 \mathbf{p}}{(2\pi)^{\frac{3}{2}}} Q_{\lambda ,\lambda _1}^{h^{(2)}_{st}}(\mathbf{k},\mathbf{p})\mathcal{I}^{h^{(2)}}_{st}(\eta , p , |\mathbf{k}-\mathbf{p}|) h ^{\lambda _1} _{\mathbf{p}}\mathcal{R} _{\mathbf{k}-\mathbf{p}} \,, \label{solh2st}
        \\
        \begin{split}
            h^{(2)tt}_{\lambda}(\eta , \mathbf{k}) &= 4 \sum _{\lambda _1, \lambda _2} \int \frac{{\rm } ^3 \mathbf{p}}{(2\pi)^{\frac{3}{2}}} \bigg ( Q_{\lambda ,\lambda _1,\lambda _2}^{h^{(2)}_{tt1}}(\mathbf{k},\mathbf{p})\mathcal{I}^{h^{(2)}}_{tt1}(\eta , p , |\mathbf{k}-\mathbf{p}|) \\
            &+ Q_{\lambda ,\lambda _1,\lambda _2}^{h^{(2)}_{tt2}}(\mathbf{k},\mathbf{p})\mathcal{I}^{h^{(2)}}_{tt2}(\eta , p , |\mathbf{k}-\mathbf{p}|) \bigg ) h ^{\lambda _1} _{\mathbf{p}}h ^{\lambda _2} _{\mathbf{k}-\mathbf{p}} \, , \label{solh2tt}
        \end{split}
    \end{align}
\end{subequations}
where we have defined the polarisation functions
\begin{subequations}
    \begin{align}
        Q_{\lambda ,\lambda _1}^{h^{(2)}_{st}}(\mathbf{k},\mathbf{p}) &= \eps ^{ij}_{\lambda}(\mathbf{k}) ^* \eps _{ij}^{\lambda _1}(\mathbf{p}) \, , 
    \\
    \begin{split}
        Q_{\lambda ,\lambda _1,\lambda _2}^{h^{(2)}_{tt1}}(\mathbf{k},\mathbf{p}) &= \eps ^{ij}_{\lambda}(\mathbf{k}) ^* \bigg (  \eps^{cd}_{\lambda _1}(\mathbf{p}) \eps_{ij}^{\lambda _2} (\mathbf{k}-\mathbf{p})k_ck_d - p_dk^c  \eps_{jc}^{\lambda _1}(\mathbf{p}) \eps^{d,\lambda _2}_{i} (\mathbf{k}-\mathbf{p})  \\
        &+ \eps_{jc}^{\lambda _1}(\mathbf{p}) \eps^{c,\lambda _2}_{i} (\mathbf{k}-\mathbf{p})(k-p)_dp^d +2p_ik_c \eps^{dc}_{\lambda _1}(\mathbf{p}) \eps^{\lambda _2}_{jd} (\mathbf{k}-\mathbf{p})\\
        &+\frac{1}{2}p_ip_j \eps^{cd}_{\lambda _1}(\mathbf{p}) \eps^{\lambda _2}_{cd}(\mathbf{k}-\mathbf{p})\bigg ) \label{qtt1} \, ,
    \end{split}
    \\
    Q_{\lambda ,\lambda _1,\lambda _2}^{h^{(2)}_{tt2}}(\mathbf{k},\mathbf{p}) &= \eps ^{ij}_{\lambda}(\mathbf{k}) ^* \eps ^{c,\lambda _1}_i (\mathbf{p}) \eps _{jc}^{\lambda _2} (\mathbf{k}-\mathbf{p})  \label{qtt2}  \, ,
    \end{align}
\end{subequations}
and the kernels
\begin{subequations}
    \begin{align}
        \mathcal{I}^{h^{(2)}}_{st}(\eta , p , |\mathbf{k}-\mathbf{p}|) &= \left (\frac{3+3w}{5+3w} \right ) \int _{\eta _i}^{\infty} {\rm }  \bareta \, \frac{a(\bareta)}{a(\eta)} G^h_k(\eta , \bareta)f^{h^{(2)}}_{st}(\bareta , p , |\mathbf{k}-\mathbf{p}|)\, , \label{stnestedbase}
        \\
        \mathcal{I}^{h^{(2)}}_{tt1}(\eta , p , |\mathbf{k}-\mathbf{p}|) &= \int _{\eta _i}^{\infty} {\rm }  \bareta \, \frac{a(\bareta)}{a(\eta)} G^h_k(\eta , \bareta)f^{h^{(2)}}_{tt1}(\bareta , p , |\mathbf{k}-\mathbf{p}|)\, ,
        \\
        \mathcal{I}^{h^{(2)}}_{tt2}(\eta , p , |\mathbf{k}-\mathbf{p}|) &= \int _{\eta _i}^{\infty} {\rm }  \bareta \, \frac{a(\bareta)}{a(\eta)} G^h_k(\eta , \bareta)f^{h^{(2)}}_{tt2}(\bareta , p , |\mathbf{k}-\mathbf{p}|)\, ,
    \end{align}
\end{subequations}
with 
\begin{subequations}
    \begin{align}
        \begin{split}
            f^{h^{(2)}}_{st}(\bareta , p , |\mathbf{k}-\mathbf{p}|)&=-2p^2T_h(\eta p) T_{\Psi}(w \eta|\mathbf{k}-\mathbf{p}|)-2(\mathbf{k}-\mathbf{p})\cdot \mathbf{p}\, T_h(\eta p) T_{\Psi}(w \eta|\mathbf{k}-\mathbf{p}|) \\
            &+ \mathcal{H}(1+3w^2)T_h(\eta p) T_{\Psi}^{\prime}(w \eta|\mathbf{k}-\mathbf{p}|)-|\mathbf{k}-\mathbf{p}|^2(1-w^2)T_h(\eta p) \\
            &\times T_{\Psi}(w \eta|\mathbf{k}-\mathbf{p}|)\, , \label{fh2st}
        \end{split}
    \\
        f^{h^{(2)}}_{tt1}(\bareta , p , |\mathbf{k}-\mathbf{p}|) &= T_h(\eta|\mathbf{k}-\mathbf{p}|)T_h(\eta p)\,, \label{fh2tt1}
        \\
        f^{h^{(2)}}_{tt2}(\bareta , p , |\mathbf{k}-\mathbf{p}|) &= T^{\prime}_h(\eta|\mathbf{k}-\mathbf{p}|)T^{\prime}_h(\eta p)\,. \label{fh2tt2}
    \end{align}
\end{subequations}
The iGW solution is again structured in three distinct components: a polarisation part, a kernel that governs the time evolution, and a term representing the initial values of the perturbations. The linear tensor perturbations have been separated according to Eq.~\eqref{tensor1sol}, and we note that the primordial value depends on the GW polarisation. Unlike the scalar-scalar and scalar-tensor contributions, there is no common factor that can be extracted from the tensor-tensor solution, necessitating its decomposition into two components.

To extract the power spectrum, we correlate Eq.~\eqref{GW2split}, where we recall that we are working with Gaussian initial conditions, so that
\begin{equation}\label{2pointSOGWs}
    \begin{split}
    \langle h^{(2)}_{\lambda}(\eta , \mathbf{k})h^{(2)}_{\lambda \conf}(\eta , \mathbf{k \conf})\rangle =&  \langle h^{(2)ss}_{\lambda}(\eta , \mathbf{k})h^{(2)ss}_{\lambda \conf}(\eta , \mathbf{k \conf})\rangle + \langle  h^{(2)st}_{\lambda}(\eta , \mathbf{k}) h^{(2)st}_{\lambda \conf}(\eta , \mathbf{k \conf})\rangle \\
    + & \langle h^{(2)tt}_{\lambda}(\eta , \mathbf{k})h^{(2)tt}_{\lambda \conf}(\eta , \mathbf{k \conf})
    \rangle \,.
    \end{split}
\end{equation}
Here we have used the fact that the two-point function $\langle \mathbf{\mathcal{R}_{\mathbf{p}}}h_{\mathbf{p^{\prime}}}\rangle$ vanishes for any momenta ${\mathbf{p}}$ \& ${\mathbf{p^{\prime}}}$. This can be seen as a consequence of the SVT theorem: everything decouples at linear order. The total power spectrum $\mathcal{P}_{h^{(2)}}$, can therefore be expressed as the sum of the three separate contributions $\mathcal{P}_{ss}(\eta ,k)$, $\mathcal{P}_{st}(\eta ,k)$ and $\mathcal{P}_{tt}(\eta ,k)$, as
\begin{equation}\label{2pointgeneral}
    \langle h^{\lambda(2)}(\eta, \mathbf{k})h^{\lambda^{\prime}(2)}(\eta, \mathbf{k^{\prime}})\rangle = \delta ^{(3)}(\mathbf{k} + \mathbf{k^{\prime}}) \delta ^{\lambda \lambda ^{\prime}} \frac{2\pi ^2}{k^3} \{ \mathcal{P}_{h^{(2)}_{ss}}(\eta, k) + \mathcal{P}_{h^{(2)}_{st}}(\eta, k) + \mathcal{P}_{h^{(2)}_{tt}}(\eta, k) \} \,.
\end{equation}
The next subsections are dedicated to computing $\mathcal{P}_{h^{(2)}_{st}}(\eta, k)$ and $\mathcal{P}_{h^{(2)}_{tt}}(\eta, k)$ respectively.

%%%%%%%%%%%%%%%%%%%%%%%%%%%%%%%%%%%%%%%%%%%%%%%%%%%%%%%
\subsection{Scalar-tensor iGWs}\label{sec:stinduced}
In this section, we compute the spectral density of scalar–tensor iGWs and examine the resulting spectrum for a Dirac-delta input for both scalars and tensors. The prospect of having first-order tensors coupled to scalar perturbations as a source term was studied in Refs.~\cite{Chang:2022vlv, Bari:2023rcw, Picard:2023sbz, Yu:2023lmo}.

To extract the power spectrum of scalar-tensor iGWs, we look at the two-point function of the GW solution in Eq.~\eqref{solh2st}
\begin{equation}
    \begin{split}
        \langle  h^{(2)st}_{\lambda}(\eta , \mathbf{k}) h^{(2)st}_{\lambda \conf}(\eta , \mathbf{k \conf})\rangle &= \frac{16}{(2\pi)^3} \sum _{\lambda _1 , \lambda \conf _1} \int {\rm } ^3 \mathbf{p} \int {\rm } ^3 \mathbf{p \conf} \, ,  Q_{\lambda ,\lambda _1}^{h^{(2)}_{st}}(\mathbf{k},\mathbf{p})  Q_{\lambda \conf ,\lambda _1 \conf}^{h^{(2)}_{st}}(\mathbf{k \conf},\mathbf{p \conf}) \\
        &\times \mathcal{I}^{h^{(2)}}_{st}(\eta , p , |\mathbf{k}-\mathbf{p}|) \mathcal{I}^{h^{(2)}}_{st}(\eta , p \conf , |\mathbf{k \conf}-\mathbf{p \conf }|) \langle  h ^{\lambda _1} _{\mathbf{p}}\mathcal{R} _{\mathbf{k}-\mathbf{p}} h ^{\lambda \conf _1} _{\mathbf{p\conf}}\mathcal{R} _{\mathbf{k\conf}-\mathbf{p\conf}} \rangle \, .
    \end{split}
\end{equation}
The four-point function can be further broken down as (where we recall linear scalar and tensor perturbations do not correlate)
\begin{equation}
    \langle  h ^{\lambda _1} _{\mathbf{p}}\mathcal{R} _{\mathbf{k}-\mathbf{p}} h ^{\lambda \conf _1} _{\mathbf{p\conf}}\mathcal{R} _{\mathbf{k\conf}-\mathbf{p\conf}} \rangle = \delta ^{\lambda \lambda _1} \delta (\mathbf{p}+\mathbf{p\conf}) \delta (\mathbf{k}+ \mathbf{k \conf}+\mathbf{p} + \mathbf{p\conf}) P^{\lambda_1}_h (p)P_{\mathcal{R}} (|\mathbf{k}-\mathbf{p}|)\, ,
\end{equation}
so that 
\begin{equation}
    \begin{split}
        \langle  h^{(2)st}_{\lambda}(\eta , \mathbf{k}) h^{(2)st}_{\lambda \conf}(\eta , \mathbf{k \conf})\rangle &= \frac{16}{(2\pi)^3} \sum _{\lambda _1 } \int {\rm } ^3 \mathbf{p} \int {\rm } ^3 \mathbf{p \conf} \,   Q_{\lambda ,\lambda _1}^{h^{(2)}_{st}}(\mathbf{k},\mathbf{p})  Q_{\lambda \conf ,\lambda _1 }^{h^{(2)}_{st}}(\mathbf{-k },\mathbf{-p }) \\
        &\times \mathcal{I}^{h^{(2)}}_{st}(\eta , p , |\mathbf{k}-\mathbf{p}|) ^2 \delta (\mathbf{k}+\mathbf{k\conf}) P^{\lambda_1}_h (p)P_{\mathcal{R}} (|\mathbf{k}-\mathbf{p}|) \, .
    \end{split}
\end{equation}
Hence, the only non-trivial contribution to the scalar-tensor power spectrum is 
\begin{equation}
    \begin{split}
        P^{\lambda}_{h^{(2)}_{st}}(\eta, k) &= \frac{16}{(2\pi)^3} \sum _{\lambda _1 } \int {\rm } ^3 \mathbf{p} \int {\rm } ^3 \mathbf{p \conf} \,   Q_{\lambda ,\lambda _1}^{h^{(2)}_{st}}(\mathbf{k},\mathbf{p})  Q_{\lambda  ,\lambda _1 }^{h^{(2)}_{st}}(\mathbf{-k },\mathbf{-p }) \\
        &\times \mathcal{I}^{h^{(2)}}_{st}(\eta , p , |\mathbf{k}-\mathbf{p}|) ^2 \delta (\mathbf{k}+\mathbf{k\conf}) P^{\lambda_1}_h (p)P_{\mathcal{R}} (|\mathbf{k}-\mathbf{p}|) \, .
    \end{split}
\end{equation}
We can now compute the explicit form of the polarisation functions, where we notice that there is a sum over the polarisation of the linear tensor mode ($\lambda_1$) and, importantly, the polarisation tensors are coupled to the power spectrum of primordial GWs. We have
\begin{align}
    \begin{split}
        \sum _{\lambda _1} Q^{st}_{\lambda , \lambda _1}(\mathbf{k},\mathbf{p})&Q^{st}_{\lambda , \lambda _1}(\mathbf{-k},-\mathbf{p})P^{\lambda_1}_h(p) 
        =\sum _{\lambda _1} \left( q^{ab}_{\lambda}(\mathbf{k}) \right)^*q^{\lambda _1}_{ab}(\mathbf{p}) \left( q^{mn}_{\lambda}(-\mathbf{k}) \right)^*q^{\lambda _1}_{mn}(-\mathbf{p})P^{\lambda_1}_h(p)\,\\
       &=
       \begin{cases}
         \cos \left ( \tfrac{\theta _p}{2} \right )^8 P_h^R(p)  + \sin \left ( \tfrac{\theta _p}{2} \right )^8P_h^L(p) \, , \lambda = R  \\
         \cos \left ( \tfrac{\theta _p}{2} \right )^8P_h^L(p)  + \sin \left ( \tfrac{\theta _p}{2} \right )^8P_h^R(p) \, , \lambda = L \, 
    \end{cases}
        \\
       &=
       \begin{cases}
        \frac{\left ( u^2 - (1+v)^2 \right )^4P_h^R(kv)  + \left ( u^2 - (-1+v)^2 \right )^4P_h^L(kv) }{256v^4}\, , \lambda = R \, \\
         \frac{\left ( u^2 - (1+v)^2 \right )^4P_h^L(kv)  + \left ( u^2 - (-1+v)^2 \right )^4P_h^R(kv) }{256v^4}\, ,\lambda =L 
    \end{cases}
    \,,
    \end{split}
\end{align}
where in the last equality we switched to $v$ and $u$ coordinates defined in Eq.~\eqref{uvcoordinates}. The power spectrum can then be expressed as
\begin{equation}\label{stpowerspectrum}
    \begin{split}
        P^{R/L}_{h^{(2)}_{st}}(\eta, k) &= k^3 \frac{16}{(2\pi)^2}\int ^{\infty}_0 \rm  v \int ^{v+1}_{|v-1|}\rm u \, \,  \frac{uv}{256v^4} \bigg ( \left ( u^2 - (1+v)^2 \right )^4P_h^{R/L}(kv) \\
        &+ \left ( u^2 - (-1+v)^2 \right )^4P_h^{L/R}(kv) \bigg )  \mathcal{I}^{h^{(2)}}_{st}(x,v,u) ^2 P_{\mathcal{R}}(ku) \, .
    \end{split}
\end{equation}
We notice that, unlike the SIGWs power spectrum, the induced power spectrum emerging from scalar-tensor correlations is different for $\lambda=R$ and $\lambda=L$. This marks a major difference with SIGWs. Finally, the time-averaged power spectrum is
\begin{equation}\label{avergaepsst}
    \begin{split}
        \overline{\mathcal{P}_{h^{(2)}_{st}}(\eta, k)}  &= \frac{1}{32}\int ^{\infty}_0 \rm  v \int ^{v+1}_{|v-1|}\rm u \, \,  \frac{1}{u^2v^6} \bigg ( \left ( u^2 - (1+v)^2 \right )^4 \\
        &+ \left ( u^2 - (-1+v)^2 \right )^4 \bigg )  \overline{\mathcal{I}^{h^{(2)}}_{st}(x,v,u) ^2} \left ( \mathcal{P}^R_{h}(kv)+\mathcal{P}^L_{h}(kv) \right ) \mathcal{P}_{\mathcal{R}}(ku) \, ,
    \end{split}
\end{equation}
where the kernel 
\begin{equation}\label{kernelST}
    \mathcal{I}^{h^{(2)}}_{st}(x,v,u) =  \left (\frac{3+3w}{5+3w} \right ) \frac{1}{k^2x^b} \int _{x _i}^{x} {\rm }  \barx \, \, \barx ^{2+b} \left ( y_b(x)j_b(\barx)- j_b(x)y_b(\barx) \right )f^{h^{(2)}}_{st}(\barx ,v , u) \, ,
\end{equation}
and the function $f^{h^{(2)}}_{st}(\barx ,v , u)$ in Eq.~\eqref{fh2st} reads
\begin{equation}
    \begin{split}
            f^{h^{(2)}}_{st}(\barx ,v , u) &= -2v^2k^2T_h(\barx v) T_{\Psi}(\barx u)+k^2(-1+u^2+v^2) T_h(\barx v) T_{\Psi}(\barx u) \\
            &+ (1+3w^2)\frac{k^2}{\barx}T_h(\barx v) \frac{\partial T_{\Psi}(\barx u)}{\partial \barx}-u^2k^2(1-w^2)T_h(\barx v) T_{\Psi}(\barx u) \, .
    \end{split}
\end{equation}

In a RD universe the kernel defined in Eq.~\eqref{kernelST} is computed to be \cite{Picard:2023sbz, Yu:2023lmo}
\begin{equation}
    \begin{split}\label{kernelSTRD}
        \overline{\mathcal{I}^{h^{(2)}}_{st}(v,u)^2}&=x^2\overline{\mathcal{I}^{h^{(2)}}_{st}(x\rightarrow \infty,v,u)^2} = \frac{1}{1152v^2u^6} \bigg [ \pi ^2 (u^2-3(v-1)^2)^2(u^2-3(v+1)^2)^2  \\
        &\times \Theta \left(v+\frac{u}{\sqrt{3}}-1\right) + \bigg ( 4uv \left (9-9v^2 +u^2 \right )-\sqrt{3}(u^2-3(v-1)^2) \\
        &\times (u^2-3(v+1)^2) \log \bigg | \frac{\left (\sqrt{3}v-u \right )^2-3}{\left (\sqrt{3}v+u \right )^2-3} \bigg | \bigg )^2 \bigg ] \, .
    \end{split}
\end{equation}
Hence, in an RD universe, the spectral density is
\begin{equation}
    \begin{split}\label{omegaSTRD}
        \Omega (\eta , k) &= \frac{1}{384}\int ^{\infty}_0 \rm  v \int ^{v+1}_{|v-1|}\rm u \, \,  \frac{1}{u^2v^6} \bigg ( \left ( u^2 - (1+v)^2 \right )^4 \\
        &+ \left ( u^2 - (-1+v)^2 \right )^4 \bigg ) \overline{\mathcal{I}^{h^{(2)}}_{st}(v,u)^2}\left ( \mathcal{P}^R_{h}(kv)+\mathcal{P}^L_{h}(kv) \right ) \mathcal{P}_{\mathcal{R}}(ku) \, .
    \end{split} 
\end{equation}
For the case of SIGWs, certain values of $v$ and $u$ can cause a logarithmic resonance, so it is important to check whether the same arises for this contribution. We note that when $v = 1 \pm \sqrt{3}$, the logarithmic term in Eq.~\eqref{kernelSTRD} diverges. However, the prefactor ensures that the kernel itself remains finite at this scale, and thus no divergence occurs for scalar-tensor iGWs. Furthermore, as pointed out in Ref.~\cite{Bari:2023rcw}, scalar-tensor iGWs do not exhibit a logarithmic slope in the IR sector, since scalar and tensor fluctuations propagate at different speeds. Combined with the distinctive way polarisation functions couple to the polarisation of primordial tensor modes, these features mark a fundamental departure from the behaviour of SIGWs. Additionally, we can verify how the integrand in Eq.~\eqref{omegaSTRD} behaves in IR and UV limits. We find that the integrand is finite in the IR limit. When taking the $k\rightarrow \infty$ (UV) limit, we now have to distinguish between two different, distinct limits: $v\rightarrow 1$ and $u\rightarrow 0$, which couples large wavelength scalar perturbations with short wavelength tensor modes and $v\rightarrow 0$ and $u\rightarrow 1$, which couples short wavelength scalar perturbations with long wavelength tensor modes. In the first of these two limits, the integrand blows up as $u^{-4}$ and in the opposite UV limit, the integrand blows up as $v^{-2}$ \cite{Bari:2023rcw, Picard:2023sbz}. The integrand also diverges in the $u\rightarrow 0$ limit when $v = 1 \pm \sqrt{3}$. For these reasons, we will exclusively use peaked power spectra as input, which we do next, by first looking at the induced waves from monochromatic power spectra. Having a sufficiently sharp peak means that all the power in the power spectrum will be at a particular scale and hence will not encompass UV scales, where the rest of the integrand diverges. Moreover, enhancements of primordial tensor fluctuations during inflation often lead to peaked primordial spectra \cite{Freese:1990rb, Adams:1992bn, Pajer:2013fsa, Anber:2009ua, Garretson:1992vt, Garcia-Bellido:2016dkw, Namba:2015gja, Thorne:2017jft, Shiraishi:2016yun}. We will come back to the UV properties of the scalar-tensor contribution and their consequences on the spectral density in the final chapter of this thesis. In what follows, we take $\mathcal{P}^R_{h}(kv)=\mathcal{P}^L_{h}(kv)$, the possibility of having an initial parity violation was discussed in Ref.~\cite{Bari:2023rcw} and will be explored for the case of non-Gaussian scalar fluctuations in the following chapter. 

\begin{figure}[h!]
    \centering
    \begin{subfigure}[b]{0.49\textwidth}
        \includegraphics[width=\linewidth]{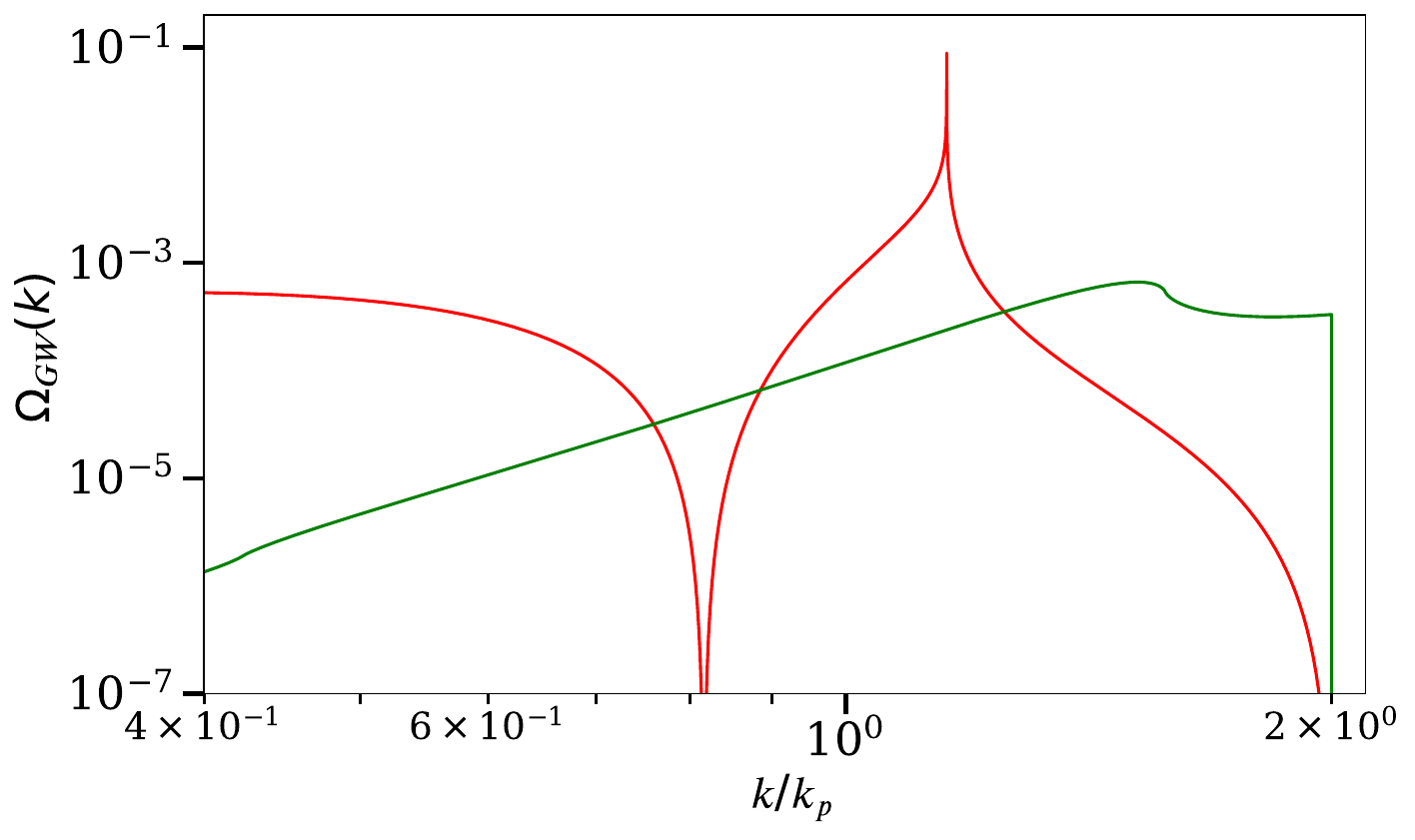}
    \end{subfigure}
    \begin{subfigure}[b]{0.49\textwidth}
        \includegraphics[width=\linewidth]{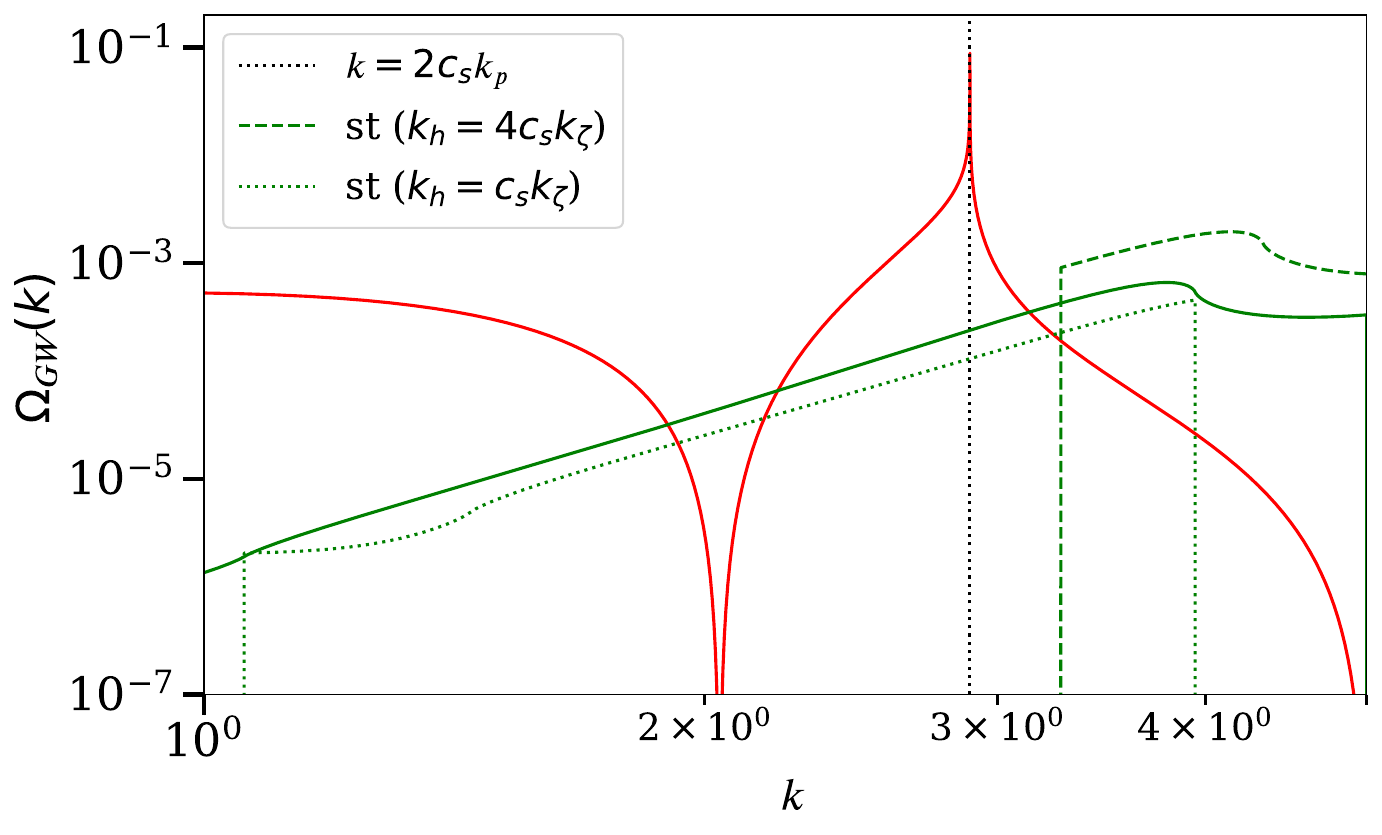}
    \end{subfigure}
    
    \caption{\footnotesize{(\textit{Left}). The spectral density of second-order iGWs induced by scalar-tensor mixing (solid green), where the primordial power spectra peak at the same scale $k_p$. We have also plotted the spectral density of SIGWs (solid red). There is no resonance in the scalar-tensor iGWs. The amplitude spanned by the scalar-tensor induced waves is comparable to the scalar-scalar case and is higher at smaller scales. (\textit{Right}). Scalar-tensor iGWs where we present three different scenarios: when the primordial power spectra peak at the same scale (solid line), when the primordial tensor peaks on larger scales than the scalar-scalar resonant peak (dotted line), and when the primordial tensor peaks on smaller scales than the scalar-scalar resonant peak (dashed line). The `impulse-like' behaviour of the dashed line can be understood from momentum conservation encoded in the Heaviside step functions.}}
    \label{fig:STDirac}
\end{figure}

We first model the primordial peak of tensors and scalars as monochromatic spectra. For tensors, we have
\begin{equation}\label{diracdelatatensors}
    \mathcal{P}^{R/L}_{h }(k)= \mathcal{A}^{R/L}_{h} \frac{k_{h}}{k} \delta (\log \frac{k}{k_h} )\, ,
\end{equation}
whilst for scalars we reuse the spectrum from Eq.~\eqref{diracdeltascalars}. We choose $\mathcal{A}^{R}_{h}=\mathcal{A}^{L}_{h}=\mathcal{A}_{h}=0.1\mathcal{A_R}$. First we set $k_\mathcal{R}=k_h=k_p$, which reduces Eq.~\eqref{omegaSTRD} to 
\begin{equation}\label{STDDsp}
     \Omega (\eta , k) = \mathcal{A}_\mathcal{R}\mathcal{A}_h \frac{1}{96} \tilde{k}^2 (16+24\tilde{k}^2+\tilde{k}^4)  \overline{\mathcal{I}^{h^{(2)}}_{st}(v=\tilde{k}^{-1},u=\tilde{k}^{-1})^2} \Theta(2k_p-k)\, .
\end{equation}
The resulting spectral density is plotted on the left of Fig.~\ref{fig:STDirac}. As expected, no resonant peak is observed. More interestingly, however, we find that while the spectral density of SIGWs decreases after the resonant peak, the scalar–tensor contribution remains comparatively larger, indicating that it can dominate the production of second-order iGWs at this order in perturbation theory. 

If the primordial peaks of scalars and tensors are at different scales then we have
\begin{equation}\label{STdiracdp}
    \begin{split}
        \Omega (\eta , k) &= \mathcal{A}_\mathcal{R}\mathcal{A}_h \frac{((k-k_h)^2-k^2_\mathcal{R} )^4+((k+k_h)^2-k^2_\mathcal{R} )^4}{192k^2k^5_hk_\mathcal{R}} \overline{\mathcal{I}^{h^{(2)}}_{st}(v=\tfrac{k_h}{k},u=\tfrac{k_\mathcal{R}}{k})^2} \\
        &\times \Theta (k-|k_{\mathcal{R}}-k_h|) \Theta (-k+k_{\mathcal{R}}+k_h)\,, 
    \end{split}
\end{equation}
where the Heaviside function comes from taking the correct limits of the integrands into account:
\begin{equation}
    \begin{cases}|1-v|<u<1+v\\ 0<v<\infty\end{cases}  =\begin{cases}|k-k_h|<k_{\mathcal{R}}<k+k_h\\ 0<k_h/k<\infty\end{cases}  =
    |k_{\mathcal{R}}-k_h|<k<k_{\mathcal{R}}+k_h \, .
\end{equation}
We look at two different informative cases: first, when the tensor power spectrum peaks at a scale higher than the resonance from the scalar-scalar contribution ($k_h\approx c_sk_{\zeta}$), and then at a scale smaller than the resonance ($k_h\approx 4c_sk_{\zeta}$). The results are presented in the right of Fig.~\ref{fig:STDirac}. We see that when $k_h\approx c_sk_{\zeta}$ the scalar-tensor contribution is smaller than when the power spectra peak at the same scale, in fact we chose this value for $k_h$ because this value corresponds to an approximate threshold: when $k_h \leq c_sk_{\zeta}$ the signal will always be smaller than the case when $k_h=k_\zeta$. Secondly, we chose the peak scale $k_h\approx 4c_sk_{\zeta}$ to show that we can get higher amplitude signals than the previous setup. The short `burst' like signal is due to the allowed $k$ values encoded in the Heaviside step functions in Eq.~\eqref{STdiracdp} (outside of this range, no incoming modes can satisfy momentum conservation and hence there is no outgoing mode). Indeed, the closer $k_{h}$ is to $k_{\zeta}$, the more the range of $k$ increases. 

Before turning to an input peak with finite width, like a Log-normal power spectrum, we first compute the spectral density of tensor–tensor iGWs.

%%%%%%%%%%%%%%%%%%%%%%%%%%%%%%%%%%%%%%%%%%%%%%%%%%%%%%%
\subsection{Tensor-tensor iGWs}\label{sec:ttinduced}

Linear tensor-tensor interactions can source gravitational waves at second-order Refs.~\cite{Chang:2022vlv, Picard:2023sbz, Yu:2023lmo}. In Eq.~\eqref{solh2tt}, we presented the solution to the equation of motion of tensor-tensor iGWs. The objective of this section is to extract the power spectrum resulting from these types of interactions.

The two-point function reads
\begin{equation}
    \begin{split}
        &\langle  h^{(2)tt}_{\lambda}(\eta , \mathbf{k}) h^{(2)tt}_{\lambda \conf}(\eta , \mathbf{k \conf})\rangle = \frac{16}{(2\pi)^3} \sum _{\lambda _1 , \lambda _2 , \lambda _1 ^{\prime}, \lambda _2 ^{\prime}} \int {\rm } ^3 \mathbf{p} \int {\rm } ^3 \mathbf{p \conf} \, \bigg ( Q_{\lambda ,\lambda _1,\lambda _2}^{h^{(2)}_{tt1}}(\mathbf{k},\mathbf{p}) \\
        &\times \mathcal{I}^{h^{(2)}}_{tt1}(\eta , p , |\mathbf{k}-\mathbf{p}|)+ Q_{\lambda ,\lambda _1,\lambda _2}^{h^{(2)}_{tt2}}(\mathbf{k},\mathbf{p})\mathcal{I}^{h^{(2)}}_{tt2}(\eta , p , |\mathbf{k}-\mathbf{p}|) \bigg ) \bigg ( Q_{\lambda \conf ,\lambda _1 \conf,\lambda \conf _2}^{h^{(2)}_{tt1}}(\mathbf{k\conf},\mathbf{p\conf}) \\
        &\times \mathcal{I}^{h^{(2)}}_{tt1}(\eta , p\conf , |\mathbf{k\conf}-\mathbf{p\conf}|) + Q_{\lambda \conf ,\lambda \conf _1,\lambda \conf _2}^{h^{(2)}_{tt2}}(\mathbf{k\conf},\mathbf{p\conf})\mathcal{I}^{h^{(2)}}_{tt2}(\eta , p\conf , |\mathbf{k\conf}-\mathbf{p\conf}| \bigg ) \langle  h^{\lambda _1} _{\mathbf{p}}h ^{\lambda _2} _{\mathbf{k}-\mathbf{p}} h^{\lambda \conf _1} _{\mathbf{p\conf}}h ^{\lambda \conf _2} _{\mathbf{k\conf}-\mathbf{p\conf}} \rangle  \ , 
    \end{split}
\end{equation}
and the four-point function is broken down as 
\begin{equation}\label{tensortensor2point}
    \begin{split}
        \langle h ^{\lambda _1}_{\mathbf{p}}h ^{\lambda _2}_{\mathbf{k-p}}h ^{\lambda _1^{\prime}}_{\mathbf{p^{\prime}}}h ^{\lambda _2^{\prime}}_{\mathbf{p^{\prime}-k^{\prime}}}\rangle  &= \langle h ^{\lambda _1}_{\mathbf{p}}h ^{\lambda _1^{\prime}}_{\mathbf{p^{\prime}}} \rangle \langle h ^{\lambda _2}_{\mathbf{k-p}}h ^{\lambda _2^{\prime}}_{\mathbf{p^{\prime}-k^{\prime}}}\rangle + \langle h ^{\lambda _1}_{\mathbf{p}}h ^{\lambda _2^{\prime}}_{\mathbf{p^{\prime}-k^{\prime}}} \rangle \langle h ^{\lambda _2}_{\mathbf{k-p}}h ^{\lambda _1^{\prime}}_{\mathbf{p^{\prime}}}\rangle \\
        &=\delta (\mathbf{k}+ \mathbf{k^{\prime}}) \delta (\mathbf{p}+\mathbf{p^{\prime}}) \left (\delta ^{\lambda_1 \lambda_1^{\prime}}\delta ^{\lambda_2 \lambda_2^{\prime}} +\delta ^{\lambda_1 \lambda_2^{\prime}}\delta ^{\lambda_2 \lambda_1^{\prime}} \right )  \\
        &\times P^{\lambda _1}_{h}(p)P^{\lambda _2}_h (|\mathbf{k}-\mathbf{p}|)\, ,
    \end{split}
\end{equation}
where we have kept only non-trivial contributions and used the symmetry of the convolution, see Ref.~\cite{Picard:2023sbz} for details. We note that, even if we eventually will sum over the $\lambda _1$ and $\lambda _2$, it is not possible to switch the polarisation labels on the Kronecker deltas in Eq.~\eqref{tensortensor2point} to further simplify the computation.  Inserting the above result into the tensor-tensor two-point function, we arrive at
\begin{equation}
    \begin{split}
        &P^{\lambda}_{h^{(2)}_{tt}}(\eta, k) = \frac{16}{(2\pi)^3} \sum _{\lambda _1 , \lambda _2 , \lambda _1 ^{\prime}, \lambda _2 ^{\prime}} \int {\rm } ^3 \mathbf{p}  \, \bigg ( Q_{\lambda ,\lambda _1,\lambda _2}^{h^{(2)}_{tt1}}(\mathbf{k},\mathbf{p})Q_{\lambda ,\lambda \conf _1,\lambda \conf _2}^{h^{(2)}_{tt1}}(\mathbf{-k},\mathbf{-p}) \\
        &\times \mathcal{I}^{h^{(2)}}_{tt1}(\eta , p , |\mathbf{k}-\mathbf{p}|)^2+ Q_{\lambda ,\lambda _1,\lambda _2}^{h^{(2)}_{tt2}}(\mathbf{k},\mathbf{p})Q_{\lambda ,\lambda \conf _1,\lambda \conf _2}^{h^{(2)}_{tt2}}(\mathbf{-k},\mathbf{-p})\mathcal{I}^{h^{(2)}}_{tt2}(\eta , p , |\mathbf{k}-\mathbf{p}|)^2 \\
        &+ Q_{\lambda ,\lambda _1,\lambda _2}^{h^{(2)}_{tt1}}(\mathbf{k},\mathbf{p})Q_{\lambda ,\lambda \conf _1,\lambda \conf _2}^{h^{(2)}_{tt2}}(\mathbf{-k},\mathbf{-p})\mathcal{I}^{h^{(2)}}_{tt1}(\eta , p , |\mathbf{k}-\mathbf{p}|)\mathcal{I}^{h^{(2)}}_{tt2}(\eta , p , |\mathbf{k}-\mathbf{p}|) \\
        &+ Q_{\lambda ,\lambda _1,\lambda _2}^{h^{(2)}_{tt2}}(\mathbf{k},\mathbf{p})Q_{\lambda ,\lambda \conf _1,\lambda \conf _2}^{h^{(2)}_{tt1}}(\mathbf{-k},\mathbf{-p})\mathcal{I}^{h^{(2)}}_{tt1}(\eta , p , |\mathbf{k}-\mathbf{p}|)\mathcal{I}^{h^{(2)}}_{tt2}(\eta , p , |\mathbf{k}-\mathbf{p}|)  \bigg )\\
        &\times \left (\delta ^{\lambda_1 \lambda_1^{\prime}}\delta ^{\lambda_2 \lambda_2^{\prime}} +\delta ^{\lambda_1 \lambda_2^{\prime}}\delta ^{\lambda_2 \lambda_1^{\prime}} \right ) P^{\lambda _1}_{h}(p)P^{\lambda _2}_h (|\mathbf{k}-\mathbf{p}|) \, .
    \end{split}
\end{equation}
The evaluation of the polarisation functions defined in Eqs.~\eqref{qtt1} and \eqref{qtt2} is long and tedious; it is useful to switch to the coordinates $v$ and $u$ and define the following functions to reduce clutter
{
\allowdisplaybreaks
\begin{align}
    A(v,u)&= (1+u-v)^2(1-u+v)^2(-1+u+v)^2(1+u+v)^6 \, ,
        \\
        B(v,u)&= (1+u-v)^2(1-u+v)^6(-1+u+v)^2(1+u+v)^2 \, ,
        \\
        C(v,u)&= (1+u-v)^6(1-u+v)^2(-1+u+v)^2(1+u+v)^2 \, ,
        \\
        D(v,u)&= (1+u-v)^2(1-u+v)^2(-1+u+v)^6(1+u+v)^2 \, ,
        \\
        E(v,u)&= 9u^2-2u(v-1)+(v+3)(v-1)\, ,
        \\
        F(v,u)&= 9u^2+2u(v-1)+(v+3)(v-1)\, ,
        \\
        G(v,u)&= 9u^2+2u(v+1)+(v-3)(v+1)\, ,
        \\
        H(v,u)&= 9u^2-2u(v+1)+(v-3)(v+1)\, ,
        \\
        I(v,u)&= \left ( (u-v)^2-1 \right )^4 (-1+u+v)^2(1+u+v)^2 \, .
\end{align}
}

It follows
\begin{equation}
    \begin{split}
        &P^{R/L}_{h^{(2)}_{tt}}(\eta, k) = \frac{16}{(2\pi)^2} k^3\int ^{\infty}_0 \rm  v \int ^{v+1}_{|v-1|}\rm u \, \, (uv) \bigg [  \frac{k^4}{65536} \bigg( A(v,u)E^2(v,u) P^{R/L}_h(kv)P^{R/L}_h(ku) \\
        &+ B(v,u)F^2(v,u) P^{R/L}_h(kv)P^{L/R}_h(ku)+ C(v,u)G^2(v,u) P^{L/R}_h(kv)P^{R/L}_h(ku) \\
        &+ D(v,u)H^2(v,u) P^{L/R}_h(kv)P^{L/R}_h(ku)\bigg )  \mathcal{I}^{h^{(2)}}_{tt1}(x,v,u)^2 +  \frac{k^4}{65536} \bigg( A(v,u)E^2(v,u)\\
        &\times  P^{R/L}_h(kv) P^{R/L}_h(ku) + I(v,u)G(v,u)F(v,u) \big \{  P^{R/L}_h(kv)P^{L/R}_h(ku) +  P^{L/R}_h(kv)\\
        &\times P^{R/L}_h(ku)  \big \}  + D(v,u)H^2(v,u) P^{L/R}_h(kv)P^{L/R}_h(ku)  \bigg )  \mathcal{I}^{h^{(2)}}_{tt1}(x,v,u)^2 -\frac{2k^2}{16384u^4v^4} \bigg ( \\
        &A(v,u) E(v,u) P^{R/L}_h(kv)P^{R/L}_h(ku) + B(v,u)F(v,u) P^{R/L}_h(kv)P^{L/R}_h(ku) + C(v,u)G(v,u)\\
        &\times P^{L/R}_h(kv)P^{R/L}_h(ku) + D(v,u)H(v,u) P^{L/R}_h(kv)P^{L/R}_h(ku)  \bigg )\mathcal{I}^{h^{(2)}}_{tt1}(x,v,u)\mathcal{I}^{h^{(2)}}_{tt2}(x,v,u)   \\
        &-\frac{k^2}{16384u^4v^4} \bigg (A(v,u) E(v,u) P^{R/L}_h(kv)P^{R/L}_h(ku) + I(v,u)F(v,u) P^{R/L}_h(kv)P^{L/R}_h(ku) \\
        & + I(v,u)G(v,u) P^{L/R}_h(kv)P^{R/L}_h(ku) + D(v,u)H(v,u)P^{L/R}_h(kv)P^{L/R}_h(ku) \bigg ) \mathcal{I}^{h^{(2)}}_{tt1}(x,v,u) \\
        &\times \mathcal{I}^{h^{(2)}}_{tt2}(x,v,u) -  \frac{k^2}{16384u^4v^4} \bigg ( A(v,u) E(v,u) P^{R/L}_h(kv)P^{R/L}_h(ku) + I(v,u)G(v,u)\\
        &\times P^{R/L}_h(kv)P^{L/R}_h(ku) +  I(v,u)F(v,u) P^{L/R}_h(kv)P^{R/L}_h(ku) + D(v,u)H(v,u)P^{L/R}_h(kv)\\
        &\times P^{L/R}_h(ku)\bigg ) \mathcal{I}^{h^{(2)}}_{tt1}(x,v,u) \mathcal{I}^{h^{(2)}}_{tt2}(x,v,u) + \frac{1}{4056u^4v^4} \bigg ( A(v,u)P^{R/L}_h(kv)P^{R/L}_h(ku) \\
        &+ B(v,u)P^{R/L}_h(kv)P^{L/R}_h(ku) + C(v,u)P^{L/R}_h(kv)P^{R/L}_h(ku) + D(v,u)P^{L/R}_h(kv)\\
        &\times P^{L/R}_h(ku)  \bigg ) \mathcal{I}^{h^{(2)}}_{tt2}(x,v,u)^2 \frac{1}{4056u^4v^4} \bigg ( A(v,u)P^{R/L}_h(kv)P^{R/L}_h(ku) + I(v,u) \\
        &\times \big \{  P^{R/L}_h(kv)P^{L/R}_h(ku) +  P^{L/R}_h(kv) P^{R/L}_h(ku)  \big \}  +  D(v,u)P^{L/R}_h(kv) P^{L/R}_h(ku) \bigg ) \\
        &\times \mathcal{I}^{h^{(2)}}_{tt2}(x,v,u)^2  \bigg ] \, .
    \end{split}
\end{equation}
And finally the dimensionless time-averaged power spectrum can be computed to be
\begin{equation}
    \begin{split}
        &\overline{\mathcal{P}_{h^{(2)}_{tt}}(\eta, k)} = 8 \int ^{\infty}_0 \rm  v \int ^{v+1}_{|v-1|}\rm u \, \, \frac{1}{u^6v^6} \bigg [ \bigg ( \mathcal{P}^R_h(kv)\mathcal{P}^R_h(ku)+\mathcal{P}^L_h(kv)\mathcal{P}^L_h(ku) \bigg )\\
        &\times \bigg ( \frac{k^4}{32768} \bigg \{ A(v,u)E^2(v,u) + D(v,u)H^2(v,u) \bigg \}\overline{\mathcal{I}^{h^{(2)}}_{tt1}(x,v,u)^2} - \frac{k^2}{4096} \bigg \{ \\
        &  A(v,u)E(v,u) + D(v,u)H(v,u)  \bigg \} \overline{\mathcal{I}^{h^{(2)}}_{tt1}(x,v,u)}\overline{\mathcal{I}^{h^{(2)}}_{tt2}(x,v,u)} + \frac{1}{2028} \bigg \{ A(v,u)\\
        &+D(v,u) \bigg \}\overline{\mathcal{I}^{h^{(2)}}_{tt2}(x,v,u)^2}  \bigg ) +  \bigg ( \mathcal{P}^R_h(kv)\mathcal{P}^L_h(ku)+\mathcal{P}^L_h(kv)\mathcal{P}^R_h(ku) \bigg ) \bigg ( \frac{k^4}{65536} \\
        &\times \bigg \{ B(v,u)F^2(v,u)+ C(v,u)G^2(v,u) + 2I(v,u)G(v,u)F(v,u) \bigg \}\overline{\mathcal{I}^{h^{(2)}}_{tt1}(x,v,u)^2}\\
        &-\frac{k^2}{8192} \bigg \{ B(v,u)F(v,u)+C(v,u)G(v,u)+ I(v,u)F(v,u)+I(v,u)G(v,u) \bigg \} \\
        &\times \overline{\mathcal{I}^{h^{(2)}}_{tt1}(x,v,u)}\overline{\mathcal{I}^{h^{(2)}}_{tt2}(x,v,u)} +\frac{1}{4056} \bigg \{ B(v,u) + C(v,u) + I(v,u) \bigg \} \overline{\mathcal{I}^{h^{(2)}}_{tt2}(x,v,u)^2}  \bigg )    \bigg ] \, ,
    \end{split}
\end{equation}
with the kernels defined as ($i=1,2$)
\begin{equation}\label{kernelTT}
    \mathcal{I}^{h^{(2)}}_{tti}(x,v,u) =   \frac{1}{k^2x^b} \int _{x _i}^{x} {\rm }  \barx \, \, \barx ^{2+b} \left ( y_b(x)j_b(\barx)- j_b(x)y_b(\barx) \right )f^{h^{(2)}}_{tti}(\barx ,v , u) \, ,
\end{equation}
and the functions $f^{h^{(2)}}_{tt1}(\barx ,v , u)$ and $f^{h^{(2)}}_{tt2}(\barx ,v , u)$ defined respectively in Eqs.~\eqref{fh2tt1} and \eqref{fh2tt2} are
\begin{subequations}
    \begin{align}
        f^{h^{(2)}}_{tt1}(\barx ,v , u) &= T_h(\barx v) T_h(\barx u) \, ,\\
        f^{h^{(2)}}_{tt2}(\barx ,v , u) &= k^2 \frac{\partial T_h(\barx v)}{\partial \barx} \frac{\partial T_h(\barx u)}{\partial \barx} \, .
    \end{align}
\end{subequations}

In an RD universe, the kernels are evaluated to be
\cite{Picard:2023sbz}
{
\allowdisplaybreaks
\begin{align}
    &\overline{\mathcal{I}_{tt,1}^2(v,u)} = x^2\overline{\mathcal{I}_{tt,1}^2(x\rightarrow \infty,v,u)} = \frac{1}{32k^4u^2v^2}\left(\pi ^2 \Theta (u+v-1) + \log \bigg | \frac{1-(u+v)^2}{1-(u-v)^2} \bigg |^2 \right) \,,\\
    &\overline{\mathcal{I}_{tt,2}^2(v,u)} = x^2\overline{\mathcal{I}_{tt,2}^2(x\rightarrow \infty,v,u)} = \frac{k^4}{128u^2v^2}\bigg ( \pi ^2(u^2+v^2-1)^2\Theta (u+v-1) \nonumber \\
    &+ \left (\log \bigg | \frac{1-(u+v)^2}{1-(u-v)^2} \bigg |-4vu \right )^2 \bigg ) \,,\\ 
    &\overline{\mathcal{I}_{tt,1}(v,u)\mathcal{I}_{tt,2}(v,u)} = \frac{k^2}{64u^2v^2}\bigg (\pi ^2(u^2+v^2-1)\Theta (u+v-1) + \log \bigg | \frac{1-(u+v)^2}{1-(u-v)^2} \bigg | \nonumber \\
    & \bigg (-4vu+\log \bigg | \frac{1-(u+v)^2}{1-(u-v)^2} \bigg |\bigg ) \bigg )\,.
\end{align}
}
If we have no initial primordial parity violation, the spectral density in an RD universe is
\begin{equation}\label{omegaTTRD}
    \begin{split}
        &\Omega (\eta , k) = \frac{4}{3} \int ^{\infty}_0 \rm  v \int ^{v+1}_{|v-1|}\rm u \, \, \frac{1}{u^6v^6} \bigg [  \frac{k^4}{32768} \bigg ( A(v,u)E^2(v,u) + D(v,u)H^2(v,u) \bigg ) \\
        & \times \overline{\mathcal{I}^{h^{(2)}}_{tt1}(x,v,u)^2} - \frac{k^2}{4096} \bigg ( A(v,u)E(v,u) + D(v,u)H(v,u)  \bigg ) \overline{\mathcal{I}^{h^{(2)}}_{tt1}(x,v,u)}\\
        &\times \overline{\mathcal{I}^{h^{(2)}}_{tt2}(x,v,u)} + \frac{1}{2028} \bigg ( A(v,u)+D(v,u) \bigg )\overline{\mathcal{I}^{h^{(2)}}_{tt2}(x,v,u)^2}   +    \frac{k^4}{65536} \\
        &\times \bigg ( B(v,u)F^2(v,u)+ C(v,u)G^2(v,u) + 2I(v,u)G(v,u)F(v,u) \bigg )\overline{\mathcal{I}^{h^{(2)}}_{tt1}(x,v,u)^2}\\
        &-\frac{k^2}{8192} \bigg ( B(v,u)F(v,u)+C(v,u)G(v,u)+ I(v,u)F(v,u)+I(v,u)G(v,u) \bigg ) \\
        &\times \overline{\mathcal{I}^{h^{(2)}}_{tt1}(x,v,u)}\overline{\mathcal{I}^{h^{(2)}}_{tt2}(x,v,u)} +\frac{1}{4056} \bigg ( B(v,u) + C(v,u) + I(v,u) \bigg ) \overline{\mathcal{I}^{h^{(2)}}_{tt2}(x,v,u)^2}     \bigg ] \\
        &\times \mathcal{P}_h(kv)\mathcal{P}_h(ku) \, .
    \end{split}
\end{equation}
The kernels could enter a logarithmic divergence when $v+u=1$; however, we can show that this is not the case. By considering momentum conservation, we can extract a range of allowed values for $v$ and $u$. The wave vectors in Fourier space of the three modes form a triangle, with sides: $\mathbf{k}$, $\mathbf{p}$ and $\mathbf{k}-\mathbf{p}$, where the magnitude of the vectors satisfy the triangle inequality, which in terms of the variables $v$ and $u$ is given by: $|u-v|<1<u+v$. Hence, we see that the tensor-tensor spectral density will not enter a logarithmic resonance since the values of $v$ and $u$ needed do not satisfy the triangle inequality. We also note that the tensor-tensor integrand is finite in the IR limit but in the UV limit goes as $v^{-4}$, hence, once again, we will only consider an input peaked tensor power spectrum. 

\begin{figure}
    \centering
    \includegraphics[width=0.8\linewidth]{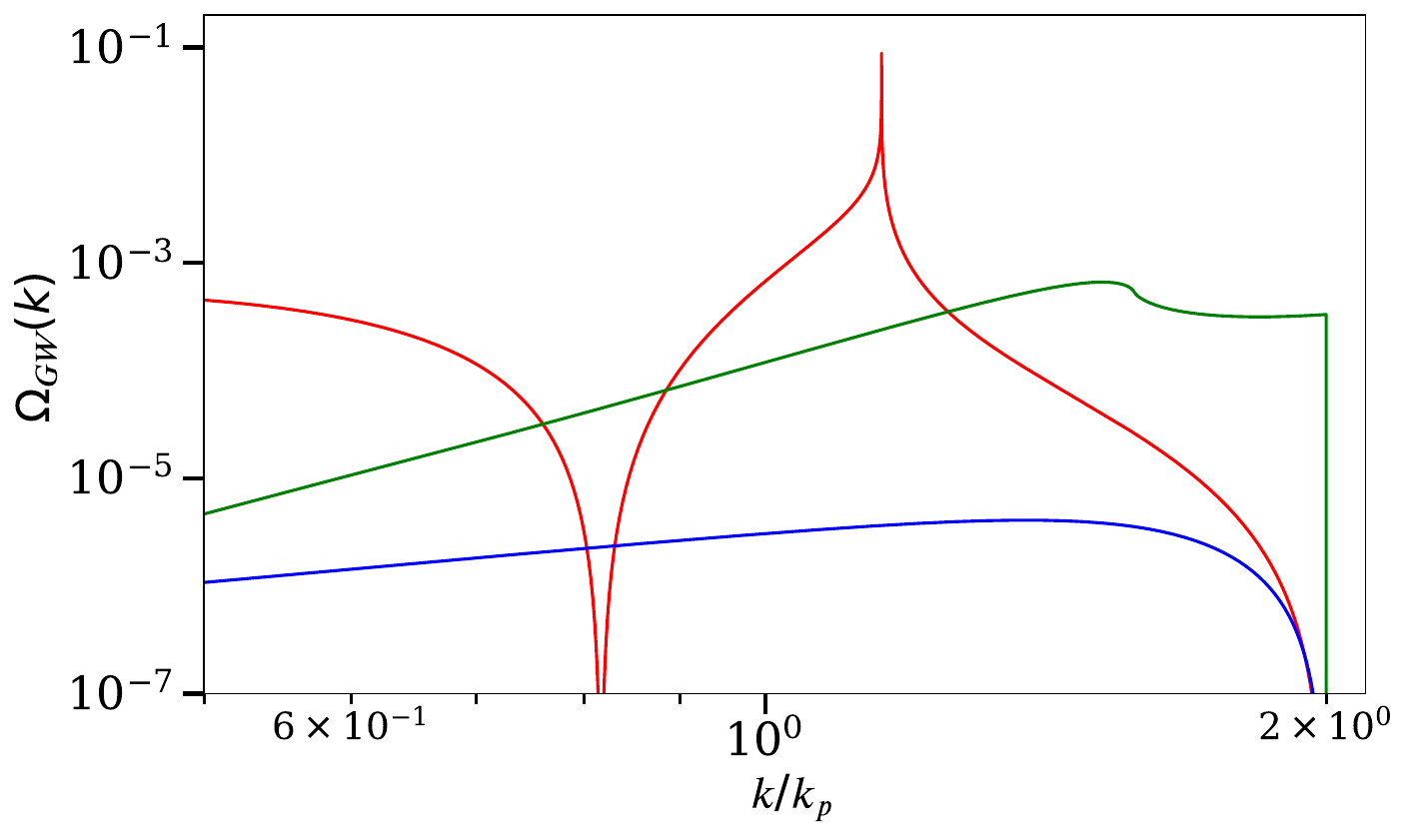}
    \caption{\footnotesize{The spectral density tensor-tensor iGWs (solid blue) in a RD universe with input Dirac delta power spectra  SIGWs against a range of $k$. For reference, we have included the spectral density of SIGWs (solid red) and scalar-tensor iGWs (solid green). The tensor-tensor is sub-dominant in amplitude at all scales.}}
    \label{fig:DiracDeltaTT}
\end{figure}

For the Dirac peaked spectra, the tensor-tensor contribution reduces to 
\begin{equation}
    \begin{split}
        \Omega (\eta , k) &= \frac{4}{3}A_h^2 \tilde{k}^{10} \bigg ( \frac{(-4+\tilde{k}^2)^2 (512+1536\tilde{k}^2-336\tilde{k}^4-40\tilde{k}^6 + 9\tilde{k}^8)}{8192 \tilde{k}^{12}}\\
        &\times k^4  \overline{\mathcal{I}^{h^{(2)}}_{tt1}(v=\tilde{k}^{-1},u=\tilde{k}^{-1})^2} + \frac{(-4+\tilde{k}^2)^2 (64+96\tilde{k}^2+7\tilde{k}^4)}{4056 \tilde{k}^{8}} \\
        &\times \overline{\mathcal{I}^{h^{(2)}}_{tt2}(v=\tilde{k}^{-1},u=\tilde{k}^{-1})^2} +k^2\overline{\mathcal{I}^{h^{(2)}}_{tt1}(v=\tilde{k}^{-1},u=\tilde{k}^{-1})} \\
        &\times \overline{\mathcal{I}^{h^{(2)}}_{tt2}(v=\tilde{k}^{-1},u=\tilde{k}^{-1})} \frac{(-2+\tilde{k})^2(2+\tilde{k})^2(-64-136\tilde{k}^2+10\tilde{k}^4+3\tilde{k}^6)}{1024\tilde{k}^{10}}  \bigg ) \\
        &\times \Theta(2k_p-k) \, ,
    \end{split}
\end{equation}
and is plotted in Fig.~\ref{fig:DiracDeltaTT}. 
We conclude that the tensor-tensor term is subdominant and has no special features (such as a resonant peak).

%%%%%%%%%%%%%%%%%%%%%%%%%%%%%%%%%%%%%%%%%%%%%%%%%%%%%%%
\subsection{All contributions for peaked input spectra in an RD universe}

\begin{figure}[t]
    \centering
    \begin{subfigure}[b]{0.49\textwidth}
        \includegraphics[width=\linewidth]{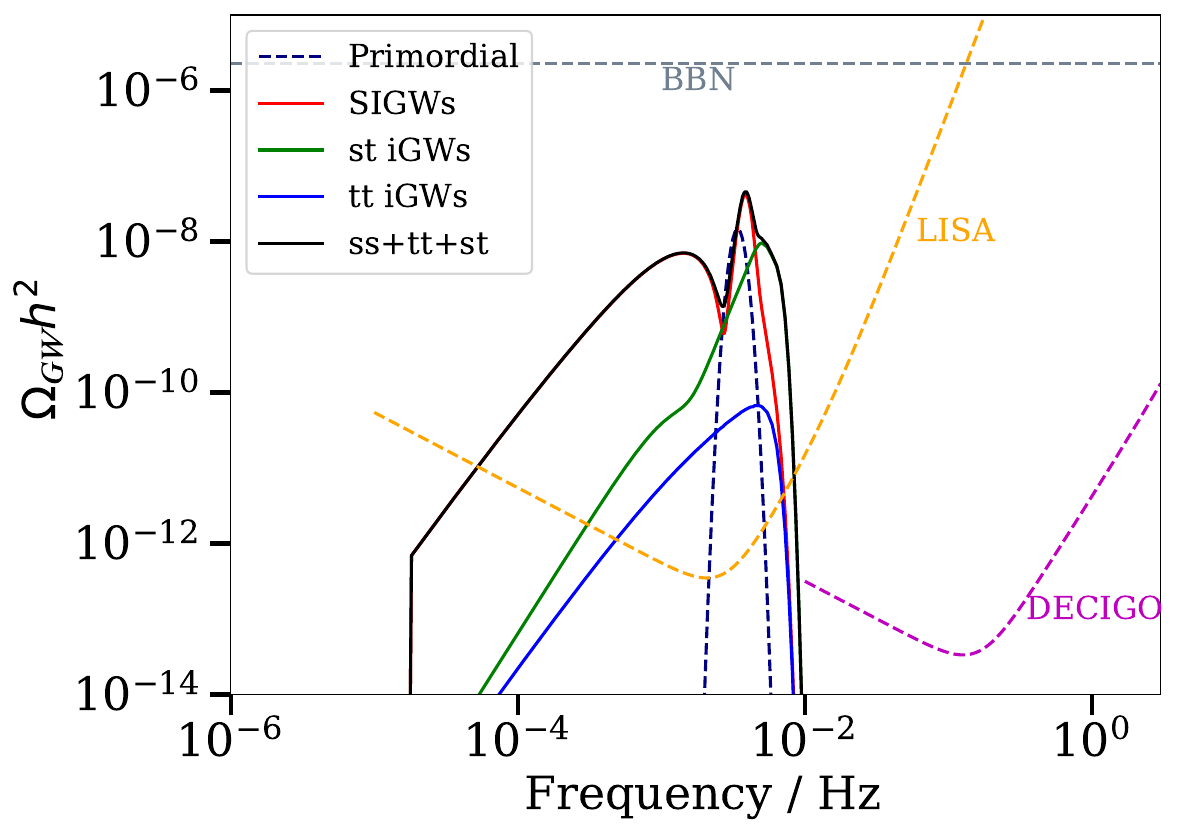}
    \end{subfigure}
    \begin{subfigure}[b]{0.49\textwidth}
        \includegraphics[width=\linewidth]{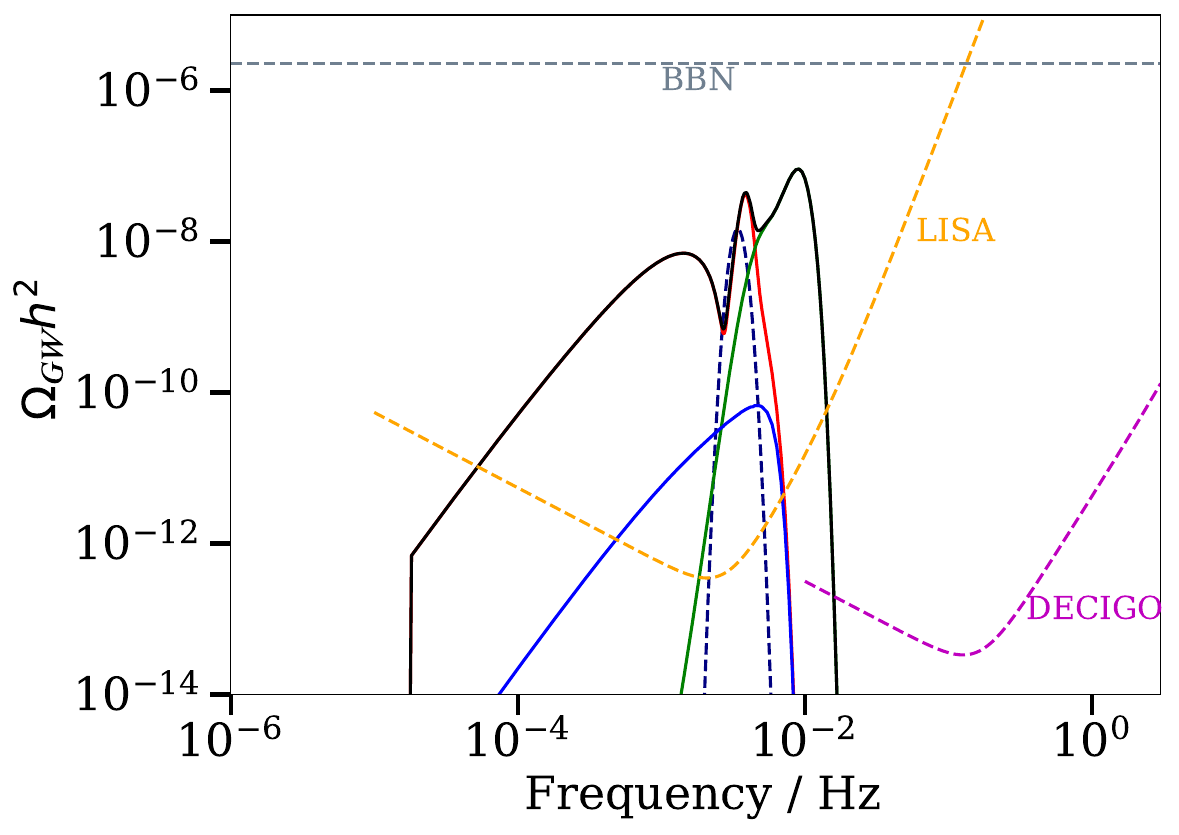}
    \end{subfigure}
 
    \caption{\footnotesize{Plots of the spectral density against frequency. Each graph corresponds to a different setup in the location of the primordial tensor: $k_\zeta=k_h$ (\textit{Left}) $k_\zeta=4c_sk_h$ (\textit{Right}). The dashed dark blue line is the primordial tensor power spectrum. We see that the scalar-tensor contribution is non-negligible. The sharp signals on larger scales and below the threshold for detection are artefacts of numerical integration.}}
    \label{fig:lognormal_all}
\end{figure}

We conclude this chapter by looking at all contributions for a peaked Log-normal input power spectrum for both scalar and tensor perturbations
\begin{equation}
\label{lognorm_def}
    \mathcal{P}_{\zeta , h}= \frac{\mathcal{A}_{\zeta , h}}{\sqrt{2\pi}\sigma}\exp \left ( {-\frac{\log ^2(k/k_{\zeta , h})}{2\sigma ^2}}\right )\,.
\end{equation}
For the reasons outlined previously, we choose to work with a narrowly peaked input power spectra: $\sigma=0.1$. In Fig.~\ref{fig:lognormal_all} we present the results of our toy model for different scenarios. 
When the primordial scalar and tensor spectra peak at the same scale, the scalar-tensor contribution becomes the dominant contribution. For the case when $k_h = 4c_sk_{\zeta}$, we can still distinguish the resonant peak and also have an extra contribution coming from the scalar-tensor terms, in the form of a peak.

In summary, we have shown in this chapter how lower-order terms can couple to generate second-order gravitational waves. Specifically, we looked at how adding first-order tensor perturbations affects the spectral density of GWs. Two new sources appear: terms quadratic in tensor fluctuations and terms that couple a scalar and a tensor fluctuation. We have shown that the tensor-tensor contribution is subdominant due to its inherent structure, and for this reason, we will neglect this term in what follows. On the other hand, the scalar-tensor contribution is non-negligible at smaller scales. In particular, we showed that if there is a peak in the first order tensor power spectrum, then the overall spectral density will be different to waves induced by quadratic scalar fluctuations, particularly on small scales where the scalar-tensor waves can become dominant. Since the power spectrum for first-order scalars and tensors is directly linked to inflation, these results can be used to constrain models of inflation which enhance the primordial scalar and tensor power spectrum.

Due to the behaviour of source terms including tensors in the UV limit, we limited our analysis to waves induced by peaked spectra. We argued that by putting all the power at one scale, any divergent behaviour in the UV limit will not be a problem, since non-peak scales are sufficiently suppressed. There are other contributions to the spectral density at the same order in perturbation theory coming from the correlation of third-order tensor modes correlated with primordial tensors. These extra contributions will be the object of Chap.~\ref{Chap:Third}, where we will study these contributions to determine whether they cancel out UV enhancements.

So far, our analysis has been centred around Gaussian initial conditions. There has also been plenty of interest in the possible impact of primordial non-Gaussianity (PNG) on any potentially observable SIGW signal~\cite{Cai:2018dig,Unal:2018yaa,Yuan:2020iwf,Atal:2021jyo,Adshead:2021hnm,Ragavendra:2020sop,Zeng:2024ovg,Pi:2024jwt,Perna:2024ehx,Li:2023xtl,Yuan:2023ofl,He:2024luf,Papanikolaou:2024kjb}. PNG arises when the higher-order correlations of the curvature perturbation become non-zero, and can be seen as a measure of the interactions of the inflaton field beyond the free-field dynamics. This means that any detection of PNG would be particularly enlightening for cosmology, since it would tell us a wealth of information about the inflationary action beyond second order. The impact of PNG on potentially observable GW signals is particularly pertinent since models of inflation capable of generating enhanced peaks, and therefore PBHs and SIGWs, typically involve curvature perturbations with non-Gaussian statistics, see e.g. Refs.~\cite{Cai:2021zsp,Davies:2021loj,Taoso:2021uvl,Rezazadeh:2021clf}. So far, the majority of works considering non-Gaussian scalar curvature perturbations have modelled non-Gaussianity as a local expansion\footnote{See e.g. Refs~\cite{Garcia-Saenz:2022tzu,Ragavendra:2021qdu} for works that consider PNG beyond a local expansion.}. It has been found that the introduction of scalar non-Gaussianity typically results in the smoothing of peaks and the appearance of `knees' at higher frequencies in the GW spectrum~\cite{Unal:2018yaa,Atal:2021jyo,Adshead:2021hnm,Pi:2024jwt,Perna:2024ehx,Li:2023xtl}. Since we would expect sizeable scalar non-Gaussianity to accompany an enhanced scalar power spectrum, the contribution to the iGW spectrum from scalar-tensor interactions is likely to be affected by PNG. Inclusion of PNG yields a new non-Gaussian term in the scalar-tensor induced GW spectrum. In the following Chapter, we will analyse the behaviour of the new term and compare the effect of this contribution to the non-Gaussian SIGWs terms studied previously.

% % % % % % % % % % % % % % % % % % % % % % % % % % % % % % % % 
\chapter{Non-Gaussian scalar contributions to scalar-tensor induced gravitational waves}
\label{Chap:NG}
In this Chapter, we will analyse the effects of scalar PNG on the spectral density of scalar-tensor iGWS. Our starting point is Eq.~\eqref{2pointSOGWs}, without the tensor-tensor contribution. If we now don't assume that primordial scalar fluctuations are drawn from some Gaussian distribution, then schematically 
\begin{equation}\label{PSbreakup}
    \langle h^{(2)}_{\lambda}(\eta, \mathbf{k})h^{(2)}_{\lambda ^{\prime}}(\eta, \mathbf{k^{\prime}}) \rangle \sim  \langle \mathcal{R}_{\mathbf{p}} \mathcal{R}_{\mathbf{k}-\mathbf{p}} \mathcal{R}_{\mathbf{p^{\prime}}} \mathcal{R}_{\mathbf{k^{\prime}}-\mathbf{p^{\prime}}} \rangle + \langle h_{\mathbf{p}}^{\lambda _1} \mathcal{R}_{\mathbf{k}-\mathbf{p}}h^{\lambda _1 ^{\prime}}_{\mathbf{p^{\prime}}}\mathcal{R}_{\mathbf{k^{\prime}}-\mathbf{p^{\prime}}}  \rangle\,.
\end{equation}
Since the scalar fluctuations are no longer Gaussian, the four-point function involving purely scalar perturbations can no longer be decomposed into products of two-point functions. Instead, it gives rise to a `connected' contribution (the trispectrum) and a `disconnected' contribution\footnote{These names can be interpreted in terms of the corresponding interactions when represented through Feynman diagrams, see Ref.~\cite{Adshead:2021hnm, Picard:2024ekd} for more details.}
\begin{equation}
\label{condiscon}
    \langle \mathcal{R}_{\mathbf{p}} \mathcal{R}_{\mathbf{k}-\mathbf{p}} \mathcal{R}_{\mathbf{p^{\prime}}} \mathcal{R}_{\mathbf{k^{\prime}}-\mathbf{p^{\prime}}} \rangle = \langle \mathcal{R}_{\mathbf{p}} \mathcal{R}_{\mathbf{k}-\mathbf{p}} \mathcal{R}_{\mathbf{p^{\prime}}} \mathcal{R}_{\mathbf{k^{\prime}}-\mathbf{p^{\prime}}} \rangle _c + \langle \mathcal{R}_{\mathbf{p}} \mathcal{R}_{\mathbf{k}-\mathbf{p}} \mathcal{R}_{\mathbf{p^{\prime}}} \mathcal{R}_{\mathbf{k^{\prime}}-\mathbf{p^{\prime}}} \rangle _d \,.
\end{equation}
The connected part is
\begin{equation}
    \langle \mathcal{R}_{\mathbf{p}} \mathcal{R}_{\mathbf{k}-\mathbf{p}} \mathcal{R}_{\mathbf{p^{\prime}}} \mathcal{R}_{\mathbf{k^{\prime}}-\mathbf{p^{\prime}}} \rangle _c = \delta ^{3}\left ( \mathbf{k}+\mathbf{k^{\prime}} \right ) \mathcal{T}\left (\mathbf{k},\mathbf{k}-\mathbf{p}, \mathbf{k^{\prime}},\mathbf{k^{\prime}}-\mathbf{p^{\prime}} \right ) \,,
\end{equation}
with $\mathcal{T}(\mathbf{k}_1,\mathbf{k}_2,\mathbf{k}_3,\mathbf{k}_4)$ the connected trispectrum, whilst the disconnected part can be broken down further and simplified as
\begin{equation}
    \langle \mathcal{R}_{\mathbf{p}} \mathcal{R}_{\mathbf{k}-\mathbf{p}} \mathcal{R}_{\mathbf{p^{\prime}}} \mathcal{R}_{\mathbf{k^{\prime}}-\mathbf{p^{\prime}}} \rangle _d = \langle \mathcal{R}_{\mathbf{p}} \mathcal{R}_{\mathbf{p^{\prime}}} \rangle\langle \mathcal{R}_{\mathbf{k}-\mathbf{p}} \mathcal{R}_{\mathbf{k^{\prime}}-\mathbf{p^{\prime}}} \rangle + \langle \mathcal{R}_{\mathbf{p}} \mathcal{R}_{\mathbf{k^{\prime}}-\mathbf{p^{\prime}}}  \rangle \langle \mathcal{R}_{\mathbf{k}-\mathbf{p}} \mathcal{R}_{\mathbf{p^{\prime}}} \rangle \,.
\end{equation}
The scalar-tensor correlation is inherently disconnected and so the second term in Eq.~\eqref{PSbreakup} is broken down as 
\begin{equation}
  \langle h_{\mathbf{p}}^{\lambda _1} \mathcal{R}_{\mathbf{k}-\mathbf{p}}h^{\lambda _1 ^{\prime}}_{\mathbf{p^{\prime}}}\mathcal{R}_{\mathbf{k^{\prime}}-\mathbf{p^{\prime}}}  \rangle _d = \langle h_{\mathbf{p}}^{\lambda _1} h_{\mathbf{p^{\prime}}}^{\lambda _1 ^{\prime}} \rangle \langle \mathcal{R}_{\mathbf{k}-\mathbf{p}} \mathcal{R}_{\mathbf{k^{\prime}}-\mathbf{p^{\prime}}} \rangle \,.
\end{equation}

We model the primordial scalar non-Gaussianity as a local-type expansion of the scalar curvature perturbation field in real space
\begin{equation}
\label{localNGexpansion}
    \mathcal{R}(\mathbf{x}) = \mathcal{R}_\textrm{G}(\mathbf{x}) + F_{\textrm{NL}} \left (\mathcal{R}^2_\textrm{G}(\mathbf{x}) - \langle \mathcal{R}^2_\textrm{G}(\mathbf{x}) \rangle \right ) \, ,
\end{equation}
where $\mathcal{R}_\textrm{G}(\mathbf{x})$ is a scalar curvature perturbation with Gaussian statistics (the one we used in the previous chapter), and $F_{\textrm{NL}}$ is the local non-Gaussianity parameter. Other works have considered the effects of expanding the curvature perturbation to higher orders ($\mathcal{O}(G_{\textrm{NL}})$) and beyond~\cite{Perna:2024ehx,Zeng:2024ovg,Pi:2024jwt,Li:2023xtl,Yuan:2023ofl}, but we will restrict ourselves here to just $\mathcal{O}(F_{\textrm{NL}})$\footnote{Technically, this is an infinite series expansion
\begin{equation*}
    \begin{split}
        \mathcal{R}(\mathbf{x}) = \mathcal{R}_\textrm{G}(\mathbf{x}) + F_{\textrm{NL}} \left (\mathcal{R}^2_\textrm{G}(\mathbf{x}) - \langle \mathcal{R}^2_\textrm{G}(\mathbf{x}) \rangle \right ) + G_{\textrm{NL}}\,\mathcal{R}^3_\textrm{G}(\mathbf{x}) + H_{\textrm{NL}} \left (\mathcal{R}^4_\textrm{G}(\mathbf{x}) - 3\langle \mathcal{R}^2_\textrm{G}(\mathbf{x}) \rangle ^2 \right ) + ... 
    \end{split}
\end{equation*}}. Including higher-order non-Gaussian parameters leads to the generation of many extra contributions to the iGW spectrum.  At present, the constraints on non-Gaussianity are fairly weak; the most recent Planck data from CMB measurements give the local non-Gaussianity as $f_{\textrm{NL}}^{\textrm{local}}=-0.9\pm5.1$~\cite{Planck:2019kim}\footnote{This is related to the local non-Gaussianity parameter as $F_{\textrm{NL}}=\frac{3}{5}f_{\textrm{NL}}^{\textrm{local}}$.}. However, there is a lot of freedom in its magnitude on short scales. Indeed, since we will consider toy models which are peaked with large amplitudes on short scales, we expect that the magnitude of PNG may be considerably larger than CMB constraints on these scales. In Fourier space, the local-type expansion of $\mathcal{R}(\mathbf{x})$ shows up as a correction to the Gaussian power spectrum
\begin{equation}
\label{NGPS}
    P_{\mathcal{R}}(k) = P_{\mathcal{R},\textrm{G}}(k) + 2F_{\textrm{NL}}^2 \int \frac{\rm ^3\mathbf{p}}{(2\pi)^{3}}P_{\mathcal{R},\textrm{G}}(p)P_{\mathcal{R}, \textrm{G}}(|\mathbf{k}-\mathbf{p}|)  \,,
\end{equation}
with $ P_{\mathcal{R},\textrm{G}}(k)$ being the power spectrum of the Gaussian scalar perturbations from inflation\footnote{In this chapter, we distinguish between the power spectrum, $P_{\mathcal{R}}(k)$, and the Gaussian power spectrum, $P_{\mathcal{R},\textrm{G}}(k)$, we used in the previous chapter}. The effects of PNG on SIGWs and all the relevant expressions can be found in Ref.~\cite{Adshead:2021hnm}. In what follows, we are only concerned with computing the effect of PNG on scalar-tensor iGWs. 

%%%%%%%%%%%%%%%%%%%%%%%%%%%%%%%%%%%%%%%%%%%%%%%%%%%%%%%%%%%%%%%%%%%%%%%%%%%%%%%%%%%%%%%%%%%%%%%%%%%%%%%%%%%%%%%%%%%%%%
\section{The non-Gaussian scalar-tensor contribution}

If we substitute the expansion in Eq.~\eqref{NGPS} into the expression for the power spectrum of scalar-tensor iGWs from Eq.~\eqref{stpowerspectrum}, we will get two contributions: the Gaussian contribution, which was the object of focus in the last chapter, and a new contribution proportional to $F_{\text{NL}}^2$
\begin{align} \label{stHS}
    \begin{split}
        P^{\lambda}_{h^{(2)}_{st}}(\eta , k)_{\textrm{HST}}&= 32 F_{\textrm{NL}}^2 \sum_{\lambda _1} \int \frac{\rm ^3\mathbf{p}}{(2\pi)^{3}}Q^{st}_{\lambda , \lambda _1}(\mathbf{k},\mathbf{p}) Q^{st}_{\lambda , \lambda _1}(-\mathbf{k},-\mathbf{p})I^{st}(p,|\mathbf{k}-\mathbf{p}|,\eta)^2 \\
        &\times  P_{h}^{\lambda _1}(p) \int \frac{\rm ^3\mathbf{q}}{(2\pi)^{3}} P_{\mathcal{R},\textrm{G}}(q)P_{\mathcal{R},\textrm{G}}(|\mathbf{k}-\mathbf{p}-\mathbf{q}|) \, .
    \end{split}
\end{align}
The subscript “HST” denotes hybrid scalar–tensor and is consistent with the nomenclature commonly used in the literature \cite{Adshead:2021hnm,Unal:2018yaa}. The evaluation of the polarisation functions and kernel is identical to the case of Gaussian initial conditions since only the shape of the scalar power spectrum has been modified. The integral over the unconstrained momenta $\mathbf{p}$ is also identical, but for the integral over $\mathbf{q}$, we have to introduce extra coordinates. For this reason, the $v$ and $u$ coordinates defined in Eq.~\eqref{uvcoordinates} are relabelled respectively as $v_1$ and $u_1$, and we further introduce
\begin{equation}
    v_2=\frac{q}{|\mathbf{k}-\mathbf{p}|}\,, \quad u_2=\frac{|\mathbf{k}-\mathbf{p}-\mathbf{q}|}{|\mathbf{k}-\mathbf{p}|}\,.
\end{equation}
We transform to the coordinates $(u_1,v_1,u_2,v_2, \phi_p,\phi_q)$ and integrate over the azimuthal angles immediately, since the scalar-tensor NG term is disconnected and has no dependence on them. This generates a factor of $(2\pi)^2$. We then express the polarisation functions in terms of $u_1$ and $v_1$, as we did for the Gaussian scalar-tensor term, and arrive at 
\begin{align} \label{srHSspherical}
    \begin{split}
        P_{R/L}^{st}(\eta , k)_{\textrm{HST}}&= \frac{1}{8} F_{\textrm{NL}}^2 \frac{k^3}{(2\pi)^2} \int ^{\infty}_0 \,\rm v_1\, \int ^{v_1+1}_{|v_1-1|} \,\rm u_1 \, \frac{u_1}{v_1^3} \bigg [ \left (u_1^2-(v_1+1)^2 \right )^4 P^{R/L}_{h}(kv_1) \\ 
        &+ \left (u_1^2-(v_1-1)^2 \right )^4 P^{L/R}_{h}(kv_1)  \bigg ] \mathcal{I}^{st}(x,v_1,u_1)^2 \\
        &\times  \frac{k^3}{(2\pi)^2} \int ^{\infty}_0 \,\rm v_2 \int ^{v_2+1}_{|v_2-1|} \, \rm u_2 \, u_1^3v_2u_2 P_{\mathcal{R},\textrm{G}}(ku_1v_2)P_{\mathcal{R},\textrm{G}}(ku_1u_2) \, .
    \end{split}
\end{align}
Since we will evaluate these terms numerically, we transform coordinates (see Eq.~\eqref{stcoordinates}) and introduce
\begin{subequations}
    \begin{align}
        s_1&=u_1-v_1 \, , \quad t_1 = u_1+v_1+1 \, , \\
        s_2&=u_2-v_2 \, , \quad t_2 = u_2+v_2+1 \, 
    \end{align}
\end{subequations}
so that the power spectrum can be re-expressed as 
\begin{equation} \label{srHSspherical}
    \begin{split}
        P_{R/L}^{st}(\eta , k)_{\textrm{HST}}&= \frac{1}{32} F_{\textrm{NL}}^2 \frac{k^3}{(2\pi)^2} \int ^{\infty}_0 \,\rm t_1\, \int ^{1}_{-1} \,\rm s_1 \, \frac{u_1}{v_1^3} \bigg [ \left (u_1^2-(v_1+1)^2 \right )^4 P^{R/L}_{h}(kv_1) \\ 
        &+ \left (u_1^2-(v_1-1)^2 \right )^4 P^{L/R}_{h}(kv_1)  \bigg ] \mathcal{I}^{st}(x,v_1,u_1)^2 \\
        &\times  \frac{k^3}{(2\pi)^2} \int ^{\infty}_0 \,\rm t_2 \int ^{1}_{-1} \, \rm s_2 \, u_1^3v_2u_2 P_{\mathcal{R},\textrm{G}}(ku_1v_2)P_{\mathcal{R},\textrm{G}}(ku_1u_2) \, ,
    \end{split}
\end{equation}
where the integrand is written in terms of the coordinates $(u_1,v_1,u_2,v_2)$ for simplicity. Finally, the time-averaged power spectrum is 
\begin{equation}
\label{HybridScalar1}
    \begin{split}
       \overline{\mathcal{P}^{st}(\eta , k)_{\textrm{HST}}}= \frac{F_{\textrm{NL}}^2}{128}  &\int ^{\infty}_0 \, \rm t_1 \, \int ^{1}_{-1} \, \rm s_1 \, \int ^{\infty}_0 \, \rm t_2 \, \int ^{1}_{-1} \, \rm s_2 \, \frac{1}{v_1^6u_1^2v_2^2u_2^2}\\
       &\times \left[ \left (u_1^2-(v_1+1)^2 \right )^4 + \left (u_1^2-(v_1-1)^2 \right )^4 \right] \\ 
       &\times \left ( \mathcal{P}^{R}_{h}(kv_1)  + \mathcal{P}^{L}_{h}(kv_1) \right ) \overline{\mathcal{I}^{st}(x,v_1,u_1)^2} \mathcal{P}_{\mathcal{R},\textrm{G}}(ku_1v_2)\mathcal{P}_{\mathcal{R},\textrm{G}}(ku_1u_2)  \,.
    \end{split}
\end{equation}
Eq.~\eqref{HybridScalar1} is the new contribution, and we shall study its contribution to the spectral density in the following section. 

%%%%%%%%%%%%%%%%%%%%%%%%%%%%%%%%%%%%%%%%%%%%%%%%%%%%%%%%%%%%%%%%%%%%%%%%%%%%%%%%%%%%%%%%%%%%%%%%%%%%%%%%%%%%%%%%%%%
\section{Results in a radiation-dominated universe for peaked input spectra}
In this section, we present numerically evaluated examples of iGW spectra, including Gaussian and non-Gaussian scalar-scalar and scalar-tensor contributions. Through this, we may ascertain the impact of including non-Gaussian scalar-tensor contributions relative to the non-Gaussian scalar-scalar contributions that have already been studied. Again, we specialise in the concrete case of peaked scalar and tensor primordial power spectra in an RD universe, since we recall this is the case most relevant for PBH production and the observation prospects of iGWs. Strongly peaked sources are also the easiest cases to gain a semi-analytic handle on. We highlight the key features of the GW spectrum contribution it generates by combining numerical evaluations of the term for Gaussian-shaped primordial power spectra with semi-analytic results for the limiting case of a monochromatic spectrum. These features include: a sharp peak when the scalar and tensor primordial spectra peak at the same scale, a distinctive IR scaling, and a cut-off of its UV tail. We also compare the new term's GW signature to some of those studied already in the literature. To this end, in addition to computing the Gaussian and $\mathcal{O}(F_{\textrm{NL}}^2)$ scalar-tensor contributions to the second-order iGW spectrum, we also evaluate all terms to order $\mathcal{O}(F_{\textrm{NL}}^4)$ in the scalar-scalar sector. Expressions for these terms have been derived and studied extensively already in Refs.~\cite{Adshead:2021hnm, Perna:2024ehx, Picard:2024ekd}. 

We recall that in an RD universe, the kernel we need to input in Eq.~\eqref{HybridScalar1} is the same as the Gaussian case\footnote{Since the kernels and polarisation functions remain the same as in the scalar–tensor Gaussian case, the UV problem also persists in this new contribution. This further motivates our focus on peaked spectra in this chapter. Further details can be found in Ref.~\cite{Picard:2024ekd}.}, i.e. Eq.~\eqref{kernelSTRD}.

%%%%%%%%%%%%%%%%%%%%%%%%%%%%%%%%%%%%%%%%%%%%%%%%%%%%%%%%%%
\subsection{The case of a monochromatic input power spectrum} \label{UVcutoff}

The behaviour of this contribution in the UV can be derived from the consideration of its form in the limiting case of monochromatic primordial scalar and tensor power spectra. 

We assume a monochromatic peak at the scale $k_p$ for both the scalar and tensor power spectra, see Eq.~\eqref{diracdeltascalars} and Eq.~\eqref{diracdelatatensors}. For now, in the following analysis, we will assume parity is not violated, i.e. $\mathcal{A}^{R}_h=\mathcal{A}^{L}_h=\mathcal{A}_h$. For this case, the Gaussian contribution is in Eq.~\eqref{STDDsp}. While an analytic form for the Gaussian scalar-tensor contribution can be found for the case of monochromatic spectra, this is not the case for the non-Gaussian scalar-tensor terms. This is because, while in the Gaussian case we had two integrals over two Dirac-deltas, we now have four integrals over just three Dirac-deltas. It is still possible, however, to simplify the expression for the scalar-tensor NG term by performing three integrals and then studying the behaviour of what remains. 

Starting with Eq.~\eqref{HybridScalar1}, we see that, for monochromatic input spectra, the integrand will contain a product of three Dirac-deltas
\begin{equation}
    \delta\left(\ln\frac{k(1-s_1+t_1)}{2k_p}\right)\delta\left(\ln\frac{k(1+s_1+t_1)(1-s_2+t_2)}{4k_p}\right)\delta\left(\ln\frac{k(1+s_1+t_1)(1+s_2+t_2)}{4k_p}\right)\,.
\end{equation}
By integrating over the variables $t_1,t_2,s_2$, and keeping in mind the limits of the integration, it can be shown that the contribution acquires an overall factor of a product of Heaviside functions
\begin{equation}
\label{GSTcutoff}
    \Theta\left(\frac{k_p-k s_1}{k_p+k s_1}\right)\Theta\left(-1+\frac{2k_p}{k}+s_1\right)\,.
\end{equation}
We can find the largest value of $k$ at which this combination is non-zero by setting the arguments of each Heaviside function to zero and solving for $k$
\begin{equation}
    k=\frac{2k_p}{1-s_1}\, \quad \textrm{or}\, \quad k=\frac{k_p}{s_1}\,.
\end{equation}
The maximum value of $k$ will occur for the value of $s_1$ at which these equations are simultaneously satisfied. This is when $s_1=1/3$. For this value of $s_1$, the Heaviside functions both become zero at $k= 3k_p$. This means that the scalar-tensor NG term will vanish for $k \geq 3k_p$. This is a different UV cutoff to the Gaussian scalar-tensor term~\eqref{GSTcutoff}, but coincides with the non-Gaussian scalar-scalar contributions' cutoffs~\cite{Adshead:2021hnm}. For broader input power spectra, the contribution to the spectrum from these terms will typically extend past their limiting-case UV cutoffs, but maintain their hierarchy. Terms with a UV cutoff at larger momenta in the monochromatic case will dominate at higher $k$-values over terms with a smaller UV cutoff.

%%%%%%%%%%%%%%%%%%%%%%%%%%%%%%%%%%%%%%%%%%%%%%%%%%%%%%%%%%%%
\subsection{Features of the scalar-tensor NG contribution: Gaussian peaked spectra}
We now consider the more physical case of peaked primordial scalar and tensor power spectra with non-zero width. As a concrete example, we work with Gaussian spectra of the form
\begin{align}
    \mathcal{P}_{\mathcal{R},\textrm{G}}(k)&=\mathcal{A}_s\left(\frac{k}{k_s}\right)^3\frac{1}{\sqrt{2\pi (\sigma_s /k_s)^2}} \exp{\left(-\frac{(k-k_s)^2}{2\sigma_s^2}\right)}\,, \\
    \mathcal{P}^{R/L}_{h}(k)&=\mathcal{A}^{R/L}_h\left(\frac{k}{k_h}\right)^3\frac{1}{\sqrt{2\pi (\sigma_h /k_h)^2}} \exp{\left(-\frac{(k-k_h)^2}{2\sigma_h^2}\right)}\,,
\end{align}
where $\mathcal{A}_{\textrm{s/h}}$ are the peak amplitudes, $\sigma_{\textrm{s/h}}$ are the widths and $k_{\textrm{s/h}}$ are the positions of the peaks of the primordial scalar and tensor power spectra, respectively. The power spectra are normalised such that $\int \rm \ln k \, \mathcal{P}_{\mathcal{R},\textrm{G}}(k)=\mathcal{A}_s$ and $\int \rm \ln k \, \mathcal{P}^{R/L}_{h}(k)=\mathcal{A}^{R/L}_h$. In the following, we typically position the scalar and tensor peaks at a scale accessible to LISA, again assuming that there is no parity violation and that tensor perturbations are generically smaller than scalar ones.

\begin{figure}
    \centering
    \includegraphics[width=\linewidth]{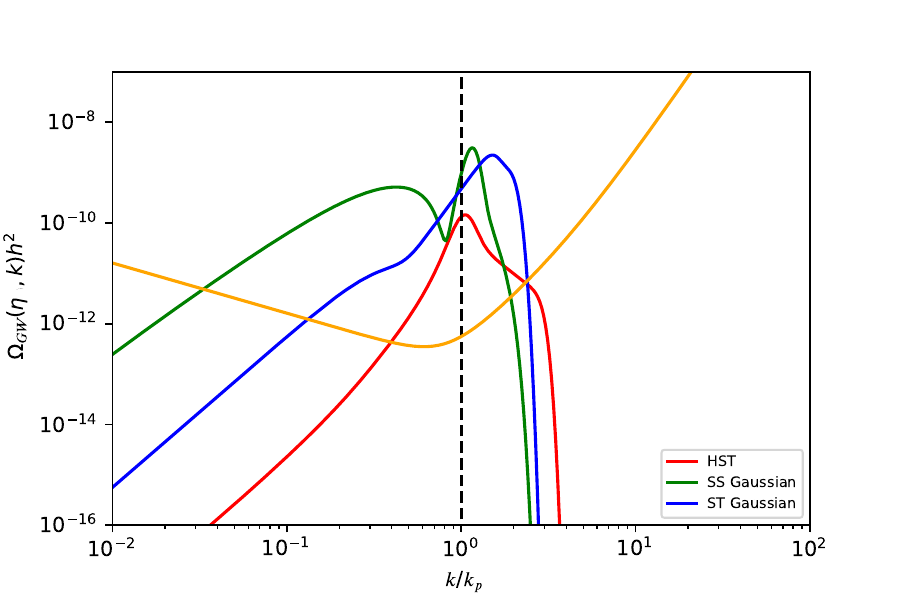}
    \caption{\footnotesize{The present-day spectral density $\Omega_{\textrm{GW}}(\eta,k)h^2$ of GWs by scalar-scalar (green) and scalar-tensor (blue) modes at Gaussian order and scalar-tensor modes at $\mathcal{O}(F^2_{\textrm{NL}})$ (red) for Gaussian primordial scalar and tensor power spectra with amplitude $\mathcal{A}_s=10^{-2}$, $\mathcal{A}^{R/L}_h=0.1 \mathcal{A}_s$, with peak at $k=k_p$ and width $\sigma_s=\sigma_h=k_p/10$. The solid orange curve denotes the LISA sensitivity, and the dashed black line marks the location of the peaks in the scalar and tensor power spectra $k_p$. The non-Gaussianity parameter is set to $F_{\textrm{NL}}=1$.}}
    \label{fig:HybridScalar}
\end{figure}

In Fig.~\ref{fig:HybridScalar}, we plot the contributions from the Gaussian scalar-scalar\footnote{All the NG scalar-scalar contributions can be found in Refs.~\cite{Adshead:2021hnm,Perna:2024ehx}.}, Gaussian scalar-tensor and NG scalar-tensor terms in the default scenario just described. This gives us an impression of the sort of effect introducing scalar non-Gaussianity might have on the iGW spectrum through the scalar-tensor sector. Immediately, we notice that the NG scalar-tensor term peaks at a scale closer to $k_p$ than either of the Gaussian contributions. In the last chapter, we showed that for a delta-function primordial spectrum, the Gaussian scalar-scalar term peaks at $k=2/\sqrt{3}k_p$. We have verified numerically that, in the limit of monochromatic spectra, our new term peaks at $k=k_p$. This behaviour is unique to the new term under consideration; other contributions from the scalar-scalar and scalar-tensor sectors generally peak at $k$-values after $k=k_p$. The scalar-tensor Gaussian peaks just a little before $k=(1+\frac{1}{\sqrt{3}})k_p$, which is the point at which the Heaviside function in the kernel~\eqref{kernelSTRD} cuts out. Non-Gaussian scalar-scalar contributions typically peak either at the peak of the Gaussian, or at later `knees' around $k=2k_p$, which is the cutoff of the Gaussian contributions. The fact that the NG scalar-tensor term peaks close to $k_p$ contributes further to the typical influence of non-Gaussianity on iGW spectra --- the smoothing out of peaks in the second-order iGW spectrum. 

The additional smoothing effect of the new term is unlikely to be distinguishable from that of other non-Gaussian terms. A more distinctive feature of this term, however, is its large amplitude at shorter scales. At scales just beyond $k_p$, the scalar-tensor Gaussian term dominates over the scalar-scalar, but for scales even smaller than this, eventually the NG scalar-tensor term takes over. This is a result of the conservation of momentum discussed in the previous section~\ref{UVcutoff}. In the limit of monochromatic primordial spectra, the Gaussian terms have a UV cutoff at $k=2k_p$, whereas the new contribution does not vanish until later at $k=3k_p$. When we consider Gaussian-shaped primordial spectra with finite width, the Gaussian and NG scalar-tensor terms may extend a little beyond these cutoffs, but the dominance of the new term at these scales is preserved.

This suggests that the distinctive feature of scalar non-Gaussianity in the scalar-tensor sector will be the presence of so-called `knees' at short scales in the spectrum, just like in the case of scalar-scalar non-Gaussianity. This presents us with two questions: is it plausible that LISA could detect such a knee, and can the knee from the NG scalar-tensor term be distinguished from the knee generated by scalar-scalar non-Gaussian terms? In the following section, we analyse the effect of the new term on an iGW second-order spectrum when all scalar-scalar non-Gaussian contributions up to $\mathcal{O}(F^4_{\textrm{NL}})$ are also included. 

We also notice that the NG scalar-tensor term has a similar infra-red running as the Gaussian scalar-tensor term in the limit $k\rightarrow0$. Interestingly, though, although it tends to the same scaling as the Gaussian scalar-tensor in the far IR, it does so with an unusual concavity. 

%%%%%%%%%%%%%%%%%%%%%%%%%%%%%%%%%%%%%%%%%%%%%%%%%%%%%%%%%%%
\subsection{Comparison to the case of SIGWs}
\label{sec:compare}
\begin{figure}
\centering
\begin{subfigure}{.49\textwidth}
  \centering
  \includegraphics[width=\linewidth]{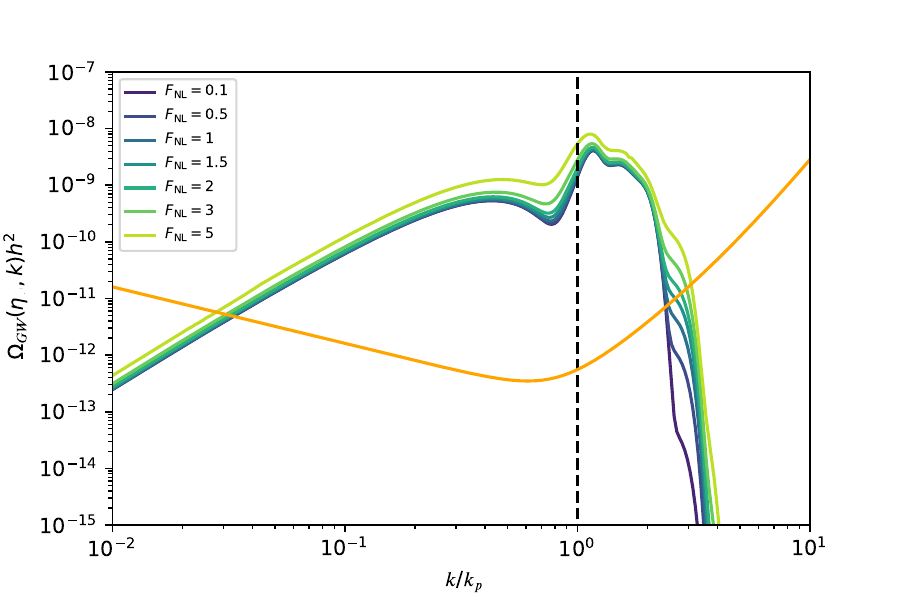}
\end{subfigure}
\begin{subfigure}{.49\textwidth}
  \centering
  \includegraphics[width=\linewidth]{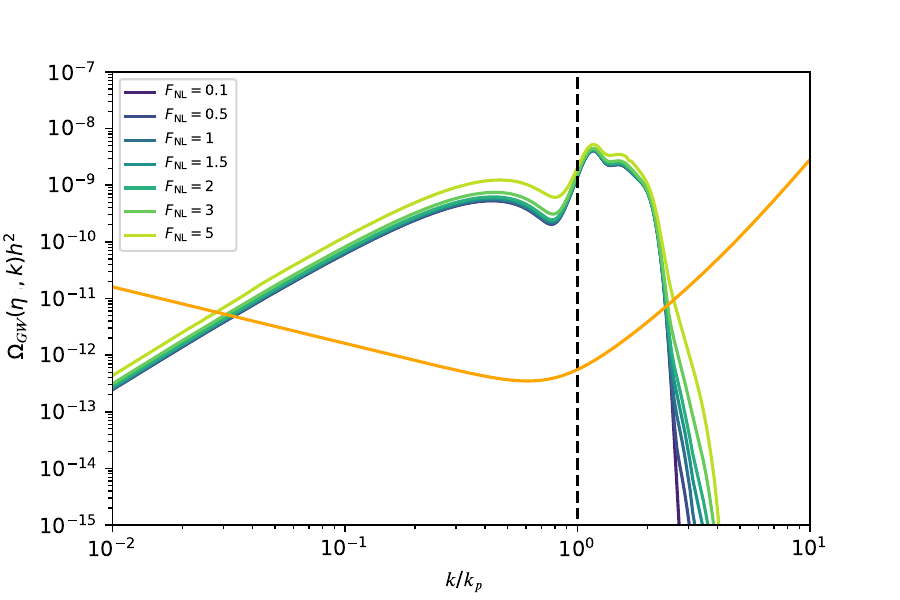}
\end{subfigure}
\begin{subfigure}{.49\textwidth}
  \centering
  \includegraphics[width=\linewidth]{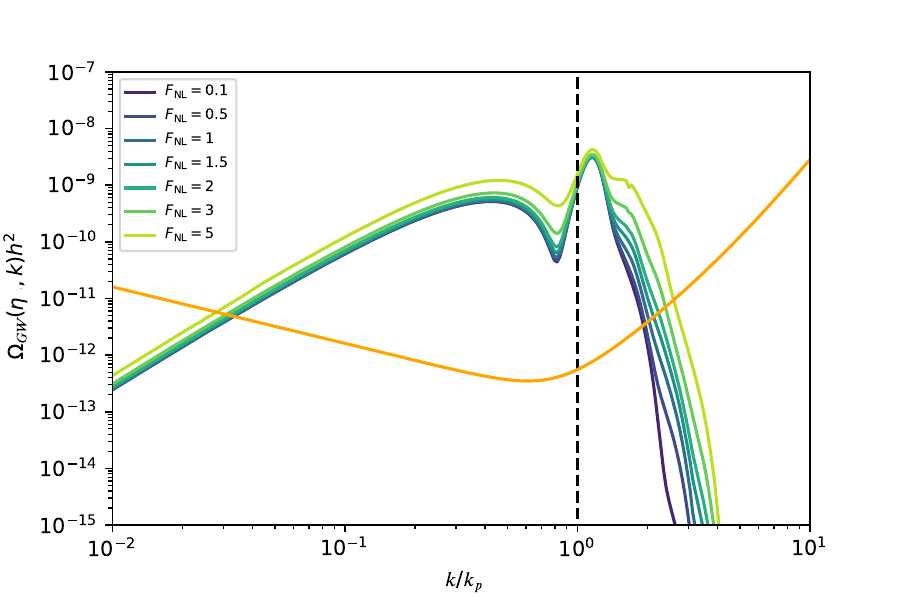}
\end{subfigure}
\caption{\footnotesize{The present-day spectral density $\Omega_{\textrm{GW}}(\eta,k)h^2$ of induced second-order GWs by scalar-scalar and scalar-tensor modes for varying values of $F_{\textrm{NL}}$ and Gaussian scalar and tensor power spectra centred at $k_p$ with width $1/10$ and $\mathcal{A}^{R/L}_h=0.1\mathcal{A}_s$. The solid orange line represents the LISA sensitivity band. The vertical black dashed line marks the scale $k_p$. (Top left) All scalar-scalar Gaussian and non-Gaussian contributions to order $\mathcal{O}(F^4_{\textrm{NL}})$ as well as the scalar-tensor Gaussian and NG scalar-tensor terms are included. (Top right) Same as the previous, but the new term has been excluded. (Bottom) same as the previous, but with the scalar-tensor Gaussian removed in addition.}}
\label{fig:SS-STNGcompare}
\end{figure}
The key question we wish to address is whether the inclusion of scalar non-Gaussianity in the scalar-tensor sector can have a noticeable effect on the overall second-order iGW spectrum beyond the contributions from scalar-scalar Gaussian and non-Gaussian terms studied previously in the literature. To this end, in Fig.~\ref{fig:SS-STNGcompare} we plot the present-day spectral density $\Omega_{\textrm{GW}}(\eta_0,k)h^2$ of second-order iGWs with varying $F_{\textrm{NL}}$ parameters for three different cases:
\begin{enumerate}
    \item Including all Gaussian and non-Gaussian contributions, to $\mathcal{O}(F^4_{\textrm{NL}})$, from both scalar-scalar and scalar-tensor sectors
    \item Excluding just the non-Gaussian scalar-tensor contribution
    \item Excluding the Gaussian and non-Gaussian scalar-tensor sector contributions
\end{enumerate}
This allows us to gauge the effect of the scalar-tensor sector on the overall spectrum, as well as to determine whether non-Gaussian scalar-tensor effects are distinguishable from all the other contributions.

Firstly, we point out that the existence of a Gaussian scalar-tensor contribution can be masked almost entirely by the non-Gaussian contributions from the scalar-scalar sector, but only for fairly large values of $F_{\textrm{NL}}$, i.e. $F_{\textrm{NL}}\geq 5$. The distinctive feature of the Gaussian scalar-tensor contribution, relative to the Gaussian scalar-scalar, is the continuation of the spectral density peak beyond $k=2k_p$. The amplitude of the Gaussian scalar-scalar term declines sharply around $k=2k_p$, whereas the Gaussian scalar-tensor term maintains a large amplitude until almost $k=3k_p$.
Regardless of the value of $F_{\textrm{NL}}$, if the scalar-tensor sector is significant, the amplitude does not fall off rapidly until past this threshold. If, on the other hand, the primordial tensor power spectrum does not have a large amplitude peak (i.e. if the scalar-tensor sector becomes negligible), then the scalar-scalar terms will dominate, and the spectral density tends to drop substantially before the threshold. This is not the case, though, if $F_{\textrm{NL}}$ begins to approach $\mathcal{O}(10)$. Already at $F_{\textrm{NL}}=5$, we can see that the scalar-scalar sector alone can mimic the behaviour of the scalar-tensor Gaussian contribution, raising the power at scales around $k=2k_p$ to an amplitude comparable to the peak.

Turning to the non-Gaussian scalar-tensor contribution, we notice that it doesn't have any significant impact on the peak of the iGW spectrum. This would suggest that the presence of a scalar-tensor sector is unlikely to be determined by LISA observations. Even if LISA detected an extended peak in the spectral density, whether it was caused by a scalar-tensor contribution or purely scalar non-Gaussianity would be uncertain. There is, however, one clear difference between the spectrum including scalar-tensor non-Gaussian terms and the others. In the top-left panel of Fig.~\ref{fig:SS-STNGcompare}, the NG scalar-tensor term clearly introduces an additional `kick' for scales $3k_p < k < 4k_p$. If $F_{\textrm{NL}}$ is of $\mathcal{O}(1)$ or larger, then the spectrum's terminal drop off in amplitude is delayed until almost $k=4k_p$. It is possible that this feature could uniquely signal the presence of a significant scalar-tensor sector, or even sizeable scalar non-Gaussianity. For our choice of $k_p$, it would be difficult for LISA to spot this extension for $F_{\textrm{NL}}$ values smaller than about 5, since the kick occurs right at the bottom of the LISA sensitivity. But, if $k_p$ is suitably smaller than the LISA scale $k_{\textrm{LISA}}$, then more of the drop-off in amplitude may be seen. In this case, $F_{\textrm{NL}}$ values as low as unity may produce kicks that LISA could detect.

%%%%%%%%%%%%%%%%%%%%%%%%%%%%%%%%%%%%%%%%%%%%%%%%%%%%%%%%%%%%%%%%%%%%%%%%%%%%%%%%%%%%%%%%%%%%
\subsection{Primordial parity violation}
%%%%%%%%%%%%%%%%%%%%%%%%%%%%%%%%%%%%%%%%%%%%%%%%%%%%%%%%%%%%%%%%%%%%%%%%%%%%%%%%%%%%%%%%%%%%
We conclude this chapter by examining the case of initial parity violation in the primordial tensor power spectrum. Certain models of inflation are capable of generating degrees of parity violation in the primordial tensor perturbations~\cite{Obata:2016tmo,Bartolo:2018elp,Bartolo:2017szm}. While primordial parity violation has no effect on the spectrum of SIGWs (since they involve the interactions of primordial scalar perturbations only), it may be transmitted to GWs induced by scalar-tensor interactions through the affected primordial tensor perturbations. Ref.~\cite{Bari:2023rcw} analysed the effects of parity violation on the Gaussian scalar-tensor contribution. In this section we give a brief review of their results and take a step forward to consider the effects of parity violation on the non-Gaussian term. For simplicity, we restrict our analysis to the right-handed spectral density of the scalar-tensor waves. This is an arbitrary choice, and the same analysis could be performed for the left-handed case.  

\begin{figure}
    \centering
    \includegraphics[width=\linewidth]{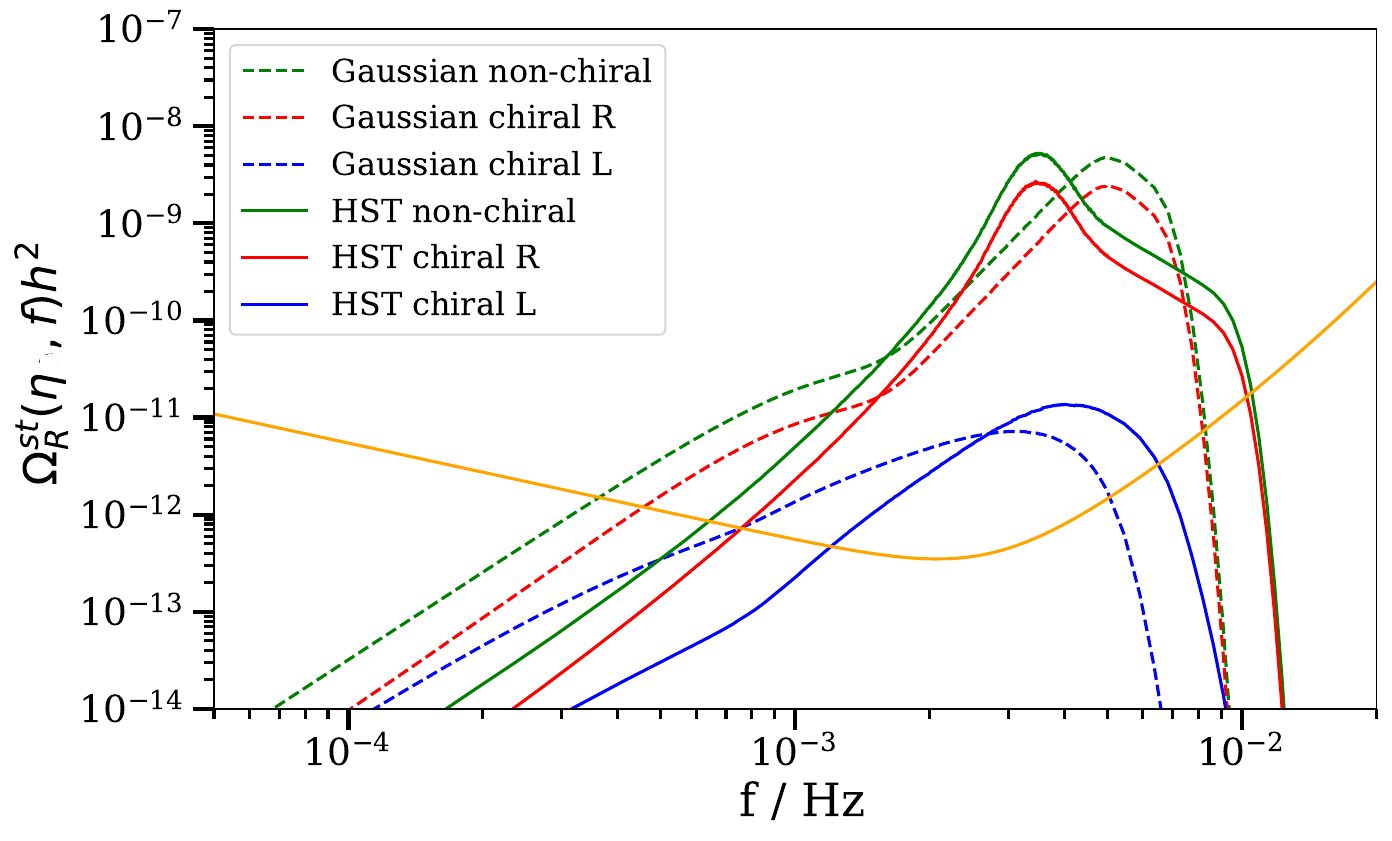}
    \caption{\footnotesize{The present-day right-handed spectral density of scalar-tensor induced SOGWs, $\Omega^{st}_{R}(\eta,f)h^2$, as a function of frequency, $f$. Dashed lines correspond to the Gaussian scalar-tensor contribution, solid lines to the non-Gaussian scalar-tensor contribution. In green we have the non-chiral contributions, in red the contribution from only right-handed primordial tensor perturbations and in blue the contribution from only left-handed primordial tensor perturbations. We have fixed $F_{\textrm{NL}}=3$. The solid orange line is the projected LISA sensitivity curve.}}
    \label{fig:chiralityplot}
\end{figure}

The right-handed spectral density of the Gaussian scalar-tensor contribution can be found straightforwardly using Eq.~\eqref{stpowerspectrum}
\begin{equation}
    \begin{split}
        \Omega _R^{st}(\eta , k)_\textrm{Gaussian} &= \frac{1}{768} \int _0^{\infty}{\rm } t_1 \int _{-1}^{1}{\rm } s_1 \, \frac{1}{v_1^6u_1^2} \bigg [ \left (u_1^2-(v_1+1)^2 \right )^4 \mathcal{P}^{R}_{h}(kv_1) ^4 \\
        &+ \left (u_1^2-(v_1-1)^2 \right ) \mathcal{P}^{L}_{h}(kv_1)  \bigg ]\mathcal{P}_{\mathcal{R},G}(ku_1) x^2\overline{\mathcal{I}^{st}(x,v_1,u_1)^2}\,. 
    \end{split}
\end{equation}
In Fig.~\ref{fig:chiralityplot}, we consider what the right-handed spectral density looks like in the case of having either no primordial left-handedness (dashed red) or no primordial right-handedness (dashed blue). For comparison, we also include the non-chiral (i.e. no initial parity violation, so Eq.~\eqref{avergaepsst}) case (dashed green). As explained in Ref.~\cite{Bari:2023rcw}, the IR behaviour of the spectral density for the case of having an initial right- or left-handed parity is similar. On shorter scales, however, the spectral density corresponding to the no left-handedness scenario exhibits a large-amplitude peak, in contrast to its no right-handedness counterpart. 

We can extend this analysis to the non-Gaussian scalar-term contribution in Eq.~\eqref{srHSspherical}, and derive an expression for the right-handed spectral density it generates
\begin{align}
    \begin{split}
        \Omega _R^{st}(\eta , k)_{\textrm{HST}} &= \frac{F_{\textrm{NL}}^2}{1536} \int _0^{\infty}{\rm } t_1 \int _{-1}^{1}{\rm } s_1 \int _0^{\infty}{\rm } t_2 \int _{-1}^{1}{\rm } s_2 \, \frac{1}{v_1^6u_1^2v_2^2u_2^2} \bigg [ \left (u_1^2-(v_1+1)^2 \right )^4   \\
        &\times \mathcal{P}^{R}_{h}(kv_1) + \left (u_1^2-(v_1-1)^2 \right )^4 \mathcal{P}^{L}_{h}(kv_1)  \bigg ]x^2\overline{\mathcal{I}^{st}(k,v_1,u_1)^2} \\
        &\times \mathcal{P}_{\mathcal{R},G}(ku_1v_2)\mathcal{P}_{\mathcal{R},G}(ku_1u_2) \text{ .} 
    \end{split}
\end{align}
The contributions from this term for different primordial parity scenarios are displayed as solid lines in Fig.~\ref{fig:chiralityplot}. One can observe in Fig.~\ref{fig:chiralityplot} that the conclusions drawn for the Gaussian case in Ref.~\cite{Bari:2023rcw} and above remain applicable to our new contribution. This fact is consistent with expectations, since any effects of primordial parity violation will be transmitted to scalar-tensor iGWs through the tensor perturbations, and our inclusion of scalar non-Gaussianity has not affected these.

\usetikzlibrary{calc}
\usetikzlibrary{arrows.meta}

% % % % % % % % % % % % % % % % % % % % % % % % % % % % % % % % 
\chapter{Third-order induced gravitational waves and the suppression of the SGWB}
\label{Chap:Third}

In Chap.~\ref{Chap:SOGWs}, we examined how linear tensor fluctuations can imprint themselves at second-order through their interactions with first-order scalar modes. We found that when the power spectrum of both scalar and tensor fluctuations is enhanced during inflation, the scalar-tensor contribution becomes non-negligible. Our analysis was restricted to sufficiently narrow peaked spectra in an RD universe, since broader spectra could artificially enhance the observable. This effect arises from the limiting behaviour of the scalar-tensor induced wave integrand (Eq.~\eqref{omegaSTRD}) in the UV limit. Because this divergence is entirely classical in origin, it is natural to ask whether other source terms, particularly those appearing at the same perturbative order, might cancel it.

Ref.~\cite{Chen:2022dah} identified `missing contributions' to the SGWB which contribute at the same perturbative order as scalar-tensor iGWs: correlations of third-order induced and primordial GWs. The authors looked at the behaviour of some source terms in the IR limit and found that on large scales, they are scale-invariant and negative. However, since these new contributions are at the same order as second-order iGWs, it is worth investigating \textit{all} possible source terms at \textit{all scales}, since we are also interested in their properties on small scales. Hence, in this chapter, we will derive these contributions and study their contribution to the SGWB, whilst also analysing their properties on small scales. Unlike in the previous chapter, we now assume for simplicity that the scalar fluctuations are Gaussian.

%%%%%%%%%%%%%%%%%%%%%%%%%%%%%%%%%%%%%%%%%%%%%%%%%%%%%%%%%%%%%%%%%%%%%%%%%%%%%%%%%%%%%%%%%%%%
%%%%%%%%%%%%%%%%%%%%%%%%%%%%%%%%%%%%%%%%%%%%%%%%%%%%%%%%%%%%%%%%%%%%%%%%%%%%%%%%%%%%%%%%%%%%
\section{Third order perturbation theory}
%%%%%%%%%%%%%%%%%%%%%%%%%%%%%%%%%%%%%%%%%%%%%%%%%%%%%%%%%%%%%%%%%%%%%%%%%%%%%%%%%%%%%%%%%%%%%%%%%%%%%%%%%%%%%%%%%%%%%%%%%%%%%%%%%%%%%%%%%%%%%%%%%%%%%%%%%%%%%%%%%%%%%%%%%%%%%%%%%%%%%%%%

We need to go to third-order in perturbation theory, and like at second-order, we pick the conformal Newtonian gauge. Following the expansion in Eq.~\eqref{metricexpansion}, we have
\begin{subequations}\label{metricdef}
    \begin{align}
        g_{00} &= -a^2(\eta)(1+2\Phi ^{(1)} + \Phi ^{(2)}) \, , \\
        g_{0i} &= a^2(\eta)\frac{1}{2}B_i^{(2)}=g_{i0}\, , \\
        g_{ij} &= a^2(\eta) \left ( (1-2\Psi ^{(1)} - \Psi ^{(2)})\delta _{ij}  +2h^{(1)}_{ij}+h^{(2)}_{ij}+\frac{1}{3}h^{(3)}_{ij} \right )\, .
    \end{align}
\end{subequations}
$h^{(3)}_{ij}$ are the third-order TT tensor modes, and we note that we have not included third-order scalar nor vector perturbations since they will not contribute to the GW equation at third order. To justify why we are going up to third-order, we will generalise the definition of the power spectrum in Eq.~\eqref{defpsh}, to the two-point function of a tensor mode of order $(n)$ and $(m)$, giving rise to the power spectrum $P_{\lambda}^{\text{$(nm)$}} (\eta , k)$, defined by 
\begin{equation}\label{generalcorrelationoftensors}
     \langle h_{\mathbf{k},\lambda}^{(n)}h_{\mathbf{k \conf} ,\lambda \conf}^{(m)} \rangle= \delta ^{\lambda \lambda ^{\prime}} \delta (\mathbf{k}+ \mathbf{k^{\prime}}) P_{\lambda}^{\text{$(nm)$}} (\eta , k) = \delta ^{\lambda \lambda ^{\prime}} \delta (\mathbf{k}+ \mathbf{k^{\prime}}) \frac{2\pi ^2}{k^3}\mathcal{P}_{\lambda}^{\text{$(nm)$}} (\eta , k) \, .
\end{equation}
By focusing on the power series expansion of the tensor perturbation in the spatial part of the metric 
\begin{equation}
    h_{ij}(\eta,\mathbf{x}) = h^{(1)}_{ij}(\eta,\mathbf{x})+\frac{1}{2}h^{(2)}_{ij}(\eta,\mathbf{x})+\frac{1}{6}h^{(3)}_{ij}(\eta,\mathbf{x}) \, ,
\end{equation}
and then by considering the two-point function in Fourier space of $\langle h_{\mathbf{k},\lambda}h_{\mathbf{k \conf},\lambda \conf} \rangle$, we can see what contributes to the GW background up to fourth order in the correlator
\begin{align}\label{tensortotalcorrelation}
    \begin{split}
        \langle h_{\mathbf{k},\lambda}h_{\mathbf{k \conf} ,\lambda \conf} \rangle &= \underbrace{ \langle h_{\mathbf{k},\lambda}^{(1)}h_{\mathbf{k \conf} ,\lambda \conf}^{(1)} \rangle }_{\text{Primordial}} + \frac{1}{4} \underbrace{ \langle h_{\mathbf{k},\lambda}^{(2)}h_{\mathbf{k \conf} ,\lambda \conf}^{(2)} \rangle }_{\text{iSOGWs}} +   \underbrace{ \frac{1}{6}\langle h_{\mathbf{k},\lambda}^{(1)}h_{\mathbf{k \conf} ,\lambda \conf}^{(3)} \rangle   + \frac{1}{6}\langle h_{\mathbf{k},\lambda}^{(3)}h_{\mathbf{k \conf} ,\lambda \conf}^{(1)} \rangle }_{\text{Additional contributions}} \\
        &=  \langle h_{\mathbf{k},\lambda}^{(1)}h_{\mathbf{k \conf} ,\lambda \conf}^{(1)} \rangle +\frac{1}{4}  \langle h_{\mathbf{k},\lambda}^{(2)}h_{\mathbf{k \conf} ,\lambda \conf}^{(2)} \rangle +  \frac{1}{3}\langle h_{\mathbf{k},\lambda}^{(3)}h_{\mathbf{k \conf} ,\lambda \conf}^{(1)} \rangle \, .
    \end{split}
\end{align}
In the first line, we have assumed that tensor modes are drawn from some Gaussian distribution, so any odd-point primordial correlator vanishes. The second equality holds due to the fact that the first-order equation of motion is even under $\mathbf{k}\rightarrow -\mathbf{k}$ (see Eq.~\eqref{tensortransfer}). On the other hand, if we recall from Eq.~\eqref{densityGWs}, the density of GWs is related to the time-averaged spectrum. We can generalise the definition for two tensors of order $(n)$ and $(m)$:
\begin{equation}
    \rho _{GW}(\eta , k) = \frac{M_{Pl}^2}{4} \frac{k^2}{a^2(\eta)} \sum _{\lambda} \overline{\mathcal{P}^{(nm)}_{\lambda}(\eta , k)} \, .
\end{equation}
Therefore, contributing to the iGW background at fourth order in the correlator is 
\begin{equation}\label{spectraldensitydef}
    \Omega (\eta , k) =\frac{k^2}{\mathcal{H}^2(\eta)}\sum _{\lambda} \left ( \frac{1}{12}\overline{\mathcal{P}^{(22)}_{\lambda}(\eta ,k )} + 2\times \frac{1}{18}\overline{\mathcal{P}^{(13)}_{\lambda}(\eta ,k )}\right ) \, .
\end{equation}
The second term in this equation is the focus of this chapter and was partially discussed in Ref.~\cite{Chen:2022dah}.

To know how third-order GWs are sourced, we need the third-order Einstein equations. For the third-order Einstein tensor, after the perturbations have been separated into scalar, vector and tensor degrees of freedom, there are approximately 500 terms. We will disregard some terms, namely: pure first-order tensor contributions, second-order perturbations coupled to first-order tensors, terms that couple quadratic first-order tensor perturbations to a first-order scalar fluctuation and terms made of cubic first-order scalars. We will see in the next section that the first two terms are subdominant and outside the scope of our analysis, and the last two disregarded terms do not contribute to the correlation of interest to this work. The third-order Einstein tensor is 
{
\allowdisplaybreaks
\begin{align}\label{thirdordereinstein}
        G_{ij}^{(3)} &= h_{ij}^{(3)''} + 2h_{ij}^{(3)'} \mathcal{H} - 2h_{ij}^{(3)} \mathcal{H}^2 - 4h_{ij}^{(3)} \mathcal{H}' - 6h_{ij}^{(2)''} \Phi - 12h_{ij}^{(2)'} \mathcal{H} \Phi + 12h_{ij}^{(2)} \mathcal{H}^2 \Phi  \nonumber \\ 
        &+ 24h_{ij}^{''} \Phi^2+ 48h_{ij}^{'} \mathcal{H} \Phi^2- 48h_{ij} \mathcal{H}^2 \Phi^2- 96h_{ij} \mathcal{H}' \Phi^2- 3h_{ij}^{(2)'} \Phi'+ 12h_{ij}^{(2)} \mathcal{H} \Phi' \nonumber \\
        &- 96h_{ij} \mathcal{H} \Phi \Phi' + 3h_{ij}^{(2)'} \Psi'+ 36h_{ij}^{(2)} \mathcal{H} \Psi'- 12h_{ij}^{'} \Phi \Psi'- 144h_{ij} \mathcal{H} \Phi \Psi'- 36h_{ij} \Phi' \Psi' \nonumber \\
        &+ 144h_{ij} \mathcal{H} \Psi \Psi'+ 24h_{ij} \Psi'^2+ 18h_{ij}^{(2)} \Psi''- 72h_{ij} \Phi \Psi''+72h_{ij} \Psi \Psi''- 24\Psi^2 \nabla^2 h_{ij}\nonumber \\
        &- 6\Psi \nabla^2 h_{ij}^{(2)}- \nabla^2 h_{ij}^{(3)}+ 6h_{ij}^{(2)} \nabla^2 \Phi - 24h_{ij} \Phi \nabla^2 \Phi + 24h_{ij} \Psi \nabla^2 \Phi - 6h_{ij} \Phi \nabla^2 \Phi^{(2)} \nonumber \\
        &+ 6h_{jm}^{(2)} \nabla^m \nabla_i \Psi + 48h_{jm} \Psi \nabla^m \nabla_i \Psi + 6h_{im}^{(2)} \nabla^m \nabla_j \Psi + 48h_{im} \Psi \nabla^m \nabla_j \Psi \nonumber \\
        &+ 12\Phi \nabla^m h_{ij} \nabla_m \Phi - 12\Psi \nabla^m h_{ij} \nabla_m \Phi - 3\nabla^m h_{ij}^{(2)} \nabla_m \Phi \nonumber \\
        &- 72\Psi \nabla^m h_{ij} \nabla_m \Psi - 9\nabla^m h_{ij}^{(2)} \nabla_m \Psi - 60h_{ij} \nabla^m \Psi \nabla_m \Psi + 6\mathcal{H} \Phi \nabla_i B_j^{(2)} \nonumber \\
        &+ \frac{3}{2}\Phi' \nabla_i B_j^{(2)} + \frac{3}{2}\Psi' \nabla_i B_j^{(2)} + 3\Phi \nabla_i B_j^{(2)'} - 12\Phi \nabla^m \Phi \nabla_i h_{jm}- 12h_{ij}^{(2)} \nabla^2 \Psi  \nonumber \\
        &+ 12\Psi \nabla^m \Phi \nabla_i h_{jm} + 24\Psi \nabla^m \Psi \nabla_i h_{jm} + 3\nabla^m \Phi \nabla_i h_{jm}^{(2)} + 3\nabla^m \Psi \nabla_i h_{jm}^{(2)}  \nonumber \\
        &- 6 B_j^{(2)} \mathcal{H} \nabla_i \Psi  + 12 h_{jm} \nabla^m \Phi \nabla_i \Psi  + 36 h_{jm} \nabla^m\Psi \nabla_i \Psi  - 3 B_j^{(2)} \nabla_i \Psi'  + 6 \mathcal{H} \Phi \nabla_j B_i^{(2)} \nonumber  \\
        & + \frac{3}{2} \Phi' \nabla_j B_i^{(2)} + \frac{3}{2} \Psi' \nabla_j B_i^{(2)}  + 3 \Phi \nabla_j B_i^{(2)'}  - 12 \Phi \nabla^m \Phi \nabla_j h_{im}  + 12 \Psi \nabla^m \Phi \nabla_j h_{im}\nonumber  \\
        & + 24 \Psi \nabla^m \Psi \nabla_j h_{im}  + 3 \nabla^m \Phi \nabla_j h_{im}^{(2)}  + 3 \nabla^m \Psi \nabla_j h_{im}^{(2)} + 6 \Psi \nabla_j \nabla_i \Psi^{(2)} \nonumber \\
        & + 3 \nabla_i \Phi^{(2)} \nabla_j \Phi  - 3 \nabla_i \Psi^{(2)} \nabla_j \Phi  + 3 \nabla_i \Phi \nabla_j \Phi^{(2)}  - 3 \nabla_i \Psi \nabla_j \Phi^{(2)}    \nonumber \\
        & + 12 h_{im} \nabla^m \Phi \nabla_j \Psi  + 36 h_{im} \nabla^m \Psi \nabla_j \Psi  - 3 \nabla_i \Phi^{(2)} \nabla_j \Psi  + 9 \nabla_i \Psi^{(2)} \nabla_j \Psi  - 3 B_i^{(2)} \nabla_j \Psi' \nonumber  \\
        & - 3 \nabla_i \Phi \nabla_j \Psi^{(2)} + 9 \nabla_i \Psi \nabla_j \Psi^{(2)}  + 6 \Phi^{(2)} \nabla_j \nabla_i \Phi  + 6 \Phi \nabla_j \nabla_i \Phi^{(2)}  + 6 \Psi^{(2)} \nabla_j \nabla_i \Psi \nonumber \\
        &+ 24h_{ij}^{(2)} \mathcal{H}' \Phi + 12h_{ij}^{'}+ 24h_{ij}^{'} \Phi \Phi' \Psi \Psi' - 12h_{ij} \nabla^m \Phi \nabla_m \Phi - 12h_{ij} \nabla^m \Psi \nabla_m \Phi  \nonumber \\
        &- 96h_{ij} \Psi \nabla^2 \Psi  - 6 B_i^{(2)} \mathcal{H} \nabla_j \Psi - 3 B_i^{(2)'} \nabla_j \Psi - 3B_j^{(2)'} \nabla_i \Psi  \nonumber \\
        &+ \text{(trace part)}\delta _{ij} + (hhh,\, \Psi \Psi \Psi, \, hh\Psi,\, \Psi ^{(2)}h, \, \Phi ^{(2)}h, \, B ^{(2)}h, \, h ^{(2)}h, \,\text{terms})   \, , 
\end{align}
}
whilst the third-order energy-momentum tensor reads
\begin{equation}
    \begin{split}\label{thirdorderenergymomentum}
        a^{-2}T_{ij}^{(3)} &=  2 h_{ij}^{(3)} P + 6 h_{ij}^{(2)} \delta P  + 6 h_{ij} P^{(2)} + h_{ij} P^{(3)} - 6 h_{ij} P^{(2)} \Psi - 6 h_{ij} \delta P \Psi^{(2)}- 2 h_{ij} P \Psi^{(3)} \\
        &   + 3 B^{(2)}_{j} P \partial_{i} v + 3 P v^{(2)}_{j} \partial_{i} v + 3 B^{(2)}_{j} \rho \partial_{i} v + 3 v^{(2)}_{j} \rho \partial_{i} v+ 12 h_{jm} P \partial^{m} v \partial_{i} v \\
        &   + 12 h_{jm} \rho \partial^{m} v \partial_{i} v + 3 B^{(2)}_{i} P \partial_{j} v + 3 P v^{(2)}_{i} \partial_{j} v + 3 B^{(2)}_{i} \rho \partial_{j} v + 3 v^{(2)}_{i} \rho \partial_{j} v  \\
        &+ 12 h_{im} P \partial^{m} v \partial_{j} v + 12 h_{im} \rho \partial^{m} v \partial_{j} v + 6 \delta P \partial_{i} v \partial_{j} v + 6 \delta \rho \partial_{i} v \partial_{j} v - 24 P \Psi \partial_{i} v \partial_{j} v \\
        &  - 24 \rho \Psi \partial_{i} v \partial_{j} v + 3 P \partial_{i} v^{(2)} \partial_{j} v + 3 \rho \partial_{i} v^{(2)} \partial_{j} v + 3 P \partial_{i} v \partial_{j} v^{(2)} + 3 \rho \partial_{i} v \partial_{j} v^{(2)}  \, .
    \end{split}
\end{equation}

%%%%%%%%%%%%%%%%%%%%%%%%%%%%%%%%%%%%%%%%%%%%%%%%%%%%%%%%%%%%%%%%%%%%%%%%%%%%%%%%%%%%%%%%%%%%%%
\subsection{What source terms do we need to consider?}
%%%%%%%%%%%%%%%%%%%%%%%%%%%%%%%%%%%%%%%%%%%%%%%%%%%%%%%%%%%%%%%%%%%%%%%%%%%%%%%%%%%%%%%%%%%%%%
At third order in perturbation theory, there are then various products of first and second order terms which source third order GWs. This becomes obvious when we extract the TT of the third-order spatial part of the Einstein equation from the previous section
\begin{equation}\label{thirdorderGWEQ}
    h_{ab}^{\prime \prime (3)} + 2\mathcal{H}h_{ab}^{\prime (3)} - \nabla ^2h_{ab}^{(3)}= 6\Lambda ^{ij}_{ab}S_{ij}^{(3)}\, ,
\end{equation}
We will now show that not all terms need to be considered when correlating third-order and linear tensor fluctuations.

For instance, the product of the first three-order scalar perturbations,
\begin{equation}
      -8\Psi \partial _i \Psi \partial _j \Psi \subset S_{ij}^{(3)}  \, ,
\end{equation}
does source GWs (i.e. it has a non-vanishing TT contribution). Nonetheless, when we consider $\langle h_{\mathbf{k},\lambda}^{(3)}h_{\mathbf{k \conf},\lambda \conf}^{(1)} \rangle$ this term will not contribute since, as we said previously, first-order scalars and tensors do not correlate. Extending this logic to other cubic first order terms, we see that we need to only consider terms that contain two scalars coupled to one tensor or cubic first order tensor terms\footnote{From a perturbative point of view $hhh$ will contribute as $\mathcal{P}_h\mathcal{P}_h$, whilst $h\Psi \Psi$ as $\mathcal{P}_h\mathcal{P}_{\Psi}$. Since $\mathcal{P}_{\Psi} >\mathcal{P}_h$, the $hhh$ contributions will be subdominant. This was shown in Chap.~\ref{Chap:SOGWs} for induced second-order GWs for peaked input power spectra. We therefore ignore these contributions for the rest of this chapter. Source terms which include pure first-order tensors are shown in App.~\ref{app4}}. This line of thought extends to second-order perturbations. Schematically, we will therefore only consider terms that source $h^{(3)}$ as:
\begin{equation}
S^{(3)} \sim h  \Psi  \Psi  \text{ or } h \underbrace{X^{(2)}} _{\Psi   \Psi  } \text{ or } \Psi \underbrace{X^{(2)}}_{ h \Psi } \, ,
\end{equation}
where $X^{(2)}$ can be a second-order scalar, vector, or tensor. In the second-to-last term, we see that $X^{(2)}$ can be sourced by two first-order scalars, while in the last term, it is sourced by one scalar and one tensor. To proceed, we require second-order equations of motion for second-order scalars, vectors, and tensors, sourced by first-order scalars and tensors. 

Using momentum conservation, a closer inspection of second-order perturbations sourced by two scalars coupled to a first-order tensor shows that the correlation $\langle h_{\mathbf{k}\conf , \lambda \conf}h^{(3)}_{\mathbf{k},\lambda}\rangle$ is proportional to
\begin{equation}
    \langle  h_{\mathbf{k}\conf , \lambda \conf} h_{\mathbf{k}-\mathbf{p}} \underbrace{X^{(2)}_{\mathbf{p}}} _{\Psi _{\mathbf{q}} \Psi _{\mathbf{p}-\mathbf{q}}} \rangle \propto \langle  h_{\mathbf{k}\conf , \lambda \conf} h_{\mathbf{k}-\mathbf{p}} \Psi _{\mathbf{q}} \Psi _{\mathbf{p}-\mathbf{q}} \rangle \propto \delta (\mathbf{k}\conf  + \mathbf{k}-\mathbf{p}) \delta (\mathbf{q}+\mathbf{p}-\mathbf{q}) \, ,
\end{equation}
which implies $\mathbf{p}=0$. Therefore, we see that we do not need to consider second-order scalars, vectors and tensors sourced by terms quadratic in first-order scalar fluctuations, when said second-order perturbations are sourcing third-order iGWs.

Bringing all of this together, the source terms in Eq.~(\ref{thirdorderGWEQ}) we need to consider are
\begin{equation}
    S^{(3)}_{ij} =  S^{h\Psi \Psi}_{ij}  + S^{h^{(2)}\Psi}_{ij} + S^{B^{(2)}\Psi}_{ij} + S^{\Psi^{(2)}\Psi}_{ij} \, ,
\end{equation}
where for clarity: $S^{h\Psi \Psi}_{ij}$ is made of linear terms with one tensor perturbation and quadratic scalar perturbations, $S^{h^{(2)}\Psi}_{ij}$ a second order tensor with a first order scalar, $S^{B^{(2)}\Psi}_{ij}$ a second order vector with a first order scalar, and $S^{\Psi^{(2)}\Psi}_{ij}$ a second order scalar with a first order scalar. We now proceed to give the exact form of these source terms.

%%%%%%%%%%%%%%%%%%%%%%%%%%%%%%%%%%%%%%%%%%%%%%%%%%%%%%%%%%%%%%%%%%%%%%%%%%%%%%%%%%%%%%%%%%%%%%
\subsection{Relevant source terms}
%%%%%%%%%%%%%%%%%%%%%%%%%%%%%%%%%%%%%%%%%%%%%%%%%%%%%%%%%%%%%%%%%%%%%%%%%%%%%%%%%%%%%%%%%%%%%%
In this subsection, we present in real space all the source terms that we need to consider throughout our work. These source terms have been simplified as far as possible, and we have extensively used the fact that perturbations are adiabatic and that the equation of state parameter is constant. For source terms which couple second-order perturbations to first-order ones, we also give equations of motion for induced second-order quantities, as well as their solutions in Fourier space. 
%%%%%%%%%%%%%%%%%%%%%%%%%%%%%%%%%%%%%%%%%%%%%%
\subsubsection{Source term made of first-order fluctuations}
%%%%%%%%%%%%%%%%%%%%%%%%%%%%%%%%%%%%%%%%%%%%%%
There is one source term made purely of first-order perturbations. The scalar-scalar-tensor source is
\begin{align}\label{sourcetermssh}
    \begin{split}
        S^{h\Psi \Psi}_{ij} &= -4h_{ij}\conf \Psi  \Psi \conf -\frac{1}{3(1+w)}h_{ij} \bigg ( 3(1+4w+3w^2)\Psi \conf \Psi \conf + 24(w^2-1)\Psi \nabla ^2 \Psi \\
        &+(-17-16w+9w^2)\partial _m \Psi \partial ^m \Psi \bigg ) - 16 h_j^m\Psi \partial _m \partial _i \Psi + 12 \Psi \partial _m h_{ij} \partial ^m \Psi \\
        &-8\Psi \partial ^m \Psi \partial _i h_{jm} - 16 h_{jm}\partial ^m \Psi \partial _i \Psi - 8\hubble h_{ij} \Psi \Psi \conf + \frac{8}{3(1+w)\hubble} \bigg ((w-1)h_{ij} \partial _m \Psi \partial ^m \Psi \conf \\
        &-h_{jm}\conf \partial ^m \Psi \partial _i \Psi +  h_{jm} \partial ^m \Psi \conf \partial _i \Psi   \bigg ) + \frac{4}{3(1+w)\hubble ^2} \bigg ((w-1)h_{ij}\partial _m \Psi \conf \partial ^m \Psi \conf \\
        &-2h_{jm}  \conf \partial ^m \Psi \partial _ i \Psi \conf + 2h_{jm} \partial ^m \Psi \conf \partial _i \Psi \conf  \bigg ) \, . 
    \end{split}
\end{align}
%%%%%%%%%%%%%%%%%%%%%%%%%%%%%%%%%%%%%%%%%%%%%%
\subsubsection{Source terms consisting of second-order scalar perturbations coupled to first-order perturbations}\label{inducedscalar}
%%%%%%%%%%%%%%%%%%%%%%%%%%%%%%%%%%%%%%%%%%%%%%
The source term which couples second-order scalar perturbations to first-order scalars, reads
\begin{align}
    \begin{split}\label{realspacesourcepsi2psi}
        S^{\Psi^{(2)}\Psi}_{ij} &= \frac{4}{3(1+w)\mathcal{H}^2} \partial _i \Psi ^{\prime} \partial _j \Psi ^{(2)\prime} + \frac{4}{3(1+w)\mathcal{H}} \left ( \partial _i \Psi ^{\prime} \partial _j \Phi ^{(2)} + \partial _i \Psi \partial _j \Psi ^{(2)\prime} \right )  \\
        &+\frac{2}{3}\frac{(5+3w)}{(1+w)}\partial _i \Psi \partial _j \Phi ^{(2)} \, .
    \end{split}
\end{align}
This term was considered in Ref.~\cite{Zhou:2021vcw} in the context of third-order SIGWs, and our expression above is identical in an RD universe. To proceed, we need equations of motion for $\Psi ^{(2)}$ and $\Phi ^{(2)}$, which was the focus of Sec.~\ref{sec:inducedscalars}.

From the trace Eq.~\eqref{eqtracescalarsrealspace}, we are only interested in the scalar-tensor source term, which, after simplification, is 
\begin{align}
    \begin{split}
        S^{\Psi^{(2)}_{st}} & =2(1-3w)h^{ij}\partial _i \partial _j \Psi \text{ .}
    \end{split}
\end{align}
The other equation, Eq.~\eqref{eqanisrealspace}, obtained by extracting the symmetric trace-free part of the spatial second-order Einstein equations, encodes the difference between $\Phi ^{(2)}$ and $\Psi ^{(2)}$, sourced from
\begin{align}
    \begin{split}
        \Pi^{(2),st}_{ij} & = 12h_i^m \partial _j \partial _m \Psi -6(1+3w)\hubble h_{ij}\Psi \conf +6(w-1)h_{ij} \nabla ^2 \Psi -12\partial ^m \Psi \partial _m h_{ij} \\
    &+12 \partial ^m \Psi \partial _j h_{im}- 12 \Psi \nabla ^2 h_{ij} \, .
    \end{split}
\end{align}
In Ref.~\cite{Chen:2022dah}, the authors set $\Phi ^{(2)}$ and $\Psi ^{(2)}$ to be the same, since they focus on the IR limit of the iGW background.

To solve the third-order equations of motion, we then need the second-order equations of motion for $\Phi ^{(2)}$ and $\Psi ^{(2)}$. We proceed to extract these in Fourier space. The second-order anisotropic stress equation sourced by first-order tensor and scalar fluctuations, Eq.~\eqref{eqanisrealspace}, is given in Fourier space by
\begin{equation} \label{anis}
    \Phi ^{(2)}(\eta , \mathbf{k}) - \Psi ^{(2)}(\eta , \mathbf{k}) = \frac{1}{k^4} \Pi ^{(2)}_{st} (\eta , \mathbf{k}) \, ,
\end{equation}
with
\begin{align}\label{anisinducedst}
    \begin{split}
        \Pi ^{(2)}_{st} (\eta , \mathbf{k}) & = \left (\frac{3+3w}{5+3w} \right ) \sum _{\lambda _1 }  \int \frac{d^3\mathbf{p}}{(2\pi)^{\frac{3}{2}}}  \text{ } Q_{\lambda _1}^{\Psi ^{(2)} _{st}}(\mathbf{k},\mathbf{p})f^{\Pi ^{(2)}}_{st}(\eta ,k,p,|\mathbf{k}-\mathbf{p}|) h^{\lambda _1}_{\mathbf{p}}\mathcal{R} _{\mathbf{k}-\mathbf{p}} \, .
    \end{split}
\end{align}
In the above, we factored out the primordial values and defined for compactness
\begin{align}\label{polinducedscalars}
    Q_{\lambda _1}^{\Psi ^{(2)} _{st}}(\mathbf{k},\mathbf{p}) &= k^ik^j\epsilon_{ij}^{\lambda _1}(\mathbf{p}),
\end{align}
and
\begin{align}\label{fpistscalar}
    \begin{split}
        &f^{\Pi ^{(2)}}_{st}(\eta ,k,p,|\mathbf{k}-\mathbf{p}|) = -12 \left (k^ip_i-\frac{2}{3}k^2 \right ) T_h(\eta p) T_{\Psi}(\eta|\mathbf{k}-\mathbf{p}|)-12 p^2  T_h(\eta p)T_{\Psi}(\eta|\mathbf{k}-\mathbf{p}|)  \\
        &+6(1+3w)\hubble T_h(\eta p) T_{\Psi}\conf(\sqrt{w}\eta|\mathbf{k}-\mathbf{p}|)+6(w-1)|\mathbf{k}-\mathbf{p}|^2 T_h(\eta p) T_{\Psi}(\eta|\mathbf{k}-\mathbf{p}|) \\
        &- 12 (k^ip_i-p^2)T_h(\eta p) T_{\Psi}(\eta|\mathbf{k}-\mathbf{p}|) + 12 \, k^ip_i T_h(\eta p)T_{\Psi}(\eta|\mathbf{k}-\mathbf{p}|) \, .
    \end{split}
\end{align}
The second order trace equation for scalars, Eq.~\eqref{eqtracescalarsrealspace}, in Fourier space is given by
\begin{align}
    \begin{split}
        &3\Psi ^{\prime \prime (2)}(\eta, \mathbf{k}) + 3(2+3w)\hubble \Psi ^{\prime (2)}(\eta, \mathbf{k}) + (3w+1)k^2 \Psi ^{(2)}(\eta, \mathbf{k}) + 3\hubble \Phi ^{\prime (2)}(\eta, \mathbf{k}) \\
        &- k^2 \Phi ^{(2)}(\eta, \mathbf{k}) =  \mathcal{S}^{\Psi ^{(2)}}_{st}(\eta , \mathbf{k})
    \end{split}
\end{align}
where
\begin{align}
    \begin{split}
        \mathcal{S}^{\Psi ^{(2)}}_{st}(\eta , \mathbf{k}) & = \left (\frac{3+3w}{5+3w} \right ) \sum _{\lambda _1 } \int \frac{d^3\mathbf{p}}{(2\pi)^{\frac{3}{2}}} Q_{\lambda _1}^{\Psi ^{(2)} _{st}}(\mathbf{k},\mathbf{p})f^{\Psi ^{(2)}}_{st}(\eta , p,|\mathbf{k}-\mathbf{p}|) h^{\lambda _1}_{\mathbf{p}}\mathcal{R} _{\mathbf{k}-\mathbf{p}} \, ,
    \end{split}
\end{align}
\begin{equation}
    f^{\Psi ^{(2)}}_{st}(\eta , p,|\mathbf{k}-\mathbf{p}|) = -2  (1-3w)T_h(\eta p)T_{\Psi} (\eta |\mathbf{k}-\mathbf{p}|) \, .
\end{equation}
We can now solve for $\Psi ^{(2)} (\eta , \mathbf{k})$ by making the substitution in Eq.~(\ref{anis}) to get rid of $\Phi ^{(2)} (\eta , \mathbf{k})$. This leaves us with
\begin{align} \label{eqscalarfourier}
    \begin{split}
        &\Psi ^{\prime \prime (2)}(\eta, \mathbf{k}) + 3(1+w)\hubble \Psi ^{\prime (2)}(\eta, \mathbf{k}) +wk^2 \Psi ^{(2)}(\eta, \mathbf{k}) = \frac{1}{3}  \mathcal{S}^{st}(\eta , \mathbf{k}) - \frac{\hubble}{k^4} \timeder   \Pi ^{(2)}_{st} (\eta , \mathbf{k})    \\
        &+ \frac{1}{3k^2} \Pi ^{(2)}_{st} (\eta , \mathbf{k})= \sum _{\lambda _1 , \lambda _2} \int \frac{d^3\mathbf{p}}{(2\pi)^{\frac{3}{2}}} Q_{\lambda _1}^{\Psi ^{(2)} _{st}}(\mathbf{k},\mathbf{p})\Sigma ^{st} (\eta , p,|\mathbf{k}-\mathbf{p}|)h^{\lambda _1}_{\mathbf{p}}\mathcal{R} _{\mathbf{k}-\mathbf{p}} \, ,
    \end{split}
\end{align}
where for convenience we defined
\begin{equation}\label{Sigmapsi2}
        \Sigma ^{st} (\eta , p,|\mathbf{k}-\mathbf{p}|) =\frac{1}{3}f^{\Psi ^{(2)}}_{st}(\eta , p,|\mathbf{k}-\mathbf{p}|) - \frac{\hubble}{k^4}\timeder  f^{\Pi ^{(2)}}_{st}(\eta ,k , p,|\mathbf{k}-\mathbf{p}|) + \frac{1}{3k^2}f^{\Pi ^{(2)}}_{st}(\eta ,k , p,|\mathbf{k}-\mathbf{p}|) \, .
\end{equation}
Eq.~\eqref{eqscalarfourier} can be solved via Green's method. After applying the substitution $z=\Psi a^{\tfrac{3(1+w)}{2}}$ and using the initial conditions:  $\Psi^{(2)}(\eta _i ,k)=0$ and $\partial _{\eta}\Psi^{(2)}(\eta _i ,k)=0$, we find the Green's function 
\begin{equation}
    G^s_k(\eta,\bareta) = \frac{(k\bareta)(k\eta)\sqrt{w}}{k} \left (j_{b+1}(k\bareta\sqrt{w})y_{b+1}(k\eta\sqrt{w}) - y_{b+1}(k\bareta\sqrt{w})j_{b+1}(k\eta\sqrt{w}) \right ) \Theta (\eta - \bareta) \, ,
\end{equation}
which is valid for $w>0$ and $b$ we recall is defined in Eq.~\eqref{defb}. The solution to the equation of motion for $\Psi ^{(2)}(\eta , \mathbf{k})$ is then given by
\begin{equation}\label{psiinducedst}
    \Psi ^{(2)}_{st}(\eta , \mathbf{k}) = \left (\frac{3+3w}{5+3w} \right )\sum _{\lambda _1 }\int \frac{d^3\mathbf{p}}{(2\pi)^{\frac{3}{2}}} Q_{\lambda _1}^{\Psi ^{(2)} _{st}}(\mathbf{k},\mathbf{p}) I^{\Psi ^{(2)}}_{st}(\eta ,p,|\mathbf{k}-\mathbf{p}|)h_{\mathbf{p}}^{\lambda _1} \mathcal{R} _{\mathbf{k}-\mathbf{p}} \, ,
\end{equation}
where we have further defined
\begin{equation} \label{kernelinducedstscalars}
    I^{\Psi ^{(2)}}_{st}(\eta ,p,|\mathbf{k}-\mathbf{p}|)= \int _{\eta _i}^{\infty} {\rm}  \bareta \, \left ( \frac{a(\bareta)}{a(\eta)} \right )^{\frac{3(1+w)}{2}} G^s_k(\eta , \bareta) \Sigma ^{st}(\bareta , p,|\mathbf{k}-\mathbf{p}|) \, .
\end{equation}

%%%%%%%%%%%%%%%%%%%%%%%%%%%%%%%%%%%%%%%%%%%%%%
\subsubsection{Source terms consisting of second-order vector perturbations coupled to first-order perturbations}\label{inducedvector}
%%%%%%%%%%%%%%%%%%%%%%%%%%%%%%%%%%%%%%%%%%%%%%
Another source term which we need to consider contains second-order vector perturbations, $B^{(2)}_{i}$, coupled to first-order scalars  
\begin{align}\label{realspacesourcevectorpsi}
    \begin{split}
        S^{B^{(2)} \Psi}_{ij}\ &= -\Psi ^{\prime} \partial _i B_j^{(2)} - \Psi \partial _i B^{(2)\prime}_j + B^{(2)\prime}_j\partial _i \Psi - 2\mathcal{H}\Psi \partial _i B_j^{(2)} + 2\mathcal{H} B_j^{(2)} \partial _i \Psi + B_j^{(2)} \partial _i \Psi ^{\prime} \\
        &-\frac{1}{3(1+w)\mathcal{H}^2}\nabla ^2 B^{(2)}_i \left (\mathcal{H} \partial _j \Psi + \partial _j \Psi ^{\prime} \right ) \, ,
    \end{split}
\end{align}
also considered in Ref.~\cite{Zhou:2021vcw} and identical to ours. The second-order equation of motion for induced vectors is Eq.~\eqref{eqinducedvectorsrealspace}, where, after some simplifications using linear-order equations of motion, the source term is given by
\begin{align}\label{sourcetermsV}
    \begin{split}
        S_{ij}^{B^{(2)}_{st}} & = -4h_j^m\partial _m \partial _i \Psi + 4 \partial _m h_{ij} \partial ^m \Psi + 4 \Psi \nabla ^2 h_{ij} -4\partial ^m \Psi \partial _j h_{im} \\
    &+2h_{ij} \left ((1+3w)\mathcal{H}\Psi ^{\prime} + (1-w) \nabla ^2 \Psi  \right ) \, .
    \end{split}
\end{align}

In Fourier space, the second-order vector equation can be solved
\begin{align} \label{inducedvectorequation}
    \begin{split}
         B_r^{\prime (2)}(\eta , \mathbf{k}) + 2\mathcal{H}B_r^{(2)}(\eta , \mathbf{k}) =   \mathcal{S}^{B ^{(2)}_{st}}_{r}(\eta , \mathbf{k})\, ,
    \end{split}
\end{align}
where the source terms read
\begin{align}
    \begin{split}
        \mathcal{S}^{B ^{(2)}_{st}}_{r}(\eta , \mathbf{k}) & = \left ( \frac{3+3w}{5+3w} \right ) \sum _{\lambda _1} \frac{i}{k^2}\int \frac{d^3\mathbf{p}}{(2\pi)^{\frac{3}{2}}} \bigg ( Q^{B ^{(2)}_{st1}} _{r,\lambda _1 } (\mathbf{k},\mathbf{p}) f^{B ^{(2)}}_{st1}(\eta , p ,|\mathbf{k}-\mathbf{p}| ) + Q^{B ^{(2)}_{st2}} _{r,\lambda _1 } (\mathbf{k},\mathbf{p})  \\
    &\times f^{B ^{(2)}}_{st2}(\eta , p ,|\mathbf{k}-\mathbf{p}| ) \bigg ) h^{\lambda _1}_{\mathbf{p}} \mathcal{R}_{\mathbf{k}-\mathbf{p}} \, .
    \end{split}
\end{align}
We have defined
\begin{subequations}
    \begin{align}
       \label{pol2ndvecst1}
        Q^{ B^{(2)}_{st1}}_{r,\lambda _1 }(\mathbf{k},\mathbf{p}) &= (k^2-k^ip_i)k^me^j_r(\mathbf{k})^*\eps_{mj}^{\lambda _1} (\mathbf{p}) + e^j_r(\mathbf{k})^*p_jk^ik^m\eps_{im}^{\lambda _1} (\mathbf{p}) \, ,
        \\ \label{pol2ndvecst2}
        Q^{ B^{(2)}_{st2}}_{r,\lambda _1 }(\mathbf{k},\mathbf{p}) &= k^ie^j_r(\mathbf{k})^*\eps_{ij}^{\lambda _1} (\mathbf{p}) \, ,
    \end{align}
\end{subequations}
and
\begin{subequations}
    \begin{align}
        \label{fun2ndvecst1}
        f^{B^{(2)}}_{st1}(\eta , p,|\mathbf{k}-\mathbf{p}|) & = -4 T_h(\eta p)T_\Psi( \eta |\mathbf{k}- \mathbf{p}|) \, ,
        \\ \label{fun2ndvecst2}
        \begin{split} 
            f^{B^{(2)}}_{st2}(\eta , p,|\mathbf{k}-\mathbf{p}|) &=  4(\mathbf{p}\cdot \mathbf{k}-p^2)T_h(\eta p)T_\Psi (\eta |\mathbf{k}-\mathbf{p}|) +4p^2T_h(\eta p)T_\Psi (\eta |\mathbf{k}-\mathbf{p}|) \\
            &-2(1+3w)\hubble T_h(\eta p)T^{\prime}_{\Psi}  (\eta |\mathbf{k}-\mathbf{p}|) \\
            &+ 2(1-w)|\mathbf{k}-\mathbf{p}|^2T_h(\eta p)T_\Psi (\eta |\mathbf{k}-\mathbf{p}|) \, .
        \end{split}
    \end{align}
\end{subequations}
The general solution to Eq.~\eqref{inducedvectorequation} can be found via Green's method (with initial condition $B^{(2)}(\eta_i,k)=0$)
\begin{equation} \label{greensvectors}
    G^v(\eta , \bareta) = \left( \frac{\bareta}{\eta} \right )^{2(1+b)} \Theta(\eta - \bareta) \, .
\end{equation}
The Green's function for induced vectors does not depend on $k$ as for the case of induced scalars or tensors, i.e. there is no propagating mode. The solution to the equation is then
\begin{align}\label{2ndordervecst}
    \begin{split}
        B_{r}^{(2)st}(\eta , \mathbf{k}) &= \left ( \frac{3+3w}{5+3w} \right ) \sum _{\lambda _1} \frac{i}{k^2}\int \frac{d^3\mathbf{p}}{(2\pi)^{\frac{3}{2}}} \bigg ( Q^{B ^{(2)}_{st1}} _{r,\lambda _1 } (\mathbf{k},\mathbf{p}) I^{B^{(2)}}_{st1}(\eta,p,|\mathbf{k}-\mathbf{p}|) + Q^{B ^{(2)}_{st2}} _{r,\lambda _1 } (\mathbf{k},\mathbf{p})  \\
            &\times I^{B^{(2)}}_{st2}(\eta,p,|\mathbf{k}-\mathbf{p}|) \bigg ) h^{\lambda _1}_{\mathbf{p}} \mathcal{R}_{\mathbf{k}-\mathbf{p}} \, ,
    \end{split}
\end{align}
where
\begin{subequations}
    \begin{align}
          I^{B^{(2)}}_{st1}(\eta,p,|\mathbf{k}-\mathbf{p}|) &=  \int _{\eta _i}^{\infty} {\rm}  \bareta \,  G^v(\eta , \bareta) f^{B^{(2)}}_{st1}(\overline{\eta},  p,|\mathbf{k}-\mathbf{p}|) \label{kernelvectorst1}  \, ,
          \\
          I^{B^{(2)}}_{st2}(\eta,p,|\mathbf{k}-\mathbf{p}|) &=  \int _{\eta _i}^{\infty} {\rm}  \bareta \,  G^v(\eta , \bareta) f^{B^{(2)}}_{st2}(\overline{\eta},  p,|\mathbf{k}-\mathbf{p}|) \label{kernelvectorst2}  \, .
    \end{align}
\end{subequations}
Second-order induced vectors were previously studied in Ref.~\cite{Lu:2007cj}.
%%%%%%%%%%%%%%%%%%%%%%%%%%%%%%%%%%%%%%%%%%%%%%
\subsubsection{Source terms consisting of second-order tensor perturbations coupled to first-order perturbations}\label{inducedtensor}
%%%%%%%%%%%%%%%%%%%%%%%%%%%%%%%%%%%%%%%%%%%%%%
Finally, the last source term couples second-order tensor perturbations, $h^{(2)}_{ij}$, to first-order scalars
\begin{align}
    \begin{split}\label{realspaceh2psi}
        S^{h^{(2)} \Psi}_{ij} &= \left ( h_{ij}^{(2)\prime \prime} + 2\mathcal{H}h_{ij}^{(2)\prime} - \nabla ^2 h_{ij}^{(2)} \right )\Psi + 2\Psi \nabla ^2 h^{(2)}_{ij}+ (3w+1)\mathcal{H}h^{(2)}_{ij}\Psi ^{\prime} \\
        & - (w-1)h^{(2)}_{ij}\nabla ^2 \Psi  + 2 \partial ^m \Psi \partial _m h^{(2)}_{ij}  \, ,
    \end{split}
\end{align}
In an RD universe, our source term agrees with Ref.~\cite{Zhou:2021vcw}. Second-order tensors sourced by first-order scalars and tensors were the focus of Chap.~\ref{Chap:SOGWs}, and all relevant equations can be found in Sec.~\ref{sec:stinduced}.

Having listed all the contributions at third order, we have also shown how the second-order variables are sourced from first-order perturbations. We now move on to solving the third-order GW equation for each source term.

\section{Solution to the third-order induced gravitational wave equation}\label{sec3}
%%%%%%%%%%%%%%%%%%%%%%%%%%%%%%%%%%%%%%%%%%%%%%%%%%%%%%%%%%%%%%%%%%%%%%%%%%%%%%%%%%%%%%%%%%%%%%
%%%%%%%%%%%%%%%%%%%%%%%%%%%%%%%%%%%%%%%%%%%%%%%%%%%%%%%%%%%%%%%%%%%%%%%%%%%%%%%%%%%%%%%%%%%%%%
We now proceed to solving the third-order iGW equation, Eq.~\eqref{thirdorderGWEQ}, in Fourier space  
\begin{equation}\label{GW3eqfourier}
     h_{\mathbf{k}, \lambda}^{\prime \prime (3)} + 2\mathcal{H}h_{\mathbf{k}, \lambda}^{\prime (3)} + k ^2h_{\mathbf{k}, \lambda}^{(3)}=6 \epsilon ^{ij}_{\lambda}(\mathbf{k})^{*}\mathcal{S}_{ij}^{(3)}(\eta , \mathbf{k})\equiv 6\mathcal{S}_{\mathbf{k},\lambda}^{(3)} (\eta)\, .
\end{equation}
We can obtain the solution for the third-order tensor mode using Green's method; the structure of the solution is identical to the case of second-order induced tensors
\begin{equation}\label{GWsol3rdorder}
    h^{(3)}_{\mathbf{k}, \lambda} (\eta ) = 6 \int _{\eta _i}^{\infty} {\rm}\overline{\eta} \, \frac{a(\overline{\eta})}{a(\eta)}  G^h_k (\eta , \overline{\eta})\mathcal{S}_{\mathbf{k},\lambda}^{(3)} (\overline{\eta}) \, .
\end{equation}
where the Green’s function was defined in Eq.\eqref{greenstensors}.

In the following subsections, we present the solutions for each contribution in Fourier space. Since the solution to Eq.~\eqref{GW3eqfourier} depends on the equations of motion and solutions of both first- and second-order perturbations, we first provide a brief overview of how the corresponding third-order solutions are constructed.

For adiabatic initial conditions, first-order perturbations can be split up as Eq.~\eqref{tensor1sol} and Eq.~\eqref{psitransfer} for tensors and scalars, respectively. However, a more complicated decomposition applies for second-order perturbations. While it is still possible to isolate the primordial value of the perturbation, one must also account for the non-linear evolution of the second-order variables. The time evolution of second-order perturbations is encoded in a kernel, as opposed to a transfer function at linear order, and so this will give rise to nested kernels. For this reason, in what follows, we distinguish between source terms made of cubic first-order fluctuations and the ones containing second-order perturbations. 

%%%%%%%%%%%%%%%%%%%%%%%%%%%%%%%%%%%%%%%%%%%%%%%%%%%%%%%%%%%%%%%%%%%%
\subsection{Solution to the gravitational wave equation for the pure first-order contribution}
%%%%%%%%%%%%%%%%%%%%%%%%%%%%%%%%%%%%%%%%%%%%%%%%%%%%%%%%%%%%%%%%%%%%
The source term composed exclusively of first-order perturbations, corresponding to the coupling of quadratic scalar fluctuations to a tensor mode, given in Eq.~\eqref{sourcetermssh}. This leads to the following solution in Fourier space:
\begin{align}\label{solhpsipsi}
    \begin{split}
        &h^{\lambda \, (3)}_{\Psi \Psi h} (\eta , \mathbf{k}) = \frac{6}{(2\pi)^3} \int {\rm} ^3 \mathbf{q} \int {\rm} ^3 \mathbf{p} \, \, \epsilon^{ij}_{\lambda}(\mathbf{k}) ^* \sum _{\lambda _1} \bigg ( \epsilon_{ij}^{\lambda _1}(\mathbf{k}-\mathbf{p}-\mathbf{q}) I_1^{ssh}(\eta , k,p,q) \\
        &+ 16 \epsilon_{jm}^{\lambda _1}(\mathbf{k}-\mathbf{p}-\mathbf{q})p^mp_i I_2^{ssh}(\eta , k,p,q) - 8\epsilon_{jm}^{\lambda _1}(\mathbf{k}-\mathbf{p}-\mathbf{q})p^m(q_i+p_i) I_3^{ssh}(\eta , k,p,q) \\
        &+8\epsilon_{jm}^{\lambda _1}(\mathbf{k}-\mathbf{p}-\mathbf{q})q^mp_i I_4^{ssh}(\eta , k,p,q)\bigg ) h^{\lambda _1}_{\mathbf{k}-\mathbf{p}-\mathbf{q}}\mathcal{R} _{\mathbf{q}} \mathcal{R} _{\mathbf{p}} \, .
    \end{split}
\end{align}
Above, we have decomposed the first-order perturbations into their primordial values and transfer functions (and polarisation functions for tensors) according to Eq.~\eqref{tensor1sol} and Eq.~\eqref{psitransfer}. The factor of $\epsilon^{ij}_{\lambda}(\mathbf{k}) ^*$ comes from the TT projection operator, and we note that there is a sum over the two polarisation states of the first-order GW, denoted $\lambda _1$. Furthermore, we have collected all the time dependence into the kernels $I_{\textit{i}}^{ssh}(\eta , k,p,q)$ ($\textit{i}=1,2,3,4$), defined as
\begin{equation}
    I_{\textit{i}}^{\Psi \Psi h}(\eta , k,p,q) =  \int _{\eta _i}^{\infty} {\rm} \overline{\eta} \, \frac{a(\overline{\eta})}{a(\eta)} G^h_{k}(\eta , \overline{\eta}) f^{\Psi \Psi h}_{\textit{i}}(\overline{\eta},k,p,q) \, ,
\end{equation}
with
{
\allowdisplaybreaks
\begin{align}
     &f_1^{\Psi \Psi h}(\eta ,k,p,q)= \left ( \frac{3+3w}{5+3w} \right )^2 \bigg ( -4T\conf _h(\eta|\mathbf{k}-\mathbf{p}-\mathbf{q}|)T_{\Psi}(\eta q) T\conf _{\Psi}(\eta p) - \frac{1}{3(1+w)}\nonumber\\
     &\times T_h(\eta|\mathbf{k}-\mathbf{p}-\mathbf{q}|)  \bigg ( 3(1+4w+3w^2) T\conf _{\Psi}(\eta q) T \conf _{\Psi}(\eta p) -24(w^2-1)T_{\Psi}(\eta q)T_{\Psi}(\eta p) \nonumber \\
     &- (-17-16w+9w^2)\, \mathbf{p}\cdot \mathbf{q}\, T_{\Psi}(\eta q)T_{\Psi}(\eta p) \bigg ) - 12( \mathbf{k}\cdot \mathbf{p} -  \mathbf{q}\cdot \mathbf{p} -p^2)\nonumber \\
     &\times T_h(\eta|\mathbf{k}-\mathbf{p}-\mathbf{q}|)T_{\Psi}(\eta q) T_{\Psi}(\eta p) - 8 \hubble T_h(\eta|\mathbf{k}-\mathbf{p}-\mathbf{q}|)T_{\Psi}(\eta q) T\conf _{\Psi}(\eta p)\nonumber \\
     &-\frac{8(w-1)}{3(1+w)\hubble} \, \mathbf{q} \cdot \mathbf{p} \, T_h(\eta|\mathbf{k}-\mathbf{p}-\mathbf{q}|)T_{\Psi}(\eta q) T\conf _{\Psi}(\eta p)\nonumber \\
    &-\frac{4(w-1)}{3(1+w)\hubble ^2} \, \mathbf{q}\cdot \mathbf{p} \, T_h(\eta|\mathbf{k}-\mathbf{p}-\mathbf{q}|)T \conf _{\Psi}(\eta q) T\conf _{\Psi}(\eta p) \bigg ) \, , \\
    &f_2^{\Psi \Psi h}(\eta ,k,p,q) =  f_3^{\Psi \Psi h}(\eta ,k,p,q)  = \left ( \frac{3+3w}{5+3w} \right )^2 \bigg ( T_h(\eta|\mathbf{k}-\mathbf{p}-\mathbf{q}|)T_{\Psi}(\eta q) T_{\Psi}(\eta p) \bigg ) \, , \\
    &f_4^{\Psi \Psi h}(\eta ,k,p,q) =\left ( \frac{3+3w}{5+3w} \right )^2 \bigg ( 2  T_h(\eta|\mathbf{k}-\mathbf{p}-\mathbf{q}|)T_{\Psi}(\eta q) T_{\Psi}(\eta p) + \frac{1}{3(1+w)\hubble } \nonumber \\
     &\times T \conf_h(\eta|\mathbf{k}-\mathbf{p}-\mathbf{q}|) T_{\Psi}(\eta q) T_{\Psi}(\eta p) -  \frac{1}{3(1+w)\hubble }T_h(\eta|\mathbf{k}-\mathbf{p}-\mathbf{q}|)T \conf _{\Psi}(\eta q) T_{\Psi}(\eta p) \nonumber \\
     &+  \frac{1}{3(1+w)\hubble ^2} T \conf_h(\eta|\mathbf{k}-\mathbf{p}-\mathbf{q}|)T_{\Psi}(\eta q)T \conf _{\Psi}(\eta q) - \frac{1}{3(1+w)\hubble ^2} T_h(\eta|\mathbf{k}-\mathbf{p}-\mathbf{q}|) \nonumber\\
     &\times T \conf _{\Psi}(\eta q) T\conf _{\Psi}(\eta p) \bigg ) \, 
\end{align}
}
We observe that the structure of these solutions remains largely unchanged compared to the case of second-order iGWs, except that there are now two unconstrained momenta, $\mathbf{p}$ and $\mathbf{q}$. 
%%%%%%%%%%%%%%%%%%%%%%%%%%%%%%%%%%%%%%%%%%%%%%%%%%%%%%%%%%%%%%%%%%%%
\subsection{Solution to the gravitational-wave equation for source terms with second-order contributions}
%%%%%%%%%%%%%%%%%%%%%%%%%%%%%%%%%%%%%%%%%%%%%%%%%%%%%%%%%%%%%%%%%%%%
We now present the solutions for the source terms that involve second-order perturbations coupled to first-order ones. We begin with sources containing second-order scalars, then consider those coupling second-order vectors to first-order perturbations, and finally address the source terms involving second-order tensors.
%%%%%%%%%%%%%%%%%%%%%%%%%%%%%%%%%%%%%%%%%%%%%%%%%%%%%%%%%%%%%%%%%%%%
\subsubsection{Source term containing second-order scalars}
%%%%%%%%%%%%%%%%%%%%%%%%%%%%%%%%%%%%%%%%%%%%%%%%%%%%%%%%%%%%%%%%%%%%
The source term that couples second order scalars to first order scalars, Eq.~\eqref{realspacesourcepsi2psi}, in Fourier space reads
\begin{align}
    \begin{split}
        \mathcal{S}_{\lambda }^{\Psi^{(2)}\Psi}(\eta , \mathbf{k}) &= \frac{1}{(2\pi)^{\frac{3}{2}}} \int {\rm} ^3 \mathbf{p}  \, \eps ^{ij}_{\lambda}(\mathbf{k})^* p_ip_j \bigg ( \frac{4}{3(1+w)\hubble ^2} \timeder \Psi (\eta,\mathbf{k}-\mathbf{p}) \timeder \Psi ^{(2)}_{st}(\eta,\mathbf{p}) \\
        &+ \frac{4}{3(1+w)\hubble}\left (\timeder \Psi (\eta,\mathbf{k}-\mathbf{p}) \Phi ^{(2)}_{st}(\eta,\mathbf{p}) +  \Psi (\eta,\mathbf{k}-\mathbf{p}) \timeder \Psi ^{(2)}_{st}(\eta,\mathbf{p})  \right ) \\
        &+ \frac{2(5+3w)}{3(1+w)}  \Psi (\eta,\mathbf{k}-\mathbf{p})  \Phi ^{(2)}_{st}(\eta,\mathbf{p})  \bigg ) \, .
    \end{split}
\end{align}
We recall that second-order perturbations coupled to first-order scalar fluctuations will only contribute to the GW background if these second-order scalar fluctuations are induced by first-order tensors coupled to first-order scalars. We can write the above source term exclusively in terms of $\Psi^{(2)}$, via Eq.~(\ref{anis}), which leads to 
\begin{align}
    \begin{split}
        &\mathcal{S}_{\lambda }^{\Psi^{(2)}\Psi}(\eta , \mathbf{k}) = \frac{1}{(2\pi)^{\frac{3}{2}}}  \int {\rm} ^3 \mathbf{p}  \, \eps ^{ij}_{\lambda}(\mathbf{k})^* p_ip_j \bigg ( \frac{4}{3(1+w)\hubble ^2} \timeder \Psi (\eta,\mathbf{k}-\mathbf{p}) \timeder \Psi ^{(2)}_{st}(\eta,\mathbf{p}) \\
        &+ \frac{4}{3(1+w)\hubble } \bigg (\timeder \Psi (\eta,\mathbf{k}-\mathbf{p}) \Psi ^{(2)}_{st}(\eta,\mathbf{p}) + \frac{1}{p^4}\timeder \Psi (\eta,\mathbf{k}-\mathbf{p}) \Pi ^{(2)}_{st}(\eta , \mathbf{p}) +  \Psi (\eta,\mathbf{k}-\mathbf{p})  \\
        &\times \timeder \Psi ^{(2)}_{st}(\eta,\mathbf{p})  \bigg )+ \frac{2(5+3w)}{3(1+w)} \bigg (  \Psi (\eta,\mathbf{k}-\mathbf{p})  \Psi ^{(2)}_{st}(\eta,\mathbf{p})+\frac{1}{p^4}\Psi (\eta,\mathbf{k}-\mathbf{p})  \Pi ^{(2)}_{st}(\eta , \mathbf{p})\bigg)  \bigg ) \, .
    \end{split}
\end{align}
The above source term is then substituted into the solution of the third-order iGW equation, Eq.~\eqref{GWsol3rdorder}, and we arrive at 
\begin{align}
    \begin{split} \label{GWeqsolpsi2psi1}
        h_{\lambda }^{\Psi^{(2)}\Psi}(\eta , \mathbf{k}) &= \frac{6}{(2\pi)^3} \int {\rm} ^3 \mathbf{q} \int {\rm} ^3 \mathbf{p} \sum _{ \lambda _1} \epsilon^{ij}_{\lambda}(\mathbf{k})^* p_ip_jQ_{\lambda _1}^{\Psi ^{(2)} _{st}}(\mathbf{p},\mathbf{q}) I^{\Psi ^{(2)}\Psi} (\eta , k,p,q) \\
        &\times  \mathcal{R} _{\mathbf{k}-\mathbf{p}}h_{\mathbf{q}}^{\lambda _1}  \mathcal{R} _{\mathbf{p}-\mathbf{q}} \, .
    \end{split}
\end{align}
To arrive at the expression above, we have substituted in the equation of motion for $ \Psi ^{(2)}_{st}(\eta,\mathbf{p})$, given in Eq.~\eqref{psiinducedst}, the expression for $\Pi ^{(2)}_{st}(\eta , \mathbf{p})$ in Eq.~\eqref{anisinducedst} and split up the first order perturbations into their primordial values and transfer functions. $ Q_{\lambda _1}^{\Psi ^{(2)} _{st}}(\mathbf{p},\mathbf{q})$ is defined in Eq.~\eqref{polinducedscalars} and there is a sum over the polarisation of the first-order GW since they are inducing the second-order scalars. The time dependence of the iGW is in the kernel, $I^{\Psi ^{(2)}\Psi} (\eta , k,p,q)$, defined as
\begin{equation} \label{kernelpsi2psi1}
    I^{\Psi ^{(2)}\Psi}(\eta , k,p,q) =  \int _{\eta _i}^{\infty} {\rm} \overline{\eta} \, \frac{a(\overline{\eta})}{a(\eta)} G^h_k(\eta , \overline{\eta}) f^{\Psi^{(2)}\Psi}(\overline{\eta},k,p,q) \, ,
\end{equation}
where
\begin{align}\label{psi2psifunction}
    \begin{split}
        &f^{\Psi^{(2)}\Psi}(\bareta ,k,p,q) = \avant ^2 \bigg ( \frac{4}{3(1+w)\hubble ^2} T_{\Psi}\conf (\bareta |\conv | ) \timederbar I^{\Psi ^{(2)}}_{st}(\bareta , p ,q ) +  \frac{4}{3(1+w)\hubble } \\
        &\times \left ( T_{\Psi}\conf (\bareta |\conv | )  I^{\Psi ^{(2)}}_{st}(\bareta , p ,q ) + \frac{1}{p^4} T_{\Psi}\conf (\bareta |\conv | ) f^{\Pi ^{(2)}}_{st}(\bareta , p , q) + T_{\Psi}(\bareta | \conv |) \timederbar I^{\Psi ^{(2)}}_{st}(\bareta , p ,q )\right  )\\
        &  + \frac{2(5+3w)}{3(1+w)} \left (T_{\Psi}(\bareta | \conv |) I^{\Psi ^{(2)}}_{st}(\bareta , p ,q ) + \frac{1}{p^4}T_{\Psi}(\bareta | \conv |)f^{\Pi ^{(2)}}_{st}(\bareta , p , q)   \right ) \bigg ) \, .
    \end{split}
\end{align}
We now see explicitly that the third order kernel, $I^{\Psi ^{(2)}\Psi} (\eta , k,p,q)$ depends on the second order scalar kernel $I^{\Psi ^{(2)}}_{st}(\bareta , p ,q )$, defined in Eq.~\eqref{kernelinducedstscalars}, giving rise to nested kernels. The function $f^{\Pi ^{(2)}}_{st}(\bareta , p , q)$ is defined in Eq.~\eqref{fpistscalar}.

%%%%%%%%%%%%%%%%%%%%%%%%%%%%%%%%%%%%%%%%%%%%%%%%%%%%%%%%%%%%%%%%%%%%
\subsubsection{Source term containing second-order vectors}
%%%%%%%%%%%%%%%%%%%%%%%%%%%%%%%%%%%%%%%%%%%%%%%%%%%%%%%%%%%%%%%%%%%%
In this subsection, we present the solution to the third-order gravitational wave equations, where the source consists of second-order vectors coupled to first-order perturbations.

The source term, given in real space by Eq.~\eqref{realspacesourcevectorpsi}, in Fourier space reads
\begin{align}
    \begin{split}
        &\mathcal{S}_{\lambda }^{B^{(2)}\Psi}(\eta , \mathbf{k}) = -i\sum _ r\frac{1}{(2\pi)^{\frac{3}{2}}} \int {\rm} ^3 \mathbf{p}  \, \eps ^{ij}_{\lambda}(\mathbf{k})^* p_ie^r_j(\mathbf{p}) \bigg ( 2B^{(2)st}_r(\eta, \mathbf{p}) \partial _{\eta}\Psi (\eta,\mathbf{k}-\mathbf{p}) 
        \\
        &+ 2\partial _{\eta}B^{(2)st}_r(\eta, \mathbf{p})  \Psi (\eta,\mathbf{k}-\mathbf{p}) +4\hubble B^{(2)st}_r(\eta, \mathbf{p})\Psi (\eta,\mathbf{k}-\mathbf{p}) + \frac{1}{3(1+w)\hubble ^2}p^2 
        \\
        &\times B^{(2)st}_r(\eta, \mathbf{p}) \left ( \hubble \Psi (\eta,\mathbf{k}-\mathbf{p}) + \partial _{\eta} \Psi (\eta,\mathbf{k}-\mathbf{p})  \right )  \bigg ) \, ,
    \end{split}
\end{align}
where we sum over $r$, the parity of the second-order vectors. After substituting the equation of motion for second-order vectors sourced by first-order scalars coupled to tensors, i.e. Eq.~\eqref{2ndordervecst}, and splitting up first-order perturbations into transfer functions and their corresponding primordial values, we can write down the third-order solution as
\begin{align}\label{GWeqsolb2psi1}
    \begin{split} 
        h_{\lambda }^{B^{(2)}\Psi}(\eta , \mathbf{k}) &= \frac{6}{(2\pi)^3} \sum _{\lambda _1} \sum _r  \int {\rm} ^3 \mathbf{p}  \int {\rm} ^3 \mathbf{q}  \, \bigg ( Q^{B^{(2)}\Psi}_{\lambda ,r}(\mathbf{k},\mathbf{p})Q^{B^{(2)}_{st1}}_{r,\lambda _1}(\mathbf{p},\mathbf{q}) I^{B^{(2)}\Psi}_1(\eta , k,p,q) \\
        &+Q^{B^{(2)}\Psi}_{\lambda ,r}(\mathbf{k},\mathbf{p})Q^{B^{(2)}_{st2}}_{r,\lambda _1}(\mathbf{p},\mathbf{q}) I^{B^{(2)}\Psi}_2(\eta , k,p,q) \bigg ) \mathcal{R} _{\mathbf{k}-\mathbf{p}}h^{\lambda _1}_{\mathbf{q}}\mathcal{R} _{\mathbf{p}-\mathbf{q}} \, .
    \end{split}
\end{align}
We defined a polarisation function 
\begin{equation}
    Q^{B^{(2)}\Psi}_{\lambda ,r}(\mathbf{k},\mathbf{p}) =  \eps ^{ij}_{\lambda}(\mathbf{k})^* p_ie^r_j(\mathbf{p}) \, ,
\end{equation}
whilst $Q^{B^{(2)}_{st1}}_{r,\lambda _1}(\mathbf{p},\mathbf{q})$ and  $Q^{B^{(2)}_{st2}}_{r,\lambda _1}(\mathbf{p},\mathbf{q})$ are defined respectively in Eq.~\eqref{pol2ndvecst1} and Eq.~\eqref{pol2ndvecst2}. The kernels are defined as follows, ($i=1,2$),
\begin{equation}\label{kernelb2psi1}
    I^{B^{(2)}\Psi}_{\textit{i}}(\eta , k,p,q) =  \int _{\eta _i}^{\infty} {\rm} \overline{\eta} \, \frac{a(\overline{\eta})}{a(\eta)} G^h_{k}(\eta , \overline{\eta}) f^{B^{(2)\Psi}}_{\textit{i}}(\overline{\eta},k,p,q) \, ,
\end{equation}
where
\begin{subequations}
    \begin{align}
        \begin{split}
            &f^{B^{(2)\Psi}}_1(\overline{\eta},k,p,q) = \avant ^2 \bigg ( \frac{2}{p^2} I^{B^{(2)}}_{st1}(\overline{\eta} ,k,p,q) T^{\prime}_{\Psi}(\overline{\eta} |\mathbf{k}-\mathbf{p}|) + \frac{2}{p^2}\frac{\partial}{\partial _{\overline{\eta}}} I^{B^{(2)}}_{st1}(\overline{\eta} ,k,p,q) \\
            &\times T_{\Psi}(\overline{\eta} |\mathbf{k}-\mathbf{p}|) + \frac{4\hubble}{p^2} I^{B^{(2)}}_{st1}(\overline{\eta} ,k,p,q) T_{\Psi}(\overline{\eta} |\mathbf{k}-\mathbf{p}|) + \frac{1}{3(1+w)\hubble} I^{B^{(2)}}_{st1}(\overline{\eta} ,k,p,q)  \\
            &\times T_{\Psi}(\overline{\eta} |\mathbf{k}-\mathbf{p}|)+  \frac{1}{3(1+w)\hubble ^2} I^{B^{(2)}}_{st1}(\overline{\eta} ,k,p,q) T^{\prime}_{\Psi}(\overline{\eta} |\mathbf{k}-\mathbf{p}|)   \bigg ) \, ,
        \end{split}
        \\
        \begin{split}
            &f^{B^{(2)\Psi}}_2(\overline{\eta},k,p,q) =  \avant ^2 \bigg ( \frac{2}{p^2} I^{B^{(2)}}_{st2}(\overline{\eta} ,k,p,q) T^{\prime}_{\Psi}(\overline{\eta} |\mathbf{k}-\mathbf{p}|) + \frac{2}{p^2}\frac{\partial}{\partial _{\overline{\eta}}} I^{B^{(2)}}_{st2}(\overline{\eta} ,k,p,q) \\
            &\times T_{\Psi}(\overline{\eta} |\mathbf{k}-\mathbf{p}|) + \frac{4\hubble}{p^2} I^{B^{(2)}}_{st2}(\overline{\eta} ,k,p,q) T_{\Psi}(\overline{\eta} |\mathbf{k}-\mathbf{p}|) + \frac{1}{3(1+w)\hubble} I^{B^{(2)}}_{st2}(\overline{\eta} ,k,p,q)  \\
            &\times T_{\Psi}(\overline{\eta} |\mathbf{k}-\mathbf{p}|) +  \frac{1}{3(1+w)\hubble ^2} I^{B^{(2)}}_{st2}(\overline{\eta} ,k,p,q) T^{\prime}_{\Psi}(\overline{\eta} |\mathbf{k}-\mathbf{p}|)   \bigg ) \, .
        \end{split}
    \end{align}
\end{subequations}
The nested kernels $ I^{B^{(2)}}_{st1}(\overline{\eta} ,k,p,q)$ and $ I^{B^{(2)}}_{st2}(\overline{\eta} ,p,q)$ are defined in Eqs.~\eqref{kernelvectorst1} and \eqref{kernelvectorst2}, respectively.
%%%%%%%%%%%%%%%%%%%%%%%%%%%%%%%%%%%%%%%%%%%%%%%%%%%%%%%%%%%%%%%%%%%%
\subsubsection{Source term containing second-order tensors}
%%%%%%%%%%%%%%%%%%%%%%%%%%%%%%%%%%%%%%%%%%%%%%%%%%%%%%%%%%%%%%%%%%%%
Finally, the last type of source terms we need to consider contains second-order tensor perturbations.

The source term in Eq.~\eqref{realspaceh2psi} is expressed as a Fourier integral such that
\begin{align}
    \begin{split}
        &\mathcal{S}_{\lambda }^{h^{(2)}\Psi}(\eta , \mathbf{k}) = \frac{1}{(2\pi)^{\frac{3}{2}}}\sum _{\sigma }  \int {\rm} ^3 \mathbf{p}  \, \eps ^{ij}_{\lambda}(\mathbf{k})^* \eps _{ij}^{\sigma}(\mathbf{p}) \bigg ( \big ( \timeder ^2 h^{(2)st}_{\sigma}(\eta , \mathbf{p}) + 2 \hubble \timeder h^{(2)st}_{\sigma}(\eta , \mathbf{p})  \\
        &+ p^2 h^{(2)st}_{\sigma}(\eta , \mathbf{p}) \big ) \Psi (\eta,\mathbf{k}-\mathbf{p}) -2p^2h^{(2)st}_{\sigma}(\eta , \mathbf{p})\Psi (\eta,\mathbf{k}-\mathbf{p}) + (3w+1)\hubble h^{(2)st}_{\sigma}(\eta , \mathbf{p}) \\
        &\times  \timeder \Psi (\eta,\mathbf{k}-\mathbf{p})  + (w-1)|\mathbf{k}-\mathbf{p}|^2 h^{(2)st}_{\sigma}(\eta , \mathbf{p})\Psi (\eta,\mathbf{k}-\mathbf{p})-2(k_m-p_m)p^m h^{(2)st}_{\sigma}(\eta , \mathbf{p}) \\
        &\Psi (\eta,\mathbf{k}-\mathbf{p})  \bigg )\, ,
    \end{split}
\end{align}
where we have summed over the polarisation states ($\sigma$) of the second order tensor. In the first line above, we recognise the second order equation of motion for tensors and substitute in the scalar-tensor source term  and also substitute in the solution for scalar-tensor induced waves, Eq.~\eqref{solh2st}, in place of $h^{(2)st}_{\sigma}(\eta , \mathbf{p})$. We arrive at the third-order solution
\begin{align}\label{GWeqsolh2psi1}
    \begin{split}
        h_{\lambda }^{h^{(2)}\Psi}(\eta , \mathbf{k}) &= \frac{6}{(2\pi)^3} \int {\rm} ^3 \mathbf{q} \int {\rm} ^3 \mathbf{p} \sum _{\sigma , \lambda _1} \epsilon^{ij}_{\lambda}(\mathbf{k})^* \epsilon_{ij}^{\sigma}(\mathbf{p})\epsilon^{mn}_{\sigma}(\mathbf{p})^*\epsilon_{mn}^{\lambda _1}(\mathbf{q})I^{h^{(2)}\Psi}(\eta , k,p,q) \\
        &\times \mathcal{R} _{\mathbf{k}-\mathbf{p}} h_{\mathbf{q}}^{\lambda _1} \mathcal{R} _{\mathbf{p}-\mathbf{q}} \, ,
    \end{split}
\end{align}
where the kernel, $I^{h^{(2)}\Psi}(\eta  , k,p,q)$, is defined as  
\begin{equation} \label{kernelh2psi1}
    I^{h^{(2)}\Psi}(\eta  , k,p,q) =  \int _{\eta _i}^{\infty} {\rm} \overline{\eta} \, \frac{a(\overline{\eta})}{a(\eta)} G^h_{k}(\eta , \overline{\eta}) f^{h^{(2)\Psi}}(\overline{\eta},k,p,q) \, ,
\end{equation}
and
\begin{align}\label{ffuntionh2psi1}
    \begin{split}
        &f^{h^{(2)\Psi}}(\overline{\eta},k,p,q) = \left ( \frac{3+3w}{5+3w} \right )^2 \bigg ( -8q^2 T_{\Psi} (\overline{\eta} |\mathbf{p}-\mathbf{q}|) T_h(\overline{\eta} q) T_{\Psi} (\overline{\eta} |\mathbf{k}-\mathbf{p}|)  - 8 (p_i - q_i)q^i \\
        &\times T_{\Psi} (\overline{\eta} |\mathbf{p}-\mathbf{q}|) T_h(\overline{\eta} q) T_{\Psi} (\overline{\eta} |\mathbf{k}-\mathbf{p}|) + 4\hubble (1+3w) \partial _{\bareta} T_{\Psi} (\overline{\eta} |\mathbf{p}-\mathbf{q}|) T_h(\overline{\eta} q) T_{\Psi} (\overline{\eta} |\mathbf{k}-\mathbf{p}|) \\
        &-4(1-w)|\mathbf{p}-\mathbf{q}|^2 T_{\Psi} (\overline{\eta} |\mathbf{p}-\mathbf{q}|) T_h(\overline{\eta} q) T_{\Psi} (\overline{\eta} |\mathbf{k}-\mathbf{p}|) -8p^2I^{h^{(2)}}_{st}(\overline{\eta} , p ,q)T_{\Psi} (\overline{\eta} |\mathbf{k}-\mathbf{p}|) \\
        &+4(3w+1)\hubble I^{h^{(2)}}_{st}(\overline{\eta} , p ,q)\partial _{\bareta} T_{\Psi} (\overline{\eta} |\mathbf{k}-\mathbf{p}|) + 4(w-1)|\mathbf{k}-\mathbf{p}|^2 I^{h^{(2)}}_{st}(\overline{\eta} , p ,q) \\
        &\times T_{\Psi} (\overline{\eta} |\mathbf{k}-\mathbf{p}|) -8(k_i-p_i)p^iI^{h^{(2)}}_{st}(\overline{\eta} , p ,q)T_{\Psi} (\overline{\eta} |\mathbf{k}-\mathbf{p}|)  \bigg ) \, .
    \end{split}
\end{align}
The nested kernel, $I^{h^{(2)}}_{st}(\overline{\eta} , p ,q)$, is defined in Eq.~\eqref{stnestedbase}. 

This concludes our discussion of all terms and their solutions in Fourier space; we now turn to their contributions to the power spectrum. 

%%%%%%%%%%%%%%%%%%%%%%%%%%%%%%%%%%%%%%%%%%%%%%%%%%%%%%%%%%%%%%%%%%%%%%%%%%%%%%%%%%%%%%%%%%%%%%
%%%%%%%%%%%%%%%%%%%%%%%%%%%%%%%%%%%%%%%%%%%%%%%%%%%%%%%%%%%%%%%%%%%%%%%%%%%%%%%%%%%%%%%%%%%%%%
\section{Power spectra of the different third-order contributions}
%%%%%%%%%%%%%%%%%%%%%%%%%%%%%%%%%%%%%%%%%%%%%%%%%%%%%%%%%%%%%%%%%%%%%%%%%%%%%%%%%%%%%%%%%%%%%%
%%%%%%%%%%%%%%%%%%%%%%%%%%%%%%%%%%%%%%%%%%%%%%%%%%%%%%%%%%%%%%%%%%%%%%%%%%%%%%%%%%%%%%%%%%%%%%
In this section, we compute the power spectrum for each contribution we have considered. In order to do this, we substitute in the third-order solution to Eq.~\eqref{generalcorrelationoftensors} and correlate it with a first-order GW generated in inflation. Explicitly, we are after $\mathcal{P}^{\text{$(13)$}} (\eta , k)$, which can be found by inserting our third-order iGW equation solutions into 
\begin{equation}\label{correlation31}
     \langle h_{\lambda}(\mathbf{k})h_{\lambda \conf}^{(3)}(\mathbf{k \conf}) \rangle = \delta ^{\lambda \lambda ^{\prime}} \delta (\mathbf{k}+ \mathbf{k^{\prime}}) \frac{2\pi ^2}{k^3}\mathcal{P}_{\lambda}^{\text{$(13)$}} (\eta , k) \, .
\end{equation}
There are four different contributions, specifically:
\begin{align}
    \begin{split}
        \mathcal{P}^{(13)}(\eta,k) &=  \mathcal{P}_{\Psi \Psi h}^{(13)}(\eta,k) + \mathcal{P}_{\Psi ^{(2)}\Psi}^{(13)}(\eta,k) + \mathcal{P}_{B ^{(2)}\Psi}^{(13)}(\eta,k)  + \mathcal{P}_{h^{(2)}\Psi}^{(13)}(\eta,k)  \, .
    \end{split}
\end{align}
In what follows, we have dedicated a subsection to each of these terms. 

Beforehand, we recall that the power spectrum of the comoving curvature perturbation is defined in Eq.~\eqref{defpsR}. Additionally, we again parametrise vectors in spherical coordinates and align the wave-vector $\mathbf{k}$ (i.e. the wave-vector of the third-order iGW) with the $z$-axis ($\theta _k=\phi _k=0$). Moreover, we will see that the correlation constrains the ingoing wave-vectors to be $\mathbf{p}$ and $\mathbf{k}-\mathbf{p}$ (through conservation of momentum), and so it is useful to use the $v$ and $u$ coordinates defined in Eq.~\eqref{uvcoordinates}. For the nested kernels, it is useful to introduce $\tilde{y}=p\tilde{\eta}$. The subsequent relations are helpful
\begin{equation}\label{yrelations}
    \quad p\overline{\eta}=v\overline{x}, \quad |\mathbf{k}-\mathbf{p}|\tilde{\eta}=\frac{\tilde{y}u}{v}\quad \text{and} \quad k\tilde{\eta} = \frac{\tilde{y}}{v}\, .
\end{equation}

We now proceed to compute each contribution. The results in this section hold for a general equation of state parameter $w \neq  -1/3 $. 

%%%%%%%%%%%%%%%%%%%%%%%%%%%%%%%%%%%%%%%%%%%%%%%%%%%%%%%%%%%%%%%%%%%%%%%%%%%%%%%%%%%%%%%%%%%%%%
\subsection{The power spectrum of the scalar-scalar-tensor first-order contribution: 
\texorpdfstring{$\mathcal{P}^{(13)}_{\Psi \Psi h}$}{P(13)PsiPsiH}}
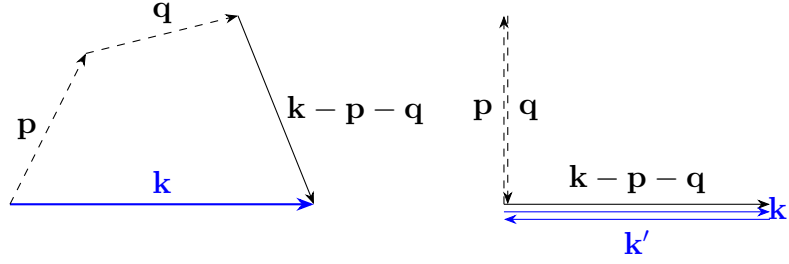
\begin{figure}[t]
  \centering
  \begin{tikzpicture}

    \coordinate (A) at (0,0);
    \coordinate (B) at (1,2);
    \coordinate (C) at (3,2.5);
    \coordinate (D) at (4,0);

    % Label the corners
    %\node[below left] at (A) {A};
    %\node[below right] at (B) {B};
    %\node[above] at (C) {C};
    %\node[below left] at (D) {D};

    \draw[thick, -{Stealth}, blue] (A) -- (D);
    \draw[ -{Stealth}, black] (C) -- (D);
    \draw[dashed, -{Stealth}, black] (A) -- (B);
    \draw[dashed, -{Stealth}, black] (B) -- (C);

    \node[above, text=blue] at ($(A)!0.5!(D)$) {$\mathbf{k}$};
    \node[left] at ($(A)!0.5!(B)$) {$\mathbf{p}$};
    \node[above] at ($(B)!0.5!(C)$) {$\mathbf{q}$};
    \node[right] at ($(C)!0.5!(D)$) {$\mathbf{k}-\mathbf{p}-\mathbf{q}$};

    \coordinate (E) at (6.5,0);
    \coordinate (F) at (6.55,0);
    \coordinate (G) at (6.5,2.5);
    \coordinate (H) at (6.55,2.5);
    \coordinate (I) at (10,0);
    \coordinate (J) at (6.5,-0.1);
    \coordinate (K) at (10,-0.1);
    \coordinate (L) at (6.5,-0.2);
    \coordinate (M) at (10,-0.2);

    \draw[dashed, -{Stealth}, black] (E) -- (G);
    \draw[dashed, -{Stealth}, black] (H) -- (F);
    \draw[ -{Stealth}, black] (E) -- (I);
    \draw[ -{Stealth}, blue] (J) -- (K);
    \draw[ -{Stealth}, blue] (M) -- (L);
    %\draw[thick, -{Stealth}, blue] (F) -- (E);
    %\draw[ -{Stealth}, black] (A) -- (C);
    %\draw[-{Stealth}, black] (B) -- (D);
    %\draw[dashed, -{Stealth}, black] (D) -- (A);
    %\draw[thick] (A) -- (B) -- (C) -- cycle;

    % Label the corners
    %\node[below left] at (E) {E};
    %\node[below right] at (B) {B};
    %\node[above] at (C) {C};
    %\node[below left] at (D) {D};

    \node[text=blue] at (10.1,-0.05) {$\mathbf{k}$};
    \node[below, text=blue] at ($(M)!0.5!(L)$) {$\mathbf{k^{\prime}}$};
    \node[left] at ($(E)!0.5!(G)$) {$\mathbf{p}$};
    \node[right] at ($(F)!0.5!(H)$) {$\mathbf{q}$};
    \node[above] at ($(E)!0.5!(I)$) {$\mathbf{k}-\mathbf{p}-\mathbf{q}$};
  \end{tikzpicture}
  
  \caption{\footnotesize{(\textit{Left}). Momentum configuration of two scalar modes (dashed black) and a tensor mode (solid black), inducing a third-order tensor mode (solid blue). (\textit{Right}). Geometrical configuration of the scalar-scalar-tensor two-point function when contracted with a first-order tensor mode. The two scalar modes must have opposite momentum, and the first-order tensor mode has equal momentum to the outgoing (third-order) one.}}
  \label{fig:ssh2point}
\end{figure}

%%%%%%%%%%%%%%%%%%%%%%%%%%%%%%%%%%%%%%%%%%%%%%%%%%%%%%%%%%%%%%%%%%%%%%%%%%%%%%%%%%%%%%%%%%%%%%
In this subsection, we compute the correlation of first and third-order tensor modes sourced by scalar-scalar-tensor interactions. We substitute Eq.~\eqref{solhpsipsi} into Eq.~\eqref{correlation31}
\begin{align}\label{h1h3psipsih}
    \begin{split}
            &\langle h_{\lambda ^{\prime}}(\mathbf{k^{\prime}})h^{\Psi \Psi h}_{\lambda }(\mathbf{k}) \rangle = \frac{6}{(2\pi)^3} \int {\rm} ^3 \mathbf{q} \int {\rm} ^3 \mathbf{p} \, T_h(\eta k \conf ) \epsilon^{ij}_{\lambda}(\mathbf{k})^*  \sum _{\lambda _1} \bigg ( \epsilon_{ij}^{\lambda _1}(\mathbf{k}-\mathbf{p}-\mathbf{q})  \\
            &\times I_1^{ssh}(\eta , k,p,q) + 16 \epsilon_{jm}^{\lambda _1}(\mathbf{k}-\mathbf{p}-\mathbf{q})p^mp_i I_2^{ssh}(\eta , k,p,q) - 8\epsilon_{jm}^{\lambda _1}(\mathbf{k}-\mathbf{p}-\mathbf{q})p^m(q_i+p_i)  \\
            &\times I_3^{ssh}(\eta , k,p,q) +8\epsilon_{jm}^{\lambda _1}(\mathbf{k}-\mathbf{p}-\mathbf{q})q^mp_i I_4^{ssh}(\eta , k,p,q)\bigg )  \langle h^{\lambda \conf} _{\mathbf{k \conf}}h^{\lambda _1}_{\mathbf{k}-\mathbf{p}-\mathbf{q}}\mathcal{R} _{\mathbf{q}} \mathcal{R} _{\mathbf{p}} \rangle .
    \end{split}
\end{align}
The four-point function is broken down as
\begin{equation}
     \langle h^{\lambda \conf} _{\mathbf{k \conf}}h^{\lambda _1}_{\mathbf{k}-\mathbf{p}-\mathbf{q}}\mathcal{R} _{\mathbf{q}} \mathcal{R} _{\mathbf{p}} \rangle = \delta ^{\lambda \conf \lambda _1} \delta (\mathbf{k \conf} + \mathbf{k} - \mathbf{p} - \mathbf{q}) \delta (\mathbf{p}+ \mathbf{q}) P^{\lambda \conf}_h(\eta k\conf) P_{\mathcal{R}} (\eta q) \, ,
\end{equation}
where we recall that we are disregarding any connected contributions that could arise from inflation. The above suggests that, physically, out of the three modes responsible for inducing the third-order gravitational wave, one must share the same momentum as the outgoing third-order wave (or the first-order gravitational wave we are correlating with), while the other two modes must have opposite momenta, see Fig.~\ref{fig:ssh2point}. This physical picture contrasts with the usual second-order production mechanism encountered in SIGWs, which was depicted in Fig.~\ref{fig:sigwscontraction}. We now substitute the expression for the four-point function into Eq.~\eqref{h1h3psipsih} and integrate over $\mathbf{p}$ 
\begin{align}
    \begin{split}
            \langle h_{\mathbf{k^{\prime}},\lambda ^{\prime}}h^{\Psi \Psi h}_{\mathbf{k},\lambda } \rangle &= \frac{6}{(2\pi)^3} \int {\rm} ^3 \mathbf{q}\sum _{\lambda _1} \bigg ( \epsilon^{ij}_{\lambda}(\mathbf{k})^* \epsilon_{ij}^{\lambda _1}(\mathbf{k})  T_h(\eta k \conf )I_1^{ssh}(\eta , k,q) \\
            &+ 16\epsilon^{ij}_{\lambda}(\mathbf{k}) ^*\epsilon_{jm}^{\lambda _1}(\mathbf{k})q^mq_i  T_h(\eta k \conf )I_2^{ssh}(\eta , k,q)   \\
            & -8\epsilon^{ij}_{\lambda}(\mathbf{k}) ^*\epsilon_{jm}^{\lambda _1}(\mathbf{k})q^mq_i  T_h(\eta k \conf )I_4^{ssh}(\eta , k,q)\bigg )  \delta ^{\lambda \conf \lambda _1} \delta (\mathbf{k} + \mathbf{k \conf} ) P^{\lambda \conf}_h(\eta k\conf) P_{\mathcal{R}} (\eta q) \, .
    \end{split}
\end{align}
Using the normalisation condition of the polarisation tensors and the result
\begin{equation}
    \epsilon^{ij}_{\lambda}(\mathbf{k}) ^*\epsilon_{jm}^{\lambda _1}(\mathbf{k})q^mq_i  = \frac{1}{2}q^2 \sin ^2\theta _q \delta ^{\lambda  \lambda _1} \, ,
\end{equation}
the two-point function finally reduces to 
\begin{align}
    \begin{split}
            &\langle h_{\mathbf{k^{\prime}},\lambda ^{\prime}}h^{(3)}_{\mathbf{k},\lambda } \rangle = \frac{6}{(2\pi)^3} \int _0^{\infty} {\rm} q \int _0^{\pi} {\rm} \theta _q \int _0^{2\pi} {\rm} \phi _q  \, q^2 \sin \theta _q \bigg (  T_h(\eta k \conf )I_1^{ssh}(\eta , k,q)  \\
            &+ 8q^2 \sin ^2\theta _q  T_h(\eta k \conf ) I_2^{ssh}(\eta , k,q) -4q^2 \sin ^2\theta _q  T_h(\eta k \conf )I_4^{ssh}(\eta , k,q)\bigg ) \\
            &\times \delta ^{\lambda \conf \lambda _1}\delta ^{\lambda  \lambda _1} \delta (\mathbf{k} + \mathbf{k \conf} ) P^{\lambda \conf}_h(k\conf) P_{\mathcal{R}} ( q) \, .
    \end{split}
\end{align}
An important point to specify, relevant for source terms involving pure linear perturbations, is that once we have integrated over one of the internal momenta, there is only a trivial angular dependence remaining. This means that after expressing the remaining internal momenta in spherical coordinates, we can integrate over both the polar and azimuthal angles, as the only angular dependence will reside in the polarisation functions, see Fig.~\ref{fig:ssh2point} for a geometrical interpretation.  

Furthermore, unlike second-order iGWs, the primordial tensor power spectrum appears outside the integral and does not depend on the modes inducing the third-order perturbations. Therefore, these contributions act as a modulation to the primordial gravitational wave power spectrum. By integrating over the angular directions in spherical coordinates, relabelling $\mathbf{q}$ to $\mathbf{p}$, we can extract the power spectrum as follows
\begin{align}
    \begin{split}
        P^{(13)}_{R/L}(\eta , k) &= k^3\frac{6}{(2\pi )^2} P^{(11)}_{R/L}( k) \int _0 ^{\infty} {\rm}v \, v^2 \bigg (\mathcal{I}^{ssh}_1(x,k,v) + \frac{32}{3}v^2k^2\mathcal{I}^{ssh}_2(x,k,v) \\
        &- \frac{16}{3}v^2k^2 \mathcal{I}^{ssh}_4(x,k,v) \bigg ) P_{\mathcal{R}}( vk) \, ,
    \end{split}
\end{align}
where we have further defined 
\begin{equation}
    \mathcal{I}^{ssh}_{\textit{i}}(x,k,v) = T_h(x )  I_{\textit{i}}^{ssh}(x,k,v) \, .
\end{equation}
Finally, the dimensionless power spectrum is given by:
\begin{align}\label{powerspectrumsshfinal}
    \begin{split}
        \mathcal{P}^{(13)}_{\Psi \Psi h}(\eta , k) &= 3 \left ( \mathcal{P}^R_h(k) + \mathcal{P}^L_h(k) \right ) \int _0 ^{\infty} {\rm}v \, \frac{1}{v} \bigg (\mathcal{I}^{ssh}_1(x , k , v) + \frac{32}{3}v^2k^2\mathcal{I}^{ssh}_2(x , k , v) \\
        &- \frac{16}{3}v^2k^2 \mathcal{I}^{ssh}_4(x , k , v) \bigg ) \mathcal{P}_{\mathcal{R}}( vk) \, ,
    \end{split}
\end{align}
and the functions $f_1^{\Psi \Psi h}(\barx , k ,v)$, $f_2^{\Psi \Psi h}(\barx , k ,v)$ and $f_4^{\Psi \Psi h}(\barx , k ,v)$ in the appropriate coordinates are
{
\allowdisplaybreaks
\begin{align}
    &f_1^{\Psi \Psi h}(\barx , k ,v) = \left ( \frac{3+3w}{5+3w} \right )^2 \bigg ( -4k^2 \frac{\partial}{\partial \barx}T _h(\barx)T_{\Psi}(\barx v) \frac{\partial}{\partial \barx}T_{\Psi}(\barx v) - \frac{1}{3(1+w)}T_h(\barx )\nonumber \\
    &\times   \bigg ( 3k^2(1+4w+3w^2) \frac{\partial}{\partial \barx}T_{\Psi}(\barx v)\frac{\partial}{\partial \barx}T _{\Psi}(c_s\barx v) -24(w^2-1)T_{\Psi}(\barx v)T_{\Psi}(\barx v) \nonumber\\
    &+ (-17-16w+9w^2)\, v^2k^2\, T_{\Psi}(\barx v)T_{\Psi}(\barx v) \bigg ) - 8 \frac{k^2}{\barx}T_h(\barx)T_{\Psi}(\barx v)\frac{\partial}{\partial \barx}T_{\Psi}(\barx v)\nonumber\\
    &+\frac{8(w-1)}{3(1+w)} \barx v^2k^2 \, T_h(\barx)T_{\Psi}(\barx v) \frac{\partial}{\partial \barx}T_{\Psi}(\barx v)+\frac{4(w-1)}{3(1+w)} \, \barx ^2 v^2 k^2 \, \frac{\partial}{\partial \barx}T_h(\barx)\nonumber \\
    &\times T_{\Psi}(\barx v)\frac{\partial}{\partial \barx}  T_{\Psi}(\barx v)\bigg ) \, , \label{sshfunctionsfinal1} \\
    &f_2^{\Psi \Psi h}(\barx , k ,v) =  \left ( \frac{3+3w}{5+3w} \right )^2  T_h(\barx)T_{\Psi}(\barx v) T_{\Psi}(\barx v) \, , \label{sshfunctionsfinal2}\\
    &f_4^{\Psi \Psi h}(\barx , k ,v) =\left ( \frac{3+3w}{5+3w} \right )^2 \bigg ( 2 T_h(\barx)T_{\Psi}(\barx v) T_{\Psi}(\barx v) + \frac{1}{3(1+w) }\barx \frac{\partial}{\partial \barx}T_h(\barx) \nonumber \\
    & \times T_{\Psi}(\barx v) T_{\Psi}(\barx v) -  \frac{1}{3(1+w) }\barx T_h(\barx) \frac{\partial}{\partial \barx}T_{\Psi}(\barx v) T_{\Psi}(\barx v) +  \frac{1}{3(1+w)} \barx ^2   \nonumber \\
    &\times \frac{\partial}{\partial \barx}T_h(\barx) T_{\Psi}(\barx v) \frac{\partial}{\partial \barx}T_{\Psi}(\barx v) - \frac{1}{3(1+w)\hubble ^2}\barx ^2 T_h(\barx)\frac{\partial}{\partial \barx} T_{\Psi}(\barx v) \frac{\partial}{\partial \barx}T_{\Psi}(\barx v) \bigg ) \,  . \label{sshfunctionsfinal3}
\end{align}
}

%%%%%%%%%%%%%%%%%%%%%%%%%%%%%%%%%%%%%%%%%%%%%%%%%%%%%%%%%%%%%%%%%%%%%%%%%%%%%%%%%%%%%%%%%%%%%%
\subsection{The power spectrum of the second-order scalars coupled to first-order scalars: 
\texorpdfstring{$\mathcal{P}^{(13)}_{\Psi ^{(2)} \Psi}$}{P(13)Psi2Psi}}
\begin{figure}[t]
  \centering
  \begin{tikzpicture}

    \coordinate (A) at (0,0);
    \coordinate (B) at (1,2);
    \coordinate (C) at (3,2.5);
    \coordinate (D) at (4,0);

    % Label the corners
    %\node[below left] at (A) {A};
    %\node[below right] at (B) {B};
    %\node[above] at (C) {C};
    %\node[below left] at (D) {D};

    \draw[thick, -{Stealth}, blue] (A) -- (D);
    \draw[dashed, -{Stealth}, black] (C) -- (D);
    \draw[ -{Stealth}, black] (A) -- (B);
    \draw[dashed, -{Stealth}, black] (B) -- (C);
    \draw[ -{Stealth}, red] (A) -- (C);

    \node[above, text=blue] at ($(A)!0.5!(D)$) {$\mathbf{k}$};
    \node[left] at ($(A)!0.5!(B)$) {$\mathbf{q}$};
    \node[above] at ($(B)!0.5!(C)$) {$\mathbf{p}-\mathbf{q}$};
    \node[right] at ($(C)!0.5!(D)$) {$\mathbf{k}-\mathbf{p}$};
    \node[below, text=red] at ($(A)!0.5!(C)$) {$\mathbf{p}$};

    \coordinate (E) at (6.5,0);
    \coordinate (F) at (6.55,0);
    \coordinate (G) at (6.5,2.5);
    \coordinate (H) at (6.55,2.5);
    \coordinate (I) at (10,0);
    \coordinate (J) at (6.5,-0.1);
    \coordinate (K) at (10,-0.1);
    \coordinate (L) at (6.5,-0.2);
    \coordinate (M) at (10,-0.2);
    \coordinate (O) at (9.9,0.1);
    \coordinate (P) at (9.9,2.5);
    \coordinate (Q) at (10,0.1);
    \coordinate (R) at (10,2.5);

    \draw[dashed, -{Stealth}, black] (O) -- (P);
    \draw[dashed, -{Stealth}, black] (R) -- (Q);
    \draw[ -{Stealth}, black] (E) -- (I);
    \draw[ -{Stealth}, blue] (J) -- (K);
    \draw[ -{Stealth}, blue] (M) -- (L);
    %\draw[thick, -{Stealth}, blue] (F) -- (E);
    %\draw[ -{Stealth}, black] (A) -- (C);
    %\draw[-{Stealth}, black] (B) -- (D);
    %\draw[dashed, -{Stealth}, black] (D) -- (A);
    %\draw[thick] (A) -- (B) -- (C) -- cycle;

    % Label the corners
    %\node[below left] at (E) {E};
    %\node[below right] at (B) {B};
    %\node[above] at (C) {C};
    %\node[below left] at (D) {D};

    \node[text=blue] at (10.1,-0.05) {$\mathbf{k}$};
    \node[below, text=blue] at ($(M)!0.5!(L)$) {$\mathbf{k^{\prime}}$};
    \node[left] at ($(O)!0.5!(P)$) {$\mathbf{p}-\mathbf{q}$};
    \node[right] at ($(Q)!0.5!(R)$) {$\mathbf{k}-\mathbf{p}$};
    \node[above] at ($(E)!0.5!(I)$) {$\mathbf{q}$};
  \end{tikzpicture}
  
  \caption{\footnotesize{(\textit{Left}). Momentum configuration of a second-order mode (solid red) and a scalar mode (dashed black), inducing a third-order tensor mode (solid blue). The second-order perturbation is itself induced by a scalar and a first-order tensor perturbation (solid black). (\textit{Right}). Geometrical configuration of the two-point function resulting from a second-order perturbation coupled to a first-order scalar perturbation with a first-order tensor mode. In this configuration, the two scalar modes depend on the angle between the incoming second-order mode and the outgoing third-order mode, making the dependence on the coordinate $u$ non-trivial.}}
  \label{fig:secondorderfirstorderscalar2point}
\end{figure}
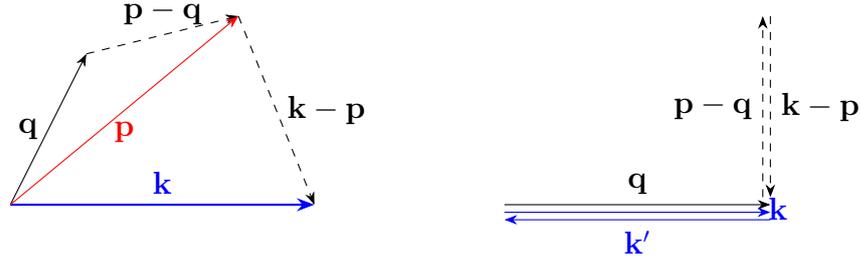

%%%%%%%%%%%%%%%%%%%%%%%%%%%%%%%%%%%%%%%%%%%%%%%%%%%%%%%%%%%%%%%%%%%%%%%%%%%%%%%%%%%%%%%%%%%%%%
In this section, we utilise the solution given by Eq.~\eqref{GWeqsolpsi2psi1}, which describes third-order gravitational waves sourced by second-order scalar perturbations coupled to first-order scalars. The corresponding two-point function is given by
\begin{align}
    \begin{split}
            \langle h_{\lambda ^{\prime}}(\mathbf{k^{\prime}})h^{\Psi ^{(2)}\Psi}_{\lambda }(\mathbf{k}) \rangle &= \frac{6}{(2\pi)^3} \int {\rm} ^3 \mathbf{q} \int {\rm} ^3 \mathbf{p} \, T_h(\eta k \conf )\sum _{\lambda _1}\eps ^{ij}_{\lambda}(\mathbf{k})^* p_ip_j Q^{\lambda _1}_{\pi ,st}(\mathbf{p},\mathbf{q}) \\
            &\times   I^{\Psi ^{(2)}\Psi}(\eta , k,p,q) \langle h^{\lambda \conf} _{\mathbf{k \conf}} \mathcal{R} _{\mathbf{k}-\mathbf{p}}h_{\mathbf{q}}^{\lambda _1} \mathcal{R} _{\mathbf{p}-\mathbf{q}}\rangle \, .
    \end{split}
\end{align}
The four-point function can be broken down as
\begin{equation}\label{stthirdorder4point}
     \langle h^{\lambda \conf} _{\mathbf{k \conf}}h_{\mathbf{q}}^{\lambda _1} \mathcal{R} _{\mathbf{p}-\mathbf{q}} \mathcal{R} _{\mathbf{k}-\mathbf{p}}\rangle = \delta ^{\lambda \conf \lambda _1} \delta (\mathbf{k \conf}+ \mathbf{q}) \delta (\mathbf{k}-\mathbf{q}) P^{\lambda \conf} _h (k \conf) P_{\mathcal{R}} (|\mathbf{p}-\mathbf{q}|) \, ,
\end{equation}
so that we can integrate over $\mathbf{q}$ to arrive at 
\begin{align}
    \begin{split}
            \langle h_{\mathbf{k^{\prime}},\lambda ^{\prime}}h^{(3)}_{\mathbf{k},\lambda } \rangle &= \frac{6}{(2\pi)^3}  \int {\rm} ^3 \mathbf{p} \,   T_h(\eta k \conf )\sum _{\lambda _1}\eps ^{ij}_{\lambda}(\mathbf{k})^* p_ip_j Q^{\lambda _1}_{\pi ,st}(\mathbf{p},\mathbf{k})  I^{\Psi ^{(2)}\Psi}(\eta , k,p,q=k) \\
            &\times \delta ^{\lambda \conf \lambda _1} \delta (\mathbf{k \conf}+ \mathbf{k})  P^{\lambda \conf} _h (k \conf) P_{\mathcal{R}} (|\mathbf{p}-\mathbf{k}|) \, .
    \end{split}
\end{align}
After switching to spherical coordinates, we can integrate the polarisation functions over the azimuthal direction  
\begin{equation}
    \int {\rm}\phi _p \,\eps ^{ij}_{\lambda}(\mathbf{k})^* p_ip_j Q_{\lambda _1}^{\Pi^{(2)}_{st}}(\mathbf{p},\mathbf{q}=\mathbf{k}) = \frac{\pi}{2}k^4v^4 \left (1-\frac{(1-u^2+v^2)^2}{4v^2} \right )^2 \delta ^{\lambda \lambda _1} \, ,
\end{equation}
where we have switched to $v$ and $u$ coordinates. As opposed to the source term made of pure first-order perturbations, the $u$ dependence is now non-trivial (see Fig.~\ref{fig:secondorderfirstorderscalar2point}) and this will be the case for all the source terms involving second-order perturbations we consider. 

The power spectrum can now be obtained and is given by
\begin{align}
    \begin{split}
        P^{(13)}_{R/L} &= \frac{3}{2^3\pi ^2}k^3 \int _0^{\infty} {\rm} v  \int _{|1-v|}^{1+v} {\rm}u \, uv \, k^4     \mathcal{I}^{h^{(2)}\Psi}(x , k,v,u) P^{(11)}_{R/L} (k) P_{\mathcal{R}} (ku)\\
        &\times \left (v^2-\frac{(1-u^2+v^2)^2}{4} \right )^2 \, , 
    \end{split}
\end{align}
where we have defined
\begin{equation}
     \mathcal{I}^{\Psi ^{(2)}\Psi}(x,k,v,u) = T_h(x)  I^{\Psi ^{(2)}\Psi}(x,k,v,u) \, ,
\end{equation}
which takes into account the sub-Hubble evolution of the tensor mode we have correlated with. Finally, the dimensionless power spectrum is given by
\begin{align}\label{pspsi2psi1}
    \begin{split}
         \mathcal{P}^{(13)}_{\Psi ^{(2)}\Psi}(\eta ,k) &= \frac{3}{4} \left ( \mathcal{P}^{(11)}_R (\eta, k) +  \mathcal{P}^{(11)}_L (\eta, k) \right ) \int _0^{\infty} {\rm} v  \int _{|1-v|}^{1+v} {\rm} u \, \frac{v}{u^2} k^4 \mathcal{I}^{\Psi ^{(2)}\Psi}(x,k,v,u)    \\
         &\times \left (v^2-\frac{(1-u^2+v^2)^2}{4} \right )^2   P_{\mathcal{R}} (ku) \, .
    \end{split}
\end{align}
We emphasise once again that the kernel and the primordial scalar power spectrum exhibit angular dependence, making the polar integral (the integral over $u$) non-trivial to evaluate—unlike the case of third-order waves sourced purely by first-order contributions, where the integration is trivial. The kernel $ I^{\Psi ^{(2)}\Psi}(x,k,v,u)$ was defined in Eq.~\eqref{kernelpsi2psi1} and we now give $f^{\Psi^{(2)}\Psi}(\barx,k,v,u)$ in terms of $v$ and $u$
\begin{align}\label{fpsi2psi1final}
    \begin{split}
        &f^{\Psi^{(2)}\Psi}(\barx,k,v,u) = \avant ^2 \bigg ( \frac{(1+3w)^2}{3(1+w)} \barx ^2 \frac{\partial}{\partial \barx}T_{\Psi} (\barx u )\frac{\partial}{\partial \barx} I^{\Psi ^{(2)}}_{st}(\barx , k ,v,u) \\
        &+  \frac{2(1+3w)}{3(1+w)} \barx \bigg (  \frac{\partial}{\partial \barx}T_{\Psi} (\barx u )  I^{\Psi ^{(2)}}_{st}(\barx  , k ,v,u) + \frac{1}{k^4v^4} \frac{\partial}{\partial \barx}T_{\Psi} (\barx u ) f^{\Pi ^{(2)}}_{st}(\barx , k, v,u)  \\
        &+ T_{\Psi} (\barx u ) \frac{\partial}{\partial \barx} I^{\Psi ^{(2)}}_{st}(\barx , k ,v,u) \bigg  )  + \frac{2(5+3w)}{3(1+w)} \bigg (T_{\Psi} (\barx u ) I^{\Psi ^{(2)}}_{st}(\barx  , k ,v,u) \\
        &+ \frac{1}{k^4v^4}T_{\Psi} (\barx u ) f^{\Pi ^{(2)}}_{st}(\barx , k, v,u)   \bigg ) \bigg ) \, ,
    \end{split}
\end{align}
with
\begin{align}
    \begin{split}
        f^{\Pi ^{(2)}}_{st}(\barx , k, v,u)  &= 2k^2(1+3u^2-3v^2) T_h(\barx) T_{\Psi}(\barx u)-12 k^2   T_h(\barx) T_{\Psi}(\barx u) \\
        &+12\frac{k^2}{\barx}  T_h(\barx) \frac{\partial}{\partial \barx}T_{\Psi}(\barx u)+6(w-1)k^2u^2T_h(\barx) T_{\Psi}(\barx u) \\
        &+6k^2 (1+u^2-v^2) T_h(\barx) T_{\Psi}(\barx u) + 6k^2(1-u^2+v^2)  T_h(\barx) T_{\Psi}(\barx u)\, ,
    \end{split}
\end{align}
and
\begin{align} \label{psi2nested}
    \begin{split}
        I^{\Psi ^{(2)}}_{st}(\barx , k ,v,u) &= \frac{\sqrt{w}}{\barx ^{1+b}k^2v} \int _0^{\barx v} {\rm}  \tilde{y} \, \, \tilde{y}^{3+b} \big ( j_{b+1}(\tilde{y}\sqrt{w})y_{b+1}(v\barx\sqrt{w}) - y_{b+1}(\tilde{y}\sqrt{w})\\
        &\times j_{b+1}(v\barx\sqrt{w}) \big ) \Sigma ^{st}(\tilde{y},v,u) \, .
    \end{split}
\end{align}
We have defined $\Sigma ^{st}(\tilde{y},v,u)$ in Eq.~\eqref{Sigmapsi2} and it now reads
\begin{align}
    \begin{split}
        &\Sigma ^{st}(\tilde{y},v,u) = \frac{2}{3(1+3w)v^3\tildey ^3} \bigg (6\tildey \frac{\partial}{\partial \tildey} T_h\left (\frac{\tildey}{v}\right ) \bigg ( (3+v^2-3u^2w)\tildey T_{\Psi}\left (\frac{\tildey u}{v}\right ) - 6uv \\
        &\times \frac{\partial}{\partial \tildey} T_{\Psi}\left (\frac{\tildey u}{v}\right ) \bigg ) +  T_h\left (\frac{\tildey}{v}\right ) \bigg [v(1+3w) (-3-2v^2+3(u^2+v^2)w)\tildey ^3 T_{\Psi}\left (\frac{\tildey u}{v}\right ) \\
        &+ 6u \bigg ( \big[6v^2 + (3-3u^2w+v^2(2+3w))\tildey ^2\big] \frac{\partial}{\partial \tildey} T_{\Psi}\left (\frac{\tildey u}{v}\right ) -6uv\tildey \frac{\partial ^2}{\partial \tildey ^2}T_{\Psi}\left (\frac{\tildey u}{v}\right )  \bigg ) \bigg] \bigg ) \, ,
    \end{split}
\end{align}
where we have used the relations shown in Eq.~\eqref{yrelations}.

%%%%%%%%%%%%%%%%%%%%%%%%%%%%%%%%%%%%%%%%%%%%%%%%%%%%%%%%%%%%%%%%%%%%%%%%%%%%%%%%%%%%%%%%%%%%%%
\subsection{The power spectrum of second-order vectors coupled to first-order scalars: 
\texorpdfstring{$\mathcal{P}^{(13)}_{B^{(2)} \Psi}$}{P(13)B2Psi}}
%%%%%%%%%%%%%%%%%%%%%%%%%%%%%%%%%%%%%%%%%%%%%%%%%%%%%%%%%%%%%%%%%%%%%%%%%%%%%%%%%%%%%%%%%%%%%%
In this subsection, we compute the contribution coming from second-order vectors coupled to first-order scalars. This amounts to substituting in the solution Eq.~\eqref{GWeqsolb2psi1} to the two-point function,
\begin{align}
    \begin{split}
            \langle h_{\lambda ^{\prime}}(\mathbf{k^{\prime}})h^{B^{(2)\Psi}}_{\lambda }(\mathbf{k}) \rangle &= \frac{6}{(2\pi)^3} \int {\rm} ^3 \mathbf{q} \int {\rm} ^3 \mathbf{p} \, T_h(\eta k \conf )\sum _{r}\sum _{\lambda _1} \bigg ( Q^{B^{(2)}\Psi}_{\lambda ,r}(\mathbf{k},\mathbf{p})Q^{B^{(2)}_{st1}}_{r,\lambda _1}(\mathbf{p},\mathbf{q})  \\
            &\times I^{B^{(2)}\Psi}_1(\eta , k,p,q)+Q^{B^{(2)}\Psi}_{\lambda ,r}(\mathbf{k},\mathbf{p})Q^{B^{(2)}_{st2}}_{r,\lambda _1}(\mathbf{p},\mathbf{q}) I^{B^{(2)}\Psi}_2(\eta , k,p,q) \bigg )\\
            &\times \langle h^{\lambda \conf} _{\mathbf{k \conf}}h_{\mathbf{q}}^{\lambda _1} \mathcal{R} _{\mathbf{p}-\mathbf{q}} \mathcal{R} _{\mathbf{k}-\mathbf{p}}\rangle \, .
    \end{split} 
\end{align}
Inserting the expression for the four-point function given in Eq.~\eqref{stthirdorder4point}, we arrive at
    \begin{align}
    \begin{split}
            &\langle h_{\lambda ^{\prime}}(\mathbf{k^{\prime}})h^{B^{(2)\Psi}}_{\lambda }(\mathbf{k}) \rangle = \frac{6}{(2\pi)^3} \int {\rm} ^3 \mathbf{q} \int {\rm} ^3 \mathbf{p} \, T_h(\eta k \conf )\sum _{r}\sum _{\lambda _1} \bigg ( Q^{B^{(2)}\Psi}_{\lambda ,r}(\mathbf{k},\mathbf{p})Q^{B^{(2)}_{st1}}_{r,\lambda _1}(\mathbf{p},\mathbf{q}=\mathbf{k})  \\
            &\times I^{B^{(2)}\Psi}_1(\eta , k,p,q=k)+Q^{B^{(2)}\Psi}_{\lambda ,r}(\mathbf{k},\mathbf{p})Q^{B^{(2)}_{st2}}_{r,\lambda _1}(\mathbf{p},\mathbf{q}=\mathbf{k}) I^{B^{(2)}\Psi}_2(\eta , k,p,q=k) \bigg )\\
            &\times \delta ^{\lambda ^{\prime} \lambda _1} \delta (\mathbf{k}+\mathbf{k^{\prime}}) P^{(11)}_{\lambda ^{\prime}}(k)P_{\mathcal{R}}(|\mathbf{p}-\mathbf{k}|) \, .
    \end{split}
\end{align}
We can now integrate over $\mathbf{q}$ and perform the azimuthal integral of the wave-vector $\mathbf{p}$ over the polarisation functions
\begin{subequations}\label{evalpolb2psi1}
    \begin{align}
        \int {\rm} \phi_p \, \sum _{r}\sum _{\lambda _1}Q^{B^{(2)}\Psi}_{\lambda ,r}(\mathbf{k},\mathbf{p})Q^{B^{(2)}_{st1}}_{r,\lambda _1}(\mathbf{p},\mathbf{q}=\mathbf{k}) &=-\frac{k^4\pi}{32v^2} a^{B^{(2)}\Psi}(v,u)b^{B^{(2)}\Psi}(v,u)\delta ^{\lambda \lambda _1} \, , \\ 
        \int {\rm} \phi_p \, \sum _{r}\sum _{\lambda _1}Q^{B^{(2)}\Psi}_{\lambda ,r}(\mathbf{k},\mathbf{p})Q^{B^{(2)}_{st2}}_{r,\lambda _1}(\mathbf{p},\mathbf{q}=\mathbf{k}) &=-\frac{k^2\pi}{32v^2} a^{B^{(2)}\Psi}(v,u)c^{B^{(2)}\Psi}(v,u)\delta ^{\lambda \lambda _1} \, ,
    \end{align}
\end{subequations}
where we have further defined
\begin{subequations}
    \begin{align}
        a^{B^{(2)\Psi}}(v,u) &= (-1+u-v)(1+u-v)(-1+u+v)(1+u+v) \, , \\
        b^{B^{(2)\Psi}}(v,u) &= -3+u^4+2v^2+v^4-2u^2(v^2-1) \, , \\
        c^{B^{(2)\Psi}}(v,u) &= 1+u^4+6v^2+v^4-2u^2(1+v^2) \, .
    \end{align}
\end{subequations}
The power spectrum is then given by
\begin{align}
    \begin{split}
           P^{(13)}_{R/L} &= \frac{3}{(2\pi) ^2}k^3P^{(11)}_{R/L}(k)  \int _0^{\infty} {\rm} v  \int _{|1-v|}^{1+v} {\rm}u \, uv\bigg ( -\frac{k^4}{32v^2} a^{B^{(2)\Psi}}(v,u)b^{B^{(2)\Psi}}(v,u)  \\
            &\times \mathcal{I}^{B^{(2)}\Psi}_1(x,k,v,u) -\frac{k^2}{32v^2} a^{B^{(2)\Psi}}(v,u)c^{B^{(2)\Psi}}(v,u) \mathcal{I}^{B^{(2)}\Psi}_2(x,k,v,u) \bigg ) P_{\mathcal{R}}(ku) \, ,
    \end{split}
\end{align}
with
\begin{equation}
     \mathcal{I}^{B^{(2)}\Psi}_{\textit{i}}(x,k,v,u) = T_h(x)  I^{B ^{(2)}\Psi}_{\textit{i}}(x,k,v,u) \, .
\end{equation}
Finally, the dimensionless power spectrum corresponding to iGWs sourced by second-order vectors coupled to first-order scalar perturbations is given by:
\begin{align}\label{finalpsb2psi1}
    \begin{split}
        \mathcal{P}^{(13)}_{B ^{(2)}\Psi}(\eta ,k) &=- \frac{3}{64} \left ( \mathcal{P}^{(11)}_R (\eta, k) +  \mathcal{P}^{(11)}_L (\eta, k) \right ) \int _0^{\infty} {\rm} v  \int _{|1-v|}^{1+v} {\rm} u \, \frac{1}{vu^2}   \\
        &\times \bigg ( a^{B^{(2)\Psi}}(v,u)b^{B^{(2)\Psi}}(v,u) k^4\mathcal{I}^{B^{(2)}\Psi}_1(x,k,v,u) + a^{B^{(2)\Psi}}(v,u)c^{B^{(2)\Psi}}(v,u)\\
        &\times k^2\mathcal{I}^{B^{(2)}\Psi}_2(x,k,v,u) \bigg )  P_{\mathcal{R}} (ku) \, .
    \end{split}
\end{align}
The two kernel's $I^{B ^{(2)}\Psi}_{\textit{i}}(x,k,v,u)$, defined in Eq.~\eqref{kernelb2psi1} are time integrals over the functions
\begin{subequations} \label{fb2psi1final}
    \begin{align}
        \begin{split}
            &f^{B^{(2)\Psi}}_{\textit{i}}(\overline{x},k,p,q) = \avant ^2 \bigg ( \frac{2}{v^2k} I^{B^{(2)}}_{sti}(\overline{x} ,k,v,u) \frac{\partial}{\partial \barx}T_{\Psi}(\barx u) + \frac{2}{v^2k}\frac{\partial}{\partial \barx}I^{B^{(2)}}_{sti}(\overline{x} ,k,v,u) \\
            &\times T_{\Psi}(\barx u) + \frac{8}{v^2k\barx (1+3w)} I^{B^{(2)}}_{sti}(\overline{x} ,k,v,u) T_{\Psi}(\barx u) +\frac{\barx (1+3w)}{6k(1+w)} I^{B^{(2)}}_{sti}(\overline{x} ,k,v,u) T_{\Psi}(\barx u) \\
            &+  \frac{\barx ^2(1+3w)^2}{12k(1+w)}  I^{B^{(2)}}_{sti}(\overline{x} ,k,v,u) \frac{\partial}{\partial \barx} T_{\Psi}(\barx u)   \bigg ) \, .
        \end{split}
    \end{align}
\end{subequations}
The expression for the nested kernels ($i=1,2$) are
\begin{equation}
    I^{B^{(2)}}_{sti}(\overline{x} ,k,v,u) = \frac{1}{(kv)^{3+2b}\barx^{2(1+b)}} \int _0^{\barx v} {\rm}  \tilde{y} \, \, \tilde{y}^{2(1+b)} f^{B^{(2)}}_{sti}(\tildey , k, v,u) \, ,
\end{equation}
with
\begin{subequations}
    \begin{align}
        f^{B^{(2)}}_{st1}(\tildey , k, v,u) &= -4 T_h\left (\frac{\tilde{y}}{v} \right )T_{\Psi} \left (\frac{\tilde{y}u}{v} \right )  \, ,
        \\
        \begin{split}
         f^{B^{(2)}}_{st2}(\tildey , k, v,u) &= 2k^2(-1+v^2-u^2)T_h\left (\frac{\tilde{y}}{v} \right )T_{\Psi} \left (\frac{\tilde{y}u}{v} \right )+4k^2T_h\left (\frac{\tilde{y}}{v} \right )T_{\Psi} \left (\frac{\tilde{y}u}{v} \right ) \\
         &-\frac{4v^2k^2}{\tildey} T_h\left (\frac{\tilde{y}}{v} \right )\frac{\partial}{\partial \tildey}T_{\Psi} \left (\frac{\tilde{y}u}{v}  \right )+2(1-w)u^2k^2T_h\left (\frac{\tilde{y}}{v} \right )T_{\Psi} \left (\frac{\tilde{y}u}{v} \right ) \, .
        \end{split}
    \end{align}
\end{subequations}

%%%%%%%%%%%%%%%%%%%%%%%%%%%%%%%%%%%%%%%%%%%%%%%%%%%%%%%%%%%%%%%%%%%%%%%%%%%%%%%%%%%%%%%%%%%%%%
\subsection{The power spectrum of second-order tensors coupled to first-order scalars: 
\texorpdfstring{$\mathcal{P}^{(13)}_{h^{(2)} \Psi}$}{P(13)h2Psi}}
%%%%%%%%%%%%%%%%%%%%%%%%%%%%%%%%%%%%%%%%%%%%%%%%%%%%%%%%%%%%%%%%%%%%%%%%%%%%%%%%%%%%%%%%%%%%%%
The last term to consider is the two-point function involving third-order iGWs sourced by second-order tensors and first-order scalars. The equation of motion for these iGWs was derived in Eq.~\eqref{GWeqsolh2psi1}. So we have
\begin{align}
    \begin{split}
            \langle h_{\lambda ^{\prime}}(\mathbf{k^{\prime}})h^{h^{(2)\Psi}}_{\lambda }(\mathbf{k}) \rangle &= \frac{6}{(2\pi)^3} \int {\rm} ^3 \mathbf{q} \int {\rm} ^3 \mathbf{p} \, T_h(\eta k \conf )\sum _{\sigma ,\lambda _1}\eps^{ij}_{\lambda}(\mathbf{k})^* \eps _{ij}^{\sigma}(\mathbf{p})\eps ^{mn}_{\sigma}(\mathbf{p})^*\eps _{mn}^{\lambda _1}(\mathbf{q})   \\
            &\times I_{st}^{h^{(2)}\Psi}(\eta , k,p,q) \langle h^{\lambda \conf} _{\mathbf{k \conf}}h_{\mathbf{q}}^{\lambda _1} \mathcal{R} _{\mathbf{p}-\mathbf{q}} \mathcal{R} _{\mathbf{k}-\mathbf{p}}\rangle \, ,
    \end{split}
\end{align}
and following the same steps outlined in the previous subsections, we arrive at 
\begin{align}
    \begin{split}
            \langle h_{\mathbf{k^{\prime}},\lambda ^{\prime}}h^{(3)}_{\mathbf{k},\lambda } \rangle &= \frac{6}{(2\pi)^3}  \int {\rm} ^3 \mathbf{p}  T_h(\eta k \conf )\sum _{\sigma ,\lambda _1}\eps ^{ij}_{\lambda}(\mathbf{k})^* \eps _{ij}^{\sigma}(\mathbf{p})\eps ^{mn}_{\sigma}(\mathbf{p})^*\eps _{mn}^{\lambda _1}(\mathbf{k})  I_{st}^{h^{(2)}\Psi}(\eta , k,p,q=k) \\
            &\times \delta ^{\lambda \conf \lambda _1} \delta (\mathbf{k \conf}+ \mathbf{k})  P^{\lambda \conf} _h (k \conf) P_{\mathcal{R}} (|\mathbf{p}-\mathbf{k}|) \, .
    \end{split}
\end{align}
Contracting the polarisation tensors and integrating over the azimuthal direction, the only non-trivial contribution happens when $\lambda =\lambda _1$, which leads to 
\begin{align}
    \begin{split}
            \langle h_{\mathbf{k^{\prime}},\lambda ^{\prime}}h^{(3)}_{\mathbf{k},\lambda } \rangle &= \frac{6}{(2\pi)^2}  \int _0^{\infty} {\rm}  p \int _0^{\pi} {\rm}  \theta _p \, p^2 \sin \theta _p \left (\cos{\left (\frac{\theta _p}{2}\right )}^8 + \sin{\left (\frac{\theta _p}{2}\right )}^8 \right )  \\
            &\times T_h(\eta k \conf )I^{h^{(2)}\Psi}(\eta , k,p,q=k) \delta ^{\lambda \conf \lambda } \delta (\mathbf{k \conf}+ \mathbf{k})  P^{\lambda \conf} _h (k \conf) P_{\mathcal{R}} (|\mathbf{p}-\mathbf{k}|) \, .
    \end{split}
\end{align}
Switching to the ($v,u$) coordinate system, the power spectrum is now given by
\begin{align}
    \begin{split}
        P^{(13)}_{R/L}(\eta , k) &= \frac{6}{(2\pi)^2}k^3 \int _0^{\infty} {\rm} v  \int _{|1-v|}^{1+v} {\rm} u \, \frac{u}{256v^3} \big [ \left (u^2-(v+1)^2 \right )^4 + \left (u^2-(v-1)^2 \right )^4  \big ] \\
        &\times \mathcal{I}^{h^{(2)}\Psi}(x,k,v,u)P^{(11)}_{R/L} (k) P_{\mathcal{R}} (ku) \, ,
    \end{split}
\end{align}
with the kernel defined as
\begin{equation}\label{kernelh2psi1}
      \mathcal{I}^{h^{(2)}\Psi}(x,k,v,u) = T_h(x)  I^{h^{(2)}\Psi}(x , k,v,u) \, .
\end{equation}
Finally, the dimensionless power spectrum reads
\begin{align}\label{psh2psi1}
    \begin{split}
         \mathcal{P}^{(13)}_{h^{(2)}\Psi}(\eta ,k) &= \frac{3}{128} \left ( \mathcal{P}^{R}_h (\eta, k) +  \mathcal{P}^{L}_h (\eta, k) \right ) \int _0^{\infty} {\rm} v  \int _{|1-v|}^{1+v} {\rm} u \, \frac{1}{v^3u^2} \bigg [ \left (u^2-(v+1)^2 \right )^4 \\
         &+ \left (u^2-(v-1)^2 \right )^4  \bigg ]  \mathcal{I}^{h^{(2)}\Psi}(x,k,v,u)  P_{\mathcal{R}} (ku) \, .
    \end{split}
\end{align}
The kernel $I^{h^{(2)}\Psi}(x , k,v,u)$, defined in Eq.~\eqref{kernelh2psi1}, is an integral over the function $f^{h^{(2)\Psi}}(\overline{x},k,v,u)$, which was previously defined in Eq.~\eqref{ffuntionh2psi1}. In our new coordinates, it is given by
\begin{align} \label{fh2psi1final}
    \begin{split}
        &f^{h^{(2)\Psi}}(\overline{x},k,v,u) = \left ( \frac{3+3w}{5+3w} \right )^2 \bigg ( -8k^2 T_{\Psi} ( \barx u) T_h(\barx) T_{\Psi} ( \barx u)  +4k^2(1+u^2-v^2) \\
        &\times  T_{\Psi} ( \barx u) T_h(\barx) T_{\Psi} ( \barx u) + 4 \frac{k^2}{\barx}\frac{\partial}{\partial \barx} T_{\Psi} ( \barx u) T_h(\barx) T_{\Psi} ( \barx u) \\
        &-4(1-w)k^2u^2 T_{\Psi} ( \barx u) T_h(\barx) T_{\Psi} ( \barx u) -8v^2k^2I^{h^{(2)}}_{st}(\barx,k,v,u)T_{\Psi} ( \barx u) \\
        &+8\frac{k^2}{\barx} I^{h^{(2)}}_{st}(\barx,k,v,u)\frac{\partial}{\partial \barx}T_{\Psi} ( \barx u) + 4(w-1)k^2u^2 I^{h^{(2)}}_{st}(\barx,k,v,u)T_{\Psi} ( \barx u) \\
        &+4k^2(-1+u^2-v^2)I^{h^{(2)}}_{st}(\barx,k,v,u)T_{\Psi} ( \barx u)  \bigg ) \, .
    \end{split}
\end{align}
Finally, the nested kernels can be expressed as 
\begin{align}
    \begin{split}
        I^{h^{(2)}}_{st}(\barx , k ,v,u) &= \frac{v^{b-2}}{\barx ^{b}k^2} \int _0^{\barx v} {\rm}  \tilde{y} \, \, \tilde{y}^{2+b} \big ( j_b(\tilde{y})y_b(v\barx) - y_b(\tilde{y}) j_b(v\barx) \big )f^{h^{(2)}}_{st}(\tilde{y},k,v,u) \, ,
    \end{split}
\end{align}
with
\begin{align}
    \begin{split}
    f^{h^{(2)}}_{st}(\tilde{y} ,k, v,u)&=-2k^2T_h\left (\frac{\tilde{y}}{v} \right ) T_{\Psi}\left (\frac{\tilde{y}u}{v} \right )+k^2(1+u^2-v^2) T_h\left (\frac{\tilde{y}}{v} \right ) T_{\Psi}\left (\frac{\tilde{y}u}{v} \right ) \\
    &+ 2\frac{v^2k^2}{\tilde{y}}T_h\left (\frac{\tilde{y}}{v} \right ) T_{\Psi}\left (\frac{\tilde{y}u}{v} \right )-k^2u^2(1-w)T_h\left (\frac{\tilde{y}}{v} \right ) T_{\Psi}\left (\frac{\tilde{y}u}{v} \right )\, .
    \end{split}
\end{align}

%%%%%%%%%%%%%%%%%%%%%%%%%%%%%%%%%%%%%%%%%%%%%%%%%%%%%%%%%%%%%%%%%%%%%%%%%%%%%%%%%%%%%%%%%%%%%%
%%%%%%%%%%%%%%%%%%%%%%%%%%%%%%%%%%%%%%%%%%%%%%%%%%%%%%%%%%%%%%%%%%%%%%%%%%%%%%%%%%%%%%%%%%%%%%
\section{Results in a radiation-dominated universe for a peaked input power spectrum}\label{sec5}
%%%%%%%%%%%%%%%%%%%%%%%%%%%%%%%%%%%%%%%%%%%%%%%%%%%%%%%%%%%%%%%%%%%%%%%%%%%%%%%%%%%%%%%%%%%%%%
%%%%%%%%%%%%%%%%%%%%%%%%%%%%%%%%%%%%%%%%%%%%%%%%%%%%%%%%%%%%%%%%%%%%%%%%%%%%%%%%%%%%%%%%%%%%%%
In the previous section, we derived expressions for the various contributions to the power spectrum arising from the correlation between third-order iGWs and first-order GWs. Our results, so far, hold for a general constant equation of state $w \neq  -1/3$. We now proceed to derive the spectral density of these GWs in an RD universe, setting $w=1/3$ or $b=0$ henceforth. The spectral density of the iGWs, at the time of creation, defined in Eq.~\eqref{spectraldensitydef}, consists of taking the time average of the power spectrum arising from the different contributions
\begin{equation}\label{spectraldensityallcontributions}
    \Omega ^{(13)} (\eta,k) =\frac{1}{9}x^2 \left ( \overline{\mathcal{P}^{(13)}_{\Psi \Psi h}(\eta,k)} +   \overline{\mathcal{P}^{(13)}_{\Psi ^{(2)}\Psi}(\eta,k)} + \overline{\mathcal{P}^{(13)}_{B ^{(2)}\Psi}(\eta,k)} + \overline{\mathcal{P}^{(13)}_{h ^{(2)}\Psi}(\eta,k)}\right )  \, .
\end{equation}

We therefore need to take the time average of the power spectra we have derived. Since the time dependence is entirely contained within the kernels, our task now is to compute their time-averaged values, which we will do in this section. As we will see, terms involving nested kernels will need to be computed numerically; however, some terms can still be evaluated analytically. Hence, where possible, we will split kernels into an analytical and a numerical part.   

We recall that the explicit forms of the transfer functions are given in Eq.~\eqref{psitransfer} for the first-order scalar fluctuations and Eq.~\eqref{tensortransfer} for the tensor modes. Furthermore, the Green's function for the second-order tensor equation can be found in Eq.~\eqref{greenstensors}.

The kernels we have defined for each contribution, $\mathcal{I}^i(x ,k,v,u)$ (here $i$ refers to the type of source), can be expanded as
\begin{align}\label{kernelspit}
    \begin{split}
        &\mathcal{I}^i(x ,k,v,u)=T_h(x)I^i(x,k,v,u) = \frac{\sin x}{x^2k^2} \int _0^{x } {\rm }  \barx \, \barx \sin (x-\barx) f^i(\barx,k,v,u) \\
        &=\frac{\sin ^2x}{x^2k^2}\int _0^{x} {\rm } \barx \, \barx \cos\barx f^i(\barx,k,v,u) - \frac{\sin x \cos x}{x^2k^2}\int _0^{x} {\rm } \barx \, \barx \sin\barx f^i(\barx,k,v,u) \, .
    \end{split}
\end{align}
As mentioned previously, we need to take the time average of these kernels, and they encapsulate all the time dependence. Moreover, as is the case for second-order iGWs, we are interested in these waves at present times, that is, when the iGWs are deep inside the horizon and the kernels have stabilised to a constant value. This allows us to take the limit $\eta\rightarrow \infty$ or $x\rightarrow \infty$ (for some fixed value of $k$). As a result, all remaining time dependence resides in the prefactors of the integrals. We conclude that the second term above does not contribute to the SGWB. Accordingly, what we need to calculate is 
\begin{equation}\label{finalkernel}
    x^2\overline{\mathcal{I}^i(x\rightarrow\infty ,k,v,u)} = \frac{1}{2k^2}\int _0^{x \rightarrow \infty} {\rm } \barx \, \barx \cos\barx f^i(\barx,k,v,u) \, ,
\end{equation}
where we recall the functions $f^i(\barx,k,v,u)$ depend on the source term and are given in Eqs.~\eqref{sshfunctionsfinal1}, \eqref{sshfunctionsfinal2}, \eqref{sshfunctionsfinal1}, \eqref{fpsi2psi1final}, \eqref{fb2psi1final}, and \eqref{fh2psi1final}. We note that Ref.~\cite{Chen:2022dah} examines the impact of third-order contributions on the IR part of the iGW power spectrum, whereas here we will examine the impact on the spectral density. The two terms in Eq.~\eqref{kernelspit} contribute to the power spectrum, but only the first term contributes to the spectral density. Once again, the kernels and nested kernels involve integrals of products of highly oscillatory and decaying functions, so they will involve the trigonometric integrals $\text{Ci}(z)$ and $\text{Si}(z)$ defined in Eq.~\eqref{SICIfunc}. As we will see, the nested kernels depend on products of these trigonometric integrals, making it impossible to derive closed-form analytical expressions for terms containing nested kernels. Consequently, we will evaluate such kernels numerically. This is achieved using a fourth-order Runge-Kutta method, where we have tested our code by comparing numerical kernels for SIGWs and scalar-tensor iGWs to the known analytical ones. Finally, we note a key difference from second-order iGWs: the kernels are not squared. As a result, these contributions can be negative, potentially suppressing the signal.

%%%%%%%%%%%%%%%%%%%%%%%%%%%%%%%%%%%%%%%%%%%%%%%%%%%%%%%%%%%%%%%%%%%%%%%%%%%%%%%%%%%%%%%%%%%%%%
\subsection{Kernel computations}
%%%%%%%%%%%%%%%%%%%%%%%%%%%%%%%%%%%%%%%%%%%%%%%%%%%%%%%%%%%%%%%%%%%%%%%%%%%%%%%%%%%%%%%%%%%%%%
In this subsection, we compute the various kernels needed. Firstly, we look at the kernels coming from the pure first-order contribution and then move on to the second-order kernels, which involve numerical computations.

There are three kernels we need to compute for the contribution composed of the scalar-scalar-tensor interactions. With the formalism established, the next step is simply to substitute in Eqs.~\eqref{sshfunctionsfinal1}, ~\eqref{sshfunctionsfinal2} and~ ~\eqref{sshfunctionsfinal3} into Eq.~\eqref{finalkernel}. This leads to
\begin{subequations}
    \begin{align}
        \begin{split}
            x^2\overline{\mathcal{I}^{ssh}_{s,1} (x\rightarrow \infty , k, v)} &= -\frac{1}{315v^6} \bigg (v^2 \left ( 648+1179v^2-739v^4\right ) +\sqrt{3}v^3 \big (1890-672v^2 \\
            &+11v^4 \big ) \left | \frac{\sqrt{3}-v}{\sqrt{3}+v} \right | +9 \left (216-903v^2+280v^4 \right ) \log \left | \frac{v^2-3}{3} \right | \bigg ) \, ,
        \end{split}
        \\
        \begin{split}
            x^2\overline{\mathcal{I}^{ssh}_{s,2}(x\rightarrow \infty , v)} &=\frac{1}{30v^6k^2} \bigg (3v^2 \left ( 6+11v^2\right ) -2\sqrt{3}v^3 \left (v^2-15 \right )\left | \frac{\sqrt{3}-v}{\sqrt{3}+v} \right | \\
            &+18 \left (3-5v^2 \right ) \log \left | \frac{v^2-3}{3} \right | \bigg ) \, ,
        \end{split}
        \\
        \begin{split}
            x^2\overline{\mathcal{I}^{ssh}_{s,4}(x\rightarrow \infty , v)} &=\frac{1}{120v^6k^2} \bigg (6v^2 \left ( 62v^2-3\right ) +\sqrt{3}v^3 \left (285-13v^2 \right ) \left | \frac{\sqrt{3}-v}{\sqrt{3}+v} \right | \\
            &-9 \left (6+65v^2+5v^4 \right ) \log \left | \frac{v^2-3}{3} \right | \bigg ) \, .
        \end{split}
    \end{align}
\end{subequations}
We also observe that these kernels remain well-behaved and finite in the limit $v\rightarrow\sqrt{3}$, exhibiting no logarithmic divergence. 

We shift our focus to computing the kernels from source terms that couple second-order perturbations, which will involve nested kernels. For second-order scalars coupled to first-order fluctuations, the function we have to integrate over is Eq.~\eqref{fpsi2psi1final}, which depends on the nested kernel defined in Eq.~\eqref{psi2nested} 
\begin{align}
    \begin{split}
        I^{\Psi ^{(2)}}_{st}(\barx , k ,v,u)&= \frac{1}{k^2v^3\barx ^3}   \int _0^{v\barx} {\rm } \tildey \, \, \tildey \Bigg ( 3(-v\barx+\tildey)\cos \left ( \frac{v\barx-\tildey}{\sqrt{3}} \right ) +\sqrt{3}(3+v\barx\tildey)  \\
        &\times \sin \left ( \frac{v\barx-\tildey}{\sqrt{3}} \right )  \Bigg )\Sigma ^{st}(\tilde{y},v,u)  \, , 
    \end{split}
\end{align}
in an RD universe. We further split this up as
\begin{align}
    \begin{split}
        &I^{\Psi ^{(2)}}_{st}(\barx , k ,v,u)=  \frac{1}{k^2v^3\barx ^3}  \Bigg ( 3\cos \left ( \frac{v\barx}{\sqrt{3}} \right ) I^{\Psi ^{(2)}}_{c2}(\barx ,v,u) -3v\barx \cos \left ( \frac{v\barx}{\sqrt{3}} \right ) I^{\Psi ^{(2)}}_{c1}(\barx ,v,u) \\
        &-3\sqrt{3} \cos \left ( \frac{v\barx}{\sqrt{3}} \right ) I^{\Psi ^{(2)}}_{s1}(\barx ,v,u) -\sqrt{3}v\barx  \cos \left ( \frac{v\barx}{\sqrt{3}} \right ) I^{\Psi ^{(2)}}_{s2}(\barx ,v,u) \\
        & + 3\sqrt{3} \sin \left ( \frac{v\barx}{\sqrt{3}} \right ) I^{\Psi ^{(2)}}_{c1}(\barx ,v,u)+\sqrt{3} v\barx \sin \left ( \frac{v\barx}{\sqrt{3}} \right ) I^{\Psi ^{(2)}}_{c2}(\barx ,v,u)  \\
        &+3 \sin \left ( \frac{v\barx}{\sqrt{3}} \right ) I^{\Psi ^{(2)}}_{s2}(\barx ,v,u) -  3v\barx \sin \left ( \frac{v\barx}{\sqrt{3}} \right ) I^{\Psi ^{(2)}}_{s1}(\barx ,v,u) \Bigg ) \, ,
    \end{split}
\end{align}
where we have defined
\begin{subequations}
    \begin{align}
        I^{\Psi ^{(2)}}_{c1}(\barx ,v,u)&= \int _0^{v\barx} {\rm } \tildey \, \cos \left ( \frac{\tildey}{\sqrt{3}} \right ) \tildey \,  \Sigma ^{st}(\tilde{y},v,u)  \, ,
        \\
        I^{\Psi ^{(2)}}_{s1}(\barx ,v,u)&=  \int _0^{v\barx} {\rm } \tildey \, \sin \left ( \frac{\tildey}{\sqrt{3}} \right ) \tildey \, \Sigma ^{st}(\tilde{y},v,u) \, ,
        \\
        I^{\Psi ^{(2)}}_{c2}(\barx ,v,u)&= \int _0^{v\barx} {\rm } \tildey \, \cos \left ( \frac{\tildey}{\sqrt{3}} \right ) \tildey ^2 \,  \Sigma ^{st}(\tilde{y},v,u)  \, ,
        \\
        I^{\Psi ^{(2)}}_{s2}(\barx ,v,u)&= \int _0^{v\barx} {\rm } \tildey \, \sin \left ( \frac{\tildey}{\sqrt{3}} \right ) \tildey ^2 \,  \Sigma ^{st}(\tilde{y},v,u) \, .
    \end{align}
\end{subequations}
The analytical results for $I^{\Psi ^{(2)}}_{c1}(\barx ,v,u)$, $I^{\Psi ^{(2)}}_{s1}(\barx ,v,u)$, $I^{\Psi ^{(2)}}_{c2}(\barx ,v,u)$ and $I^{\Psi ^{(2)}}_{s2}(\barx ,v,u)$ in a RD universe can be found in Eqs.~\eqref{Ic1RDfinal}, \eqref{Is1RDfinal}, \eqref{Ic2RDfinal} and \eqref{Is2RDfinal} respectively. These functions then need to be substituted back into Eq.~\eqref{fpsi2psi1final} and then integrated over in Eq.~\eqref{finalkernel}. Unfortunately, there are no analytical solutions for these functions. However, not every term in Eq.~\eqref{fpsi2psi1final} depends on these nested kernels, and we are able to compute analytical expressions for the terms that contain no nested kernels. This means we split up $ f^{\Psi ^{(2)}\Psi}(\barx,k,v,u)$ into an analytical part and numerical part:
\begin{equation}
    f^{\Psi ^{(2)}\Psi}(\barx,k,v,u) = f_a^{\Psi ^{(2)}\Psi}(\barx,k,v,u) + f_n^{\Psi ^{(2)}\Psi}(\barx,k,v,u) \, ,
\end{equation}
where
\begin{subequations}
    \begin{align}
        \begin{split}
            f_a^{\Psi ^{(2)}\Psi}(\barx,k,v,u)& =   \frac{24 \sin \barx}{k^2 u^4 v^4 \barx^7}  \sin\left(\frac{u \barx}{\sqrt{3}}\right) \Big(-\sqrt{3} u \barx (-18 + (1 + u^2 - 3 v^2) \barx^2) \cos\left(\frac{u \barx}{\sqrt{3}}\right)  \\
            &+ 3 (-18 + (1 + 3 u^2 - 3 v^2) \barx^2) \sin\left(\frac{u \barx}{\sqrt{3}}\right) \Big) \, ,
        \end{split}
        \\
        \begin{split}\label{finalkernelpsi2}
            f_n^{\Psi ^{(2)}\Psi}(\barx,k,v,u)& = \frac{4}{3 u^3 \bar{x}^2}  \bigg(6 u \bar{x} \cos\left(\frac{u \bar{x}}{\sqrt{3}}\right) + \sqrt{3} \big(-6 + u^2 \bar{x}^2\big) \sin\left(\frac{u \bar{x}}{\sqrt{3}}\right)\bigg)  \\
            &\times \frac{\partial}{\partial \barx} I^{\Psi ^{(2)}}_{st}(\barx , k ,v,u)+\frac{4}{\sqrt{3} u \bar{x}} \sin\left(\frac{u \bar{x}}{\sqrt{3}}\right)
            I^{\Psi ^{(2)}}_{st}(\barx , k ,v,u) \, .
        \end{split}
    \end{align}
\end{subequations}
The analytical contribution, once integrated, is
\begin{align}\label{analyticalkernelpsi2psi1}
    \begin{split}
        x^2&\overline{\mathcal{I}_a^{\Psi^{(2)}\Psi}(x\rightarrow\infty ,k,v,u)}= \frac{1}{30 k^4 u^4 v^4} \Bigg( 
        84 u^4 + u^2 (984 - 360 v^2) + (828 + 540 u^2 - 540 v^2)  \\
        &\times \log 3 + \big [(-828 - 540 u^2 + 540 v^2 + 
        \sqrt{3} (-630 u - 80 u^3 + 6 u^5 + 270 u v^2 - 30 u^3 v^2) \big ] 
         \\
        &\times \log \left |\sqrt{3} + u \right | + \big [-828 - 540 u^2 + 540 v^2 + 
        \sqrt{3} (630 u + 80 u^3 - 6 u^5 - 270 u v^2 \\
        &\times  + 30 u^3 v^2) \big ] 
        \log \left| \sqrt{3} - u \right| \Bigg) \, .
    \end{split}
\end{align}

For the kernels involving second-order vectors, there is no `analytical' part, and so we do everything numerically. In an RD universe, Eq.~\eqref{fb2psi1final} becomes
\begin{align}\label{kernelb2rd}
    \begin{split}
        f_{\textit{i}}^{B^{(2)\Psi}}(\barx,k,v,u) &=\frac{1}{3 k u^3 v^2 \bar{x}^4} \bigg( 6 u \bar{x} \big(4 + v^2 \bar{x}^2\big) \cos\left (\frac{u \bar{x}}{\sqrt{3}}\right ) + \sqrt{3} \big(-24 + \bar{x}^2 \big(-6 v^2  \\
        &+ u^2 \big(8 + v^2 \bar{x}^2\big)\big)\big) \sin\left(\frac{u \bar{x}}{\sqrt{3}} \right ) \bigg)
        I^{B^{(2)}}_{sti}(\overline{x} ,k,v,u)\\
        &+ \frac{1 }{k u^3 v^2 \bar{x}^3} 8 \sqrt{3} \sin\left(\frac{u \bar{x}}{\sqrt{3}}\right)-8 u \bar{x} \cos\left(\frac{u \bar{x}}{\sqrt{3}}\right) \frac{\partial}{\partial \barx} I^{B^{(2)}}_{sti}(\overline{x} ,k,v,u) \, ,
    \end{split}
\end{align}
and expressions for the nested kernels can be found in Eq.~\eqref{RDnestedb2psi11} for $f_1^{B^{(2)\Psi}}(\barx,k,v,u)$ and Eq.~\eqref{RDnestedb2psi12} for $f_2^{B^{(2)\Psi}}(\barx,k,v,u)$.

Finally, for the background generated by second-order tensors coupled to first-order scalars, we can similarly split up the kernel Eq.~\eqref{fh2psi1final} into an analytical contribution and a numerical contribution. We define the analytical and numerical parts as
\begin{subequations}
    \begin{align}
        \begin{split}
        &f_a^{h^{(2)\Psi}}(\barx,k,v,u) = \frac{48 k^2 \sin \barx }{u^6 \barx^9}\left( u \barx \cos\left(\frac{u \barx}{\sqrt{3}}\right) - \sqrt{3} \sin\left(\frac{u \barx}{\sqrt{3}}\right) \right) \bigg ( u \barx (-18 + (-3  \\
        &+ u^2 - 3 v^2)\barx^2)  \cos\left(\frac{u \barx}{\sqrt{3}}\right) + 3 \sqrt{3} (6 + (1 - u^2 + v^2)\barx^2) \sin\left(\frac{u \barx}{\sqrt{3}}\right) \bigg ) \, , 
        \end{split}
        \\
        \begin{split} \label{finalkernelh2}
         &f_n^{h^{(2)\Psi}}(\barx,k,v,u) = \frac{12 k^2}{u^3 \barx^5} \bigg ( ( u \barx \left(18 + (3 - u^2 + 9 v^2) \barx^2 \right) \cos\left(\frac{u \barx}{\sqrt{3}}\right) + 3 \sqrt{3} \\
         &\times  \big(-6 + (-1 + u^2 - 3 v^2) \barx^2 \big )\sin\left(\frac{u \barx}{\sqrt{3}}\right) \bigg ) I^{h^{(2)}}_{st}(\barx ,k,v,u) \, .
        \end{split}
    \end{align}
\end{subequations}
Substituting $f_a^{h^{(2)\Psi}}(\barx,k,v,u)$ into Eq.~\eqref{finalkernel}, we arrive at  
\begin{align}\label{analyticalkernelh2psi1}
    \begin{split}
         x^2&\overline{\mathcal{I}_a^{h^{(2)}\Psi}(x\rightarrow\infty ,v,u)}=\frac{2}{315 u^6} \Bigg( 127 u^6 + 324 (3 + 7 v^2) \log 3 
        + 9 u^4 \big (25 - 77 v^2 + 140 \log 3 \big)  \\
        &- 54 u^2 \big (3 + 28 \log 3 + v^2 (7 + 70 \log 3) \big )  + 2 \bigg[-81 (3 + 7 v^2) + u^2 \bigg ( 189 (2 + 5 v^2)   \\
        & + u \big[ 315 \sqrt{3} v^2+ u \big (-315 + \sqrt{3} u \{4 u^2 - 21 (4 + v^2)\}\big )\big]\bigg )\bigg ] 
        \log\left| 3 + \sqrt{3} u \right|\\
        &- 2 \bigg [(81 (3 + 7 v^2) 
        + u^2 \bigg(-189 (2 + 5 v^2)  + u \big [ 315 \sqrt{3} v^2  + u \big (315 + \sqrt{3} u \{4 u^2 \\ &- 21 (4 + v^2)\}\big )\big]\bigg )\bigg ] 
        \log\left| -3 + \sqrt{3} u \right| \Bigg) \, .
    \end{split}
\end{align}
The nested kernel has been computed and is shown in Eq.~\eqref{RDnestedh2psi1}.
%%%%%%%%%%%%%%%%%%%%%%%%%%%%%%%%%%%%%%%%%%%%%%%%%%%%%%%%%%%%%%%%%%%%%%%%%%%%%%%%%%%%%%%%%%%%%%
\subsection{Log-normal input power spectrum}
%%%%%%%%%%%%%%%%%%%%%%%%%%%%%%%%%%%%%%%%%%%%%%%%%%%%%%%%%%%%%%%%%%%%%%%%%%%%%%%%%%%%%%%%%%%%%%
We are now ready to compute the spectral density of these extra contributions. We consider a peaked input power spectrum for both the primordial scalar and tensor cases and study how varying the peak width affects the observable. We do this for a log-normal input power spectrum, building directly on the values introduced in Chap.~\ref{Chap:SOGWs}.  The spectral density of each contribution is then computed by evaluating the integrands numerically with the \texttt{vegas+} integration package, using the $(s,t)$ coordinate system defined in Eq.~\eqref{stcoordinates}.

\begin{figure}[h!]
    \centering
    \includegraphics[width=\linewidth]{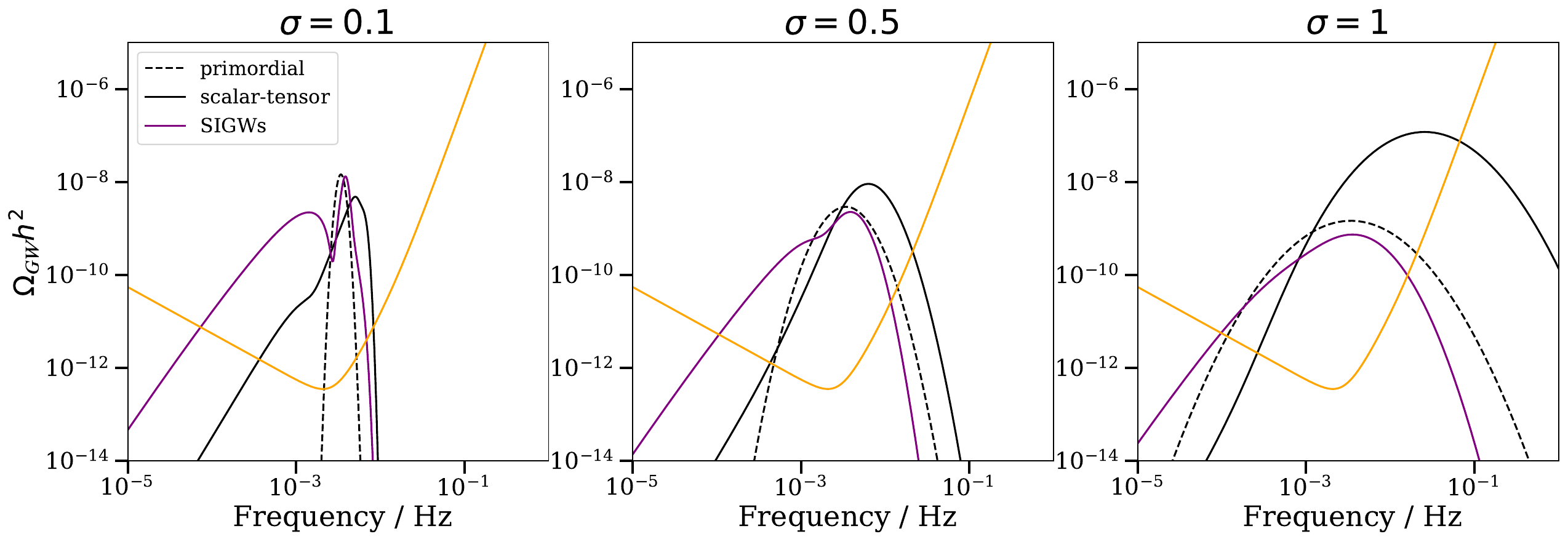}
    \caption{\footnotesize{Plot of the spectral density against frequency, where we look at SIGWs (solid purple) and scalar-tensor iGWs at second order (solid black). The dashed black lines correspond to primordial GWs, and the orange curve is the LISA sensitivity \cite{Moore:2014lga, Sathyaprakash:2009xs}. Each graph corresponds to a different value of $\sigma$, increasing as we go towards the right. Regardless of the value of $\sigma$ we would expect the higher order contributions to converge. This is not the case for the scalar-tensor contribution (as opposed to the SIGWs), which becomes bigger for increasing $\sigma$. This `unphysical enhancement' was discussed in Chap.~\ref{Chap:SOGWs}. We further recall that the integrand for SIGWs is well behaved in the IR and UV limits.}}
    \label{fig:secondorderplots}
\end{figure}

In Fig.~\ref{fig:secondorderplots}, we plot the present-day spectral density of scalar–tensor iGWs and SIGWs. While in Chap.~\ref{Chap:SOGWs} we restricted the scalar–tensor iGWs to a very narrow peak ($\sigma=0.1$), due to the UV divergence of the integrand, here we show the effect of widening the peaks. 
The expressions for the spectral density are from Eqs.~\eqref{omegaSIGWsfinal} and \eqref{omegaSTRD}, for SIGWs and scalar-tensor iGWs, respectively. Notably, as the width of the primordial peak increases, the scalar-tensor contribution seems enhanced. This is not the case for SIGWs. This `unphysical-enhancement' was attributed to a divergence of the integrand in Eq.~\eqref{omegaSTRD} when taking the UV limit of the expression. We recall that in the UV limit, as $k\rightarrow\infty$, there are two distinct limits: $v\rightarrow 1$ and $u\rightarrow 0$ (or $t\rightarrow 0$ and $s\rightarrow -1$) which couples large wavelength scalar perturbations with short wavelength tensor modes and $v\rightarrow 0$ and $u\rightarrow 1$ (or $t\rightarrow 0$ and $s\rightarrow 1$) which couples short wavelength scalar perturbations with long wavelength tensor modes. We recall that when the input primordial power spectrum is sufficiently peaked, the enhancements are not an issue since the spectrum is sufficiently suppressed away from the peak. However, as the peak of the scalar power spectrum is widened, we start integrating over more and more scales, which include the UV scales where the rest of the integrand diverges. This results in an unphysical enhancement of the observable. In App.~\ref{meshstuff} we present mesh plots of the integrands from the new numerical contributions we will consider. For reference, we also have a mesh plot of the scalar-tensor contribution in Fig.~\ref{fig:st_meshplots}. The enhancements of the integrand appear in the limits we have discussed. We would like to emphasise that, from the mesh plot, we see that only the $u^{-4}$ divergence seems to enhance the observable, which corresponds to the scalar sector.

\begin{figure}[h!]
    \centering
    \includegraphics[width=\linewidth]{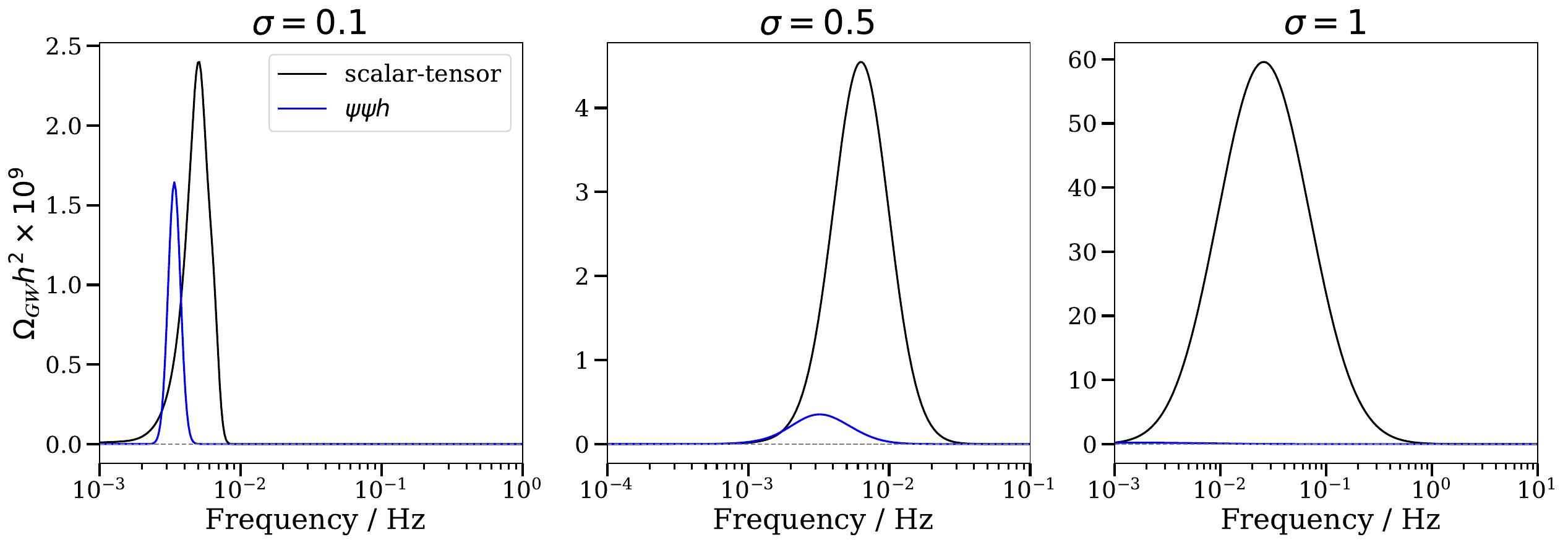}
    \caption{\footnotesize{Contribution of the scalar-scalar-tensor source term to the iGW background. We have plotted the spectral density enhanced by a factor of $10^9$ (on a linear scale) against the frequency (on a logarithmic scale). We see that these induced waves are subdominant compared to the scalar-tensor iGWs and also decrease in amplitude as $\sigma$ gets larger.}}
    \label{fig:sshvaryingsigma}
\end{figure}

\begin{figure}[h!]
    \centering
    \includegraphics[width=\linewidth]{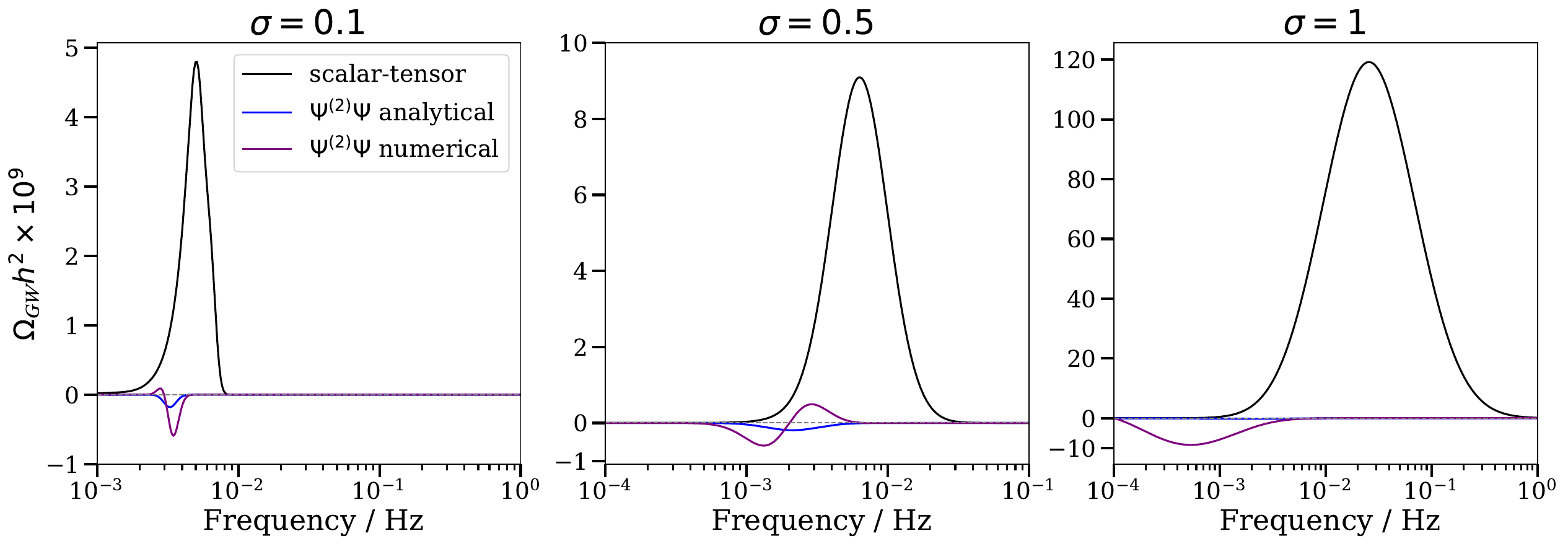}
    \caption{\footnotesize{Contribution to the iGW background originating from second-order scalar perturbations coupled to first-order scalar modes. We have plotted the numerical contribution (purple), analytical contribution (blue) and second-order scalar-tensor induced waves (black). This new contribution acts as a suppression of the signal on certain scales when $\sigma$ is small, and the suppression gets larger whilst increasing the width of $\sigma$. The numerical and analytical contributions are different, with the numerical contribution dominating in magnitude. When $\sigma =1$ the spectral density is heavily suppressed for lower frequencies.}}
    \label{fig:psi2psi1varyingsigma}
\end{figure}

\begin{figure}[h!]
    \centering
    \includegraphics[width=\linewidth]{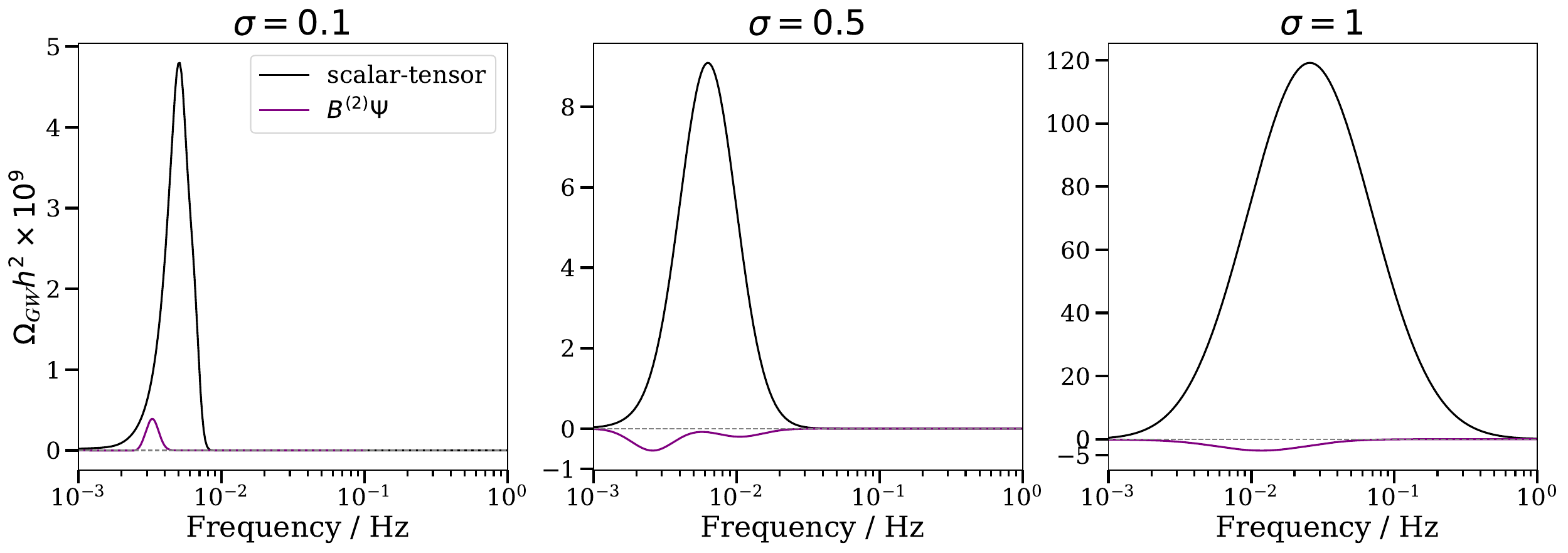}
    \caption{\footnotesize{Contribution to the iGW background originating from second-order vector perturbations coupled to first-order scalar modes (purple). For small input power spectra ($\sigma=0.1$), the contribution is positive but subdominant compared to the scalar-tensor iGWs. However, as $\sigma$ gets large, the contribution becomes negative and acts as a suppression to the spectral density. The suppression occurs at a different scale than the waves sourced by second-order scalar perturbations and is also subdominant in magnitude.}}
    \label{fig:b2psi1varyingsigma}
\end{figure}

\begin{figure}[h!]
    \centering
    \includegraphics[width=\linewidth]{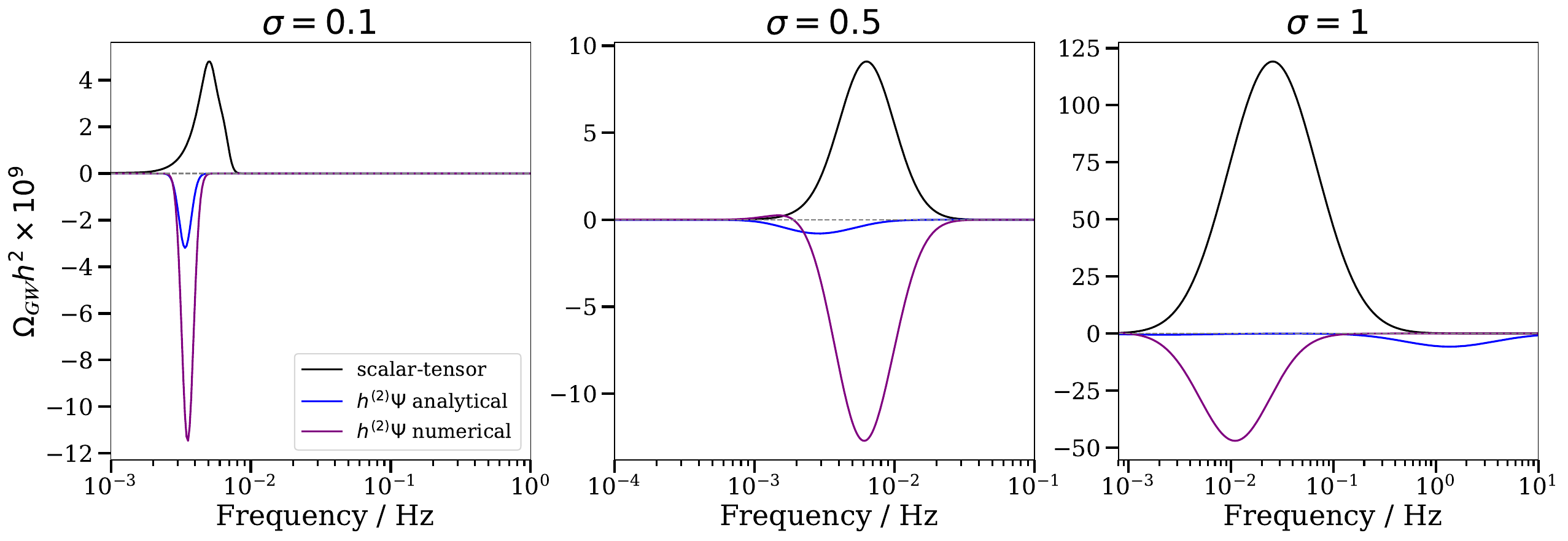}
    \caption{\footnotesize{Contribution to the iGW background originating from second-order tensor perturbations coupled to first-order scalar modes. Similar to the previous plots, we differentiate the analytical contribution (blue) from the numerical contribution (purple). Strikingly, the magnitude of the suppression is large compared to the other contributions for both the analytical and numerical contributions, particularly from the numerical contribution.}}
    \label{fig:h2psi1varyingsigma}
\end{figure}

We now move on to computing the spectral density of the new contributions we have derived. Instead of using a logarithmic scale for the spectral density, as in Fig.~\ref{fig:secondorderplots}, we plot $\Omega_{GW} h^2 \times 10^9$ on a linear scale, since the contributions can take negative values. We first show the contributions from each individual term: Fig.~\ref{fig:sshvaryingsigma} shows the contribution coming from the scalar-scalar-tensor-induced waves, Fig.~\ref{fig:psi2psi1varyingsigma} shows the contribution coming from the second-order scalars coupled to first-order scalars, Fig.~\ref{fig:b2psi1varyingsigma} shows the contribution coming from the second-order vectors coupled to first-order scalars, and Fig.~\ref{fig:h2psi1varyingsigma} shows the contribution coming from second-order tensors and scalars. Where possible, we distinguish between contributions that we have computed numerically and analytically. 

Our results show a suppression of the scalar-tensor iGWs generated at second order from all the source terms, including second-order perturbations. The contribution from pure first-order terms, i.e.~the scalar-scalar-tensor iGWs, is additive yet subdominant compared to the scalar-tensor iGWs. For iGWs generated by source terms which couple first-order scalar to second-order perturbations, we have differentiated between the `analytical contribution' (terms where we were able to compute an analytical kernel) and `numerical contribution' (terms where we were unable to compute an analytical kernel). There are two main differences between the analytical and numerical contributions: Firstly, the analytical contributions always act as a suppression to the spectral density (i.e.~always negative or zero), whereas the numerical contributions can be positive at certain scales.  Secondly, the magnitude of the numerical contribution is always larger than its analytical counterpart. Furthermore, as $\sigma$ increases, the magnitude of both numerical and analytical contributions increases. When this was the case for the scalar-tensor iGWs we concluded that this was due to how the integrand behaves in the UV limit, and so we shall carry out a similar analysis here. For terms involving analytical expressions, we can track the limits of the integrands in the UV and IR limits; however, this is not possible for the terms where we have integrated numerically. Hence, when the kernels are computed numerically, we look at the mesh plots in App.~\ref{meshstuff} and draw some conclusions by comparing with the scalar-tensor mesh plot in Fig.~\ref{fig:st_meshplots}. For the third-order contributions, we have plotted the logarithm of the absolute values of the integrand since they can be positive and negative and then we can deduce if there is a positive or negative divergence by looking at the observable\footnote{There are ways to plot on logarithmic scales negative and positive values (for example by using \texttt{symlog}) however, the data spans lots of scales in magnitude and we found that the most informative way was to plot the absolute value of the data.}. In our analysis, we have kept the width of the primordial spectrum for scalars and tensors equal; however, we insist that since we are only integrating over the primordial scalar power spectrum, if there is a divergence, it will come from there. 

For terms involving second-order scalars, we find that the analytical contribution is well behaved in the IR limit and in the UV limit. A potential problematic limit would be when $u=\sqrt{3}$, then the integrand diverges as $v^{-3}$ and then integrating over $v=0$ would cause a divergence. However, this is not a region we integrate over (this corresponds to $s=\sqrt{3}\approx1.7$). To study the numerical behaviour of the integrand, we look at Fig.~\ref{fig:psi2psi1um_meshplots}. From the mesh plots, we see that the kernel does get enhanced as $v\rightarrow0$ and $u\rightarrow 1$; nevertheless, this enhancement does not blow up as the width of the primordial scalar power spectrum gets larger. We are therefore unable to conclude that the contribution which involves second-order scalars diverges. 

The contribution coming from terms involving second-order vectors is fully numerical; the corresponding mesh plots of the integrand are shown in Fig.~\ref{fig:b2psi1um_meshplots}. Both kernels exhibit an enhancement in the limit of $t\rightarrow 0$ and $s\rightarrow -1$. We also notice diagonals in the mesh plots of the kernels, which seem to be enhanced, which could correspond to constant values of $u$ entering a resonance (for example, a logarithmic divergence like in the case of SIGWS). However, we have plotted constant lines of multiples of $u$ and cannot come to the conclusion that these lines correspond to a logarithmic resonance. Nevertheless, we see that as the primordial peak of the scalar spectrum gets wider, the enhancement in the UV limit stands out. We conclude from these mesh plots that this term diverges negatively in $v\rightarrow1$ and $u\rightarrow 0$ limit, but at a slower rate than the scalar-tensor iGWS.

The analytical term coming from second-order tensors is finite in the IR and UV limit where $v\rightarrow 0$ $u\rightarrow1$. The over UV limit, which couples large wavelength scalar modes with short wavelength tensor modes, tends to $-u^{-2}$. We recall that scalar-tensor iGWs also had a similar divergence in this limit but as $u^{-4}$, hence we cannot conclude any `cancellation' coming from these analytical terms. From the numerical mesh plots in Fig.~\ref{fig:h2psi1um_meshplots}, we can see an enhancement in the same limit which seems similar in magnitude to the one occurring in scalar-tensor iGWS; however, we cannot conclude its magnitude is of $-u^{-4}$.

Parity violation in the primordial tensor power spectrum leaves observational signatures for scalar-tensor iGWs, as was shown in Ref.~\cite{Bari:2023rcw} for Gaussian initial conditions (see also Fig.~\ref{fig:chiralityplot}). We note that for scalar-tensor iGWs, the tensor power spectrum appears in the integrand and is coupled to the polarisation tensors in a way which depends on parity, see Eq.~\eqref{stpowerspectrum}. In the case of third-order iGWs, the tensor power spectrum appears outside the integrand, and the way in which it couples to the polarisation tensors is independent of the parity (we get the same result whether $\lambda = R$ or $\lambda = L$), making the dependence on the parity of the initial spectrum trivial. For similar reasons, we do not consider peaks at different scales. 

We would like to clarify that the results presented are independent of any tunable parameters in the \texttt{vegas+}~\cite{Lepage:2020tgj} integration package used in our analysis. In particular, the final plots remain unchanged when varying the sampling strategy, the number of iterations (\texttt{nitn}), or the number of integrand evaluations per iteration (\texttt{neval}). Furthermore, we checked that our results are insensitive to integrating over more scales (i.e. changing the upper limit of the $t$ integrand in the (s,t) coordinate system) and also changing the lower limit of the $t$ integrand from $0$ to a value infinitely close to $0$ or to the value of $1$.

\begin{figure}[htbp]
    \centering
    \begin{subfigure}[b]{0.7\textwidth}
        \includegraphics[width=\linewidth]{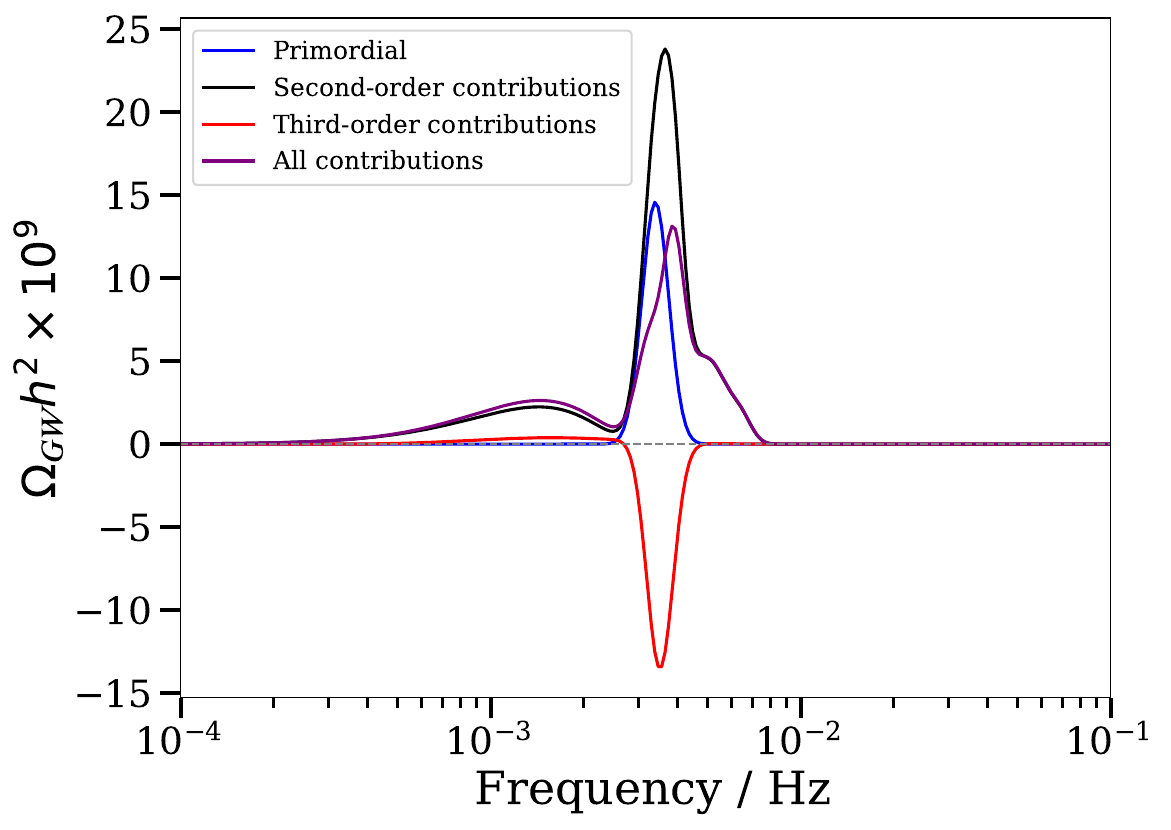}
    \end{subfigure}

    \vspace{0.5cm}
    
    \begin{subfigure}[b]{0.4\textwidth}
        \includegraphics[width=\linewidth]{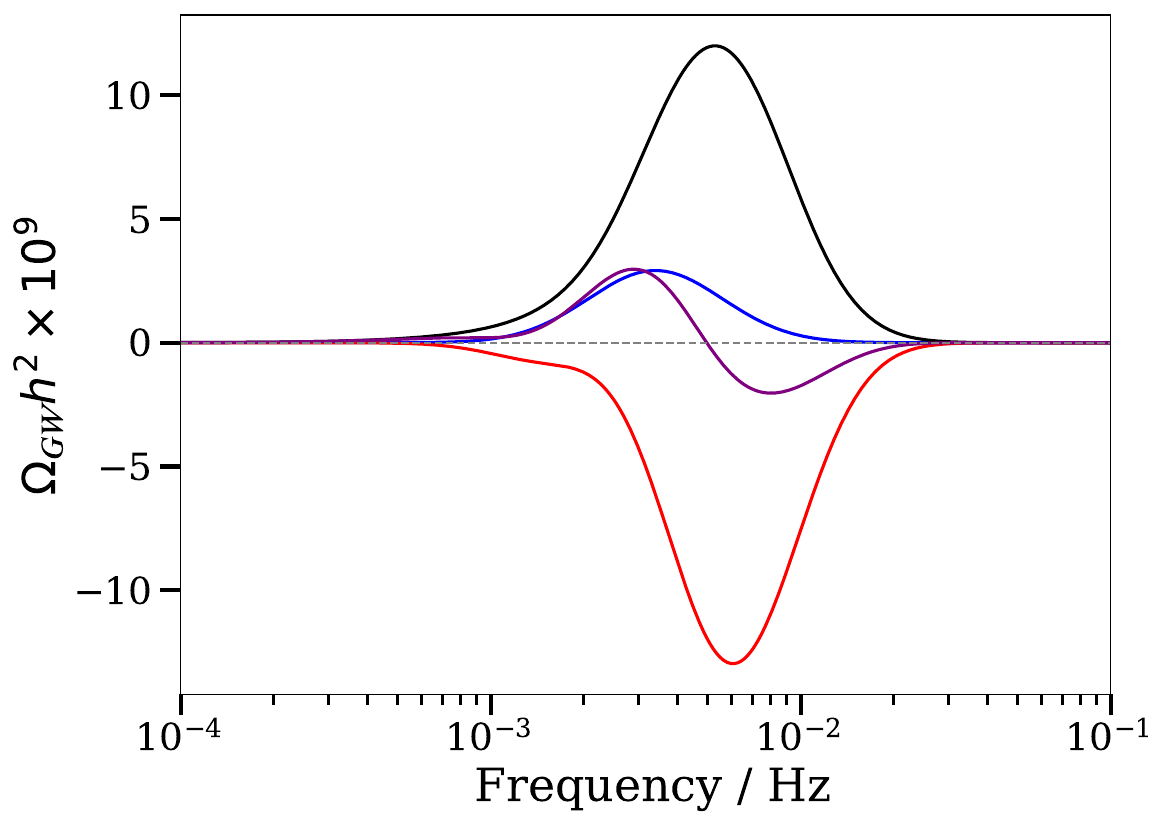}
    \end{subfigure}
    \begin{subfigure}[b]{0.4\textwidth}
        \includegraphics[width=\linewidth]{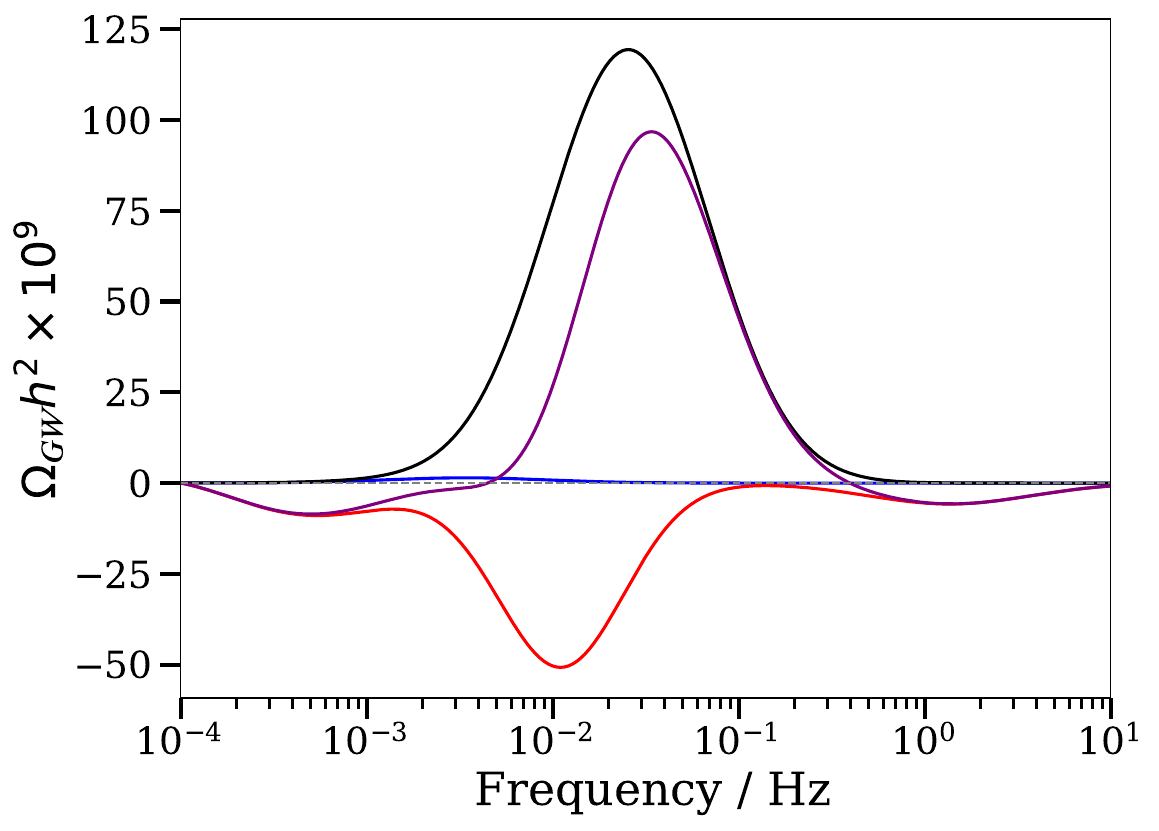}
    \end{subfigure}
   
    \caption{\footnotesize{Plot of the spectral density today for tensor modes coming from inflation (blue), second-order contributions (black), third-order contributions (red) and the total contributions (purple) for input log-normal power spectra with $\sigma=0.1$ corresponding to the top plot, $\sigma=0.5$ bottom left plot and $\sigma=1$ bottom right plot. The inclusion of third-order tensor modes drastically reduces the spectra around peak scales. As $\sigma$ gets larger, more scales are suppressed. For values of $\sigma >0.1$, the observable becomes negative, which is unphysical. Although this is partially due to the divergence, including the autocorrelations of third-order iGWs would add a positive contribution and could potentially make the overall observable positive.}}
    \label{fig:allcontirbutions}
\end{figure}

Finally, we present all contributions together in Fig.~\ref{fig:allcontirbutions} for different values of $\sigma$. We stress that only the case $\sigma = 0.1$ should be regarded as realistic, since in this regime the scalar input power spectrum is sufficiently peaked. Including third-order tensor modes correlated with linear tensor perturbations suppresses the spectral density for all $\sigma$ values. The scale at which this suppression occurs depends on $\sigma$; for $\sigma = 0.1$ it coincides with the peak, whereas for larger $\sigma$ the suppression extends to a broader range of scales, and for $\sigma = 1$ it affects the entire spectrum. Although we attribute the appearance of negative spectral densities in part to unphysical enhancements at small scales, we emphasise that a negative contribution is not inherently problematic. Correlations can be negative, which might indicate that we should also include the auto-correlation of third-order iGWs, which would yield a positive contribution, much like the auto-correlation of second-order iGWs, which is always positive\footnote{This was shown in Ref.~\cite{Zhou:2021vcw} for the case of first-order scalar-perturbations inducing second and third-order perturbations.}. Put differently, the observable (the spectral density) receives contributions from various correlations in our analysis: $\langle h^{(1)} h^{(1)} \rangle$, $\langle h^{(2)} h^{(2)} \rangle$, and $\langle h^{(1)} h^{(3)} \rangle$. If the $\langle h^{(3)} h^{(3)} \rangle$ term were also included, and the total spectral density then remained positive for all $\sigma$ values, the negative contribution from $\langle h^{(1)} h^{(3)} \rangle$ would not be problematic.

% % % % % % % % % % % % % % % % % % % % % % % % % % % % 
% chapter.tex - Ian Huston
% Sample chapter layout
% % % % % % % % % % % % % % % % % % % % % % % % % % % % 
% Redefine CVSRevision for this section. 
% If you don't want to use CVS tags comment out this line
\renewcommand{\CVSrevision}{\version$Id: chapter.tex,v 1.3 2009/12/17 18:16:48 ith Exp $}

% % % % % % % % % % % % % % % % % % % % % % % % % % % % % % % % 
% =========================================================== %
% % % % % % % % % % % % % % % % % % % % % % % % % % % % % % % % 
\chapter{Discussion and Outlook}
\label{Ch_conclusions}
% % % % % % % % % % % % % % % % % % % % % % % % % % % % % % % % 
% =========================================================== %
% % % % % % % % % % % % % % % % % % % % % % % % % % % % % % % % 
The detection of the stochastic gravitational wave background (SGWB) appears to be only a matter of time, making the computation and characterisation of its various contributions both timely and essential. In this thesis, we have focused on a particular class of gravitational waves (GWs) that may contribute to the SGWB: induced gravitational waves (iGWs). These waves arise naturally at different orders in cosmological perturbation theory. In particular, SIGWs emerge at second order, sourced by products of first-order scalar fluctuations. Waves generated by large scalar fluctuations in the early universe have attracted considerable interest, as they may provide a counterpart signal to PBH formation in the early universe. Any detection of induced GWs by future GW observatories would offer a wealth of information concerning inflationary dynamics at scales far smaller than the CMB.

In this thesis, we developed the formalism for including first-order tensor modes as a source term for GWs at second-order. This added two source terms at second-order: one which coupled first-order scalar and tensor fluctuations (scalar-tensor iGWs), and one which is made of quadratic linear tensor modes (tensor-tensor iGWs). In Chapter~\ref{Chap:SOGWs}, we derived expressions for the time-averaged power spectrum of these two new source terms, without specifying an equation of state parameter. We then computed the spectral density for these terms in a radiation-dominated (RD) universe, assuming Gaussian adiabatic initial conditions and neglecting anisotropic stress. We found that the integrand of both the new terms diverges in the UV limit when the input spectra is not sufficiently peaked, which was not the case for the SIGWs term. For this reason, we only considered narrowly peaked input power spectra for both scalars and tensors. This can be explained by the fact that when we are putting all the power at one particular scale, it sufficiently suppresses the UV scales where the integrand blows up. For peak input spectra, we found that the tensor-tensor iGWs were subdominant and thus neglected them through the rest of our work. On the other hand, we showed that the scalar-tensor iGWs were dominant at smaller scales compared to SIGWs. This showed that \emph{if} scalar \emph{and} tensor primordial perturbations are enhanced during inflation, then we should not neglect the new scalar-tensor source we computed.  

We then moved on to look at how scalar primordial non-Gaussianity (PNG) affects the scalar-tensor spectrum in Chap.~\ref{Chap:NG}. Specifically, since any observation of iGWs would require a significant amplification of either the scalar or tensor primordial power spectra on small scales, detection of PNG could suggest departures from standard single-field slow-roll inflation, multi-field inflationary scenarios, non-Gaussian perturbations or the existence of PBHs. We modelled PNG as a local expansion of the Gaussian curvature perturbation in real space up to second order. Many works had already investigated signatures of scalar local non-Gaussianity on SIGWs, and so our work focused exclusively on the scalar-tensor sector. The expansion in real space of the curvature perturbation creates the appearance of a new non-Gaussian scalar-tensor source term. We found that in the scenario with coincident peaked scalar and tensor primordial power spectra, the new term shares many similarities with the Gaussian scalar-tensor and non-Gaussian scalar-scalar terms. With a sharp peak at the location of the first-order scalar and tensor power spectrum peak, its effect on the overall iGW spectrum is to smooth out the peaks in the Gaussian terms. This is a typical effect of non-Gaussian contributions. The new term also suffers from the same divergence issue as the scalar-tensor Gaussian term for wide primordial power spectra and is affected by parity violation in a similar way. One feature that does seem to distinguish the new non-Gaussian term from others coming from scalar-scalar interactions, however, is the presence of a pronounced `knee'. This knee typically has a much larger amplitude than the knees generated by scalar-scalar non-Gaussian terms and, for suitably large values of $F_{\textrm{NL}}$ or a fortuitous position of the power spectrum peak, could lie within LISA's sensitivity. There is a possibility, therefore, that features in the UV tail of any detected iGW signature could be used to indicate the existence of scalar PNG.

In the final Chapter of this thesis, we studied extra contributions to the SGWB. These appear at the same order as the second-order correlations we studied in Chap.~\ref{Chap:SOGWs}. Indeed, in Chap.~\ref{Chap:Third} we derived the third-order GW equation and found that there were non-trivial contributions when some of the source terms were correlated with primordial tensor modes. We did not include third-order perturbations sourced by pure first-order tensor fluctuations, as we previously showed that they were subdominant. These terms can be found in App.~\ref{app4}. The divergence of the scalar–tensor term at second order motivated us to investigate whether additional contributions from scalar–tensor interactions could cancel this behaviour. More fundamentally, since these terms contribute at the same perturbative order, their computation is necessary regardless. After providing the time-averaged power spectra of these source terms, we computed the spectral density of these terms in an RD universe with an input log-normal power spectrum for the first-order curvature and tensor perturbations. For very peaked spectra, we found that these new contributions suppress the spectral density, particularly around peak scales. As the peak of the input spectra gets wider, the suppression happens over more scales. However, some of the suppression was due to the divergent nature of the integrands. We showed that some of the source terms diverge in the UV limit, but we cannot claim that these divergences cancel out the ones seen in the scalar-tensor iGWs. Indeed, source terms made of second-order perturbations gave rise to nested kernels, which made it impossible to determine analytically how these terms behave. Numerically, our results show that there is no cancellation of the divergence. 

Several promising directions remain beyond the scope of this work. When we considered scalar PNG in Chap.~\ref{Chap:NG}, we truncated the expansion at second-order. A straightforward extension of our work would be to include higher-order non-Gaussian parameters. Moreover, we did not consider tensor PNG. This would have an impact on the scalar-tensor term investigated in this work and also on the tensor-tensor sector. From a phenomenological perspective, however, it is more challenging -- and arguably less general -- to justify the local expansion of the tensor field in the same way we did for the scalar case. It would nevertheless be a good starting point as these analyses remain absent from the literature.

Furthermore, in Chap.~\ref{Chap:Third} we argued that although the suppression of the spectral density happens in part due to the divergent nature of the integrand, it might also be because we are not including the autocorrelation of third-order tensor modes. Additionally, for the third-order terms, which we could only compute numerically due to the nested integrals, taking the UV limits of the nested kernels may allow for an analytical evaluation. This, in turn, could clarify the nature of the divergence. It would also be worth investigating how the spectral density of the scalar-tensor and third-order contributions we have computed are in a different gauge than the conformal Newtonian gauge. If the divergence does not appear in a different gauge, this would indicate that the gauge we are using is not suitable for this computation. This, in turn, may have deeper implications for the gauge issue. Finally, since scalar PNG can leave an imprint on the spectral density of scalar-tensor iGWs, one should check whether this is the case for the new source terms we have computed. 

Throughout our work on second and third-order GWs, we used toy model input power spectra, such as the Log-normal or Gaussian spectra. A natural next step is therefore to use more realistic input power spectra, both for the scalar and tensor perturbations. In most cases, this will require the use of numerical methods. Luckily, there are already numerical packages to calculate the power spectrum for any single or multi-field model available, such as \textrm{PyTransport} \cite{Mulryne:2016mzv}. This would allow us to further constrain the parameter space of models of inflation using gravitational waves.

% % % % % % % % % % % % % % % % % % % % % % % % % % % % 
% chapter.tex - Ian Huston
% Sample chapter layout
% % % % % % % % % % % % % % % % % % % % % % % % % % % % 
% Redefine CVSRevision for this section. 
% If you don't want to use CVS tags comment out this line
\renewcommand{\CVSrevision}{\version$Id: chapter.tex,v 1.3 2009/12/17 18:16:48 ith Exp $}

% % % % % % % % % % % % % % % % % % % % % % % % % % % % % % % % 
% =========================================================== %
% % % % % % % % % % % % % % % % % % % % % % % % % % % % % % % % 
\begin{appendices}

\chapter{Fourier space}\label{appFourierspace}

In this Appendix, we present our Fourier space conventions in the case of flat spatial curvature and the spherical parametrisation we use throughout our work.

\section{Definitions and conventions}

A scalar perturbation can be expressed as a Fourier integral
\begin{equation}
    S(\eta , \mathbf{x}) = \int \frac{\rm ^3\mathbf{k}}{(2\pi)^{\frac{3}{2}}} S(\eta , \mathbf{k}) e^{i\mathbf{k}\cdot \mathbf{x}} \, ,
\end{equation}
and its Fourier inverse
\begin{equation}
    S(\eta , \mathbf{k}) = \int \frac{\rm ^3\mathbf{x}}{(2\pi)^{\frac{3}{2}}} S(\eta , \mathbf{x}) e^{-i\mathbf{k}\cdot \mathbf{x}} \, .
\end{equation}
The normalisation condition which ensures the basis functions are orthonormal is
\begin{equation}
    \int \frac{\rm ^3 \mathbf{x}}{(2\pi)^3}  e^{i(\mathbf{k}-\mathbf{k\conf})\cdot \mathbf{x}} = \delta (\mathbf{k}-\mathbf{k\conf}) \, .
\end{equation}
The reality condition on $S(\eta , \mathbf{x})$ implies $S^*(\eta , \mathbf{-k})=S(\eta , \mathbf{k})$.

A vector can also be expressed as a Fourier integral, where it decouples into its two parities $r$:
\begin{equation}
    V_i(\eta , \mathbf{x}) = \sum _r \int \frac{d^3\mathbf{k}}{(2\pi)^{\frac{3}{2}}}  V^r(\eta , \mathbf{k}) e^r_i({\mathbf{k}}) e^{i\mathbf{k}\cdot \mathbf{x}} \, ,
\end{equation}
and the Fourier inverse is given by
\begin{equation}
    V^r(\eta , \mathbf{k}) =  \int \frac{d^3\mathbf{k}}{(2\pi)^{\frac{3}{2}}}  V_i(\eta , \mathbf{x}) e_r^i({\mathbf{k}}) e^{-i\mathbf{k}\cdot \mathbf{x}} \, .
\end{equation}
The normalisation of the basis functions is
\begin{equation}
    \int \frac{\rm ^3 \mathbf{x}}{(2\pi)^3}   e^r_i(\mathbf{k}) e_s^i(\mathbf{k\conf})^*  e^{i(\mathbf{k}-\mathbf{k\conf})\cdot \mathbf{x}} = \delta ^{rs} \delta (\mathbf{k}-\mathbf{k\conf}) \, .
\end{equation}
The reality condition $V_i(\eta , \mathbf{x})=V_i(\eta , \mathbf{x})^*$ implies both $e^r_i(\mathbf{-k})^*=e^r_i(\mathbf{k})$ and $V_r^*(\eta , \mathbf{-k})=V_r(\eta , \mathbf{k})$. A basis which satisfies these conditions is the circular basis of parities ($r=+,-$) defined as
\begin{subequations}\label{circularparity}
    \begin{align}
        e^+_i({\mathbf{k}}) &= \frac{1}{\sqrt{2}} \left ( e_i(\mathbf{k}) +i \overline{e}_i(\mathbf{k}) \right ) \, ,
        \\
        e^-_i({\mathbf{k}}) &= \frac{1}{\sqrt{2}} \left ( e_i(\mathbf{k}) -i \overline{e}_i(\mathbf{k}) \right ) \, ,
    \end{align}
\end{subequations}
where $\{ e_i(\mathbf{k}) , \overline{e}_i(\mathbf{k})\}$ form a subspace perpendicular to the wave-vector $\mathbf{k}$. Useful relations used throughout this work include
\begin{equation}
    e^r_i(\mathbf{k})k^i=0 \, , \quad e^r_i(\mathbf{k})^*e_s^i(\mathbf{k}) = \delta ^r_s \, \quad \text{and} \quad e^r_i(\mathbf{k})^* = e^{-r}_i(\mathbf{k}) = e^r_i(\mathbf{-k}) \, .
\end{equation}
where $-r$ refers to the opposite parity.

Finally, a tensor can also be expressed as a Fourier integral. It decouples into two polarisations: 
\begin{equation}
    T_{ab}(\eta , \mathbf{x}) = \sum _{\lambda} \int \frac{\rm ^3\mathbf{k}}{(2\pi)^{\frac{3}{2}}}  T^{\lambda}(\eta , \mathbf{k}) \epsilon ^{\lambda}_{ab}({\mathbf{k}})  e^{i\mathbf{k}\cdot \mathbf{x}} \, ,
\end{equation}
and the inverse
\begin{equation}
    T^{\lambda}(\eta , \mathbf{k}) =  \int \frac{\rm ^3\mathbf{x}}{(2\pi)^{\frac{3}{2}}}  T_{ab}(\eta , \mathbf{x}) \epsilon _{\lambda}^{ab}({\mathbf{k}})^*  e^{-i\mathbf{k}\cdot \mathbf{x}} \, .
\end{equation}
The normalisation condition of the basis function is
\begin{equation}
    \int \frac{\rm ^3 \mathbf{x}}{(2\pi)^3}   \eps ^\lambda_{ij}(\mathbf{k}) \eps _{\lambda \conf}^{ij}(\mathbf{k\conf})^*  e^{i(\mathbf{k}-\mathbf{k\conf})\cdot \mathbf{x}} = \delta ^{\lambda \lambda \conf} \delta (\mathbf{k}-\mathbf{k\conf}) \, .
\end{equation}
The reality condition $T_{ab}(\eta , \mathbf{x})=T_{ab}(\eta , \mathbf{x})^*$ implies $\eps ^\lambda _{ij}(\mathbf{-k})^*=\eps ^\lambda _{ij}(\mathbf{k})$ and $T_\lambda ^*(\eta , \mathbf{-k})=T_\lambda (\eta , \mathbf{k})$. The circular basis of tensors satisfies these conditions. Polarisation tensors $\epsilon ^{\lambda=R,L}_{ab}({\mathbf{k}})$ are constructed from a linear superposition of the $\{+,\times \}$ polarisation basis
\begin{align}\label{circularpol}
    \begin{split}
        \eps ^R_{ab}(\mathbf{k}) &= \frac{1}{\sqrt{2}} \left ( q^+_{ab}(\mathbf{k}) + i q^{\times}_{ab}(\mathbf{k})  \right ) \text{ ,}  \\
        \eps ^L_{ab}(\mathbf{k}) &= \frac{1}{\sqrt{2}} \left ( q^+_{ab}(\mathbf{k}) - i q^{\times}_{ab}(\mathbf{k})  \right )\,,
    \end{split}
\end{align}
where
\begin{align}
    \begin{split}
         q_{ab}^{+}(\mathbf{k})&=\frac{1}{\sqrt{2}}\left ( e_a(\mathbf{k})e_b(\mathbf{k}) - \overline{e}_a(\mathbf{k}) \overline{e}_b(\mathbf{k})\right )\text{ ,} \\
        q_{ab}^{\times}(\mathbf{k})&=\frac{1}{\sqrt{2}}\left ( e_a(\mathbf{k})\overline{e}_b(\mathbf{k}) + \overline{e}_a(\mathbf{k}) e_b(\mathbf{k})\right )\,.
    \end{split}
\end{align}
Some properties and useful relation of $\epsilon ^{\lambda}_{ab}({\mathbf{k}})$ are
\begin{equation}
    \eps _{ab}^{\lambda}(\mathbf{k})\delta ^{ab}=\eps _{ab}^{\lambda}(\mathbf{k}) k^a = 0, \quad  \eps _{ab}^{\lambda}(\mathbf{k}) ^* \eps ^{ab,\lambda ^{\prime}}(\mathbf{k}) = \delta ^{\lambda \lambda ^{\prime}}, \, \, \text{and} \, \, \left ( \eps _{ab}^{\lambda}(\mathbf{k}) \right)^* = \eps _{ab}^{-\lambda}(\mathbf{k})  =\eps _{ab}^{\lambda}(-\mathbf{k})\, ,
\end{equation}
where $-\lambda$ refers to the opposite polarisation. 

We now move on to giving explicit expressions for the vectors and tensors we will use throughout our work in spherical coordinates.

\section{Spherical parametrisation}

Finally, we give the explicit expressions for the wave-vector $\mathbf{k}$ and the orthonormal basis $\{ e_i(\mathbf{k}) , \overline{e}_i(\mathbf{k})\}$ . Due to homogeneity and isotropy, it is useful to parametrise vectors in spherical coordinates. Explicitly,
\begin{equation}\label{kwavevector}
    \mathbf{k}=k(\sin{\theta_k}\cos{\phi_k},\sin{\theta_k}\sin{\phi_k},\cos{\theta_k}) \, ,
\end{equation}
then we can construct the vectors  $\{ e_i(\mathbf{k}) , \overline{e}_i(\mathbf{k})\}$ as
\begin{subequations}
    \begin{align}
         \mathbf{e}(\mathbf{k})&=(\cos{\theta_k}\cos{\phi_k},\cos{\theta_k}\sin{\phi_k},-\sin{\theta_k})\, , \\
         \overline{\mathbf{e}}(\mathbf{k})&=(-\sin{\phi_k},\cos{\phi_k},0)\,.
    \end{align}
\end{subequations}
From these expressions, it is straightforward to build the parity vectors in Eq.~\eqref{circularparity} and the circular polarisation tensors in Eq.~\eqref{circularpol}. We can also explicitly check their corresponding properties\footnote{For example, the transformation $\mathbf{k} \rightarrow \mathbf{-k}$ can be done by rotating $\theta _k \rightarrow \pi - \theta _k$ and $\phi _k \rightarrow \phi _k + \pi$}.

Similarly, for the vector $\mathbf{p}$:
\begin{equation}
    \mathbf{p} = p \left (\sin{\theta _p} \cos{\phi _p}, \sin{\theta _p}\sin{\phi _p}, \cos{\theta _p} \right )\,.
\end{equation}
When $\mathbf{k}$ is aligned with the z-axis we have $\theta_k=\phi_k=0$ and so
\begin{equation}
    |\mathbf{k}-\mathbf{p}|^2 = k^2 + p^2 - 2kp\cos{\theta} _p\,.
\end{equation}

%%%%%%%%%%%%%%%%%%%%%%%%%%%%%%%%%%%%%%%%%%%%%%%%%%%%%%%%%%%%%%%%%%%%%%%%%%%%%%%%%%%%%%%%%%%%%%%%%%%%%%%%%%%%%%%%%%%%%%%%%%%%%%%%%%%%%%%%%%%%%%%%%%%%%%%%%%5555555

\chapter{Bessel functions}\label{appBessel}
Bessel functions, denoted by $Z_n(z)$, where $n$ is an integer, are solutions to the following differential equation:
\begin{equation}
    \frac{d^2Z_n(z)}{dz^2} + \frac{1}{z}\frac{dZ_n(z)}{dz} + \left ( 1-\frac{n^2}{z^2}\right )  Z_n(z) =0.
\end{equation}
We distinguish between two Bessel functions: the first kind, $J_n(z)$, and the second kind, $Y_n(z)$. The solution to the above equation can be expressed as a linear combination of both
\begin{equation}
    Z_n(z) = c_1 J_n(z) + c_2Y_n(z) \, ,
\end{equation}
where $c_1$ and $c_2$ are constants to be determined. The Bessel functions are defined as
\begin{subequations}
    \begin{align}
        &J_n(z)=  \frac{z^n}{2^n} \sum _{k=0}^{\infty} (-1)^k \frac{z^{2k}}{2^{2k}k!\Gamma (n+k+1)}   \, , \\
        &Y_n(z)=\lim_{a\to n} Y_a(z)\, ,
    \end{align}
\end{subequations}
where $a$ is not an integer and
\begin{equation}
    Y_a(z)= \frac{1}{\sin (az)} \left ( \cos (az) J_n(z)-J_{-a}(z) \right ) \, .
\end{equation}
$\Gamma (n)=(n-1)!$ is the Gamma function.

These functions have well-defined asymptotic properties, as $z\rightarrow 0$ we have
\begin{equation}
    J_n (z \rightarrow 0) \approx \frac{1}{\Gamma (n+1)} \left ( \frac{z}{2} \right ) ^n \, , \quad Y_a(z) \approx \frac{1}{\Gamma (-a+1) \sin (a \pi)} \left ( \frac{z}{2} \right ) ^{-a} \, .
\end{equation}
If $n>0$ then Bessel functions of the first kind are finite at the origin, whilst Bessel functions of the second kind have a singularity.

\begin{figure}
    \centering
    \includegraphics[width=0.6\linewidth]{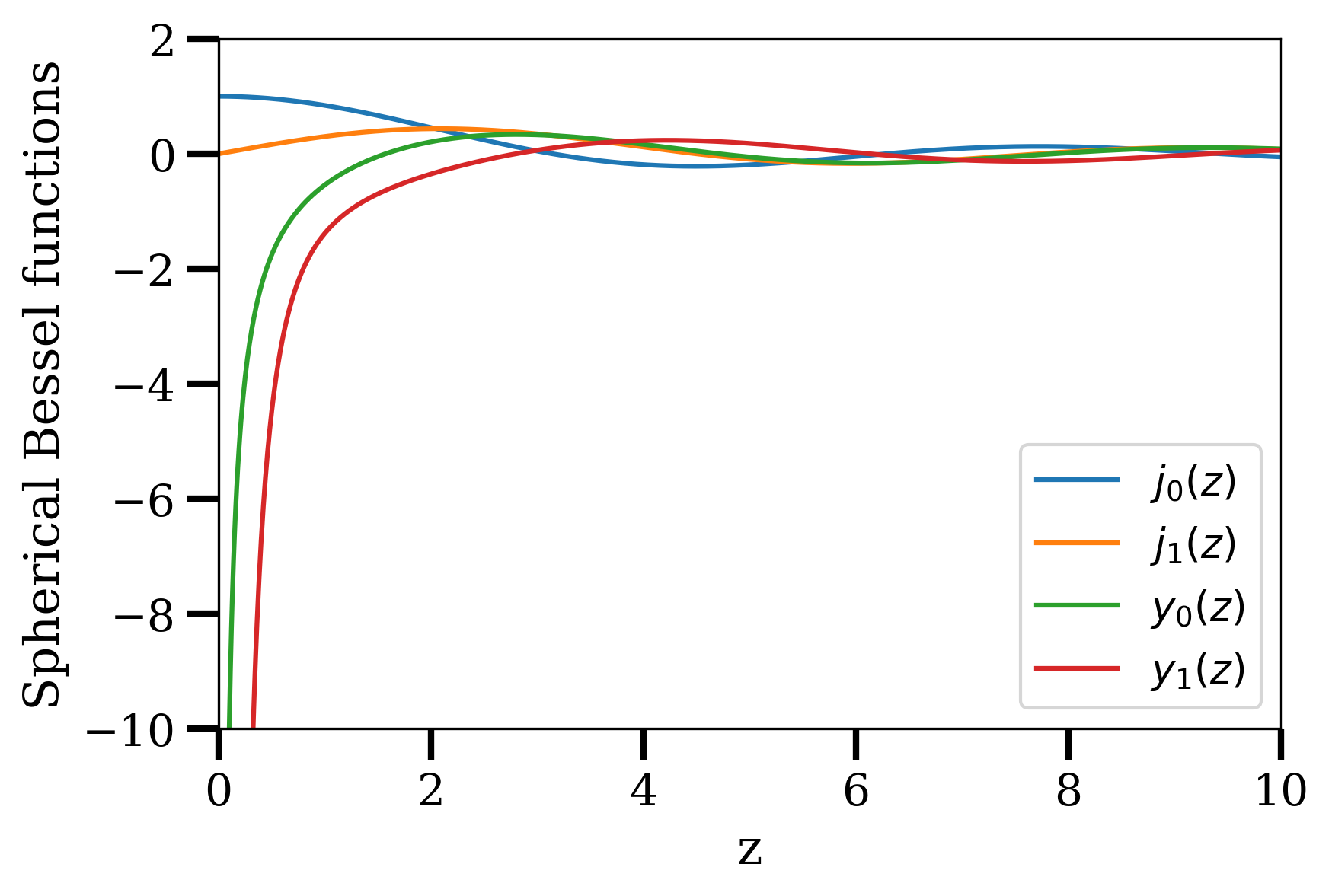}
    \caption{Some common spherical Bessel functions we use throughout our work.}
    \label{fig:bessel}
\end{figure}

Spherical Bessel function of the first kind, $j_n(z)$, are defined via
\begin{equation}
    j_n(z) = \sqrt{\frac{\pi}{2z}}J_{n + \frac{1}{2}}(z) \, ,
\end{equation}
whilst spherical Bessel function of the second kind, $y_n(z)$, are defined via
\begin{equation}
    y_n(z) = \sqrt{\frac{\pi}{2z}}Y_{n + \frac{1}{2}}(z) \, . 
\end{equation}
Some useful functions we use are
\begin{align*}
    \begin{split}
        j_0 (z) &= \frac{\sin z}{z}\, , \quad j_1(z)= \frac{\sin z}{z^2}-\frac{\cos z}{z}\, , \\
        y_0 (z) &= -\frac{\cos z}{z} \, , \quad y_1(z)= -\frac{\cos z}{z^2} -\frac{\sin z}{z} \, ,
    \end{split}
\end{align*}
which we have plotted in Fig.~\ref{fig:bessel}.

\allowdisplaybreaks
%%%%%%%%%%%%%%%%%%%%%%%%%%%%%%%%%%%%%%%%%%%%%%%%%%%%%%%%%%%%%%%%%%%%%%%%%%%%%%%%%%%%%%%%%%%%%%%%%%%%%%%%%%%%%%%%%%%%%%%%%%%%%%%%%%%%%%%%%%%%%%%%%%%%%%%%%%5555555
\chapter{Third-order nested kernels and mesh plots}
\section{Nested Kernels in a RD dominated universe}\label{nestedkernels}

In this Appendix, we present the analytical results of the nested kernels in an RD universe.

For the terms involving scalar perturbations, the relevant results are 
\begin{subequations}
\setlength{\jot}{1pt}
    \begin{align}
        \begin{split} \label{Ic1RDfinal}
        &I^{\Psi^{(2)}}_{c1}(\barx ,v,u) = \frac{1}{120 u^3 v^2 \barx^6} \bigg\{  
        - v \times a(v, u) \barx^6 \bigg[ \text{Ci} \left( \barx \left|1 + \frac{u - v}{\sqrt{3}} \right| \right)  
        - \log \left| \sqrt{3} + u - v \right| \bigg] \\  
        &+ v \times b(v, u) \barx^6 \bigg[ \text{Ci} \left( \barx \left|1 + \frac{-u + v}{\sqrt{3}} \right| \right)  
        - \log \left| \sqrt{3} - u + v \right| \bigg] \\  
        &+ v \times c(v, u) \barx^6 \bigg[ \text{Ci} \left( \barx \left|-1 + \frac{u + v}{\sqrt{3}} \right| \right)  
        - \log \left| -\sqrt{3} + u + v \right| \bigg] \\  
        &+ v \times d(v, u) \barx^6 \bigg[ \text{Ci} \left( \barx \left|1 + \frac{u + v}{\sqrt{3}} \right| \right)  
        - \log \left| \sqrt{3} + u + v \right| \bigg] \\  
        &+ 36 \sin \left( \frac{u \barx}{\sqrt{3}} \right) \bigg\{  
        - 3 \sqrt{3} \cos \left( \frac{v \barx}{\sqrt{3}} \right) 
        \bigg[ v^2 \barx^3 \bigg( -6 + \left( -\frac{7}{3} + u^2 + \frac{23 v^2}{9} \right) \barx^2 \bigg) \cos \barx \\  
        &+ 4 \bigg( -90 + (-15 + 15 u^2 - 11 v^2) \barx^2 + v^2 \left( -\frac{7}{6} + u^2 - \frac{2 v^2}{9} \right) \barx^4 \bigg) \sin \barx \bigg] \\  
        &- \frac{1}{3} v \barx \bigg[ 3 \barx \left( 54 + (21 - 9 u^2 + 31 v^2) \barx^2 \right) \cos \barx + ( 648 + 6 (21 - 18 u^2 + 22 v^2) \barx^2 \\  
        & + (-63 - 12 u^4 - 51 v^2 + 8 v^4 + u^2 (69 + 8 v^2)) \barx^4 ) \sin \barx \bigg] \sin \left( \frac{v \barx}{\sqrt{3}} \right)  
        \bigg\} \\  
        &+ 12 \barx \bigg[ u \bigg( 12 u^2 (-5 + u^2) - (21 + 17 u^2) v^2 + 23 v^4 \bigg) \barx^5 + u \cos \left( \frac{u \barx}{\sqrt{3}} \right) \\  
        &\times \bigg\{  2 \cos \left( \frac{v \barx}{\sqrt{3}} \right) \bigg[ -1620 \sin \barx  + \barx^2 \bigg( 90 (-3 + u^2) \sin \barx + v^2 \bigg( -27 \barx \cos \barx  \\  
        &+ (-198 + (-21 + 6 u^2 - 4 v^2) \barx^2) \sin \barx \bigg) \bigg) \bigg]+ \sqrt{3} v \barx \bigg[ \barx (54 + (21 - 3 u^2 + 31 v^2) \barx^2)  \\
        &\times \cos \barx + 2 (108 + (21 - 6 u^2 + 22 v^2) \barx^2) \sin \barx \bigg] \sin \left( \frac{v \barx}{\sqrt{3}} \right)  
        \bigg\}  
        \bigg]  
        \bigg\} ,
        \end{split}
        \\
        \begin{split} \label{Is1RDfinal}
        &I^{\Psi^{(2)}}_{s1}(\barx ,v,u) =\frac{1}{120 u^3 v^2 \barx^6} \bigg\{  
        36 \sin \left( \frac{u \barx}{\sqrt{3}} \right) \bigg[  
        \frac{1}{3} v \barx \cos \left( \frac{v \barx}{\sqrt{3}} \right)  
        \bigg( 3 \barx (54 + (21 - 9 u^2 + 31 v^2) \barx^2) \\  
        &\times \cos \barx+ (648 + 6 (21 - 18 u^2 + 22 v^2) \barx^2  
        + (-63 - 12 u^4 - 51 v^2 + 8 v^4 + u^2 (69 + 8 v^2)) \\  
        &\times  \barx^4) \sin \barx  
        \bigg) - 3 \sqrt{3} \bigg(  
        v^2 \barx^3 (-6 + (-\frac{7}{3} + u^2 + \frac{23 v^2}{9}) \barx^2) \cos \barx + 4 (-90 + (-15 + 15 u^2 \\  
        &- 11 v^2) \barx^2 + v^2 (-\frac{7}{6} + u^2 - \frac{2 v^2}{9}) \barx^4) \sin \barx  
        \bigg) \sin \left( \frac{v \barx}{\sqrt{3}} \right)  
        \bigg] \\  
        &+ 12 u \barx \cos \left( \frac{u \barx}{\sqrt{3}} \right) \bigg[  
        - \sqrt{3} v \barx \cos \left( \frac{v \barx}{\sqrt{3}} \right)  
        \bigg( \barx (54 + (21 - 3 u^2 + 31 v^2) \barx^2) \cos \barx \\  
        &+ 2 (108 + (21 - 6 u^2 + 22 v^2) \barx^2) \sin \barx  
        \bigg) -2 \bigg( 1620 \sin \barx  
        + \barx^2 \bigg( -90 (-3 + u^2) \sin \barx \\  
        &+ v^2 (27 \barx \cos \barx  
        + (198 + (21 - 6 u^2 + 4 v^2) \barx^2) \sin \barx)  
        \bigg) \bigg) \sin \left( \frac{v \barx}{\sqrt{3}} \right)  
        \bigg] \\  
        &- v \times b(v, u) \barx^6 \text{Si} \left(-\barx + \frac{(u - v) \barx}{\sqrt{3}} \right)  
        + v \times c(v, u) \barx^6 \text{Si} \left(-\barx + \frac{(u + v) \barx}{\sqrt{3}} \right) \\  
        &+ v \times a(v, u) \barx^6 \text{Si} \left(\barx + \frac{(u - v) \barx}{\sqrt{3}} \right)  
        + v \times d(v, u) \barx^6 \text{Si} \left(\barx + \frac{(u + v) \barx}{\sqrt{3}} \right)  
        \bigg\}  ,
        \end{split}
    \end{align}
\end{subequations}
with
\begin{subequations}
\setlength{\jot}{1pt}
    \begin{align}
        \begin{split}
            a( v, u) &= -90 u^2 \left(v^2+3\right)+v^2 \left(v \left(v \left(8 \sqrt{3} v+45\right)-120 \sqrt{3}\right)+90\right)+189 \\
            &+12 \sqrt{3} u^5+ 45 u^4-20 \sqrt{3} u^3 \left(v^2+3\right) \, ,
        \end{split}
        \\
        \begin{split}
            b(v, u) &= 90 u^2 \left(v^2+3\right)+v^2 \left(v \left(v \left(8 \sqrt{3} v-45\right)-120 \sqrt{3}\right)-90\right)-189 \\
            &+12 \sqrt{3} u^5-45 u^4-20 \sqrt{3} u^3 \left(v^2+3\right) \, ,
        \end{split}
        \\
        \begin{split}
            c( v, u) &=-90 u^2 \left(v^2+3\right)+v^2 \left(v \left(v \left(8 \sqrt{3} v+45\right)-120 \sqrt{3}\right)+90\right)+189 \\ 
            & -12 \sqrt{3} u^5+45 u^4+20 \sqrt{3} u^3 \left(v^2+3\right) \, ,
        \end{split}
        \\
        \begin{split}
            d( v, u) &= -90 u^2 \left(v^2+3\right)+v^2 \left(v \left(v \left(45-8 \sqrt{3} v\right)+120 \sqrt{3}\right)+90\right)+189 \\
            &+12 \sqrt{3} u^5+45 u^4-20 \sqrt{3} u^3 \left(v^2+3\right)\, .
        \end{split}
    \end{align}
\end{subequations}
Furthermore,
\begin{subequations}
    \begin{align}
    \begin{split} \label{Ic2RDfinal}
    I^{\Psi^{(2)}}_{c2}(\barx ,v,u) &= \frac{3}{10 u^3 v \barx^5} \bigg\{  
    \cos \left( \frac{v \barx}{\sqrt{3}} \right) \bigg[  
    3 u \barx \cos \left( \frac{u \barx}{\sqrt{3}} \right)  
    \bigg( \barx (-18 + (-7 + u^2 - \frac{31 v^2}{3}) \barx^2)  \\  
    &\times \cos \barx+ 2 (-216 + (-37 + 12 u^2 - \frac{82 v^2}{3}) \barx^2) \sin \barx  
    \bigg) + \frac{1}{\sqrt{3}} \bigg(  
    3 \barx (54 + (21 - 9 u^2  \\  
    &+ 31 v^2) \barx^2) \cos \barx+ (3888 + 6 (111 - 108 u^2 + 82 v^2) \barx^2  
    - (63 + 12 u^4 \\
    &- 39 v^2 + 22 v^4 + u^2 (-69 + 22 v^2)) \barx^4) \sin \barx  
    \bigg) \sin \left( \frac{u \barx}{\sqrt{3}} \right)  
    \bigg]- 6 v \barx  \\  
    &\times \bigg[  
    \sqrt{3} u \barx \cos \left( \frac{u \barx}{\sqrt{3}} \right) 
    \bigg( -7 \barx \cos \barx + (-18 + (-\frac{8}{3} + u^2 - \frac{41 v^2}{9}) \barx^2) \sin \barx  
    \bigg)  \\  
    &+ \frac{1}{6} \bigg( \barx (126 + (69 - 21 u^2 + 113 v^2) \barx^2) \cos \barx+ 2 (162 + (24 - 27 u^2  \\  
    &+ 41 v^2) \barx^2) \sin \barx  
    \bigg) \sin \left( \frac{u \barx}{\sqrt{3}} \right)  
    \bigg]  \sin \left( \frac{v \barx}{\sqrt{3}} \right)+ e(v, u) \barx^5 \text{Si} \left(-\barx + \frac{(u - v) \barx}{\sqrt{3}} \right)  
     \\  
    &- g(v, u) \barx^5 \text{Si} \left(-\barx + \frac{(u + v) \barx}{\sqrt{3}} \right)  + f(v, u) \barx^5 \text{Si} \left(\barx + \frac{(u - v) \barx}{\sqrt{3}} \right)  
    \\
    &+ h(v, u) \barx^5 \text{Si} \left(\barx + \frac{(u + v) \barx}{\sqrt{3}} \right)  
    \bigg\}  ,
    \end{split}
    \\
    \begin{split} \label{Is2RDfinal}
    I^{\Psi^{(2)}}_{s2}(\barx ,v,u) &=\frac{3}{10 u^3 v \barx^5} \bigg\{  
        e(v, u) \barx^5 \bigg[ \text{Ci} \left( \barx \left|-1 + \frac{u - v}{\sqrt{3}} \right| \right)  
        - \log \left| -\sqrt{3} + u - v \right| \bigg] \\  
        &+ f(v, u) \barx^5 \bigg[ \text{Ci} \left( \barx \left|1 + \frac{u - v}{\sqrt{3}} \right| \right)  
        - \log \left| \sqrt{3} + u - v \right| \bigg] \\  
        &+ g(v, u) \barx^5 \bigg[ \text{Ci} \left( \barx \left|-1 + \frac{u + v}{\sqrt{3}} \right| \right)  
        - \log \left| -\sqrt{3} + u + v \right| \bigg] \\  
        &- h(v, u) \barx^5 \bigg[ \text{Ci} \left( \barx \left|1 + \frac{u + v}{\sqrt{3}} \right| \right)  
        - \log \left| \sqrt{3} + u + v \right| \bigg] \\  
        &+ \frac{1}{3} \bigg[ \sqrt{3} u v (-69 + 7 u^2 - 113 v^2) \barx^5 + u \barx \cos \left( \frac{u \barx}{\sqrt{3}} \right) \bigg\{  
        -2 \sqrt{3} v \barx \cos \left( \frac{v \barx}{\sqrt{3}} \right)   \\  
        &\times \bigg[ 63 \barx \cos \barx + \big( 162 + (24 - 9 u^2 + 41 v^2) \barx^2 \big) \sin \barx \bigg] -3 \bigg[ \barx (54 + (21 - 3 u^2 \\  
        &  + 31 v^2) \barx^2) \cos \barx + 2 (648 + (111 - 36 u^2 + 82 v^2) \barx^2) \sin \barx \bigg]  
        \sin \left( \frac{v \barx}{\sqrt{3}} \right) \bigg\} \\  
        &+ \sin \left( \frac{u \barx}{\sqrt{3}} \right) \bigg\{  
        3 v \barx \cos \left( \frac{v \barx}{\sqrt{3}} \right)  
        \bigg[ \barx (126 + (69 - 21 u^2 + 113 v^2) \barx^2) \cos \barx  
        + 2 (162\\  
        & + (24 - 27 u^2 + 41 v^2) \barx^2) \sin \barx \bigg] + \sqrt{3} \bigg[ 3 \barx (54 + (21 - 9 u^2 + 31 v^2) \barx^2) \cos \barx \\  
        &+ (3888 + 6 (111 - 108 u^2 + 82 v^2) \barx^2  
        - (63 + 12 u^4 - 39 v^2 + 22 v^4 \\
        &+ u^2 (-69 + 22 v^2)) \barx^4) \sin \barx \bigg]  
        \sin \left( \frac{v \barx}{\sqrt{3}} \right) \bigg\}  
        \bigg]  
    \bigg\}  ,
    \end{split}
    \end{align}
\end{subequations}
with 
\begin{subequations}
    \begin{align}
        \begin{split}
            e( v, u) &=\frac{5}{2} \left(u^2-3\right)^2 v+\frac{1}{4} \sqrt{3} \left(5 \left(u^4-2 u^2 \left(v^2+3\right)+9 v^4\right)+10 v^2+21\right)+\frac{11 v^5}{6} \\
            & -u^5+\frac{5 u^3 v^2}{3}+5 u^3-5 \left(u^2-5\right) v^3\, ,
        \end{split}
        \\
        \begin{split}
            f( v, u) &=  
            -\frac{5}{2} \left(u^2-3\right)^2 v+\frac{1}{4} \sqrt{3} \left(5 \left(u^4-2 u^2 \left(v^2+3\right)+9 v^4\right)+10 v^2+21\right)-\frac{11 v^5}{6} \\
            &+u^5-\frac{5 u^3 v^2}{3}-5 u^3+5 \left(u^2-5\right) v^3\, ,
        \end{split}
        \\
        \begin{split}
            g( v, u) &=  
            \frac{5}{2} \left(u^2-3\right)^2 v-\frac{1}{4} \sqrt{3} \left(5 \left(u^4-2 u^2 \left(v^2+3\right)+9 v^4\right)+10 v^2+21\right)+\frac{11 v^5}{6}\\
            &+u^5-\frac{5 u^3 v^2}{3}-5 u^3-5 \left(u^2-5\right) v^3\, ,
        \end{split}
        \\
        \begin{split}
            h( v, u) &=  
            \frac{5}{2} \left(u^2-3\right)^2 v+\frac{1}{4} \sqrt{3} \left(5 \left(u^4-2 u^2 \left(v^2+3\right)+9 v^4\right)+10 v^2+21\right)+\frac{11 v^5}{6} \\
            &+u^5-\frac{5 u^3 v^2}{3}-5 u^3-5 \left(u^2-5\right) v^3\, .
        \end{split}
    \end{align}
\end{subequations}

For the second-order vector nested kernels 
\begin{subequations}
    \begin{align} \label{RDnestedb2psi11}
        I^{B^{(2)}}_{st1}(\barx ,k,v,u) &=\frac{18 \sqrt{3}}{k^3 u^3 \overline{x}^3} 
        \bigg\{
        2 \sin \overline{x} \sin \left(\frac{u \overline{x}}{\sqrt{3}}\right)  
        + \overline{x} \text{Si} \left( \overline{x} - \frac{u \overline{x}}{\sqrt{3}} \right)  
        - \overline{x} \text{Si} \left( \overline{x} + \frac{u \overline{x}}{\sqrt{3}} \right)  
        \bigg\} 
        \\
        \begin{split} \label{RDnestedb2psi12}
            I^{B^{(2)}}_{st2}(\barx ,k,v,u) &= \frac{3}{ku^3 \overline{x}^5} \bigg\{  
            6 u \overline{x} \cos \left(\frac{u \overline{x}}{\sqrt{3}}\right)  
            \left(\overline{x} \cos \overline{x} + 2 \sin \overline{x} \right) \\  
            &+ 2 \sqrt{3} \bigg( -3 \overline{x} \cos \overline{x}  
            + \left(-6 + ( u^2 - 3 v^2) \overline{x}^2 \right) \sin \overline{x} \bigg)  
            \sin \left(\frac{u \overline{x}}{\sqrt{3}}\right) \\  
            &+ 3\sqrt{3} v^2 
            \overline{x}^3 \bigg[  
            \text{Si} \left(- \overline{x} + \frac{u \overline{x}}{\sqrt{3}} \right)  
            + \text{Si} \left( \overline{x} + \frac{u \overline{x}}{\sqrt{3}} \right)  
            \bigg]  
            \bigg\} .
        \end{split}
    \end{align}
\end{subequations}

And finally, for the second-order tensors
\begin{align}\label{RDnestedh2psi1}
    \begin{split}
        I^{h^{(2)}}_{st}(\barx ,k,v,u) &=\frac{1}{16 u^3 v \overline{x}^5} \bigg \{ 4 \overline{x} 
        \Big( 
        18 u v \overline{x} \cos\left(\frac{u \overline{x}}{\sqrt{3}}\right) \left(\overline{x} \cos \overline{x} + \sin \overline{x} \right) 
        + 3 \sqrt{3} v \big( -6 \overline{x} \cos \overline{x} \\
        &+ \left(-6 + (3 + u^2 - 3 v^2) \overline{x}^2\right) \sin \overline{x} \big ) 
        \sin\left(\frac{u \overline{x}}{\sqrt{3}}\right) 
        + u (-9 + u^2 + 9 v^2) \overline{x}^3 \sin(v \overline{x}) 
        \Big) \\ 
        &+ \sqrt{3} (u^2 - 3(-1 + v)^2) (u^2 - 3(1 + v)^2) \overline{x}^4 
        \bigg ( \sin(v \overline{x})
        \Big[-\text{Ci} \left( \overline{x} \left|1 + \frac{u}{\sqrt{3}} - v \right| \right) \\ 
        &+ \text{Ci} \left( \overline{x} \left|1 - \frac{u}{\sqrt{3}} + v \right| \right) 
        + \text{Ci} \left( \overline{x} \left|-1 + \frac{u}{\sqrt{3}} + v \right| \right) 
        - \text{Ci} \left( \overline{x} \left|1 + \frac{u}{\sqrt{3}} + v \right| \right) \\
        &+ \log \left| \frac{(\sqrt{3} + u)^2 - 3 v^2}{(\sqrt{3} - u)^2 - 3 v^2} \right| \Big ]  
        - \cos(v \overline{x}) \bigg [ 
        \text{Si} \left( (1 - \frac{u}{\sqrt{3}} + v) \overline{x} \right) \\
        &+ \text{Si} \left( (-1 + \frac{u}{\sqrt{3}} + v) \overline{x} \right) 
        - \text{Si} \left( (1 + \frac{u}{\sqrt{3}} + v) \overline{x} \right) 
        + \text{Si} \left( \overline{x} + \frac{u \overline{x}}{\sqrt{3}} - v \overline{x} \right) 
        \bigg ] 
        \bigg) 
        \bigg \} .
    \end{split}
\end{align}

%%%%%%%%%%%%%%%%%%%%%%%%%%%%%%%%%%%%%%%%%%%%%%
%%%%%%%%%%%%%%%%%%%%%%%%%%%%%%%%%%%%%%%%%%%%%%
\section{Mesh plots of the integrands for the numerical contributions} \label{meshstuff} \vspace{-3cm}

\begin{figure}
    \centering
    \begin{subfigure}[b]{0.48\textwidth}
        \includegraphics[width=\linewidth]{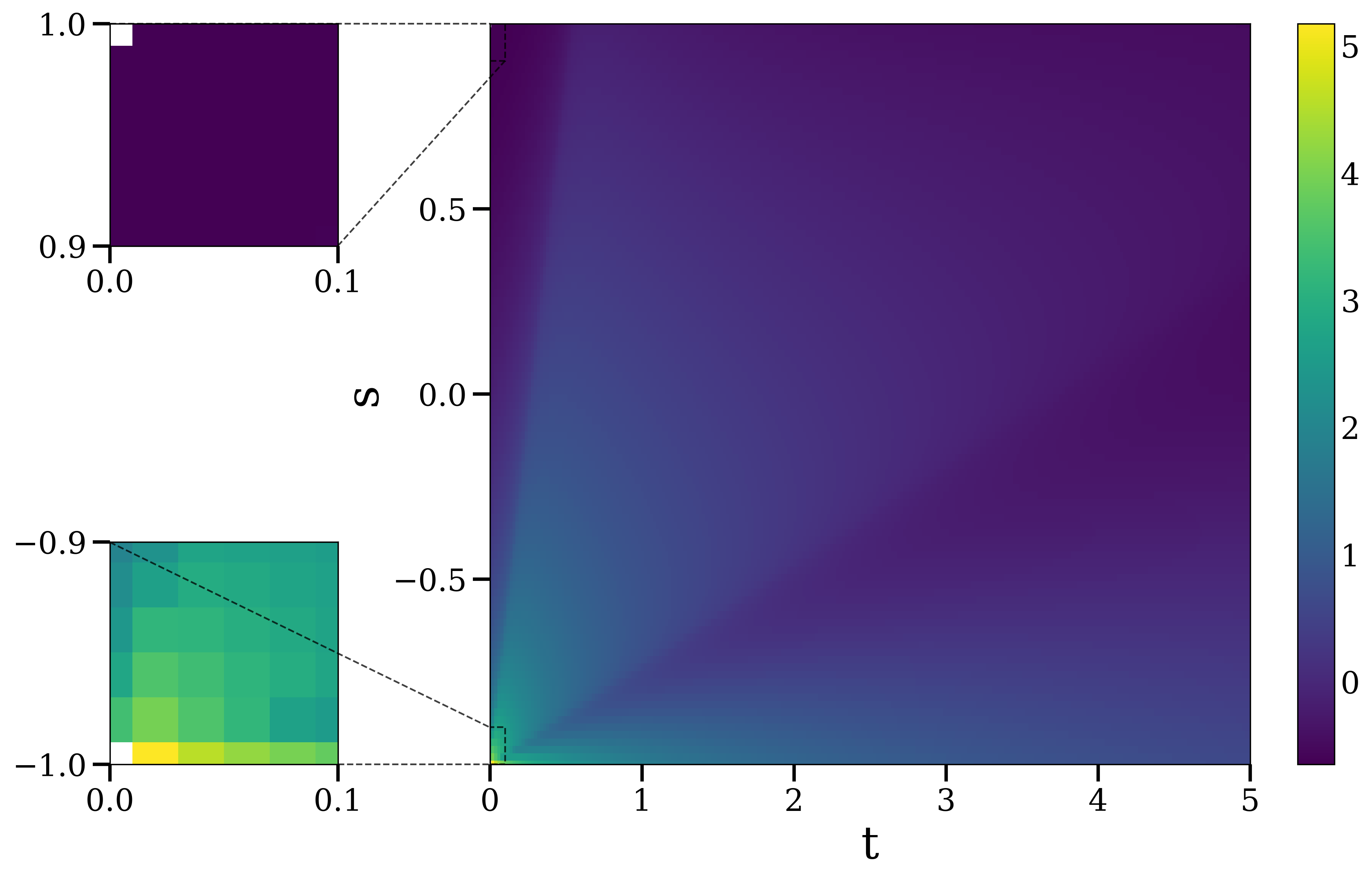}
    \end{subfigure}
    \begin{subfigure}[b]{0.48\textwidth}
        \includegraphics[width=\linewidth]{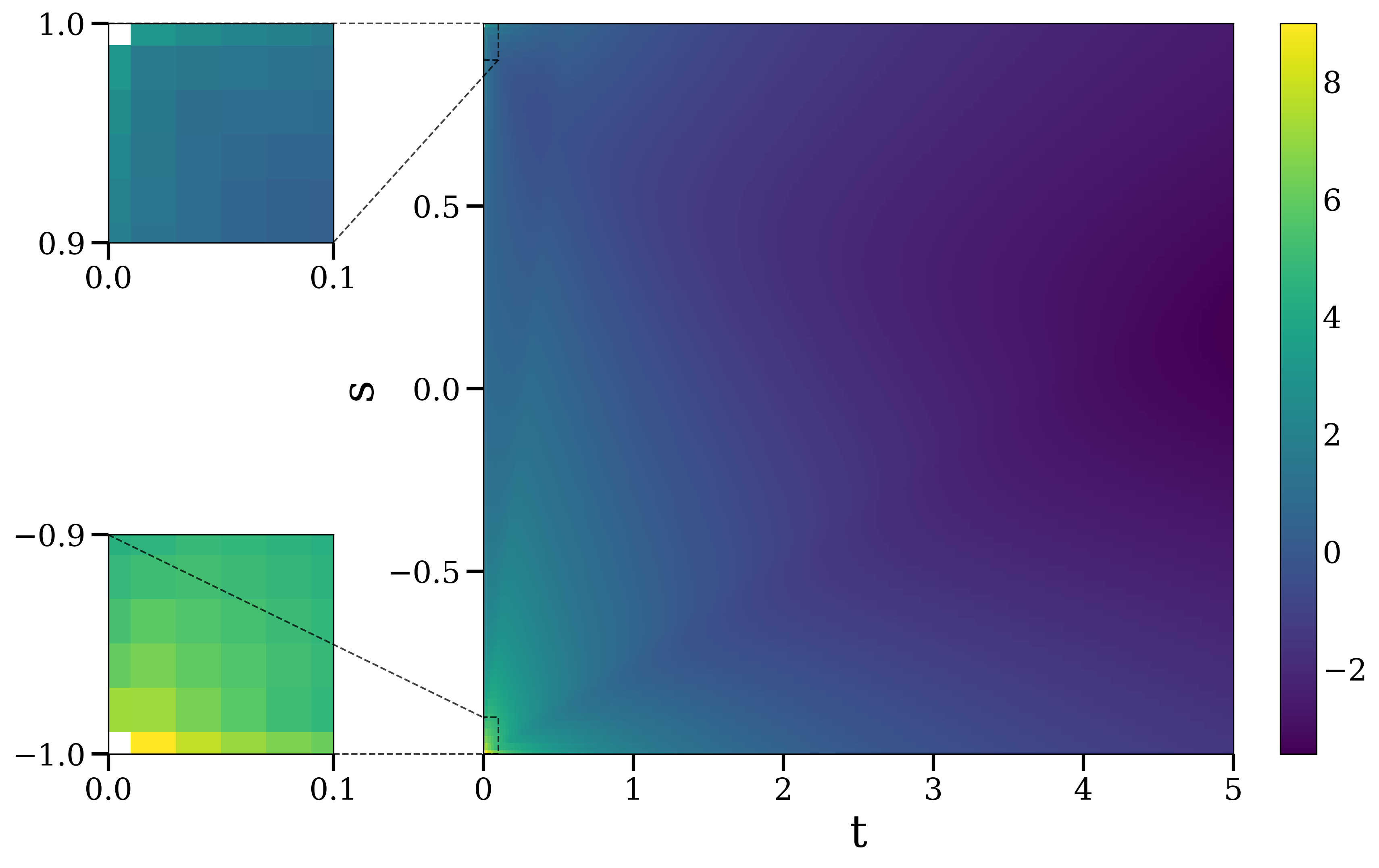}
    \end{subfigure}
    
    \begin{subfigure}[b]{0.48\textwidth}
        \includegraphics[width=\linewidth]{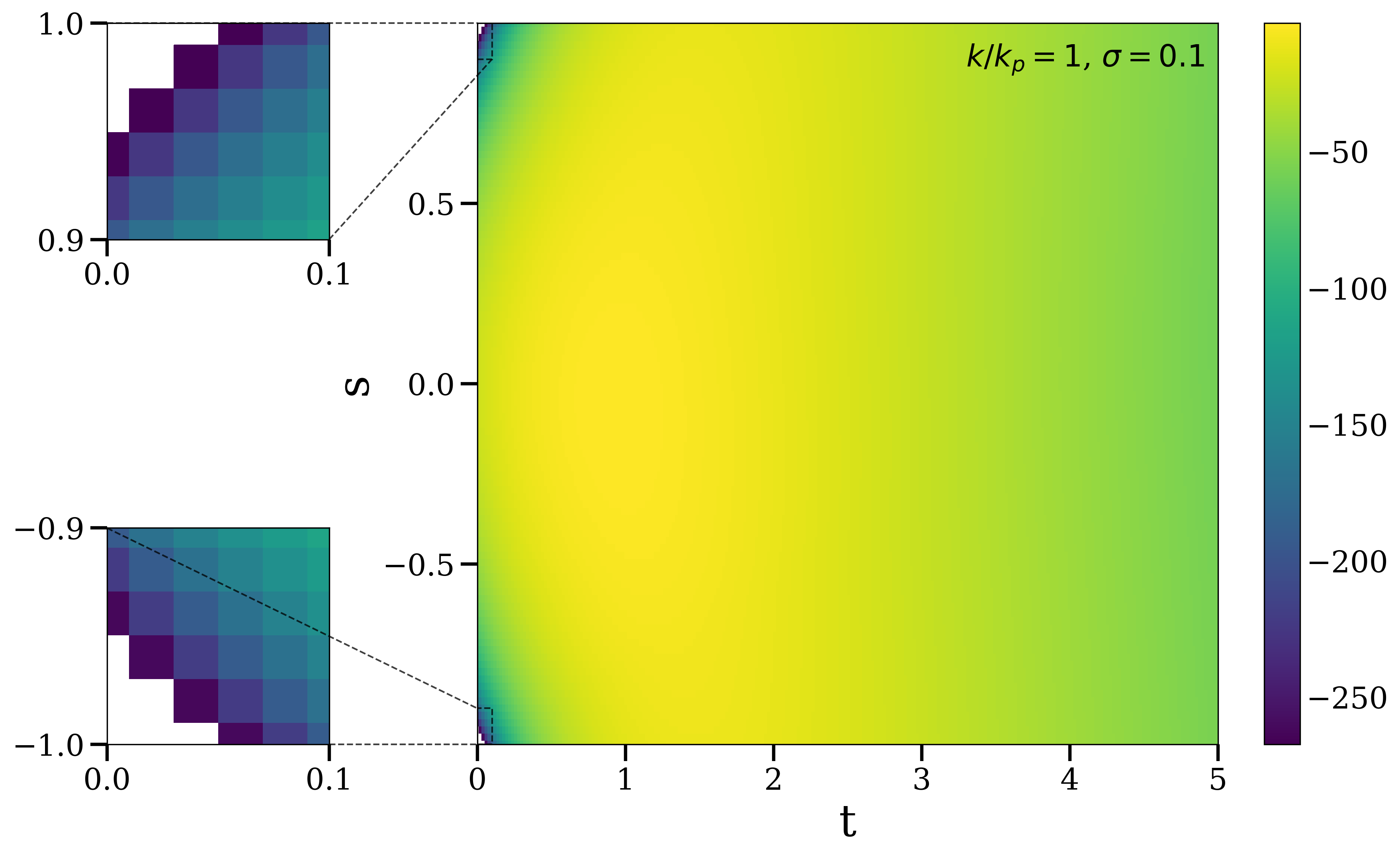}
    \end{subfigure}
    \begin{subfigure}[b]{0.48\textwidth}
        \includegraphics[width=\linewidth]{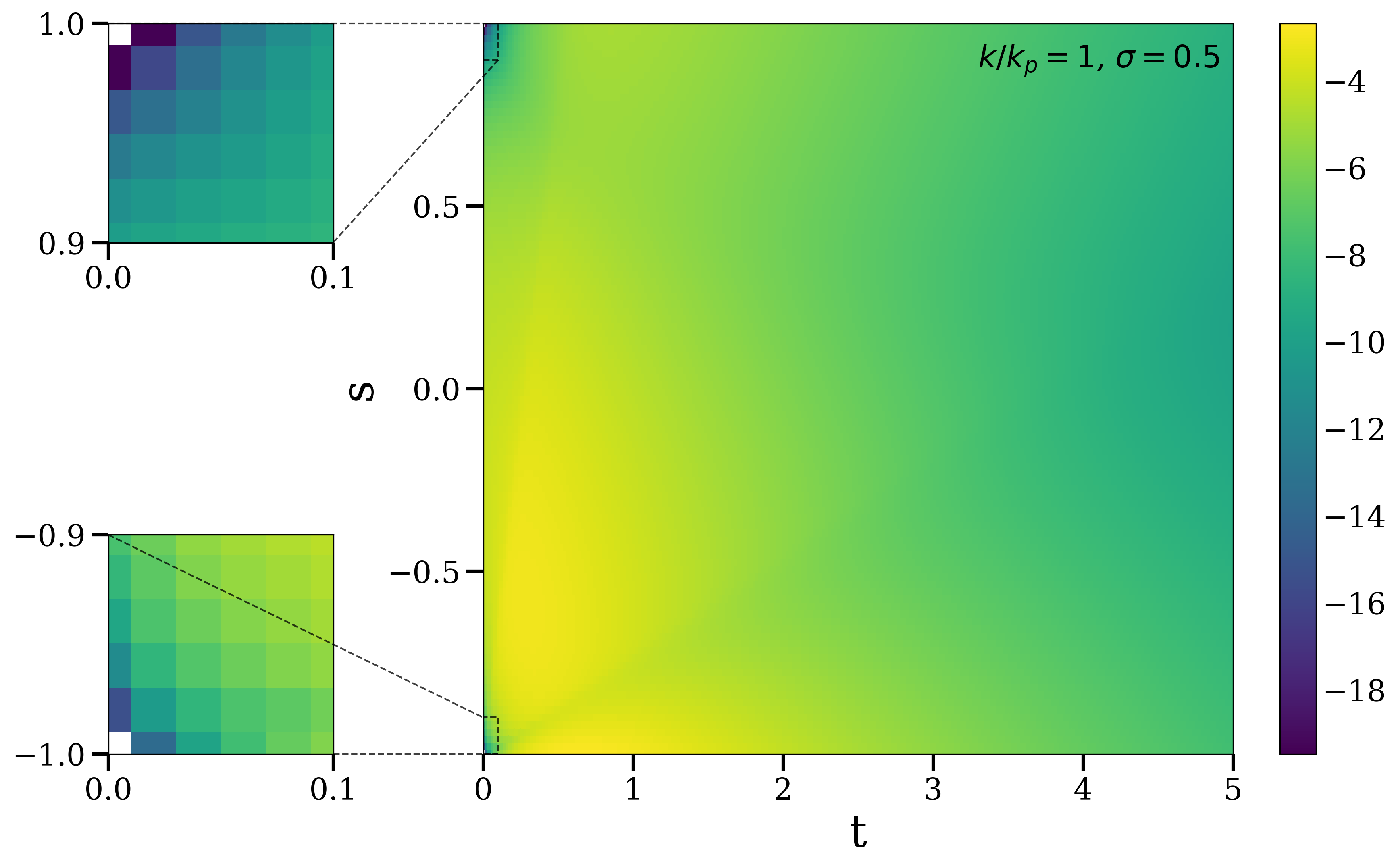}
    \end{subfigure}
    
    \begin{subfigure}[b]{0.48\textwidth}
        \includegraphics[width=\linewidth]{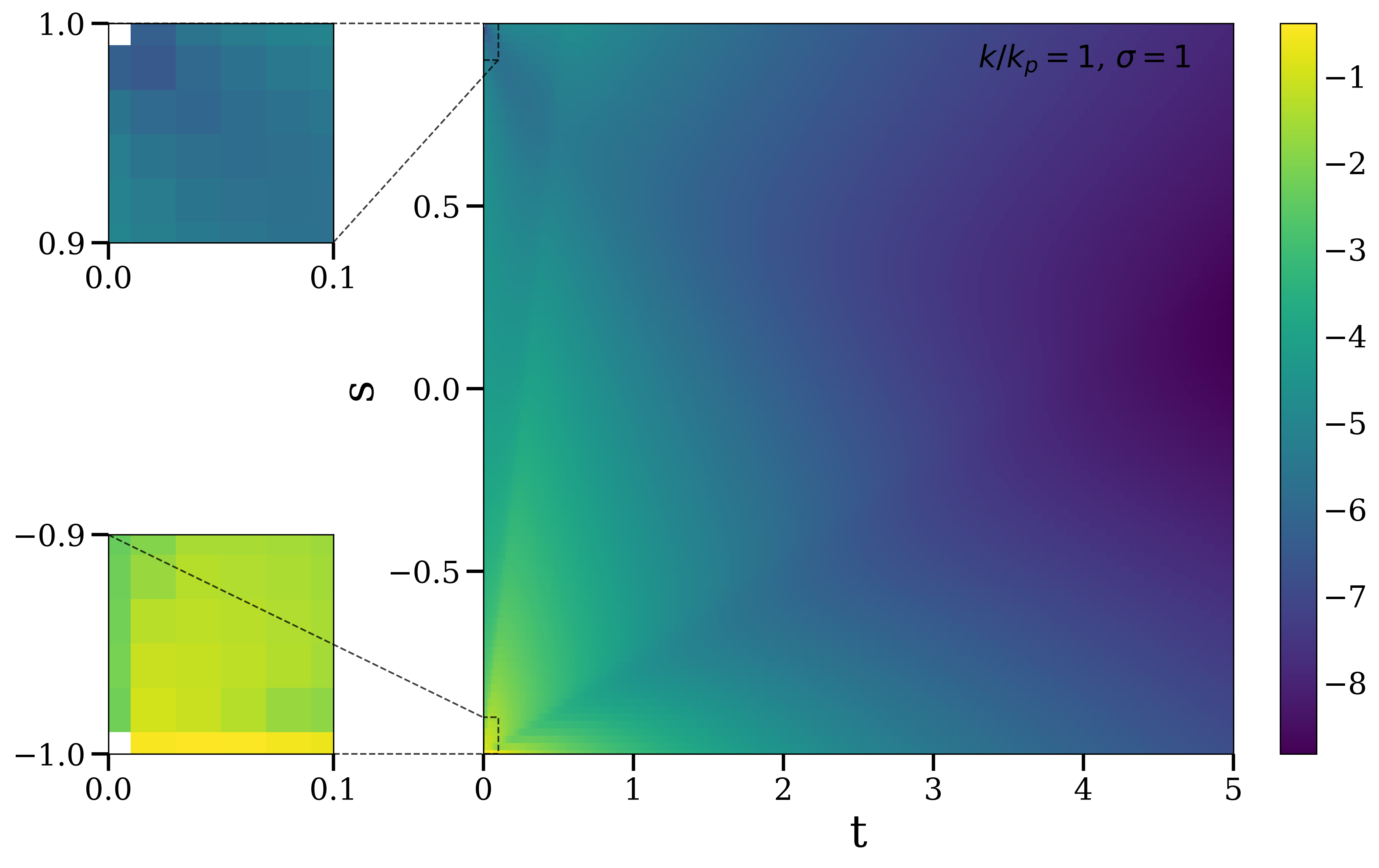}
    \end{subfigure}

    \caption{\footnotesize{Logarithmic mesh plots of the integrand of the power spectrum for the second-order scalar-tensor contribution showed in Eq.~\eqref{omegaSTRD} for different values of $\sigma$ in Eq.~\eqref{lognorm_def}. We note here that we do not show the integrand region beyond $t=5$ since we have established that the integrand is IR safe. (\textit{Top left}) Plot of the scalar-tensor kernel in an RD universe. One of the UV limits blows up. (\textit{Top right}) Plot of the integrand without the contribution of the power spectrum. We see the plot blows up in the top-left corner ($v^{-2}$ divergence) and even quicker in the bottom-left corner ($u^{-4}$ divergence). (\textit{Middle left}) Plot of the integrand this time, including the primordial scalar and tensor spectra, for $k=k_p$. The inclusion of primordial power spectra has masked the divergence. (\textit{Middle right}) Same plot as previous but with $\sigma =0.5$. The overall numerical values of the integrand increase compared to $\sigma = 0.1$ (\textit{Bottom row}). Same plot as previous but with $\sigma =1$. The divergence is no longer masked by the scalar power spectrum.}}
    \label{fig:st_meshplots}
\end{figure}

\newpage

\begin{figure}
    \centering
    \begin{subfigure}[b]{0.48\textwidth}
        \includegraphics[width=\linewidth]{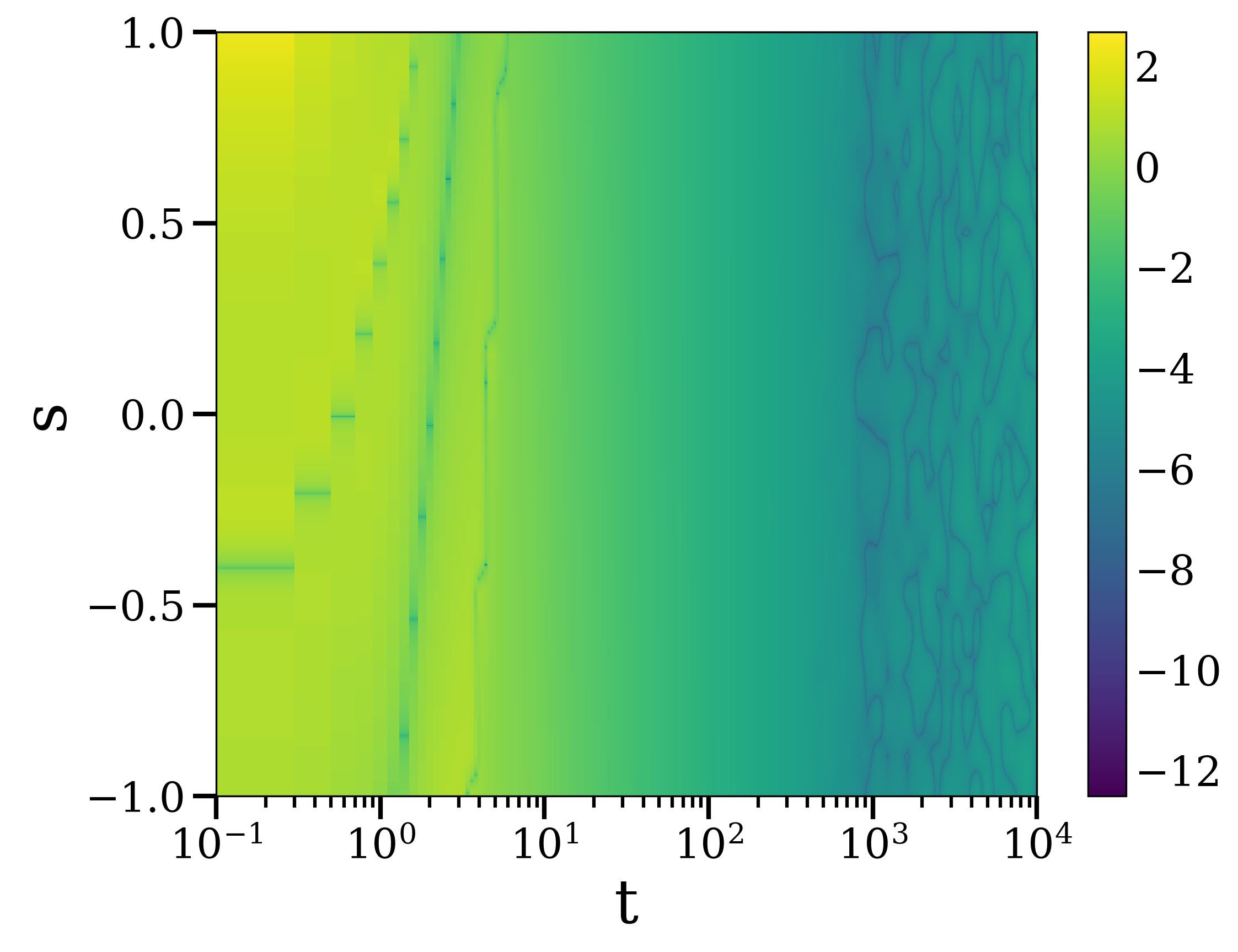}
    \end{subfigure}
    \begin{subfigure}[b]{0.48\textwidth}
        \includegraphics[width=\linewidth]{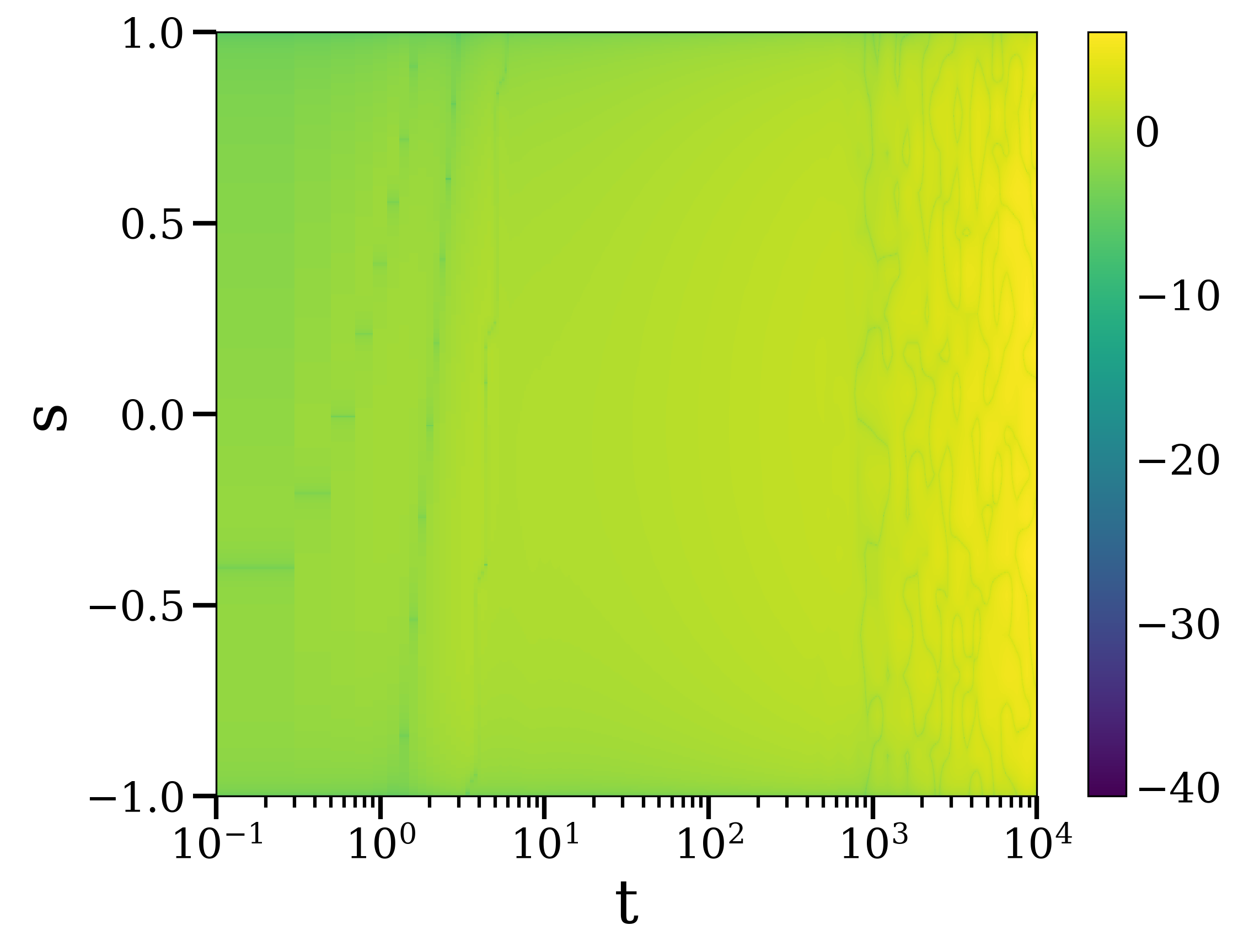}
    \end{subfigure}
    
    \begin{subfigure}[b]{0.48\textwidth}
        \includegraphics[width=\linewidth]{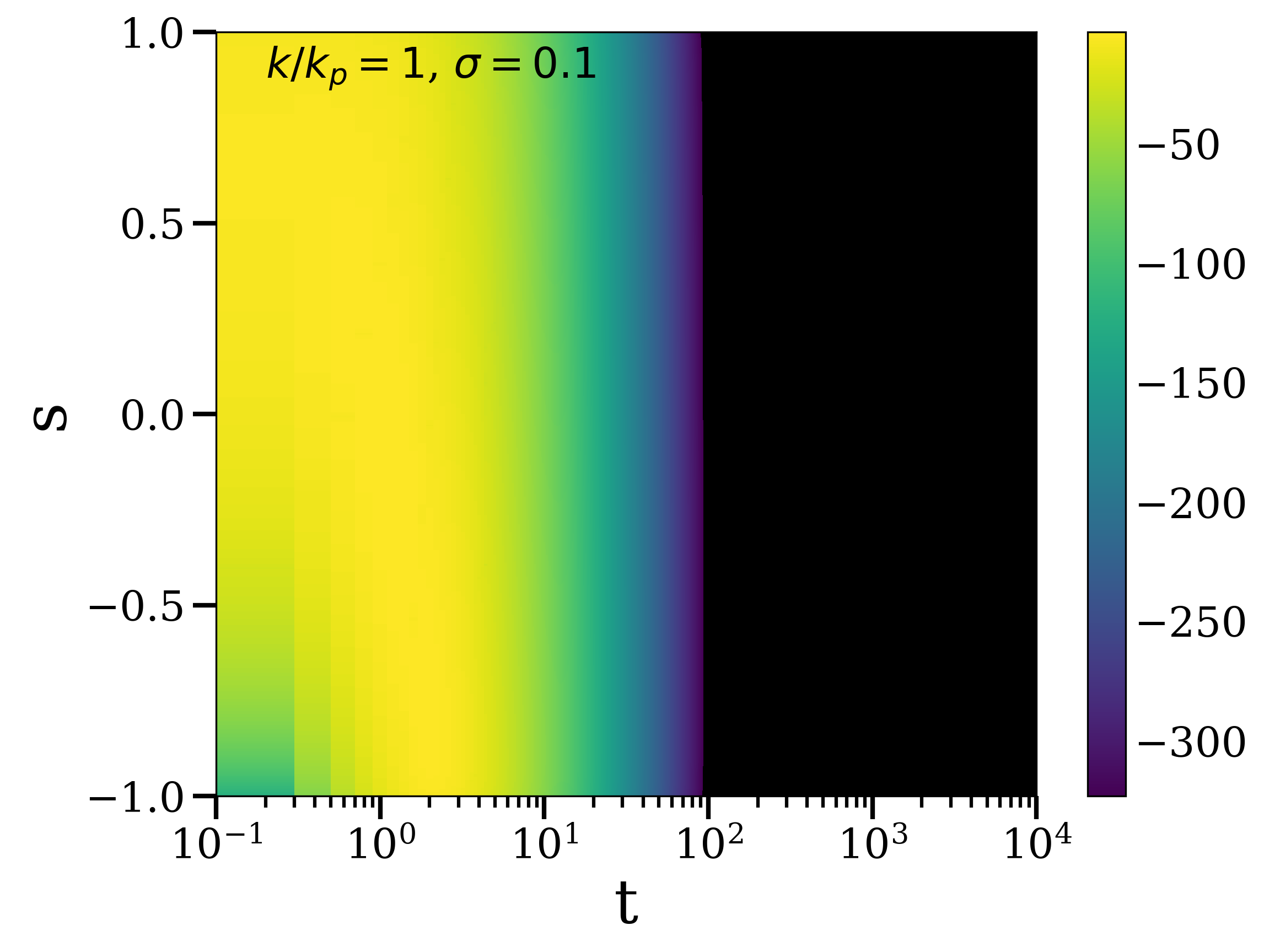}
    \end{subfigure}
    \begin{subfigure}[b]{0.48\textwidth}
        \includegraphics[width=\linewidth]{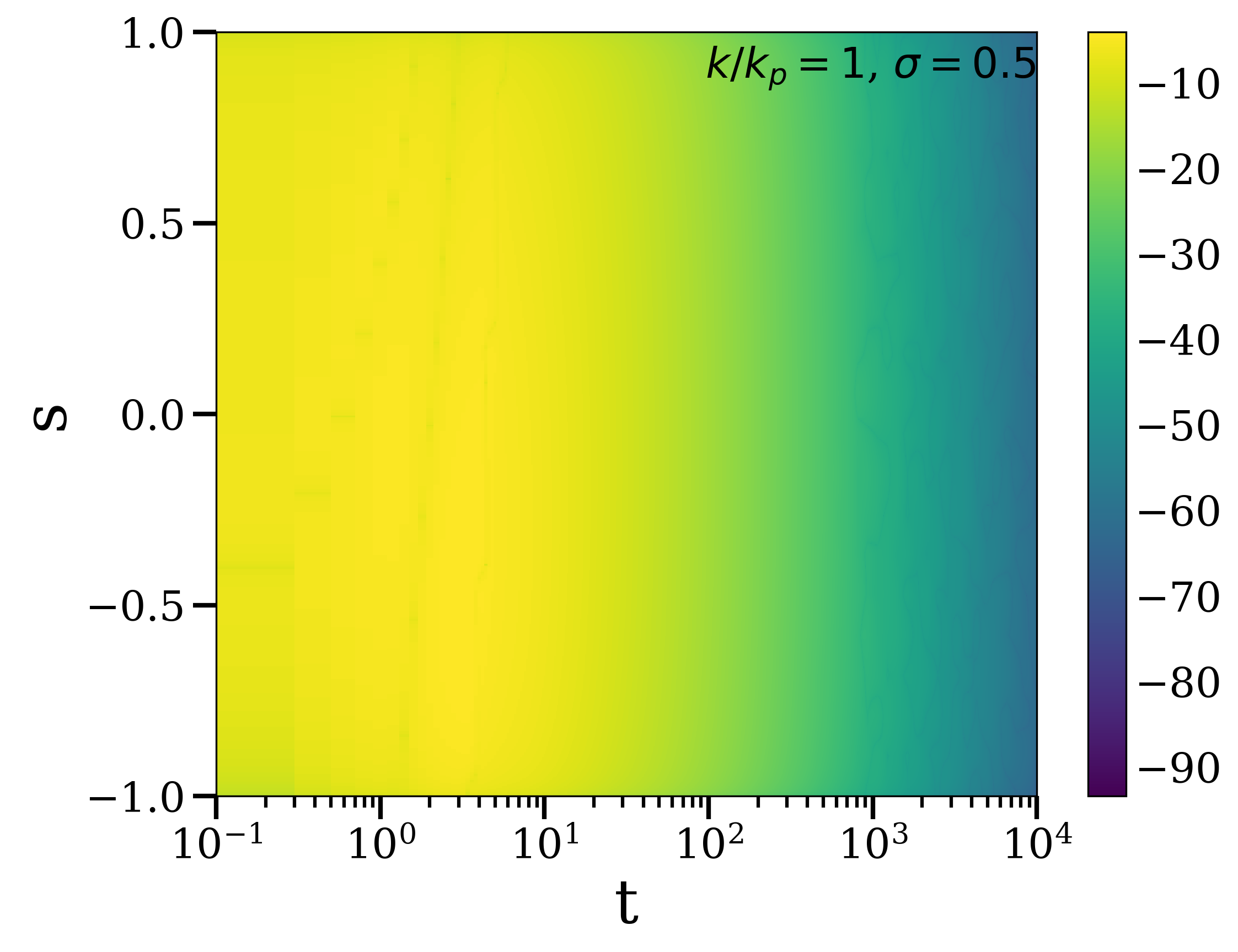}
    \end{subfigure}
    
    \begin{subfigure}[b]{0.48\textwidth}
        \includegraphics[width=\linewidth]{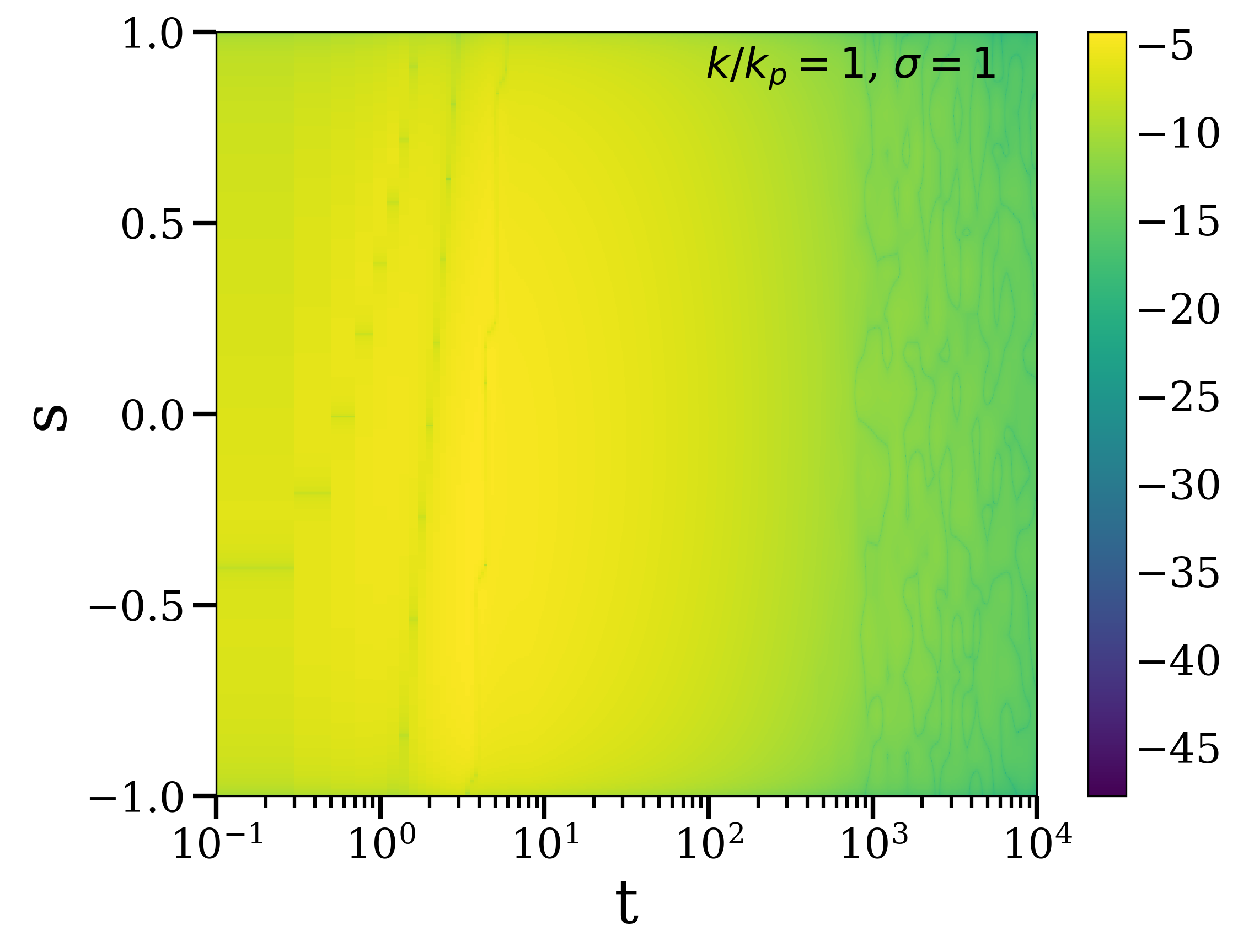}
    \end{subfigure}

    \caption{\footnotesize{Coloured mesh plots for the source term containing second-order scalars. (\textit{Top left}) Numerical kernel shown in Eq.~\eqref{finalkernelpsi2}  (\textit{Top right}). Numerical part of the integrand in Eq.~\eqref{pspsi2psi1} without the power spectra. (\textit{Middle left}) Integrand with the power spectra, with $\sigma = 0.1$ (\textit{Middle right}) $\sigma =0.5$ (\textit{Bottom row}) $\sigma =1$. Although the kernel exhibits an enhancement in the limit $v \rightarrow 1$ and $u \rightarrow 0$, we cannot conclude that there is a UV divergence as $\sigma$ increases.}}
    \label{fig:psi2psi1um_meshplots}
\end{figure}

\newpage

\begin{figure}
    \centering
    \begin{subfigure}[b]{0.48\textwidth}
        \includegraphics[width=\linewidth]{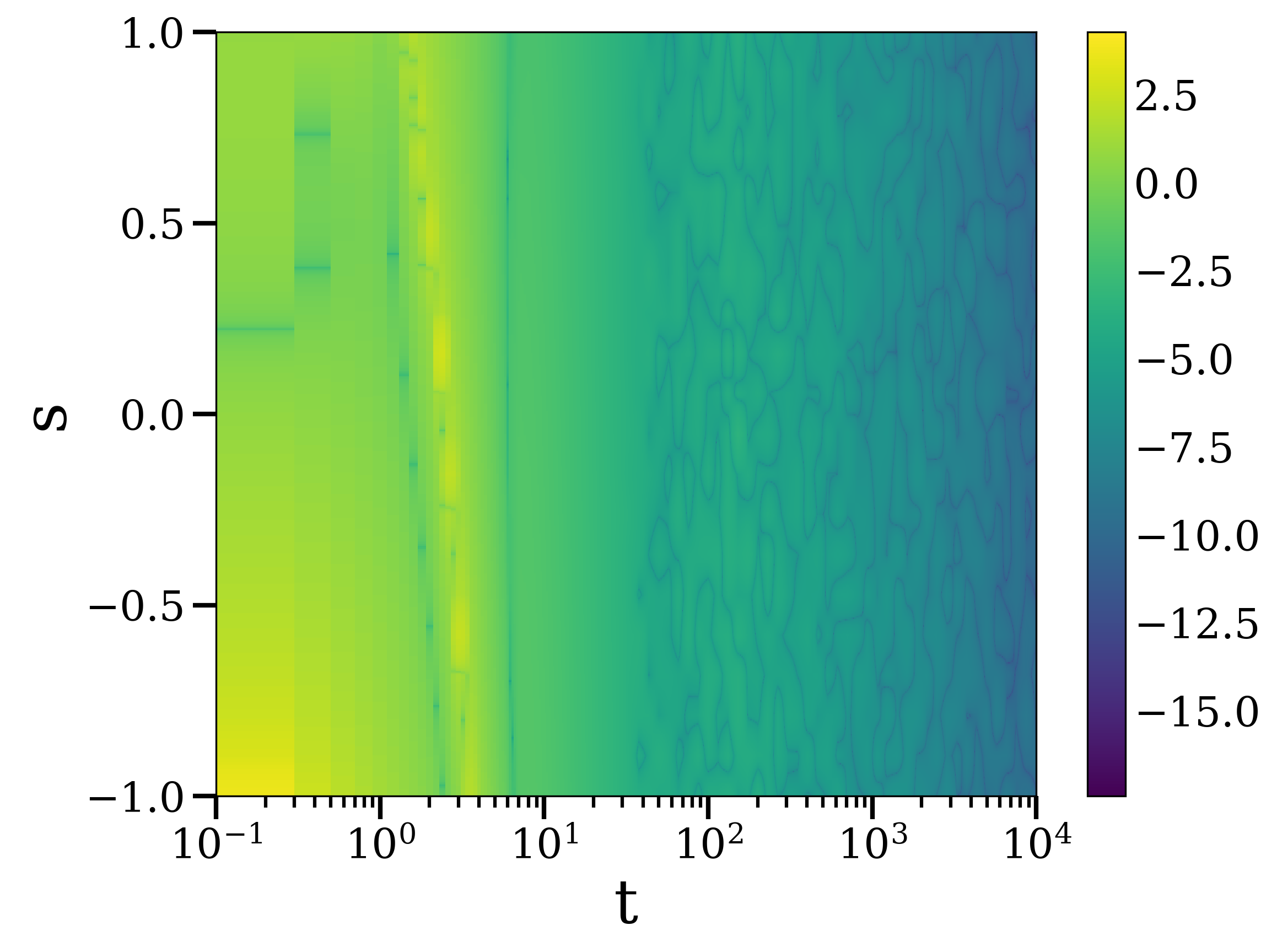}
    \end{subfigure}
    \begin{subfigure}[b]{0.48\textwidth}
        \includegraphics[width=\linewidth]{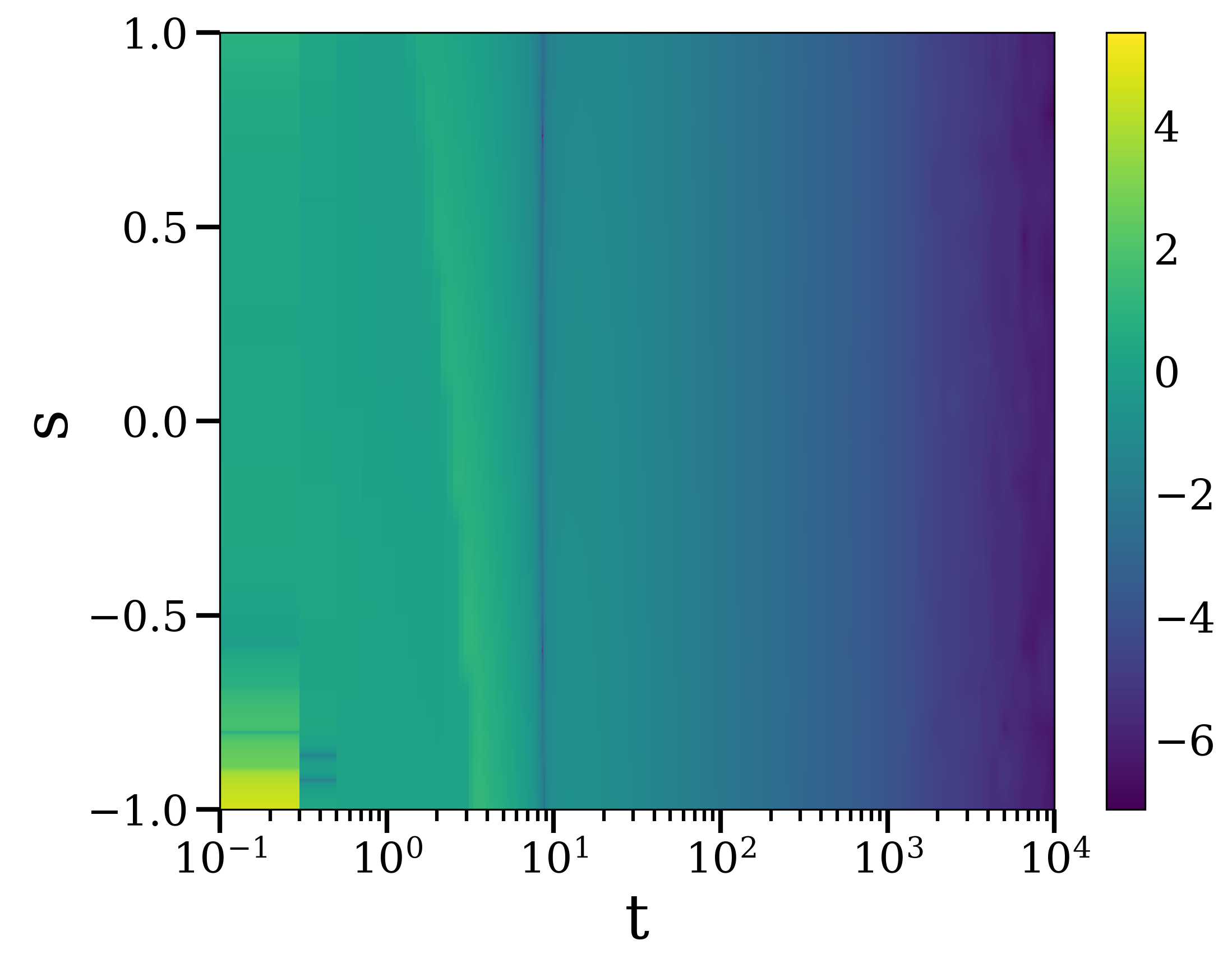}
    \end{subfigure}
    
    \begin{subfigure}[b]{0.48\textwidth}
        \includegraphics[width=\linewidth]{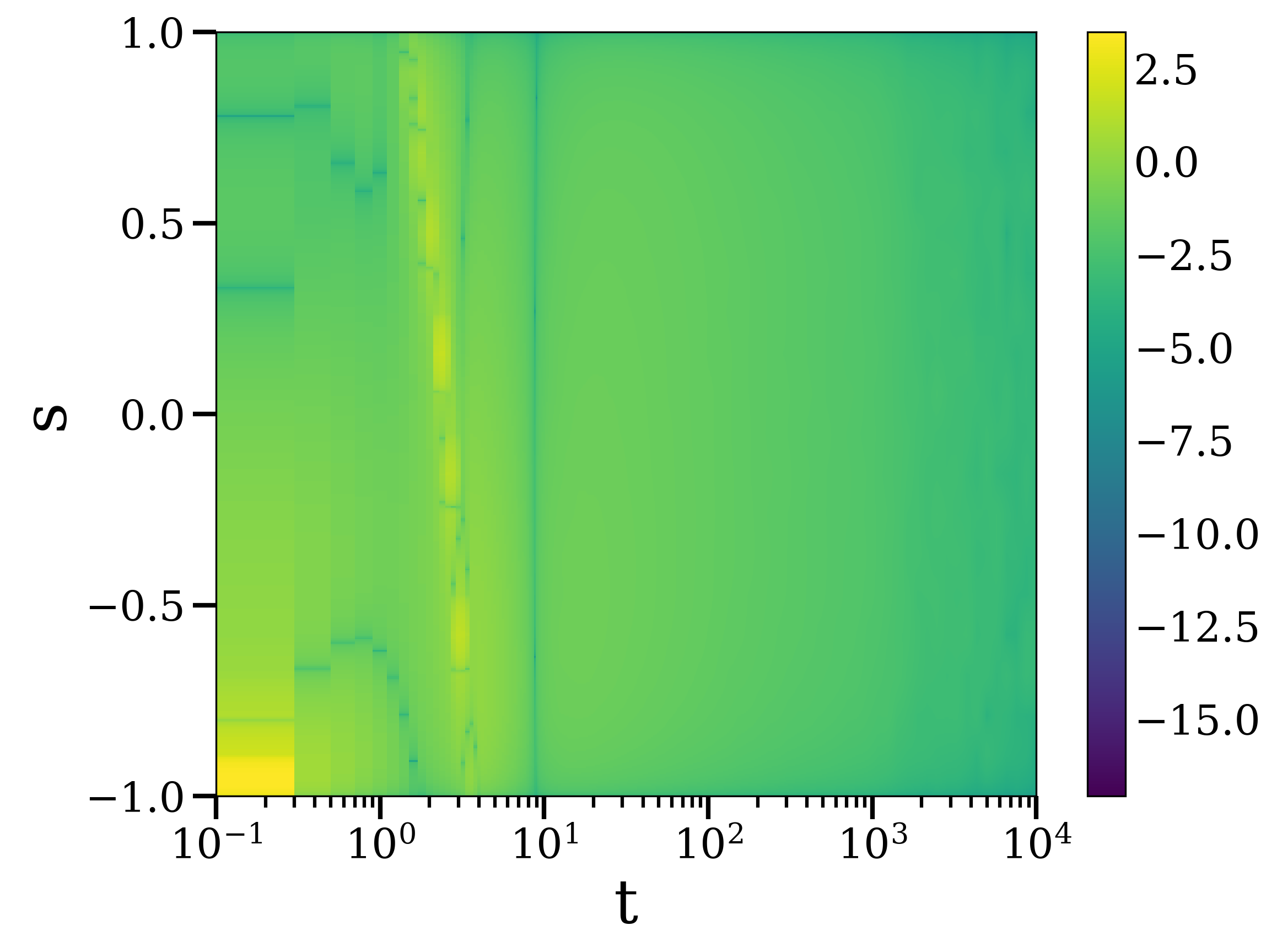}
    \end{subfigure}
    \begin{subfigure}[b]{0.48\textwidth}
        \includegraphics[width=\linewidth]{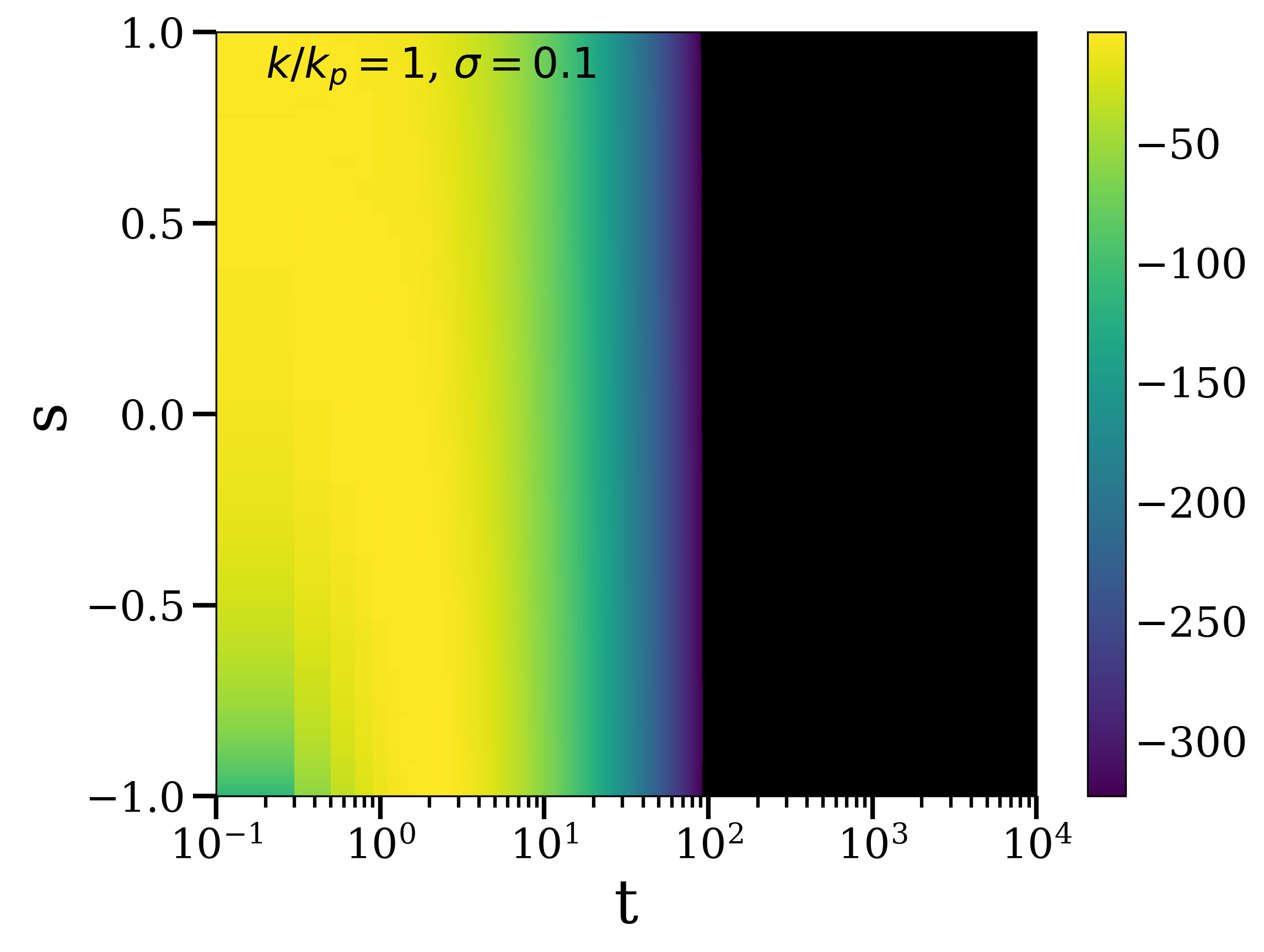}
    \end{subfigure}
    
    \begin{subfigure}[b]{0.48\textwidth}
        \includegraphics[width=\linewidth]{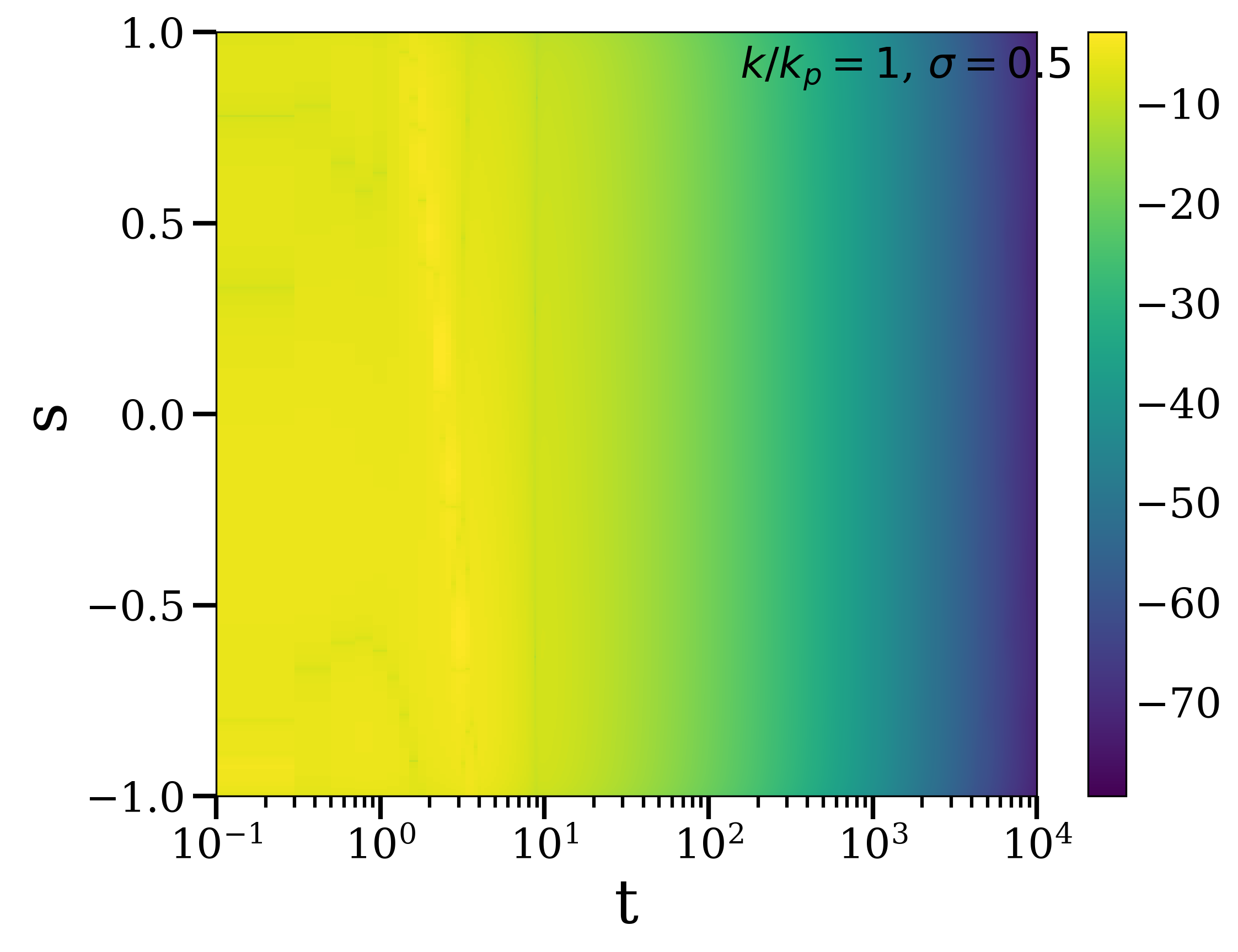}
    \end{subfigure}
    \begin{subfigure}[b]{0.48\textwidth}
        \includegraphics[width=\linewidth]{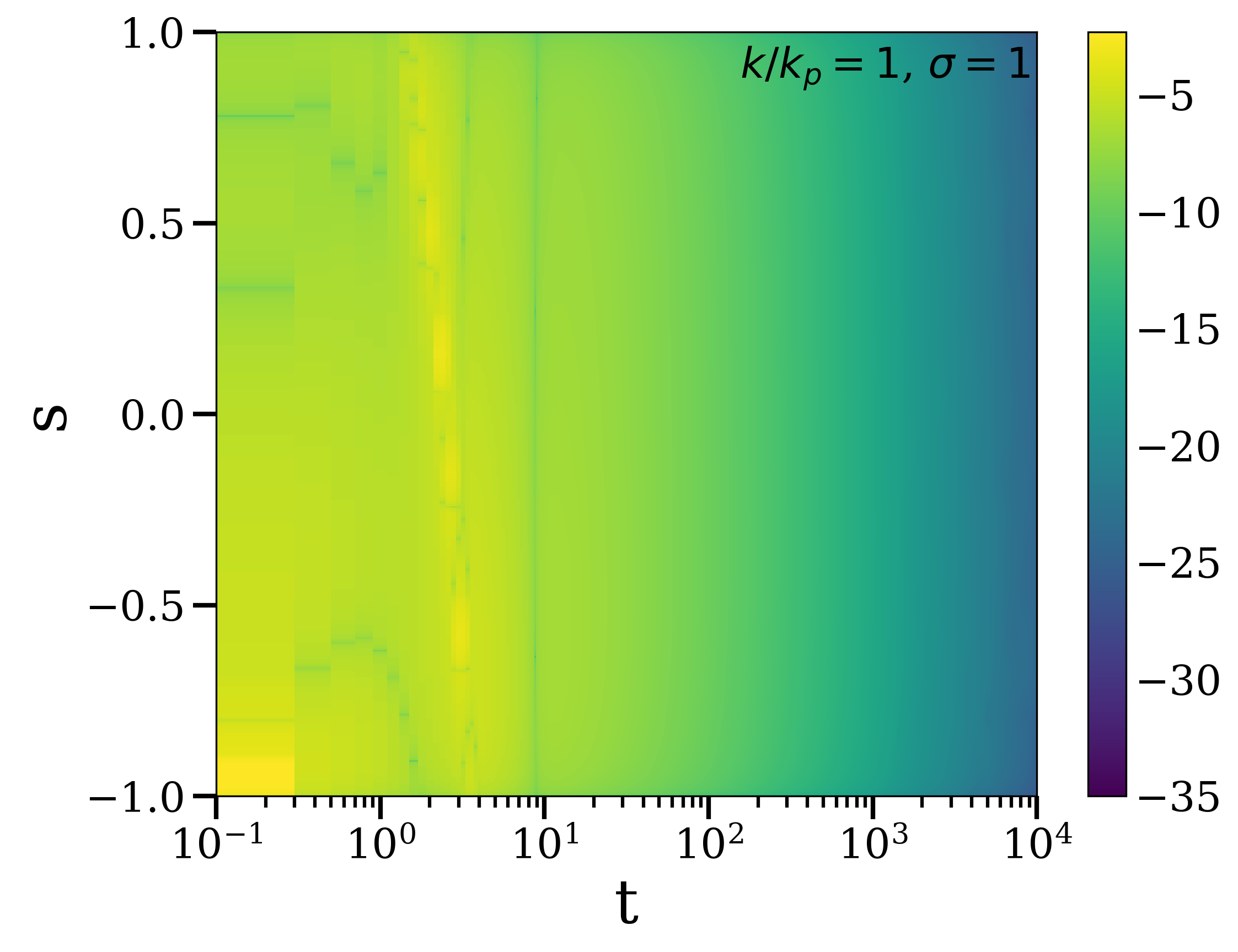}
    \end{subfigure}
    
    \caption{\footnotesize{Coloured mesh plots for the source term containing second-order vectors. (\textit{Top left}) First kernel corresponding to Eq.~\eqref{kernelb2rd}. (\textit{Top right}) Second kernel from Eq.~\eqref{kernelb2rd}. (\textit{Middle left}) Plot of the integrand in Eq.~\eqref{finalpsb2psi1} without the power spectra. (\textit{Middle right}) Plot of the integrand in Eq.~\eqref{finalpsb2psi1} with the power spectra included with $\sigma =0.1$. (\textit{Bottom left}) $\sigma = 0.5$ (\textit{Bottom right}) $\sigma = 1$. The two kernels exhibit an enhancement in the UV limit, which couples large wavelength scalar perturbations to short wavelength tensor modes. This enhancement gets larger as the width of the primordial peak of scalars gets larger; we therefore conclude that this term diverges.} }
    \label{fig:b2psi1um_meshplots}
\end{figure}

\begin{figure}
    \centering
    \begin{subfigure}[b]{0.48\textwidth}
        \includegraphics[width=\linewidth]{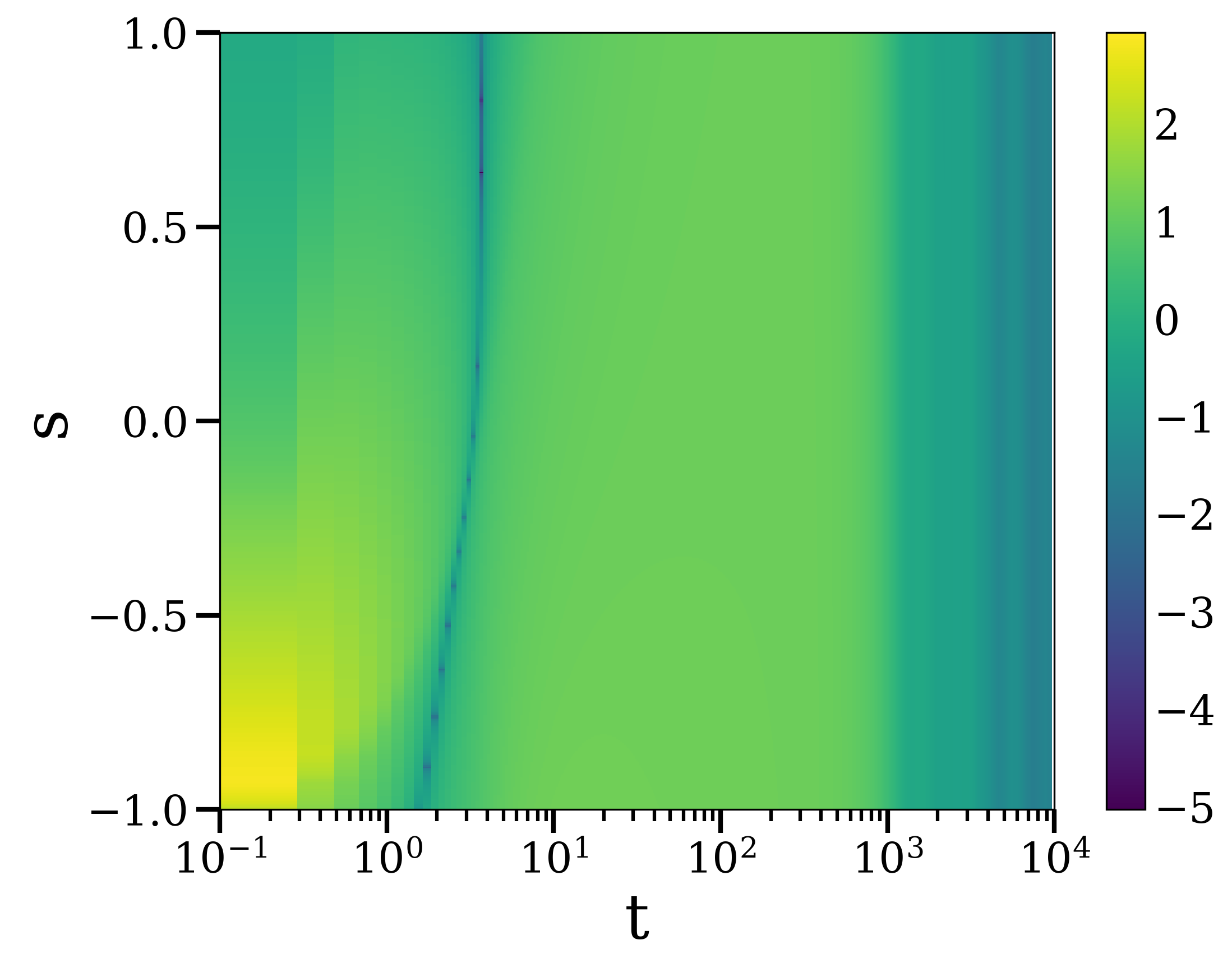}
    \end{subfigure}
    \begin{subfigure}[b]{0.48\textwidth}
        \includegraphics[width=\linewidth]{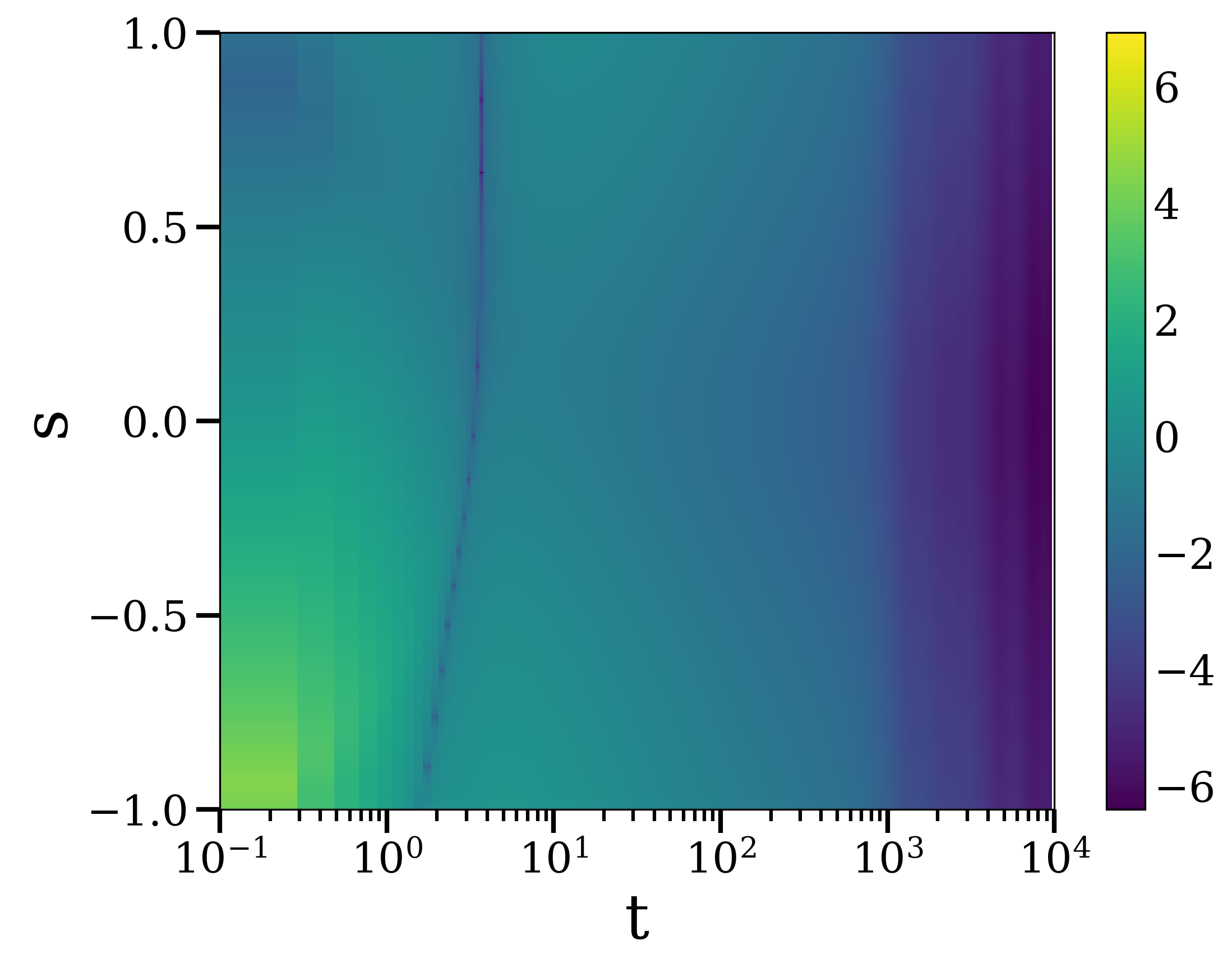}
    \end{subfigure}
    
    \begin{subfigure}[b]{0.48\textwidth}
        \includegraphics[width=\linewidth]{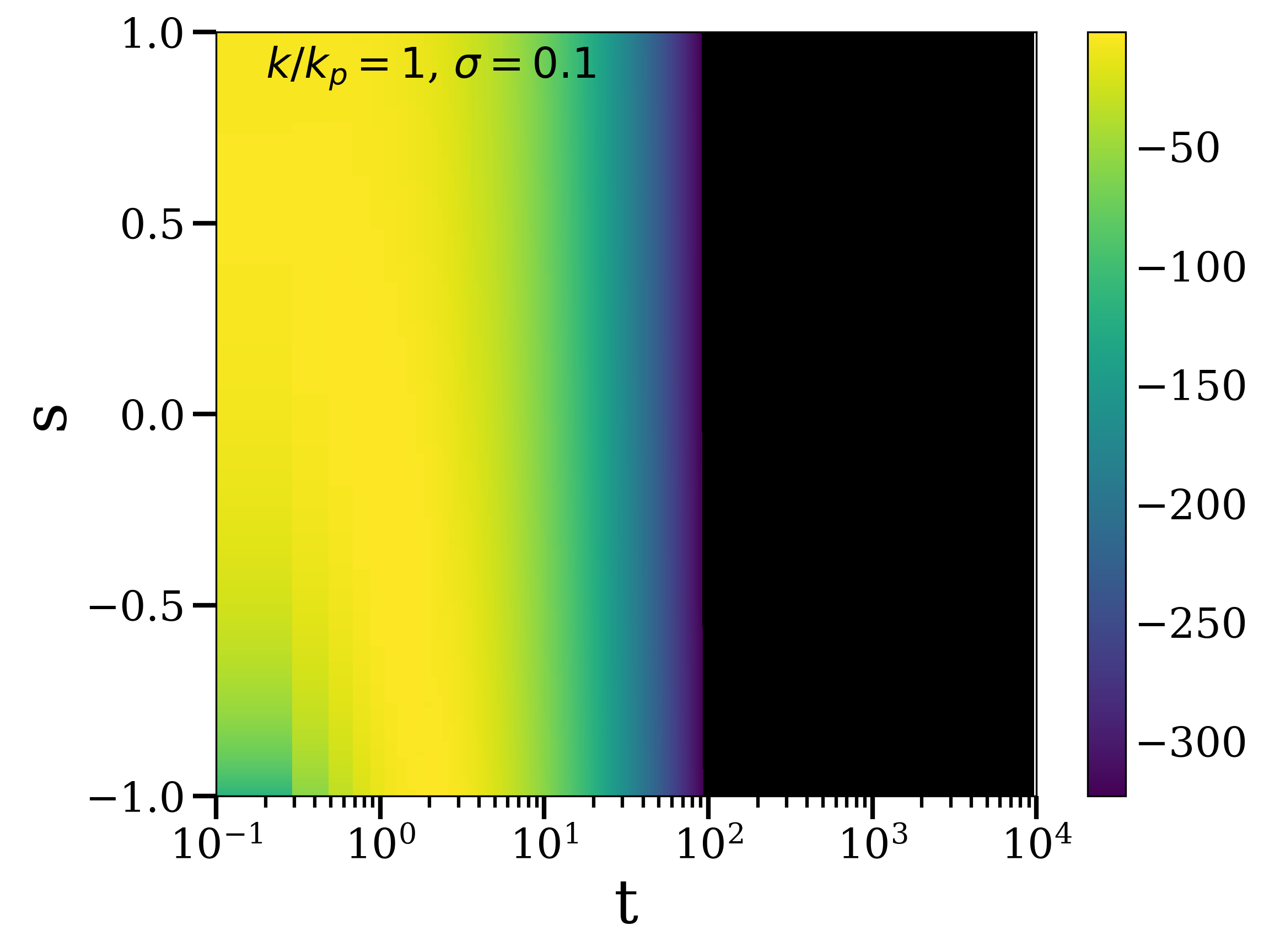}
    \end{subfigure}
    \begin{subfigure}[b]{0.48\textwidth}
        \includegraphics[width=\linewidth]{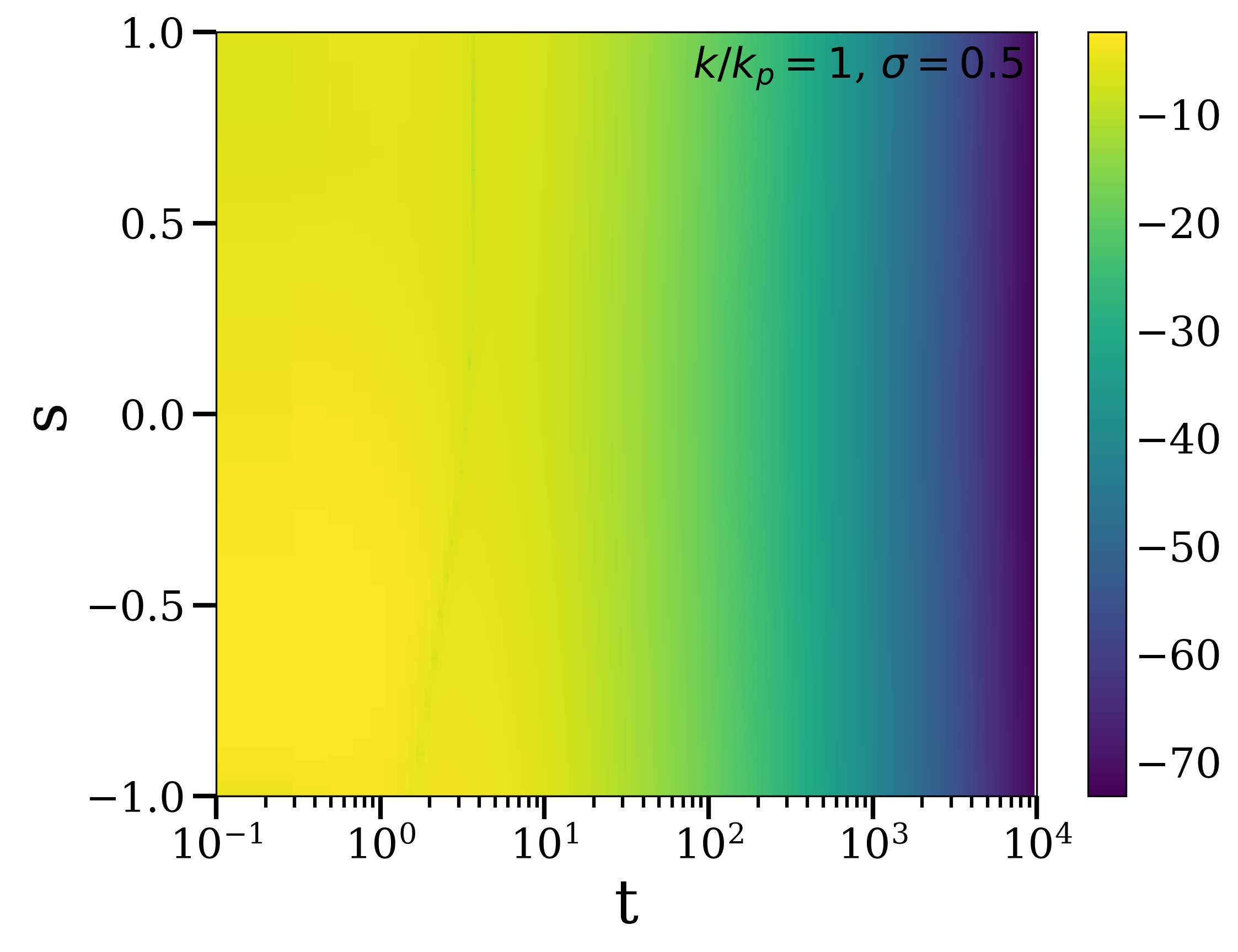}
    \end{subfigure}
    
    \begin{subfigure}[b]{0.48\textwidth}
        \includegraphics[width=\linewidth]{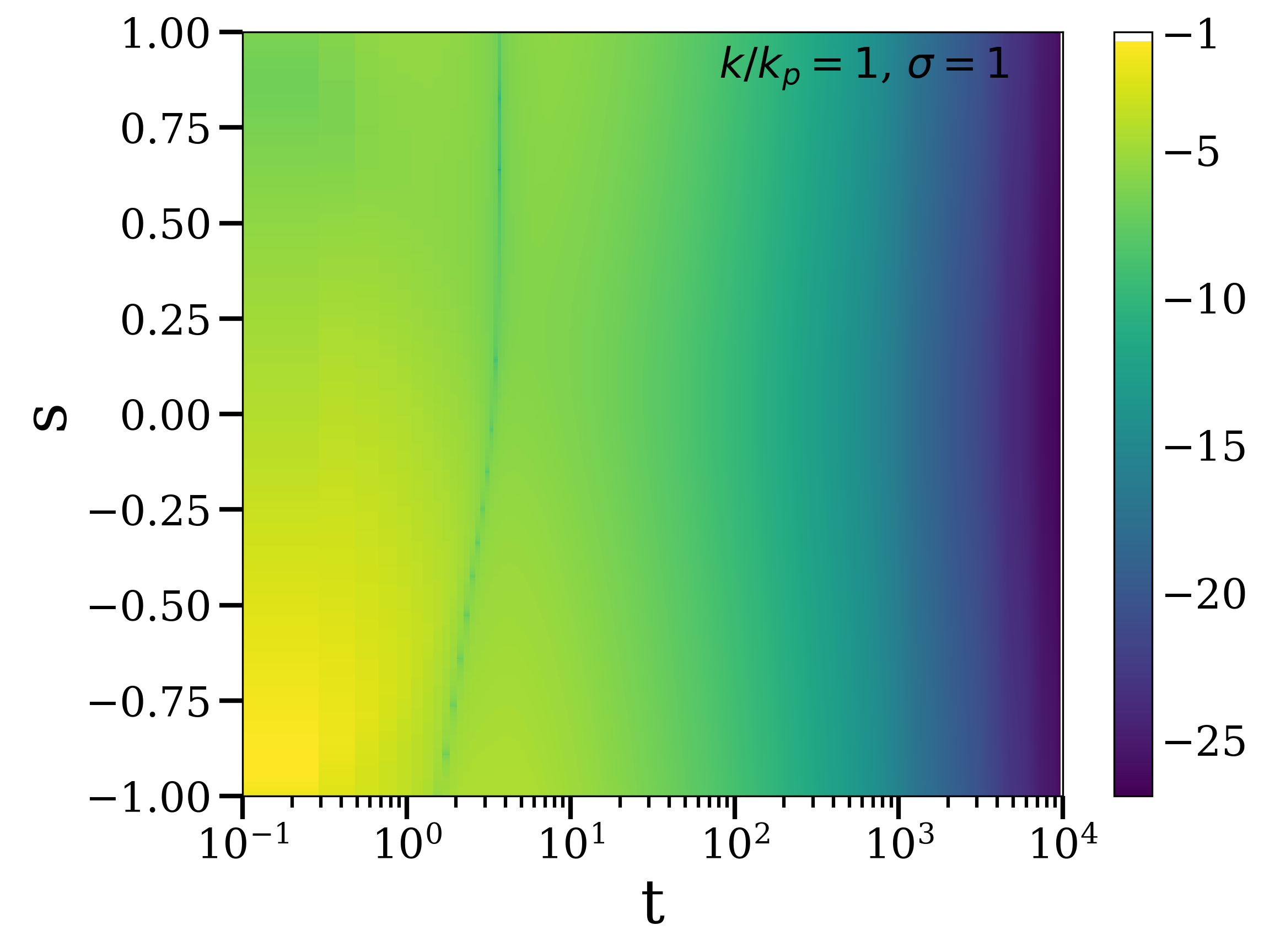}
    \end{subfigure}
    
    \caption{\footnotesize{Coloured mesh plots for the source term containing second-order tensors. (\textit{Top left}) Numerical kernel shown in Eq.~\eqref{finalkernelh2}.   (\textit{Top right}) Numerical part of the integrand shown in Eq.~\eqref{psh2psi1} without the primordial spectra. (\textit{Middle left}) Same as before but with inclusion of the first-order spectra, with $\sigma = 0.1$. (\textit{Middle right}) $\sigma =0.5$. (\textit{Bottom row}) $\sigma =1$. From these plots, we can determine that the integrand diverges in the UV limit. The kernel exhibits an enhancement in the limit $v \rightarrow 1$ and $u \rightarrow 0$, which is also present when we include the polarisation functions. This enhancement disappears when the scalar power spectrum is sufficiently peaked; however, it reappears as $\sigma$ gets larger. In fact, it diverges at a similar rate to the second-order scalar-tensor iGWs (with opposite sign); however, it is not possible to conclude whether this is also as $u^{-4}$.}}
    \label{fig:h2psi1um_meshplots}
\end{figure}

%%%%%%%%%%%%%%%%%%%%%%%%%%%%%%%%%%%%%%%%%%%%%%%%%%%%%%%%%%%%%%%%%%%%%%%%%%%%%%%%%%%%%%%%%%%%%%%%%%%%%%%%%%%%%%%%%%%%%%%%%%%%%%%%%%%%%%%%%%%%%%%%%%%%%%%%%%5555555

\chapter{Third-order tensor perturbations induced by first-order tensor perturbations}\label{app4}

In this Appendix\footnote{Hoping someone will find this useful one day.}, we present equations which describe how third-order tensors are sourced by first-order tensor modes. In Chapter~\ref{Chap:Third}, we showed that third-order tensors correlated with first-order tensors give rise to a non-trivial contribution to the iGW background, when second-order and third-order modes were sourced by couplings of first-order scalar and tensor modes. Here we present different source terms, those only made of or induced by pure first-order tensor modes.

In Eq.~\eqref{thirdorderGWEQ}, the sources we did not consider involve first-order tensors. They are
\begin{equation}
    S^{(3)}_{ij} = S^{hhh}_{ij}+ S^{\Psi^{(2)}h}_{ij}  + S^{B^{(2)}h}_{ij}  +  S^{h^{(2)}h}_{ij} \, ,
\end{equation}
where $S^{hhh}_{ij}$ contains terms cubic in first order tensor perturbations,  $S^{h^{(2)}h}_{ij}$ couples a second order tensor with a first order tensor, $S^{B^{(2)}h}_{ij}$ couples a second order vector with a first order tensor and $S^{\Psi^{(2)}h}_{ij}$. 
In each section of this Appendix, we will give these source terms and their respective solution in Fourier space. For second-order perturbations, we also give equations of motion which describe how they are sourced.
%%%%%%%%%%%%%%%%%%%%%%%%%%%%%%%%%%%%%%%%%%%%%%%%%%%%%%%%%%5555555%%%%%%%%%%%%%%%%%%%%%%%%%%%%%%%%%%%%%%%%%%%%%%%%%%%%%%%%%%5555555
\section{Source involving pure linear fluctuations}
There is one source term involving pure first-order tensor perturbations $ S^{hhh}_{ij}$, which reads
 \begin{align}\label{sourcetermhhh}
    \begin{split}
        S^{hhh}_{ij} &= -4h_{mn}h_i^{m\prime}h_j^{n\prime} + \frac{1}{2}h_{ij} \bigg \{ h_{mn}^{\prime} \big ( (w-1)h^{mn\prime} + 8\mathcal{H}c_s^2h^{mn} \big) -4h^{mn} \big ( 2 \partial _k \partial _n h_m^k \\
        &+ (w-1)\nabla ^2 h_{mn} \big) + (w-1) \big (2\partial _n h_{mk} -3\partial _k h_{mn} \big ) \partial ^k h^{mn} \bigg \} + 2 h^{mn} \bigg \{ h_{mn}^{\prime} h_{ij}^{\prime} \\
        &+ 2 \big ( \partial _n h_{jk} - \partial _k h_{jn} \big ) \big ( \partial _m h_i^k - \partial ^kh_{im} \big ) + \big (2\partial _n h_{mk} - \partial _k h_{mn} \big ) \big (\partial ^k h_{ij} -2 \partial _i h_j^k \big) \\
        &+2 \partial _i h_m^k \partial _j h_{nk} +2 h_m^k \big ( \partial _k \partial _n h_{ij} -2 \partial _i \partial _k h_{jn} + \partial _i \partial _j h_{nk} \big)  \bigg \} \, .
    \end{split}
\end{align}

In Fourier space, the solution to the third-order equation of motion is 
\begin{align}\label{solhhh}
    \begin{split}
        h^{\lambda \, (3)}_{hhh} (\eta , \mathbf{k}) &= \frac{6}{(2\pi)^3} \int {\rm } ^3 \mathbf{q} \int {\rm } ^3 \mathbf{p} \sum _{\lambda _1 , \lambda _2 ,\lambda _3} \bigg ( Q_{\lambda _1 , \lambda _2 ,\lambda _3}^{hhh,1} (\mathbf{k},\mathbf{p},\mathbf{q})I^{hhh}_1 (\eta , p,q,k) \\
        &+Q_{\lambda _1 , \lambda _2 ,\lambda _3}^{hhh,2} (\mathbf{k},\mathbf{p},\mathbf{q})I^{hhh}_2 (\eta , p,q,k) + Q_{\lambda _1 , \lambda _2 ,\lambda _3}^{hhh,3} (\mathbf{k},\mathbf{p},\mathbf{q})I^{hhh}_3 (\eta , p,q,k) \\
        &+Q_{\lambda _1 , \lambda _2 ,\lambda _3}^{hhh,4} (\mathbf{k},\mathbf{p},\mathbf{q})I^{hhh}_4 (\eta , p,q,k)+Q_{\lambda _1 , \lambda _2 ,\lambda _3}^{hhh,5} (\mathbf{k},\mathbf{p},\mathbf{q})I^{hhh}_5 (\eta , p,q,k)
        \bigg )\\
        &\times h^{\lambda _1}_{\mathbf{k}-\mathbf{p}-\mathbf{q}}h^{\lambda _2}_{\mathbf{p}}h^{\lambda _3}_{\mathbf{q}} .
    \end{split}
\end{align}
Where we have defined
\begin{equation}
    I_i^{hhh}(\eta , k,p,q) =  \int _{\eta _i}^{\infty} {\rm } \overline{\eta} \, \frac{a(\overline{\eta})}{a(\eta)} G^h_{k}(\eta , \overline{\eta}) f^{hhh}_i(\overline{\eta},k,p,q) \, 
\end{equation}
where $ f_i^{hhh}$ is made up of transfer functions
\begin{subequations} \label{hhhfunctions}
    \begin{align}
        \begin{split}
             f_1^{hhh}(\eta ,k,p,q) &= T_h\left ( \eta |\mathbf{k}-\mathbf{p}-\mathbf{q}|\right ) T_h \left ( \eta p \right ) T_h\left ( \eta q \right ) \, ,
        \end{split}
             \\
        \begin{split}
            f_2^{hhh}(\eta ,k,p,q) &= T_h\left ( \eta |\mathbf{k}-\mathbf{p}-\mathbf{q}|\right ) T_h \conf \left ( \eta p \right ) T_h \conf \left ( \eta q \right ) \, ,
        \end{split}
              \\
        \begin{split}
            f_3^{hhh}(\eta ,k,p,q) &= (w-1)T_h\left ( \eta |\mathbf{k}-\mathbf{p}-\mathbf{q}|\right ) T_h \conf \left ( \eta p \right ) T_h \conf \left ( \eta q \right ) \\
            &+ 8 \hubble w T_h\left ( \eta |\mathbf{k}-\mathbf{p}-\mathbf{q}|\right ) T_h \conf \left ( \eta p \right ) T_h\left ( \eta q \right )\, ,
        \end{split}
        \\
        \begin{split}
             f_4^{hhh}(\eta ,k,p,q) &= 2(w-1) f_1^{hhh}(\eta ,k,p,q)\, ,
        \end{split}
         \\
        \begin{split}
             f_5^{hhh}(\eta ,k,p,q) &=-\frac{1}{2}(w-1) f_1^{hhh}(\eta ,k,p,q)\, ,
        \end{split}
    \end{align}
\end{subequations}
and $Q_{\lambda _1 , \lambda _2 ,\lambda _3}^{hhh,i}(\mathbf{k},\mathbf{p},\mathbf{q})$ are the polarisation functions defined as
\begin{subequations}
    \begin{align}
        \begin{split}
            &Q_{\lambda _1 , \lambda _2 ,\lambda _3}^{hhh,1}(\mathbf{k},\mathbf{p},\mathbf{q}) = 4 \epsilon^{ij}_{\lambda}(\mathbf{k})^* q_k q_n \epsilon_{ij}^{\lambda _1}(\mathbf{k}-\mathbf{p}-\mathbf{q})\epsilon^{mn}_{\lambda _2}(\mathbf{p}) \epsilon^{k, \lambda _3}_m (\mathbf{q}) \\
            &-4 \epsilon^{ij}_{\lambda}(\mathbf{k}) ^* p_n q_m \epsilon^{mn}_{\lambda _1}(\mathbf{k}-\mathbf{p}-\mathbf{q})\epsilon_{jk}^{\lambda _2}(\mathbf{p}) \epsilon^{k, \lambda _3}_i (\mathbf{q})+ 4 \epsilon^{ij}_{\lambda}(\mathbf{k}) ^* p_n q^k \epsilon^{mn}_{\lambda _1}(\mathbf{k}-\mathbf{p}-\mathbf{q})\epsilon_{jk}^{\lambda _2}(\mathbf{p}) \epsilon^{ \lambda _3}_{im} (\mathbf{q}) \\
            &+4 \epsilon^{ij}_{\lambda}(\mathbf{k}) ^* p_k q_m \epsilon^{mn}_{\lambda _1}(\mathbf{k}-\mathbf{p}-\mathbf{q})\epsilon_{jn}^{\lambda _2}(\mathbf{p}) \epsilon^{k, \lambda _3}_i (\mathbf{q}) -4 \, \mathbf{p}\cdot \mathbf{q} \epsilon^{ij}_{\lambda}(\mathbf{k}) ^*\epsilon^{mn}_{\lambda _1}(\mathbf{k}-\mathbf{p}-\mathbf{q})\epsilon_{jn}^{\lambda _2}(\mathbf{p}) \epsilon^{ \lambda _3}_{im} (\mathbf{q}) \\
            &-4 \epsilon^{ij}_{\lambda}(\mathbf{k}) ^* p_n q_k \epsilon^{mn}_{\lambda _1}(\mathbf{k}-\mathbf{p}-\mathbf{q})\epsilon_{mk}^{\lambda _2}(\mathbf{p}) \epsilon^{ \lambda _3}_{ij} (\mathbf{q}) +8 \epsilon^{ij}_{\lambda}(\mathbf{k}) ^* p_n q_i \epsilon^{mn}_{\lambda _1}(\mathbf{k}-\mathbf{p}-\mathbf{q})\epsilon_{mk}^{\lambda _2}(\mathbf{p}) \epsilon^{k, \lambda _3}_j (\mathbf{q}) \\
            &+2 \, \mathbf{p}\cdot \mathbf{q} \epsilon^{ij}_{\lambda}(\mathbf{k}) ^*  \epsilon^{mn}_{\lambda _1}(\mathbf{k}-\mathbf{p}-\mathbf{q})\epsilon_{mn}^{\lambda _2}(\mathbf{p}) \epsilon^{ \lambda _3}_{ij} (\mathbf{q}) -4 \epsilon^{ij}_{\lambda}(\mathbf{k}) ^* p_k q_i \epsilon^{mn}_{\lambda _1}(\mathbf{k}-\mathbf{p}-\mathbf{q})\epsilon_{mn}^{\lambda _2}(\mathbf{p}) \epsilon^{k, \lambda _3}_j (\mathbf{q}) \\
            &-4 \epsilon^{ij}_{\lambda}(\mathbf{k}) ^* p_i q_j \epsilon^{mn}_{\lambda _1}(\mathbf{k}-\mathbf{p}-\mathbf{q})\epsilon_{m}^{k, \lambda _2}(\mathbf{p}) \epsilon^{ \lambda _3}_{nk} (\mathbf{q})-4 \epsilon^{ij}_{\lambda}(\mathbf{k}) ^* q_k q_n \epsilon^{mn}_{\lambda _1}(\mathbf{k}-\mathbf{p}-\mathbf{q})\epsilon_{m}^{k, \lambda _2}(\mathbf{p}) \epsilon^{\lambda _3}_{ij} (\mathbf{q}) \\
            &+8 \epsilon^{ij}_{\lambda}(\mathbf{k}) ^* q_i q_k \epsilon^{mn}_{\lambda _1}(\mathbf{k}-\mathbf{p}-\mathbf{q})\epsilon_{m}^{k, \lambda _2}(\mathbf{p}) \epsilon^{ \lambda _3}_{jn} (\mathbf{q}) -4 \epsilon^{ij}_{\lambda}(\mathbf{k}) ^* q_i q_j \epsilon^{mn}_{\lambda _1}(\mathbf{k}-\mathbf{p}-\mathbf{q})\epsilon_{m}^{k,\lambda _2}(\mathbf{p}) \epsilon^{\lambda _3}_{nk} (\mathbf{q}) \, ,
        \end{split}
        \\
        \begin{split}
            &Q_{\lambda _1 , \lambda _2 ,\lambda _3}^{hhh,2}(\mathbf{k},\mathbf{p},\mathbf{q}) = -4 \epsilon^{ij}_{\lambda}(\mathbf{k}) ^*  \epsilon_{mn}^{\lambda _1}(\mathbf{k}-\mathbf{p}-\mathbf{q})\epsilon_{i}^{m,\lambda _2}(\mathbf{p}) \epsilon^{n,\lambda _3}_{j} (\mathbf{q}) \\
            &+2 \epsilon^{ij}_{\lambda}(\mathbf{k}) ^*  \epsilon_{mn}^{\lambda _1}(\mathbf{k}-\mathbf{p}-\mathbf{q})\epsilon^{mn}_{\lambda _2}(\mathbf{p}) \epsilon^{\lambda _3}_{ij} (\mathbf{q}) \, ,
        \end{split}
        \\
        \begin{split}
             &Q_{\lambda _1 , \lambda _2 ,\lambda _3}^{hhh,3}(\mathbf{k},\mathbf{p},\mathbf{q}) = \frac{1}{2} \epsilon^{ij}_{\lambda}(\mathbf{k}) ^*  \epsilon_{ij}^{\lambda _1}(\mathbf{k}-\mathbf{p}-\mathbf{q})\epsilon_{mn}^{\lambda _2}(\mathbf{p}) \epsilon_{\lambda _3}^{mn} (\mathbf{q}) \, ,
        \end{split}
        \\
        \begin{split}
             &Q_{\lambda _1 , \lambda _2 ,\lambda _3}^{hhh,4}(\mathbf{k},\mathbf{p},\mathbf{q}) = q^2 \epsilon^{ij}_{\lambda}(\mathbf{k}) ^*  \epsilon_{ij}^{\lambda _1}(\mathbf{k}-\mathbf{p}-\mathbf{q})\epsilon_{mn}^{\lambda _2}(\mathbf{p}) \epsilon_{\lambda _3}^{mn} (\mathbf{q}) \, ,
        \end{split}
        \\
        \begin{split}
             &Q_{\lambda _1 , \lambda _2 ,\lambda _3}^{hhh,5}(\mathbf{k},\mathbf{p},\mathbf{q}) = 2\epsilon^{ij}_{\lambda}(\mathbf{k}) ^*  \epsilon_{ij}^{\lambda _1}(\mathbf{k}-\mathbf{p}-\mathbf{q})p_nq^k \epsilon_{mk}^{\lambda _2}(\mathbf{p}) \epsilon_{\lambda _3}^{mn} (\mathbf{q}) \\
             &-3 \epsilon^{ij}_{\lambda}(\mathbf{k}) ^*  \epsilon_{ij}^{\lambda _1}(\mathbf{k}-\mathbf{p}-\mathbf{q}) \, \mathbf{p} \cdot \mathbf{q} \, \epsilon_{mn}^{\lambda _2}(\mathbf{p}) \epsilon_{\lambda _3}^{mn} (\mathbf{q}) \, .
        \end{split}
    \end{align}
\end{subequations}

%%%%%%%%%%%%%%%%%%%%%%%%%%%%%%%%%%%%%%%%%%%%%%%%%%%%%%%%%%5555555%%%%%%%%%%%%%%%%%%%%%%%%%%%%%%%%%%%%%%%%%%%%%%%%%%%%%%%%%%5555555
\section{Sources involving second-order fluctuations}

%%%%%%%%%%%%%%%%%%%%%%%%%%%%%%%%%%%%%%%%%%%%%%%%%%%%%%%%%%5555555
\subsection{Second-order scalars induced by first-order tensor modes}
The source term, which couples second-order scalars to linear tensor fluctuations, $S^{\Psi^{(2)}h}_{ij}$, is given by
\begin{align}\label{realspacesourcepsi2h}
    \begin{split}
        S^{\Psi^{(2)}h}_{ij} &=  h_{ij} \left ( \mathcal{H}\Phi ^{(2)\prime} + 3\mathcal{H}w\Psi ^{(2)\prime} -(w-1)\nabla ^2 \Psi ^{(2)} \right ) +\frac{1}{2}h^{\prime}_{ij} \left (\Phi ^{(2)\prime} - \Psi ^{(2)\prime}  \right )\\
        & + \nabla ^2 h_{ij}\left (\Phi ^{(2)} + \Psi ^{(2)}  \right ) - \partial ^m \Phi ^{(2)} \partial _j h_{jm} -\partial ^m \Psi ^{(2)} \partial _i h_{jm}-2h_i^m\partial _m \partial _j \Psi ^{(2)} \\
        &+\frac{1}{2}\partial _m h_{ij}\partial ^m \Phi ^{(2)} +\frac{3}{2}\partial _m h_{ij} \partial ^m \Psi ^{(2)} , .
    \end{split}
\end{align}
Similarly to the case of second-order scalar perturbations induced by first-order scalar fluctuations, there are two evolution equations we can extract from the Einstein equations which describe how the second-order scalar fluctuations are induced by the tensor perturbations. One comes from the trace 
\begin{align}\label{eqtracerealspacett}
    \begin{split}
        &3\Psi ^{\prime \prime (2)} + 3(2+3w)\hubble \Psi ^{\prime (2)} - (3w+1)\nabla ^2 \Psi ^{(2)} + 3\hubble \Phi ^{\prime (2)} + \nabla ^2 \Phi ^{(2)} = S^{\Psi^{(2)}_{tt}} \, ,
    \end{split}
\end{align}
where the source term is given in Eq.~\eqref{rstracestt}. The second equation of motion comes from the symmetric trace-free part of the third-order Einstein equations
\begin{equation}\label{eqanisrealspacett}
    \nabla ^2 \nabla ^2 \Phi ^{(2)} - \nabla ^2 \nabla ^2 \Psi ^{(2)} = D^{ij}\Pi _{ij}^{(2),tt} \, ,
\end{equation}
where $\Pi _{ij}^{(2),tt}$ can be found in Eq.~\eqref{rsanistt}. 

Before giving the third-order solution to the iGW equation, we need expressions for the second-order equations of motion in Fourier space. The symmetric trace-free one is given by 
\begin{equation} 
    \Phi ^{(2)}(\eta , \mathbf{k}) - \Psi ^{(2)}(\eta , \mathbf{k}) = \frac{1}{k^4}  \Pi ^{(2)}_{tt} (\eta , \mathbf{k})  \, ,
\end{equation}
with
\begin{equation}
    \begin{split}\label{anisinducedtt}
    \Pi ^{(2)}_{tt} (\eta , \mathbf{k}) & =  \sum _{\lambda _1 , \lambda _2} \int \frac{\rm ^3\mathbf{p}}{(2\pi)^{\frac{3}{2}}}  \, \, \bigg (Q_{\lambda _1,\lambda _2}^{\Pi ^{(2)} _{tt1}}(\mathbf{k},\mathbf{p})f^{\Pi ^{(2)}}_{tt1}(\eta ,p,|\mathbf{k}-\mathbf{p}|) \\
    &+ Q_{\lambda _1,\lambda _2}^{\Pi ^{(2)} _{tt2}}(\mathbf{k},\mathbf{p})f^{\Pi ^{(2)}}_{tt2}(\eta ,p,|\mathbf{k}-\mathbf{p}|)\bigg ) h^{\lambda _1}_{\mathbf{p}}h^{\lambda _2}_{\mathbf{k}-\mathbf{p}}\, ,
    \end{split}
\end{equation}
where
\begin{subequations}
    \begin{align}
       \begin{split}\label{polpitt1}
            Q_{\lambda _1,\lambda _2}^{\Pi ^{(2)} _{tt1}}(\mathbf{k},\mathbf{p})&= 6k^ik^j \epsilon_i^{m ,\lambda _1}(\mathbf{p})\epsilon^{\lambda _2}_{jm}(\mathbf{k}-\mathbf{p}) - 2k^2\epsilon_{ij}^{\lambda _1}(\mathbf{p})\epsilon_{\lambda _2}^{ij}(\mathbf{k}-\mathbf{p}) \, ,
       \end{split}
        \\
        \begin{split}\label{polpitt2}
            Q_{\lambda _1,\lambda _2}^{\Pi ^{(2)} _{tt2}}(\mathbf{k},\mathbf{p}) &=  6 p_m p_n k^ik^j \epsilon_{ij}^{\lambda _1}(\mathbf{p})\epsilon^{mn}_{\lambda _2}(\mathbf{k}-\mathbf{p}) + 6 (k^ik^jp_i p_j -\frac{1}{3}k^2p^2)\epsilon_{mn}^{\lambda _1}(\mathbf{p})\epsilon^{mn}_{\lambda _2}(\mathbf{k}-\mathbf{p}) \\
            &-12p_nk^jp_jk^i\epsilon_{im}^{\lambda _1}(\mathbf{p})\epsilon^{mn}_{\lambda _2}(\mathbf{k}-\mathbf{p}) - 6k^ik^jp_mk^n\epsilon_{jn}^{\lambda _1}(\mathbf{p})\epsilon^{m, \lambda _2}_{i}(\mathbf{k}-\mathbf{p})\\
            &+2k^2p_mk^n\epsilon^{\lambda _1}_{in}(\mathbf{p})\epsilon^{im}_{\lambda _2}(\mathbf{k}-\mathbf{p}) +6 (p_nk^n-p^2)  k^ik^j\epsilon_{jm}^{\lambda _1}(\mathbf{p}) \epsilon_i^{m ,\lambda _2}(\mathbf{k}-\mathbf{p}) \\
            &-2 (p_nk^n-p^2)k^2\epsilon_{ij}^{\lambda _1}(\mathbf{p}) \epsilon^{ij}_{\lambda _2}(\mathbf{k}-\mathbf{p}) -( 3k^ik^jp_ip_j -2 k^2p^ik_i -k^2p^2  ) \epsilon^{mn}_{\lambda _1}(\mathbf{p}) \\
            &\times \epsilon_{mn}^{\lambda _2}(\mathbf{k}-\mathbf{p}) \, ,
        \end{split}
     \end{align}
\end{subequations}
and
\begin{subequations}
    \begin{align}
    \begin{split}\label{fpittscalar1}
         f^{\Pi ^{(2)}}_{tt1}(\eta ,p,|\mathbf{k}-\mathbf{p}|) &= T_{h}\conf (\eta p)T_{h}\conf(\eta|\mathbf{k}-\mathbf{p}|) \, ,
    \end{split}
    \\
    \begin{split}\label{fpittscalar2}
        f^{\Pi ^{(2)}}_{tt2}(\eta ,p,|\mathbf{k}-\mathbf{p}|) &= T_h(\eta p)T_{h}(\eta|\mathbf{k}-\mathbf{p}|) \, .
    \end{split}
    \end{align}
\end{subequations}
The second order trace equation for scalars, Eq.~\eqref{eqtracerealspacett}, in Fourier space is given by
\begin{align}
    \begin{split}
        &3\Psi ^{\prime \prime (2)}(\eta, \mathbf{k}) + 3(2+3w)\hubble \Psi ^{\prime (2)}(\eta, \mathbf{k}) + (3w+1)k^2 \Psi ^{(2)}(\eta, \mathbf{k}) + 3\hubble \Phi ^{\prime (2)}(\eta, \mathbf{k}) \\
        &- k^2 \Phi ^{(2)}(\eta, \mathbf{k}) =   \mathcal{S}^{\Psi ^{(2)}}_{tt}(\eta , \mathbf{k}) \, ,
    \end{split}
\end{align}
where the source term is defines as
\begin{equation}
    \begin{split}
    \mathcal{S}^{\Psi ^{(2)}}_{tt}(\eta , \mathbf{k}) & = \sum _{\lambda _1 , \lambda _2} \int \frac{\rm ^3\mathbf{p}}{(2\pi)^{\frac{3}{2}}}\bigg (Q_{\lambda _1 ,\lambda _2}^{\Psi ^{(2)}_{tt1}}(\mathbf{k},\mathbf{p})f^{\Psi ^{(2)}}_{tt1}(\eta , p,|\mathbf{k}-\mathbf{p}|) + Q_{\lambda _1 ,\lambda _2}^{\Psi ^{(2)}_{tt2}}(\mathbf{k},\mathbf{p})\\
    &\times f^{\Psi ^{(2)}}_{tt2}(\eta , p,|\mathbf{k}-\mathbf{p}|) \bigg ) h^{\lambda _1}_{\mathbf{p}}h^{\lambda _2}_{\mathbf{k}-\mathbf{p}} \, ,
\end{split}
\end{equation}
and we have further defined
\begin{subequations}
    \begin{align}
        Q_{\lambda _1 ,\lambda _2}^{\Psi ^{(2)}_{tt1}}(\mathbf{k},\mathbf{p}) &= \epsilon^{ij}_{\lambda _1}(\mathbf{p})\epsilon_{ij}^{\lambda _2}(\mathbf{k}-\mathbf{p}) \label{polpsitt1} \\
        Q_{\lambda _1 ,\lambda _2}^{\Psi ^{(2)}_{tt2}}(\mathbf{k},\mathbf{p}) &=p_jk^m\epsilon_{im}^{\lambda _1}(\mathbf{p})\epsilon^{ij}_{\lambda _2} (\mathbf{k}-\mathbf{p}) \label{polpsitt2}
    \end{align}
\end{subequations}
and
\begin{subequations}
    \begin{align}
        \begin{split}
            f^{\Psi ^{(2)}}_{tt1}(\eta , p,|\mathbf{k}-\mathbf{p}|) &= -\frac{1}{2} (5+3w )T_h\conf (\eta p)T_h \conf (\eta |\mathbf{k}-\mathbf{p}|) -12w \hubble T_h\conf (\eta p)T_h (\eta |\mathbf{k}-\mathbf{p}|) \\
            &-2(3w-1)p^2 T_h (\eta p)T_h (\eta |\mathbf{k}-\mathbf{p}|) -\frac{3}{2}(1+3w)  (p_mk^m-p^2)\\
            &\times T_h (\eta p)T_h (\eta |\mathbf{k}-\mathbf{p}|)
        \end{split}
        \\
        f^{\Psi ^{(2)}}_{tt2}(\eta , p,|\mathbf{k}-\mathbf{p}|) &= (1+3w)T_h(\eta p)T_h(\eta |\mathbf{k}-\mathbf{p}|) \, .
    \end{align}
\end{subequations}
Substituting $\Phi ^{(2)} (\eta , \mathbf{k})$ into the trace equation we have
\begin{align} 
    \begin{split}
        &\Psi ^{\prime \prime (2)}(\eta, \mathbf{k}) + 3(1+w)\hubble \Psi ^{\prime (2)}(\eta, \mathbf{k}) +wk^2 \Psi ^{(2)}(\eta, \mathbf{k}) = \frac{1}{3}  \mathcal{S}^{tt}(\eta , \mathbf{k}) - \frac{\hubble}{k^4} \timeder  \Pi ^{(2)}_{tt} (\eta , \mathbf{k})  \\
        &  + \frac{1}{3k^2}  \Pi ^{(2)}_{tt} (\eta , \mathbf{k}) \\
        &= \sum _{\lambda _1 , \lambda _2} \int \frac{\rm ^3\mathbf{p}}{(2\pi)^{\frac{3}{2}}} \bigg ( \bigg [ \frac{1}{3}Q_{\lambda _1 ,\lambda _2}^{\Psi ^{(2)}_{tt1}}(\mathbf{k},\mathbf{p})f^{\Psi ^{(2)}}_{tt1}(\eta , p,|\mathbf{k}-\mathbf{p}|)+\frac{1}{3}Q_{\lambda _1 ,\lambda _2}^{\Psi ^{(2)}_{tt2}}(\mathbf{k},\mathbf{p})f^{\Psi ^{(2)}}_{tt2}(\eta , p,|\mathbf{k}-\mathbf{p}|) \\
        &+Q_{\lambda _1,\lambda _2}^{\Pi ^{(2)} _{tt1}} (\mathbf{k},\mathbf{p}) \bigg \{ -\frac{\hubble}{k^4} \timeder f^{\Pi ^{(2)}}_{tt1}(\eta ,p,|\mathbf{k}-\mathbf{p}|)  + \frac{1}{3k^2}f^{\Pi ^{(2)}}_{tt1}(\eta ,p,|\mathbf{k}-\mathbf{p}|)\bigg \}  +Q_{\lambda _1,\lambda _2}^{\Pi ^{(2)} _{tt2}} (\mathbf{k},\mathbf{p})  \\
        &\times \bigg \{ -\frac{\hubble}{k^4} \timeder f^{\Pi ^{(2)}}_{tt2}(\eta ,p,|\mathbf{k}-\mathbf{p}|)  + \frac{1}{3k^2}f^{\Pi ^{(2)}}_{tt2}(\eta ,p,|\mathbf{k}-\mathbf{p}|)\bigg \} \bigg ]h^{\lambda _1}_{\mathbf{p}}h^{\lambda _2}_{\mathbf{k}-\mathbf{p}} \bigg ) \, .
    \end{split}
\end{align}
The solution to the equation of motion for $\Psi_{tt} ^{(2)}(\eta , \mathbf{k})$ is then
\begin{equation}
    \begin{split}\label{psiinducedtt}
            \Psi ^{(2)}_{tt}(\eta , \mathbf{k})&= \sum _{\lambda _1 , \lambda _2}\int \frac{\rm ^3\mathbf{p}}{(2\pi)^{\frac{3}{2}}} \bigg ( \frac{1}{3} Q_{\lambda _1 ,\lambda _2}^{\Psi ^{(2)}_{tt1}} I^{\Psi ^{(2)}}_{tt1}(\eta,p,|\mathbf{k}-\mathbf{p}|) + \frac{1}{3} Q_{\lambda _1 ,\lambda _2}^{\Psi ^{(2)}_{tt2}} I^{\Psi ^{(2)}}_{tt2}(\eta,p,|\mathbf{k}-\mathbf{p}|)\\
            &+Q^{\lambda _1 ,\lambda _2}_{\pi,tt1} I^{\Psi ^{(2)}}_{tt3}(\eta,p,|\mathbf{k}-\mathbf{p}|)+ Q^{\lambda _1 ,\lambda _2}_{\pi,tt2} I^{\Psi ^{(2)}}_{tt4}(\eta,p,|\mathbf{k}-\mathbf{p}|)\bigg ) h_{\mathbf{p}}^{\lambda _1} h_{\mathbf{k}-\mathbf{p}}^{\lambda _2} \, ,
        \end{split}
\end{equation}
with the kernels 
\begin{subequations}
    \begin{align}
        I^{\Psi ^{(2)}}_{tt1}(\eta,p,|\mathbf{k}-\mathbf{p}|)&= \int _{\eta _i}^{\infty} {\rm }  \bareta \, \left ( \frac{a(\bareta)}{a(\eta)} \right )^{\frac{3(1+w)}{2}} G^s_k(\eta , \bareta) f^{\Psi ^{(2)}}_{tt1}(\eta , p,|\mathbf{k}-\mathbf{p}|)  \, , \label{kernelinducedttscalars1}
        \\
        I^{\Psi ^{(2)}}_{tt2}(\eta,p,|\mathbf{k}-\mathbf{p}|)&= \int _{\eta _i}^{\infty} {\rm }  \bareta \, \left ( \frac{a(\bareta)}{a(\eta)} \right )^{\frac{3(1+w)}{2}} G^s_k(\eta , \bareta) f^{\Psi ^{(2)}}_{tt2}(\eta , p,|\mathbf{k}-\mathbf{p}|) \, , \label{kernelinducedttscalars2}
        \\
        \begin{split}
            I^{\Psi ^{(2)}}_{tt3}(\eta,p,|\mathbf{k}-\mathbf{p}|)&= \int _{\eta _i}^{\infty} {\rm }  \bareta \, \left ( \frac{a(\bareta)}{a(\eta)} \right )^{\frac{3(1+w)}{2}} G^s_k(\eta , \bareta) \bigg ( -\frac{\hubble}{k^4} \timeder f^{\Pi ^{(2)}}_{tt1}(\eta ,p,|\mathbf{k}-\mathbf{p}|) \\
            &+ \frac{1}{3k^2}f^{\Pi ^{(2)}}_{tt1}(\eta ,p,|\mathbf{k}-\mathbf{p}|)\bigg ) \, , \label{kernelinducedttscalars3}
        \end{split}
        \\
        \begin{split}
            I^{\Psi ^{(2)}}_{tt4}(\eta,p,|\mathbf{k}-\mathbf{p}|)&= \int _{\eta _i}^{\infty} {\rm }  \bareta \, \left ( \frac{a(\bareta)}{a(\eta)} \right )^{\frac{3(1+w)}{2}} G^s_k(\eta , \bareta) \bigg ( -\frac{\hubble}{k^4} \timeder f^{\Pi ^{(2)}}_{tt2}(\eta ,p,|\mathbf{k}-\mathbf{p}|)\\
            &+ \frac{1}{3k^2}f^{\Pi ^{(2)}}_{tt2}(\eta ,p,|\mathbf{k}-\mathbf{p}|)\bigg ) \, . \label{kernelinducedttscalars4}
        \end{split}
    \end{align}
\end{subequations}

The source term in Eq.~\eqref{realspacesourcepsi2h} is expressed as Fourier integral
\begin{equation}
    \begin{split}
        &\mathcal{S}_{\lambda }^{\Psi^{(2)}h}(\eta , \mathbf{k}) = \frac{1}{(2\pi)^{\frac{3}{2}}} \int {\rm } ^3 \mathbf{p}  \, \eps ^{ij}_{\lambda}(\mathbf{k})^* \sum_{\sigma} \bigg ( h^{\sigma}_{\mathbf{k}-\mathbf{p}}\eps^{\sigma} _{ij}(\mathbf{k}-\mathbf{p}) \bigg ( \hubble \frac{1}{p^4} \partial _{\eta}\Pi ^{(2)}_{tt}(\eta , \mathbf{p}) + \hubble (3+1w)  
        \\
        &\times \partial _{\eta}\Psi^{(2)}_{tt}(\eta,\mathbf{p}) + (w-1) p^2\Psi^{(2)}_{tt}(\eta,\mathbf{p}) \bigg ) + \frac{1}{2}\partial _{\eta}h^{\sigma}_{\mathbf{k}-\mathbf{p}}\eps^{\sigma} _{ij}(\mathbf{k}-\mathbf{p})\frac{1}{p^4}  \partial _{\eta}\Pi ^{(2)}_{tt}(\eta , \mathbf{p})- |\mathbf{k}-\mathbf{p}|^2   h^{\sigma}_{\mathbf{k}-\mathbf{p}}   
        \\
        &\times \eps^{\sigma} _{ij}(\mathbf{k}-\mathbf{p}) \left ( \frac{1}{p^4}\Pi ^{(2)}_{tt}(\eta , \mathbf{p})+2\Psi^{(2)}_{tt}(\eta,\mathbf{p})\right ) +\frac{1}{p^4}p^m(k_i-p_i)\eps ^{\sigma}_{jm}(\mathbf{k}-\mathbf{p})h^{\sigma}_{\mathbf{k}-\mathbf{p}}\Phi^{(2)}_{tt}(\eta,\mathbf{p})   
        \\
        & -\frac{1}{2p^4}(k_m-p_m)p^m \eps ^{\sigma}_{ij}(\mathbf{k}-\mathbf{p})h^{\sigma}_{\mathbf{k}-\mathbf{p}}\Pi ^{tt}_{\mathbf{p}}-2(k_m-p_m)p^m \eps ^{\sigma}_{ij}(\mathbf{k}-\mathbf{p})h^{\sigma}_{\mathbf{k}-\mathbf{p}}\Psi ^{(2)}_{tt}(\eta,\mathbf{p})  \bigg ) \, .
    \end{split}
\end{equation}
The solution to the third-order equation of motion of iGWs by second-order scalar perturbations (sourced by first-order tensors) coupled to linear tensor perturbations is then given by
\begin{equation}
    \begin{split}
        h_{\lambda }^{\Psi^{(2)}h}(\eta , \mathbf{k}) &= \frac{6}{(2\pi)^3} \int {\rm } ^3 \mathbf{q} \int {\rm } ^3 \mathbf{p} \sum _{ \sigma ,  \lambda _1 , \lambda _2}  \bigg (Q_{\lambda , \sigma}^{\Psi ^{(2)}h,1}(\mathbf{k},\mathbf{p})  Q_{\lambda _1,\lambda _2}^{\Pi ^{(2)} _{tt1}} (\mathbf{p},\mathbf{q}) \bigg ( I^{\Psi^{(2)}h}_1 (\eta ,k,p,q)
        \\
        &+ I^{\Psi^{(2)}h}_5(\eta ,k,p,q)   + I^{\Psi^{(2)}h}_9 (\eta ,k,p,q) + I^{\Psi^{(2)}h}_{11} (\eta ,k,p,q) - I^{\Psi^{(2)}h}_{13}  (\eta ,k,p,q) \\
        &-I^{\Psi^{(2)}h}_{17}  (\eta ,k,p,q) - I^{\Psi^{(2)}h}_{21}(\eta ,k,p,q)  -I^{\Psi^{(2)}h}_{25}  (\eta ,k,p,q)   \bigg ) + Q_{\lambda , \sigma}^{\Psi ^{(2)}h,1}(\mathbf{k},\mathbf{p}) 
        \\ 
        &\times Q_{\lambda _1,\lambda _2}^{\Pi ^{(2)} _{tt2}} (\mathbf{p},\mathbf{q})  \bigg ( I^{\Psi^{(2)}h}_2 (\eta ,k,p,q) + I^{\Psi^{(2)}h}_6  (\eta ,k,p,q) + I^{\Psi^{(2)}h}_{10} (\eta ,k,p,q) 
        \\
        &+I^{\Psi^{(2)}h}_{12} (\eta ,k,p,q) - I^{\Psi^{(2)}h}_{14} (\eta ,k,p,q) + I^{\Psi^{(2)}h}_{18} (\eta ,k,p,q) - I^{\Psi^{(2)}h}_{22} (\eta ,k,p,q) 
        \\
        &- I^{\Psi^{(2)}h}_{26} (\eta ,k,p,q) \bigg ) + Q_{\lambda , \sigma}^{\Psi ^{(2)}h,1}(\mathbf{k},\mathbf{p})  Q_{\lambda _1,\lambda _2}^{\Psi ^{(2)} _{tt1}} (\mathbf{p},\mathbf{q}) \bigg ( I^{\Psi^{(2)}h}_{3} (\eta ,k,p,q)
        \\
        &+ I^{\Psi^{(2)}h}_{7} (\eta ,k,p,q) - I^{\Psi^{(2)}h}_{15}  (\eta ,k,p,q) -I^{\Psi^{(2)}h}_{23} (\eta ,k,p,q) \bigg ) +   Q_{\lambda , \sigma}^{\Psi ^{(2)}h,1}(\mathbf{k},\mathbf{p}) 
        \\
        &\times Q_{\lambda _1,\lambda _2}^{\Psi ^{(2)} _{tt2}} (\mathbf{p},\mathbf{q}) \bigg ( I^{\Psi^{(2)}h}_{4} (\eta ,k,p,q) + I^{\Psi^{(2)}h}_{8} (\eta ,k,p,q) - I^{\Psi^{(2)}h}_{16} (\eta ,k,p,q)
        \\
        &- I^{\Psi^{(2)}h}_{24} (\eta ,k,p,q) \bigg ) - Q_{\lambda , \sigma}^{\Psi ^{(2)}h,2}(\mathbf{k},\mathbf{p})  Q_{\lambda _1,\lambda _2}^{\Pi ^{(2)} _{tt1}} I^{\Psi^{(2)}h}_{19} (\eta ,k,p,q) 
        \\
        &-  Q_{\lambda , \sigma}^{\Psi ^{(2)}h,2}(\mathbf{k},\mathbf{p})  Q_{\lambda _1,\lambda _2}^{\Pi ^{(2)} _{tt1}} I^{\Psi^{(2)}h}_{20} (\eta ,k,p,q)  \bigg ) \, .
    \end{split}
\end{equation}
We have defined two polarisation functions
\begin{equation}
    Q_{\lambda , \sigma}^{\Psi ^{(2)}h,1} = \eps^{ij}_{\lambda}(\mathbf{k})^* \eps _{ij}^{\sigma}(\mathbf{k} - \mathbf{p}) \quad \text{and} \quad Q_{\lambda , \sigma}^{\Psi ^{(2)}h,2} = \eps^{ij}_{\lambda}(\mathbf{k})^* p_ip^m \eps _{jm}^{\sigma}(\mathbf{k} - \mathbf{p}) \, ,
\end{equation}
whilst the polarisation functions and kernels
\begin{eqnarray}
    I^{\Psi ^{(2)}\Psi}_i(\eta , k,p,q) =  \int _{\eta _i}^{\infty} {\rm } \overline{\eta} \, \frac{a(\overline{\eta})}{a(\eta)} G^h_k(\eta , \overline{\eta}) f_i^{\Psi^{(2)}h}(\overline{\eta},k,p,q) \, ,
\end{eqnarray}
$i=1,2,...,26$, where
\begin{subequations}
    \begin{align}
    f_{1}^{\Psi^{(2)}h}(\overline{\eta},k,p,q)&=\frac{\hubble}{p^4}T_h(\bareta |\mathbf{k}-\mathbf{p}|)\partial_{\bareta}f^{\Pi}_{tt1}(\bareta , p ,|\mathbf{p}-\mathbf{q}|) \, , 
    \\
    f_{2}^{\Psi^{(2)}h}(\overline{\eta},k,p,q)&=\frac{\hubble}{p^4}T_h(\bareta |\mathbf{k}-\mathbf{p}|)\partial_{\bareta}f^{\Pi}_{tt2}(\bareta , p ,|\mathbf{p}-\mathbf{q}|)\, ,
    \\
    f_{3}^{\Psi^{(2)}h}(\overline{\eta},k,p,q)&=\frac{\hubble (3w+1)}{3}T_h(\bareta |\mathbf{k}-\mathbf{p}|)\partial_{\bareta}I^{\Psi ^{(2)}}_{tt1}(\bareta , p ,|\mathbf{p}-\mathbf{q}|)\, ,
    \\
    f_{4}^{\Psi^{(2)}h}(\overline{\eta},k,p,q)&=\frac{\hubble (3w+1)}{3}T_h(\bareta |\mathbf{k}-\mathbf{p}|)\partial_{\bareta}I^{\Psi ^{(2)}}_{tt2}(\bareta , p ,|\mathbf{p}-\mathbf{q}|)\, ,
    \\
    f_{5}^{\Psi^{(2)}h}(\overline{\eta},k,p,q)&=\hubble (3w+1)T_h(\bareta |\mathbf{k}-\mathbf{p}|)\partial_{\bareta}I^{\Psi ^{(2)}}_{tt3}(\bareta , p ,|\mathbf{p}-\mathbf{q}|)\, ,
    \\
    f_{6}^{\Psi^{(2)}h}(\overline{\eta},k,p,q)&=\hubble (3w+1)T_h(\bareta |\mathbf{k}-\mathbf{p}|)\partial_{\bareta}I^{\Psi ^{(2)}}_{tt4}(\bareta , p ,|\mathbf{p}-\mathbf{q}|)\, ,
    \\
    f_{7}^{\Psi^{(2)}h}(\overline{\eta},k,p,q)&=\frac{(w-1)}{3}p^2T_h(\bareta |\mathbf{k}-\mathbf{p}|)I^{\Psi ^{(2)}}_{tt1}(\bareta , p ,|\mathbf{p}-\mathbf{q}|)\, ,
    \\
    f_{8}^{\Psi^{(2)}h}(\overline{\eta},k,p,q)&=\frac{(w-1)}{3}p^2T_h(\bareta |\mathbf{k}-\mathbf{p}|)I^{\Psi ^{(2)}}_{tt2}(\bareta , p ,|\mathbf{p}-\mathbf{q}|)\, ,
    \\
    f_{9}^{\Psi^{(2)}h}(\overline{\eta},k,p,q)&=(w-1)p^2T_h(\bareta |\mathbf{k}-\mathbf{p}|)I^{\Psi ^{(2)}}_{tt3}(\bareta , p ,|\mathbf{p}-\mathbf{q}|)\, ,
    \\
    f_{10}^{\Psi^{(2)}h}(\overline{\eta},k,p,q)&=(w-1)p^2T_h(\bareta |\mathbf{k}-\mathbf{p}|)I^{\Psi ^{(2)}}_{tt4}(\bareta , p ,|\mathbf{p}-\mathbf{q}|)\, ,
    \\
    f_{11}^{\Psi^{(2)}h}(\overline{\eta},k,p,q)&=\frac{1}{2p^4}\partial _{\bareta}T_h(\bareta |\mathbf{k}-\mathbf{p}|)\partial _{\bareta}f^{\Pi ^{(2)}}_{tt1}(\bareta , p ,|\mathbf{p}-\mathbf{q}|)\, ,
    \\
    f_{12}^{\Psi^{(2)}h}(\overline{\eta},k,p,q)&=\frac{1}{2p^4}\partial _{\bareta}T_h(\bareta |\mathbf{k}-\mathbf{p}|)\partial _{\bareta}f^{\Pi ^{(2)}}_{tt2}(\bareta , p ,|\mathbf{p}-\mathbf{q}|)\, ,
    \\
    f_{13}^{\Psi^{(2)}h}(\overline{\eta},k,p,q)&=\frac{|\mathbf{k}-\mathbf{p}|^2}{2p^4}T_h(\bareta |\mathbf{k}-\mathbf{p}|)f^{\Pi ^{(2)}}_{tt1}(\bareta , p ,|\mathbf{p}-\mathbf{q}|)\, ,
    \\
    f_{14}^{\Psi^{(2)}h}(\overline{\eta},k,p,q)&=\frac{|\mathbf{k}-\mathbf{p}|^2}{2p^4}T_h(\bareta |\mathbf{k}-\mathbf{p}|)f^{\Pi ^{(2)}}_{tt2}(\bareta , p ,|\mathbf{p}-\mathbf{q}|) \, ,
    \\
    f_{15}^{\Psi^{(2)}h}(\overline{\eta},k,p,q)&=\frac{2}{3}|\mathbf{k}-\mathbf{p}|^2T_h(\bareta |\mathbf{k}-\mathbf{p}|)I^{\Psi ^{(2)}}_{tt1}(\bareta , p ,|\mathbf{p}-\mathbf{q}|)\, ,
    \\
    f_{16}^{\Psi^{(2)}h}(\overline{\eta},k,p,q)&=\frac{2}{3}|\mathbf{k}-\mathbf{p}|^2T_h(\bareta |\mathbf{k}-\mathbf{p}|)I^{\Psi ^{(2)}}_{tt2}(\bareta , p ,|\mathbf{p}-\mathbf{q}|)\, ,
    \\
    f_{17}^{\Psi^{(2)}h}(\overline{\eta},k,p,q)&=2|\mathbf{k}-\mathbf{p}|^2T_h(\bareta |\mathbf{k}-\mathbf{p}|)I^{\Psi ^{(2)}}_{tt3}(\bareta , p ,|\mathbf{p}-\mathbf{q}|)\, ,
    \\
    f_{18}^{\Psi^{(2)}h}(\overline{\eta},k,p,q)&=2|\mathbf{k}-\mathbf{p}|^2T_h(\bareta |\mathbf{k}-\mathbf{p}|)I^{\Psi ^{(2)}}_{tt4}(\bareta , p ,|\mathbf{p}-\mathbf{q}|)\, ,
    \\
    f_{19}^{\Psi^{(2)}h}(\overline{\eta},k,p,q)&=\frac{1}{p^4}T_h(\bareta |\mathbf{k}-\mathbf{p}|)f^{\Pi ^{(2)}}_{tt1}(\bareta , p ,|\mathbf{p}-\mathbf{q}|)\, ,
    \\
    f_{20}^{\Psi^{(2)}h}(\overline{\eta},k,p,q)&=\frac{1}{p^4}T_h(\bareta |\mathbf{k}-\mathbf{p}|)f^{\Pi ^{(2)}}_{tt2}(\bareta , p ,|\mathbf{p}-\mathbf{q}|)\, ,
    \\
    f_{21}^{\Psi^{(2)}h}(\overline{\eta},k,p,q)&=\frac{1}{2p^4}(\mathbf{k}\cdot\mathbf{p}-p^2)T_h(\bareta |\mathbf{k}-\mathbf{p}|)f^{\Pi ^{(2)}}_{tt1}(\bareta , p ,|\mathbf{p}-\mathbf{q}|)\, ,
    \\
    f_{22}^{\Psi^{(2)}h}(\overline{\eta},k,p,q)&=\frac{1}{2p^4}(\mathbf{k}\cdot\mathbf{p}-p^2)T_h(\bareta |\mathbf{k}-\mathbf{p}|)f^{\Pi ^{(2)}}_{tt2}(\bareta , p ,|\mathbf{p}-\mathbf{q}|)\, ,
    \\
    f_{23}^{\Psi^{(2)}h}(\overline{\eta},k,p,q)&=\frac{2}{3}(\mathbf{k}\cdot\mathbf{p}-p^2)T_h(\bareta |\mathbf{k}-\mathbf{p}|)I^{\Psi ^{(2)}}_{tt1}(\bareta , p ,|\mathbf{p}-\mathbf{q}|)\, ,
    \\
    f_{24}^{\Psi^{(2)}h}(\overline{\eta},k,p,q)&=\frac{2}{3}(\mathbf{k}\cdot\mathbf{p}-p^2)T_h(\bareta |\mathbf{k}-\mathbf{p}|)I^{\Psi ^{(2)}}_{tt2}(\bareta , p ,|\mathbf{p}-\mathbf{q}|)\, ,
    \\
    f_{25}^{\Psi^{(2)}h}(\overline{\eta},k,p,q)&=2(\mathbf{k}\cdot\mathbf{p}-p^2)T_h(\bareta |\mathbf{k}-\mathbf{p}|)I^{\Psi ^{(2)}}_{tt3}(\bareta , p ,|\mathbf{p}-\mathbf{q}|)\, ,
    \\
    f_{26}^{\Psi^{(2)}h}(\overline{\eta},k,p,q)&=2(\mathbf{k}\cdot\mathbf{p}-p^2)T_h(\bareta |\mathbf{k}-\mathbf{p}|)I^{\Psi ^{(2)}}_{tt4}(\bareta , p ,|\mathbf{p}-\mathbf{q}|) \, .
\end{align}
\end{subequations}
Unfortunately, there is no other way to factorise out different contributions.

%%%%%%%%%%%%%%%%%%%%%%%%%%%%%%%%%%%%%%%%%%%%%%%%%%%%%%%%%%5555555
\subsection{Second-order vectors induced by first-order tensor modes}

$S^{B^{(2)}h}_{ij}$ couples second order-vector fluctuations with first-order tensor modes
\begin{align}\label{realspacesourcevectorh}
    \begin{split}
        S^{B^{(2)} h}_{ij} &= B^{(2)}_m \left (\partial ^m h_{ij}^{\prime} + \mathcal{H}\partial ^m h_{ij} - 2\mathcal{H}\partial _i h_j^m - \partial _i h_j^{m \prime} \right )-h_{im}^{\prime} \partial ^m B^{(2)}_j\\
        &+ \frac{1}{2} B^{(2)\prime}_m \left ( \partial ^m h_{ij}-2\partial _i h_j^m \right ) \, .
    \end{split}
\end{align}
The evolution equation for second-order tensors induced by tensor-tensor interactions is
\begin{equation}\label{eqinducedvectorsrealspacett}
    B^{(2)\prime}_a + 2\mathcal{H}B^{(2)}_a= -4 V ^{ij}_{a}  S_{ij}^{B^{(2)}tt}  \, ,
\end{equation}
where the source can be found in Eq.~\eqref{rsvectortt}. In Fourier space the above equation reads
\begin{align} 
    \begin{split}
         B_r^{\prime (2)}(\eta , \mathbf{k}) + 2\mathcal{H}B_r^{(2)}(\eta , \mathbf{k}) =  \mathcal{S}^{B ^{(2)}_{tt}}_{r}(\eta , \mathbf{k})\, ,
    \end{split}
\end{align}
and the source term 
\begin{equation}
    \begin{split}
    \mathcal{S}^{B ^{(2)}_{tt}}_{r}(\eta , \mathbf{k}) & = \sum _{\lambda _1 , \lambda _2} \frac{i}{k^2}\int \frac{\rm ^3\mathbf{p}}{(2\pi)^{\frac{3}{2}}} \left ( Q^{B ^{(2)}_{tt1}} _{r,\lambda _1 ,\lambda _2} (\mathbf{k},\mathbf{p}) f^{B ^{(2)}}_{tt1}(\eta , k ,p ) + Q^{B ^{(2)}_{tt2}} _{r,\lambda _1 ,\lambda _2} (\mathbf{k},\mathbf{p}) f^{B ^{(2)}}_{tt2}(\eta , k ,p ) \right ) \\
    &\times h^{\lambda _1}_{\mathbf{p}} h^{\lambda _2}_{\mathbf{k}-\mathbf{p}} \, .
\end{split}
\end{equation}
We have defined:
\begin{subequations}
    \begin{align}
        Q^{B ^{(2)}_{tt1}} _{r,\lambda _1 ,\lambda _2} (\mathbf{k},\mathbf{p}) &= -2k^i e^j_r(\mathbf{k})^*\eps _i^{m,\lambda_1}(\mathbf{p})\eps _{jm}^{\lambda_2}(\mathbf{k-p}) \label{pol2ndvectt1} \, ,
        \\
        \begin{split} \label{pol2ndvectt2}
           Q^{B ^{(2)}_{tt2}} _{r,\lambda _1 ,\lambda _2} (\mathbf{k},\mathbf{p}) & = 2p_mk^nk^ie^j_r(\mathbf{k})^*\eps _{jn}^{\lambda _1}(\mathbf{p})\eps _i^{m,\lambda _2}(\mathbf{k}-\mathbf{p})\\
           &-2(p_nk^n-p^2)k^ie^j_r(\mathbf{k})^*\eps _{jm}^{\lambda _1}(\mathbf{p})\eps _i^{m,\lambda _2}(\mathbf{k}-\mathbf{p}) \\
           &+k^ie^j_r(\mathbf{k})^*p_ip_j\eps _{mn}^{\lambda _1}(\mathbf{p})\eps _{\lambda _2}^{mn}(\mathbf{k}-\mathbf{p})\\
           &-2 k^mk^nk^ie^j_r(\mathbf{k})^*\eps _{mn}^{\lambda _1}(\mathbf{p})\eps _{ij}^{\lambda _2}(\mathbf{k}-\mathbf{p}) \\
           &+4(k^2-k^ip_i)k_ne^j_r(\mathbf{k})^*\eps ^{mn}_{\lambda _1}(\mathbf{p})\eps _{jm}^{\lambda _2}(\mathbf{k}-\mathbf{p}) \\
           &-2(k^2-k^ip_i)e^j_r(\mathbf{k})^*p_j\eps _{mn}^{\lambda _1}(\mathbf{p})\eps _{\lambda _2}^{mn}(\mathbf{k}-\mathbf{p})\, ,
        \end{split}
    \end{align}
\end{subequations}
and
\begin{subequations}
    \begin{align}
        f^{B^{(2)}}_{tt1}(\eta, p,|\mathbf{k}-\mathbf{p}|) &= T_h'(\eta p)T_h'(\eta |\mathbf{k-p}|) \, ,
        \\
        f^{B^{(2)}}_{tt2}(\eta, p,|\mathbf{k}-\mathbf{p}|) &=T_h(\eta p)T_h(\eta |\mathbf{k-p}|) \, .
    \end{align}
\end{subequations}
The general solution for second-order vector perturbations induced by first-order tensors is then
\begin{equation}
    \begin{split} \label{2ndordervectt}
            B_{r}^{(2)tt}(\eta , \mathbf{k}) &= \sum _{\lambda _1 , \lambda _2} \frac{i}{k^2}\int \frac{\rm ^3\mathbf{p}}{(2\pi)^{\frac{3}{2}}} \bigg ( Q^{B ^{(2)}_{tt1}} _{r,\lambda _1 ,\lambda _2} (\mathbf{k},\mathbf{p}) I^{B^{(2)}}_{tt1}(\eta,p,|\mathbf{k}-\mathbf{p}|) + Q^{B ^{(2)}_{tt2}} _{r,\lambda _1 ,\lambda _2} (\mathbf{k},\mathbf{p})  \\
            &\times I^{B^{(2)}}_{tt2}(\eta,p,|\mathbf{k}-\mathbf{p}|) \bigg ) h^{\lambda _1}_{\mathbf{p}} h^{\lambda _2}_{\mathbf{k}-\mathbf{p}} \, ,
        \end{split}
\end{equation}
where
\begin{subequations}
    \begin{align}
          I^{B^{(2)}}_{tt1}(\eta,p,|\mathbf{k}-\mathbf{p}|) &=  \int _{\eta _i}^{\infty} {\rm}  \bareta \, G^v(\eta , \bareta) f^{B^{(2)}}_{tt1}(\overline{\eta}, p,|\mathbf{k}-\mathbf{p}|) \label{kernelvectortt1} \, ,
          \\
          I^{B^{(2)}}_{tt2}(\eta,p,|\mathbf{k}-\mathbf{p}|) &=  \int _{\eta _i}^{\infty} {\rm}  \bareta \,  G^v(\eta , \bareta)f^{B^{(2)}}_{tt2}(\overline{\eta}, p,|\mathbf{k}-\mathbf{p}|) \label{kernelvectortt2} \, .
    \end{align}
\end{subequations}

The source term in Eq.~\eqref{realspacesourcevectorh} in Fourier space
\begin{align}
    \begin{split}
        \mathcal{S}_{\lambda }^{B^{(2)}h}(\eta , \mathbf{k}) &= i\sum _ {r,\sigma}\frac{1}{(2\pi)^{\frac{3}{2}}} \int {\rm} ^3 \mathbf{p}  \,  \bigg ( Q^{B^{(2)h,1}}_{\lambda , r, \sigma}(\mathbf{k},\mathbf{p})B^{(2)tt}_r(\eta, \mathbf{p}) h^{\sigma} (\eta,\mathbf{k}-\mathbf{p}) \\
        &+ Q^{B^{(2)h,2}}_{\lambda , r, \sigma}(\mathbf{k},\mathbf{p})B^{(2)tt}_r(\eta, \mathbf{p}) \partial _{\eta} h^{\sigma} (\eta,\mathbf{k}-\mathbf{p}) + Q^{B^{(2)h,3}}_{\lambda , r, \sigma}(\mathbf{k},\mathbf{p})\\
        &\times \partial _{\eta}B^{(2)tt}_r(\eta, \mathbf{p})  h^{\sigma} (\eta,\mathbf{k}-\mathbf{p})  \bigg ) \, ,
    \end{split}
\end{align}
where we have defined
\begin{subequations}
  \begin{align}
      Q^{B^{(2)h,1}}_{\lambda , r, \sigma}(\mathbf{k},\mathbf{p}) &= \eps ^{ij}_{\lambda}(\mathbf{k})^* e^{r}_m(\mathbf{p})k^m\eps_{ij}^{\sigma}(\mathbf{k}-\mathbf{p}) + 2\eps ^{ij}_{\lambda}(\mathbf{k})^* e^{r}_m(\mathbf{p})p_i\eps_{j}^{m,\sigma}(\mathbf{k}-\mathbf{p}) \, ,
      \\
      \begin{split}
          Q^{B^{(2)h,2}}_{\lambda , r, \sigma}(\mathbf{k},\mathbf{p}) &= \eps ^{ij}_{\lambda}(\mathbf{k})^* e^{r}_m(\mathbf{p})k^m\eps_{ij}^{\sigma}(\mathbf{k}-\mathbf{p}) + \eps ^{ij}_{\lambda}(\mathbf{k})^* e^{r}_m(\mathbf{p})p_i\eps_{j}^{m,\sigma}(\mathbf{k}-\mathbf{p}) \\
          &- \eps ^{ij}_{\lambda}(\mathbf{k})^* e^{r}_j(\mathbf{p})\eps_{im}^{\sigma}(\mathbf{k}-\mathbf{p})p^m \, ,
      \end{split}
      \\
      Q^{B^{(2)h,3}}_{\lambda , r, \sigma}(\mathbf{k},\mathbf{p}) &= \frac{1}{2}\eps ^{ij}_{\lambda}(\mathbf{k})^* e^{r}_m(\mathbf{p})k^m\eps_{ij}^{\sigma}(\mathbf{k}-\mathbf{p}) + \eps ^{ij}_{\lambda}(\mathbf{k})^* e^{r}_m(\mathbf{p})p_i\eps_{j}^{m,\sigma}(\mathbf{k}-\mathbf{p}) \, .
  \end{align}
\end{subequations}
The third-order equation of motion admits the solution
We can now substitute in Eq.~\eqref{2ndordervectt}, the solution of the induced second-order vectors by first-order tensor terms
\begin{align}\label{GWeqsolb2h1}
    \begin{split}
        h_{\lambda }^{B^{(2)}h}(\eta , \mathbf{k}) &= -\frac{6}{(2\pi)^3} \sum _{r,\sigma, \lambda _1, \lambda _1}   \int {\rm} ^3 \mathbf{p}  \int {\rm} ^3 \mathbf{q}  \, \bigg (Q^{B^{(2)h,1}}_{\lambda , r, \sigma}(\mathbf{k},\mathbf{p})Q^{B^{(2)}_{tt1}}_{r,\lambda _1 , \lambda _2}(\mathbf{p},\mathbf{q}) I^{B^{(2)}h}_1(\eta , k,p,q) \\
        &+Q^{B^{(2)h,1}}_{\lambda , r, \sigma}(\mathbf{k},\mathbf{p})Q^{B^{(2)}_{tt2}}_{r,\lambda _1 , \lambda _2}(\mathbf{p},\mathbf{q}) I^{B^{(2)}h}_2(\eta , k,p,q) + Q^{B^{(2)h,2}}_{\lambda , r, \sigma}(\mathbf{k},\mathbf{p})Q^{B^{(2)}_{tt1}}_{r,\lambda _1 , \lambda _2}(\mathbf{p},\mathbf{q}) \\
        &\times I^{B^{(2)}h}_3(\eta , k,p,q) + Q^{B^{(2)h,2}}_{\lambda , r, \sigma}(\mathbf{k},\mathbf{p})Q^{B^{(2)}_{tt2}}_{r,\lambda _1 , \lambda _2}(\mathbf{p},\mathbf{q}) I^{B^{(2)}h}_4(\eta , k,p,q) \\
        &+Q^{B^{(2)h,3}}_{\lambda , r, \sigma}(\mathbf{k},\mathbf{p})Q^{B^{(2)}_{tt1}}_{r,\lambda _1 , \lambda _2}(\mathbf{p},\mathbf{q}) I^{B^{(2)}h}_5(\eta , k,p,q)  + Q^{B^{(2)h,3}}_{\lambda , r, \sigma}(\mathbf{k},\mathbf{p})Q^{B^{(2)}_{tt2}}_{r,\lambda _1 , \lambda _2}(\mathbf{p},\mathbf{q})\\
        &\times I^{B^{(2)}h}_6(\eta , k,p,q)  \bigg ) h^{\sigma} _{\mathbf{k}-\mathbf{p}}h^{\lambda _1}_{\mathbf{q}}h^{\lambda _2} _{\mathbf{p}-\mathbf{q}} \, .
    \end{split}
\end{align}
We have defined the kernels
\begin{equation}\label{lernelb2h1}
    I^{B^{(2)}h}_i(\eta , k,p,q) =  \int _{\eta _i}^{\infty} {\rm} \overline{\eta} \, \frac{a(\overline{\eta})}{a(\eta)} G^h_{k}(\eta , \overline{\eta}) f^{B^{(2)h}}_i (\overline{\eta},k,p,q) \, ,
\end{equation}
where $i=1,2,...,6$, and
\begin{subequations}
    \begin{align}
        f^{B^{(2)h}}_1 (\overline{\eta},k,p,q) &= T_h(\bareta|\mathbf{k}-\mathbf{p}|) I^{B^{(2)}}_{tt1}(\bareta , p , |\mathbf{p}-\mathbf{q}|)\, ,
        \\
        f^{B^{(2)h}}_2 (\overline{\eta},k,p,q) &= T_h(\bareta|\mathbf{k}-\mathbf{p}|) I^{B^{(2)}}_{tt2}(\bareta , p , |\mathbf{p}-\mathbf{q}|)\, ,
        \\
        f^{B^{(2)h}}_3 (\overline{\eta},k,p,q) &= \partial _{\bareta} T_h(\bareta|\mathbf{k}-\mathbf{p}|) I^{B^{(2)}}_{tt1}(\bareta , p , |\mathbf{p}-\mathbf{q}|)\, ,
        \\
        f^{B^{(2)h}}_4 (\overline{\eta},k,p,q) &=  \partial _{\bareta} T_h(\bareta|\mathbf{k}-\mathbf{p}|) I^{B^{(2)}}_{tt2}(\bareta , p , |\mathbf{p}-\mathbf{q}|)\, ,
        \\
        f^{B^{(2)h}}_5 (\overline{\eta},k,p,q) &=   T_h(\bareta|\mathbf{k}-\mathbf{p}|) \partial _{\bareta}I^{B^{(2)}}_{tt1}(\bareta , p , |\mathbf{p}-\mathbf{q}|)\, ,
        \\
        f^{B^{(2)h}}_6 (\overline{\eta},k,p,q) &=  T_h(\bareta|\mathbf{k}-\mathbf{p}|) \partial _{\bareta}I^{B^{(2)}}_{tt2}(\bareta , p , |\mathbf{p}-\mathbf{q}|)\, .
    \end{align}
\end{subequations}

%%%%%%%%%%%%%%%%%%%%%%%%%%%%%%%%%%%%%%%%%%%%%%%%%%%%%%%%%%5555555
\subsection{Second-order tensors induced by first-order tensor modes}

The final source term is
\begin{equation}\label{realspaceh2h}
    \begin{split}
        S^{h^{(2)} h}_{ij} &= 2h_i^{m\prime} h^{(2)\prime}_{jm} - h^{mn} \partial _m \partial _n h^{(2)}_{ij} + \partial _m h^{(2)}_{jn}\partial ^n h_i^m - 2 \partial _n h^{(2)}_{jm}\partial ^n h_i^m + \partial ^m h_{jn}\partial ^n  h^{(2)}_{im} \\
        & + h^{mn}\partial _i \partial _n h^{(2)}_{jm} - 1\partial _i h^{(2)}_{mn} \partial _j h^{mn}   + h^{mn}\partial _j \partial _n h^{(2)}_{im}- h^{mn} \partial _i \partial _j h^{(2)}_{mn} \\
        &+ h^{(2)}_{mn} \left (-\partial ^m \partial ^n h_{ij} + 2 \partial _i \partial ^n h_j^m  - \partial _i \partial _j h^{mn} \right ) \, .
    \end{split}
\end{equation}
In Sec.~\ref{sec:ttinduced}, we gave the second-order equation of motion for induced tensors by quadratic linear tensor modes, so here we give directly the solution to the third-order equation of motion. The above source term reads in Fourier space
\begin{align}
    \begin{split}
        \mathcal{S}_{\lambda }^{h^{(2)}h}(\eta , \mathbf{k}) &= \frac{1}{(2\pi)^{\frac{3}{2}}}\sum _{\sigma , \alpha }  \int {\rm} ^3 \mathbf{p}  \, \bigg ( Q_{\lambda , \sigma , \alpha}^{h^{(2)}h,1}(\mathbf{k},\mathbf{p}) \timeder h_{\alpha}(\eta, \mathbf{k}-\mathbf{p}) \timeder h^{(2)tt}_{\sigma }(\eta ,\mathbf{p})  \\
        &+ Q_{\lambda , \sigma , \alpha}^{h^{(2)}h,2}(\mathbf{k},\mathbf{p}) h_{\alpha}(\eta, \mathbf{k}-\mathbf{p})  h^{(2)tt}_{\sigma }(\eta ,\mathbf{p}) \bigg ) \, ,
    \end{split}
\end{align}
where we have defined the polarisation functions
\begin{subequations}
    \begin{align}
        \begin{split}
            &Q^{\lambda , \sigma , \alpha}_{h^{(2)}h,1}(\mathbf{k},\mathbf{p}) = \eps _{\lambda}^{ij} (\mathbf{k})^* \eps ^{\sigma} _{jm}(\mathbf{p}) \eps ^{\alpha , \,m}_i(\mathbf{k}-\mathbf{p})  \, ,
        \end{split}
        \\
        \begin{split}
           &Q^{\lambda , \sigma , \alpha}_{h^{(2)}h,2}(\mathbf{k},\mathbf{p}) =  \eps _{\lambda}^{ij} (\mathbf{k})^* \eps ^{\sigma} _{ij}(\mathbf{p}) \eps ^{mn}_{\alpha}(\mathbf{k}-\mathbf{p}) p_mp_n -  \eps _{\lambda}^{ij} (\mathbf{k})^* \eps ^{\sigma} _{jn}(\mathbf{p}) \eps ^{\alpha ,\,m}_{i}(\mathbf{k}-\mathbf{p}) p_mk^n \\
           &+2\eps _{\lambda}^{ij} (\mathbf{k})^* \eps ^{\sigma} _{jm}(\mathbf{p}) \eps ^{\alpha ,\,m}_{i}(\mathbf{k}-\mathbf{p}) p_n(k^n-p^n) - \eps _{\lambda}^{ij} (\mathbf{k})^* \eps ^{\sigma} _{im}(\mathbf{p}) \eps ^{\alpha }_{jn}(\mathbf{k}-\mathbf{p})p^nk^m \\
           &-\eps _{\lambda}^{ij} (\mathbf{k})^* \eps ^{\sigma} _{mn}(\mathbf{p}) \eps _{\alpha}^{mn}(\mathbf{k}-\mathbf{p}) p_ip_j -\eps _{\lambda}^{ij} (\mathbf{k})^* \eps ^{\sigma} _{im}(\mathbf{p}) \eps _{\alpha}^{mn}(\mathbf{k}-\mathbf{p}) p_jp_n \\
           &+\eps _{\lambda}^{ij} (\mathbf{k})^* \eps ^{\sigma} _{mn}(\mathbf{p}) \eps ^{\alpha}_{ij}(\mathbf{k}-\mathbf{p}) k^mk^n + 2\eps _{\lambda}^{ij} (\mathbf{k})^* \eps ^{\sigma} _{mn}(\mathbf{p}) \eps ^{m,\,\alpha}_{i}(\mathbf{k}-\mathbf{p}) p_jk^n \\
           &+2\eps _{\lambda}^{ij} (\mathbf{k})^* \eps ^{\sigma} _{mn}(\mathbf{p}) \eps )_{\alpha}^{mn}(\mathbf{k}-\mathbf{p}) p_ip_j \, .
        \end{split}
    \end{align}
\end{subequations}
Hence, the solution is
\begin{align} \label{solh2h1}
    \begin{split}
        h_{\lambda }^{h^{(2)}h}(\eta , \mathbf{k}) &= \frac{6}{(2\pi)^3} \int {\rm} ^3 \mathbf{q} \int {\rm} ^3 \mathbf{p} \sum _{\sigma , \alpha ,\lambda _1,\lambda _2} \bigg ( 8 Q^{\lambda , \sigma , \alpha}_{h^{(2)}h,1}(\mathbf{k},\mathbf{p}) Q^{\sigma , \lambda _1 , \lambda _2}_{h^{(2)},tt1}(\mathbf{p},\mathbf{q})I_1^{h^{(2)}h}(\eta , k,p,q)\\
        &+8 Q^{\lambda , \sigma , \alpha}_{h^{(2)}h,1}(\mathbf{k},\mathbf{p}) Q^{\sigma , \lambda _1 , \lambda _2}_{h^{(2)},tt2}(\mathbf{p},\mathbf{q}) I_2^{h^{(2)}h}(\eta , k,p,q) + 4 Q^{\lambda , \sigma , \alpha}_{h^{(2)}h,2}(\mathbf{k},\mathbf{p}) \\
        &\times Q^{\sigma , \lambda _1 , \lambda _2}_{h^{(2)},tt1}(\mathbf{p},\mathbf{q}) I_3^{h^{(2)}h}(\eta , k,p,q) + 4  Q^{\lambda , \sigma , \alpha}_{h^{(2)}h,2}(\mathbf{k},\mathbf{p}) Q^{\sigma , \lambda _1 , \lambda _2}_{h^{(2)},tt2}(\mathbf{p},\mathbf{q}) I_4^{h^{(2)}h}(\eta , k,p,q) \bigg ) \\
        &\times h^{\alpha} _{\mathbf{k}-\mathbf{p}} h_{\mathbf{q}}^{\lambda _1} h^{\lambda _2} _{\mathbf{p}-\mathbf{q}} \, .
    \end{split}
\end{align}
The time dependence is encoded in ($i=1,2,3,4$)
\begin{equation}
    I^{h^{(2)}h}_i(\eta , k,p,q) =  \int _{\eta _i}^{\infty} {\rm} \overline{\eta} \, \frac{a(\overline{\eta})}{a(\eta)} G_{\mathbf{k}}(\eta , \overline{\eta}) f^{h^{(2)h}}_i(\overline{\eta},k,p,q) \, ,
\end{equation}
where 
\begin{subequations}
    \begin{align}
        \begin{split}
            f^{h^{(2)h}}_1(\overline{\eta},k,p,q) = \frac{\partial}{\partial \bareta}T_h (\bareta |\conv|) \frac{\partial}{\partial \bareta} I_1^{tt}(\bareta ,p,q) \, ,
        \end{split}
        \\
        \begin{split}
            f^{h^{(2)h}}_2(\overline{\eta},k,p,q) = \frac{\partial}{\partial \bareta}T_h(\bareta |\conv|) \frac{\partial}{\partial \bareta} I_2^{tt}(\bareta ,p,q) \, ,
        \end{split}
        \\
        \begin{split}
            f^{h^{(2)h}}_3(\overline{\eta},k,p,q) = T_h (\bareta |\conv|) I_1^{tt}(\bareta ,p,q)\, ,
        \end{split}
        \\
        \begin{split}
            f^{h^{(2)h}}_4(\overline{\eta},k,p,q) = T_h (\bareta |\conv|) I_2^{tt}(\bareta ,p,q)\, .
        \end{split}
    \end{align}
\end{subequations}

\end{appendices}

% % % % % % % % % % % % % % % % % % % % % % % % % 

% If you want to list all the todos that you have put in the document (using
% \addtodo{}) then uncomment the next line:
% \listoftodos

% % % % % % % % % % % % % % % % % % % % % % % % % % 

% Start single space again for bibliography
\begin{singlespace}
% % % % % % % % % % % % % % % % % % % % % % % % % 

% Bibliography
% Put your bibliography file here
%\renewcommand\bibname{References} %cth added. Changes "bibliography" back to "References"
\bibliography{Thesis_Biblio}
% 
% This bibtex style file puts entries in alphabetical order and treats arxiv
% references correctly.
%\bibliographystyle{abbrvnat}
%\bibliographystyle{unsrt}
%\bibliographystyle{./utphys-ih}
%\bibliographystyle{ieeetr} %cth one I found in order of appearence
%\bibliographystyle{plainnat}

% % % % % % % % % % % % % % % % % % % % % % % % % 

\end{singlespace}

\listoffigures %cth commented
\listoftables %cth commented

% % % % % % % % % % % % % % % % % % % % % % % % % 
\end{document}